\documentclass[a4paper,10pt]{article} %
\usepackage{latexsym}
\usepackage{amssymb} %
\usepackage{amsmath} %
\usepackage{amsthm} %
\usepackage{amscd} %
\title{Equilibrium Statistical Mechanics of Fermion Lattice Systems}
\author{Huzihiro Araki \thanks{Mailing address: 
Research Institute for Mathematical Sciences, Kyoto 
 University.  Kitashirakawa-Oiwakecho, Sakyoku, Kyoto, 606-8502
 Japan} 
\;and Hajime Moriya \thanks{Mailing address:
 Institute of Particle and Nuclear Studies, 
 High Energy Accelerator Research Organization (KEK),
 1-1 Oho, Tsukuba, Ibaraki, 305-0801, Japan.}}
\date{}
\begin{document}
\maketitle
\theoremstyle{plain}%
\newtheorem{thm}{Theorem}[section]%
\newtheorem{df}[thm]{Definition}%
\newtheorem{cor}[thm]{Corollary}%
\newtheorem{lem}[thm]{Lemma}%
\newtheorem{pro}[thm]{Proposition}%
\theoremstyle{remark}%
\newtheorem*{rem}{{Remark}}%
\numberwithin{equation}{section}
\newcommand{\nonum}{\nonumber}
\newcommand{\bl}{\bigl}
\newcommand{\br}{\bigr)}%
\newcommand{\R} {{\mathbb{R}}}%
\newcommand{\Z}{{\mathbb{Z}}}%
\newcommand{\Com} {{\mathbb{C}}}%
\newcommand{\NN}{{\mathbb{N}}}%
\newcommand{\qedb}{\hbox{\rule[-2pt]{3pt}{6pt}}}%
\newcommand{\E}{{\cal E}}%
\newcommand{\M}{{\cal M}}%
\newcommand{\N}{{\cal N}}
\newcommand{\C}{{\cal C}}%
\newcommand{\Ni}{{\cal N}_1}%
\newcommand{\Me}{{\mathcal {M}}}%
\newcommand{\Ne}{{\mathcal {N}}}%
\newcommand{\EMN}{E^{\mathcal{M}}_{\mathcal{N}}}%
\newcommand{\EMNa}{\EMN(a)}%
\newcommand{\Nii}{{\cal N}_2}%
\newcommand{\Nbot}{\N^{\bot}}%
\newcommand{\Niibot}{\Nii^{\bot}}%
\newcommand{\Nibot}{\Ni^{\bot}}%
\newcommand{\Qc}{\Nibot\cap\Nii}%
\newcommand{\pot}{{\mit \Phi}}%
\newcommand{\poto}{\pot_{1}}%
\newcommand{\potzero}{\pot_{0}}%
\newcommand{\potal}{\pot_{\alpha}}%
\newcommand{\pota}{\pot_{a}}%
\newcommand{\potaI}{\pota(\I)}%
\newcommand{\pots}{\pot_{s}}%
\newcommand{\Mf}{M_{1}}%
\newcommand{\Ms}{M_{2}}%
\newcommand{\potpsi}{{\mit \Psi}}%
\newcommand{\potpsio}{\potpsi_{1}}%
\newcommand{\potpsit}{\potpsi_{2}}%
\newcommand{\potpsiA}{{\potpsi_{A}}}%
\newcommand{\Pa}{{\cal P}}%
\newcommand{\Qa}{{\cal Q}}%
\newcommand{\La}{{\cal L}}%
\newcommand{\Lai}{{\cal L}_{1}}%
\newcommand{\Laii}{{\cal L}_2}%
\newcommand{\ah}{a_{1}}%
\newcommand{\af}{a_{2}}%
\newcommand{\ap}{a_{\Pa}}%
\newcommand{\psiNi}{\psi_{\Ni}}%
\newcommand{\psiNii}{\psi_{\Nii}}%
\newcommand{\psiP}{\psi_{\Pa}}%
\newcommand{\psiI}{\potpsi_{\I}}%
\newcommand{\psiJ}{\potpsi_{\J}}%
\newcommand{\psiIuJ}{\potpsi_{\I \cup \J}}%
\newcommand{\psiIaJ}{\potpsi_{\I \cap \J}}%
\newcommand{\psiK}{\potpsi_{\K}}%
\newcommand{\psiL}{\potpsi_{\LL}}%
\newcommand{\opsiI}{\psi_{\I}}%
\newcommand{\opsiJ}{\psi_{\J}}%
\newcommand{\opsiIuJ}{\psi_{\I \cup \J}}%
\newcommand{\opsiIaJ}{\psi_{\I \cap \J}}%
\newcommand{\opsiK}{\psi_{\K}}%
\newcommand{\opsiL}{\psi_{\LL}}%
\newcommand{\Uni}{{\cal U}}%
\newcommand{\UnN}{U_{n,N}}%
\newcommand{\Unit}{\Uni_{\Theta}}%
\newcommand{\Nex}{\M \cap {\N}^{\prime}}%
\newcommand{\Niex}{\M \cap {\Ni}^{\prime}}%
\newcommand{\Pex}{\M \cap {\Pa}^{\prime}}%
\newcommand{\PexNi}{\Ni \cap {\Pa}^{\prime}}%
\newcommand{\identitybf}{{\mathbf{1} } }
\newcommand{\id}{{\mathbf{1} } }
\newcommand{\I}{{\mathrm{I}}}%
\newcommand{\J}{{\mathrm{J}}}%
\newcommand{\K}{{\mathrm{K}}}%
\newcommand{\LL}{{\mathrm{L}}}%
\newcommand{\RR}{{\mathrm{R}}}%
\newcommand{\PB}{{\cal P}} %
\newcommand{\PBI}{\PB_{\shift}} %
\newcommand{\PBIf}{\PB_{\shift}^{f}} %
\newcommand{\PBf}{\PB^{f}} %
\newcommand{\PBIdiff}{\PBI^{1}} %
\newcommand{\PBIuni}{\PBI^{unique}} 
\newcommand{\PBItil}{{\widetilde{\PB}}_{\shift}} %
\newcommand{\PBItilmone}{{\widetilde{\PB}}_{\shift,\, -1}} %
\newcommand{\PBItilone}{{\widetilde{\PB}}_{\shift,\, +1}} %
\newcommand{\PBItilzero}{{\widetilde{\PB}}_{\shift, \, 0}} %
\newcommand{\PBItilexp}{{\widetilde{\PB}}_{\shift, \,\exp \lam}} %
\newcommand{\VPBI}{\Vert_{\PBI} } %
\newcommand{\DB}{\Delta(\Alinfty)} %
\newcommand{\DBI}{\Delta_{\shift}(\Alinfty)} %
\newcommand{\IaJ}{\I \cap \J}%
\newcommand{\IuJ}{\I \cup \J}%
\newcommand{\Ja}{\J_{a}}%
\newcommand{\Jb}{\J_{b}}%
\newcommand{\Ji}{\J_{1}}%
\newcommand{\Jii}{\J_{2}}%
\newcommand{\Jiii}{\J_{3}}%
\newcommand{\Jj}{\J_{j}}%
\newcommand{\Jk}{\J_{k}}%
\newcommand{\Jai}{\J_{i}}%
\newcommand{\Jn}{\J_{n}}%
\newcommand{\Jl}{\J_{l}}%
\newcommand{\Jicapii}{\Ji \cap \Jii}%
\newcommand{\Jacapb}{\Ja \cap \Jb}%
\newcommand{\Jcirc}{\J_{\circ}}%
\newcommand{\Al}{{\cal A}}%
\newcommand{\Bl}{{\cal B}}%
\newcommand{\Bone}{{\cal B}^{1}}%
\newcommand{\AlJn}{\Al(\J_{n})}%
\newcommand{\AlJj}{\Al(\J_{j})}%
\newcommand{\AlJk}{\Al(\J_{k})}%
\newcommand{\AlJl}{\Al({\J}_{l})}
\newcommand{\Aeps}{A_{\varepsilon}}
\newcommand{\Niibi}{\Nii \cap {\Ni}^{\bot} }
\newcommand{\proofend}{{\hfill $\square$}\  \\}
\newcommand{\vp}{\varphi}
\newcommand{\lam}{\lambda}%
\newcommand{\ome}{\omega}
\newcommand{\omeone}{\ome_{1}}
\newcommand{\ometwo}{\ome_{2}}
\newcommand{\omep}{\ome_{+}}
\newcommand{\omem}{\ome_{-}}
\newcommand{\omepm}{\ome_{\pm}}
\newcommand{\vpb}{\overline{\vp}}%
\newcommand{\hb}{\overline{h}}%
\newcommand{\ub}{\overline{u}}%
\newcommand{\wb}{\overline{w}}%
\newcommand{\vpbh}{\vpb^{\hb}}%
\newcommand{\vpbu}{\vpb^{\ub}}%
\newcommand{\shift}{\tau}%
\newcommand{\shiftn}{\tau_{n}}%
\newcommand{\shiftm}{\tau_{m}}%
\newcommand{\dshift}{\tau^{\ast}}%
\newcommand{\dshiftm}{\shiftm^{\ast}}%
\newcommand{\dshiftl}{\shiftl^{\ast}}%
\newcommand{\dshiftlm}{\shift_{l+m}^{\ast}}%
\newcommand{\dshiftml}{\shift_{m+l}^{\ast}}%
\newcommand{\dshiftk}{\shiftk^{\ast}}%
\newcommand{\dshiftn}{\dshift^{\n}}%
\newcommand{\shiftk}{\tau_{k}}%
\newcommand{\shiftl}{\tau_{l}}%
\newcommand{\shiftkl}{\tau_{k+l}}%
\newcommand{\rhohat}{\hat{\rho}}%
\newcommand{\vpN}{{\vp}_{\N}}%
\newcommand{\psiN}{{\psi}_{\N}}%
\newcommand{\rhovp}{\rhohat_{\vp}}%
\newcommand{\rhodensity}{{\rhovp}}%
\newcommand{\rhodensityN}{{\rhohat_{\vpN}}}%
\newcommand{\rhohatome}{{\rhohat_{\omega}} }%
\newcommand{\rhohatomeJ}{{\rhohat_{{\omega\vert \AlJ}} } }%
\newcommand{\rhohatomeJl}{{\rhohat_{{\omega\vert \AlJl}} } }%
\newcommand{\Dome}{{D_{\omega}} }%
\newcommand{\DomeJ}{{D_{{\omega\vert \AlJ}} } }%
\newcommand{\DomeJl}{{D_{{\omega\vert \AlJl}} } }%
\newcommand{\rhoome}{{\rho_{\omega}} }%
\newcommand{\rhoomeJ}{{\rho_{{\omega\vert \AlJ}} } }%
\newcommand{\rhoomeJl}{{\rho_{{\omega\vert \AlJl}} } }%
\newcommand{\rhodensitypsi}{\rhohat_{\psi}}%
\newcommand{\rhod}{{\rho}_{\vp}}%
\newcommand{\rhodpsi}{{\rho}_{\psi}}%
\newcommand{\Tr}{\mathbf{Tr}}%
\newcommand{\Shat}{\widehat{S}}%
\newcommand{\sm}{S_{\M}}%
\newcommand{\smh}{{\hat S}_{\M}}%
\newcommand{\sn}{S_{\N}}%
\newcommand{\snh}{{\hat S}_{\N}}%
\newcommand{\sph}{{\hat S}_{\Pa}}%
\newcommand{\snih}{{\hat S}_{\Ni}}%
\newcommand{\sniih}{{\hat S}_{\Nii}}%
\newcommand{\In}{\I_{n}}%
\newcommand{\Ina}{\I_n(a)}%
\newcommand{\Ima}{\I_m(a)}%
\newcommand{\InaIma}{{\I_{\vec n}(a)}\vec {\I_{\vec m}(a)}}%
\newcommand{\aicr}{a_i^{\ast}}%
\newcommand{\ai}{a_i}%
\newcommand{\ajcr}{a_j^{\ast}}%
\newcommand{\ij}{i_{j}}%
\newcommand{\aj}{a_j}%
\newcommand{\aprim}{a^{\prime}}%
\newcommand{\ih}{i_{1}}%
\newcommand{\ik}{i_{k}}%
\newcommand{\aik}{a_{\ik}}%
\newcommand{\aikcr}{ a_{\ik}^{\ast}}%
\newcommand{\Ai}{A_{i} }%
\newcommand{\Aj}{A_{j} }%
\newcommand{\Aij}{A_{\ij} }%
\newcommand{\Aih}{A_{i_{1}} }%
\newcommand{\Ain}{A_{i_{n}} }%
\newcommand{\Aik}{A_{ i_{k}  } }%
\newcommand{\Ah}{A_{1}}%
\newcommand{\Af}{A_{2}}%
\newcommand{\Am}{A_{3}}%
\newcommand{\Ak}{A_{k}}%
\newcommand{\An}{A_{n}}%
\newcommand{\Ac}{A^{c}}%
\newcommand{\Bh}{B_{1}}%
\newcommand{\Bf}{B_{2}}%
\newcommand{\Bm}{B_{3}}%
\newcommand{\Bk}{B_{k}}%
\newcommand{\Ahp}{A_{1,+}}%
\newcommand{\Ahm}{A_{1,-}}%
\newcommand{\Ahpm}{A_{1,\pm}}%
\newcommand{\Afp}{A_{2,+}}%
\newcommand{\Afm}{A_{2,-}}%
\newcommand{\Afpm}{A_{2,\pm}}%
\newcommand{\Amp}{A_{3,+}}%
\newcommand{\Amm}{A_{3,-}}%
\newcommand{\Ampm}{A_{3,\pm}}%
\newcommand{\Bhp}{B_{1,+}}%
\newcommand{\Bhm}{B_{1,-}}%
\newcommand{\Bhpm}{B_{1,\pm}}%
\newcommand{\Bfp}{B_{2,+}}%
\newcommand{\Bfm}{B_{2,-}}%
\newcommand{\Bfpm}{B_{2,\pm}}%
\newcommand{\Bmp}{B_{3,+}}%
\newcommand{\Bmm}{B_{3,-}}%
\newcommand{\Bmpm}{B_{3,\pm}}%
\newcommand{\aial}{a_{ i_{\alpha}}}%
\newcommand{\acrial}{a^{\ast}_{ i_{\alpha}}}%
\newcommand{\Aialpha}{A_{ i_{\alpha}}}%
\newcommand{\Alialpha}{{\cal A}(  \{ i_{\alpha} \})}%
\newcommand{\Aliten}{ {\cal A}(  \{ i \}) }%
\newcommand{\ic}{ \{ i \}^{c} }%
\newcommand{\AlJisetmini}{{\cal A}(  \Ji \setminus\{i\})}%
\newcommand{\Ic}{\I^{c}}%
\newcommand{\Jc}{\J^{c}}%
\newcommand{\Kc}{\K^{c}}%
\newcommand{\Alinfty}{\Al_\circ}%
\newcommand{\AlI}{{\cal A}({\I})}%
\newcommand{\AlJ}{{\cal A}({\J})}%
\newcommand{\AlK}{{\cal A}({\K})}%
\newcommand{\AlL}{{\cal A}({\LL})}%
\newcommand{\bih}{b^{(i)}_{1}}%
\newcommand{\bif}{b^{(i)}_{2}}%
\newcommand{\bim}{b^{(i)}_{3}}%
\newcommand{\biy}{b^{(i)}_{4}}%
\newcommand{\Alisite}{{\cal A}(\{i\})}%
\newcommand{\Alisitep}{{\cal A}(\{i\})_{+}}%
\newcommand{\Alisitem}{{\cal A}(\{i\})_{-}}%
\newcommand{\Alisitecommut}{{\cal A}(\{i\})^{\prime}}%
\newcommand{\Alisitec}{{\cal A}(\{i\}^{c})}%
\newcommand{\Alisitecommutp}{{\cal A}_{+}(\{i\})^{\prime}}%
\newcommand{\Alisitecommutm}{{\cal A}_{-}(\{i\})^{\prime}}%
\newcommand{\Alic}{{\cal A}(\{i\}^{c})}%
\newcommand{\Alicp}{{\cal A}_{+}(\{i\}^{c})}%
\newcommand{\Alicm}{{\cal A}_{-}(\{i\}^{c})}%
\newcommand{\AlIc}{{\cal A}({\I}^{c})}%
\newcommand{\AlJaI}{{\cal A}({\J}\cap {\I})}%
\newcommand{\AlIaJ}{{\cal A}({\I}\cap {\J})}%
\newcommand{\AlJaIc}{{\cal A}({\J}\cap {\Ic})}%
\newcommand{\AlIcaJ}{{\cal A}({\Ic}\cap {\J})}%
\newcommand{\AlJuI}{{\cal A}({\J}\cup {\I})}%
\newcommand{\AlIuJ}{{\cal A}({\I}\cup {\J})}%
\newcommand{\AlJc}{{\cal A}({\J}^{c})}%
\newcommand{\AlKp}{{\cal A}(\K)_{+}}%
\newcommand{\AlKm}{{\cal A}(\K)_{-}}%
\newcommand{\AlKprime}{\Al(\K)^{\prime}}%
\newcommand{\AlKprimem}{\bigl(\Al(\K)^{\prime}\bigr)_{-}}%
\newcommand{\AlKprimep}{\bigl(\Al(\K)^{\prime}\bigr)_{+}}%
\newcommand{\AlKprimepm}{\bigl(\Al(\K)^{\prime}\bigr)_{\pm}}%
\newcommand{\AlKc}{\Al(\K^{c})}%
\newcommand{\AlKcp}{\Al(\K^{c})_{+}}%
\newcommand{\AlKcm}{\Al(\K^{c})_{-}}%
\newcommand{\AlKcpm}{\Al(\K^{c})_{\pm}}%
\newcommand{\Kn}{\K_{n}}%
\newcommand{\Kk}{\K_{k}}%
\newcommand{\AlKnc}{\Al((\Kn)^{c})}%
\newcommand{\AlKncp}{\Al((\Kn)^{c})_{+}}%
\newcommand{\AlKncm}{ \Al(\(Kn)^{c})_{-}}%
\newcommand{\AlKncpm}{ \Al((\Kn)^{c})_{\pm}}%
\newcommand{\AlKkc}{\Al((\Kk)^{c})}%
\newcommand{\AlKkcp}{\Al((\Kk)^{c})_{+}}%
\newcommand{\AlKkcm}{\Al((\Kk)^{c})_{-}}%
\newcommand{\AlKkcpm}{\Al((\Kk)^{c})_{\pm}}%
\newcommand{\AlIkc}{\Al((\Ik)^{c})}%
\newcommand{\AlIkcp}{\bigl(\Al(\Ik)^{c}\bigr)_{+}}%
\newcommand{\AlIkcm}{\bigl(\Al(\Ik)^{c}\bigr)_{-}}%
\newcommand{\AlIkcpm}{\bigl(\Al(\Ik)^{c}\bigr)_{\pm}}%
\newcommand{\AlLp}{{\cal A}(\LL)_{+}}%
\newcommand{\AlLm}{{\cal A}(\LL)_{-}}%
\newcommand{\Alp}{{\cal A}_{+}}%
\newcommand{\Alm}{{\cal A}_{-}}%
\newcommand{\Alpm}{{\cal A}_{\pm}}%
\newcommand{\Alminus}{{\cal A}_{-}}%
\newcommand{\AlJp}{{\cal A}(\J)_{+}}%
\newcommand{\AlJm}{{\cal A}(\J)_{-}}%
\newcommand{\AlIp}{{\cal A}(\I)_{+}}%
\newcommand{\AlIm}{{\cal A}(\I)_{-}}%
\newcommand{\AlIic}{\Al((\Ii)^{c})}%
\newcommand{\AlIip}{{\cal A}(\Ii)_{+}}%
\newcommand{\AlIim}{{\cal A}(\Ii)_{-}}%
\newcommand{\AlIsig}{{\cal A}(\I)_{\sigma}}%
\newcommand{\AlJsig}{{\cal A}(\J)_{\sigma}}%
\newcommand{\AlKsig}{{\cal A}(\K)_{\sigma}}%
\newcommand{\AlLsig}{{\cal A}(\LL)_{\sigma}}%
\newcommand{\asig}{a_{\sigma}}%
\newcommand{\bsig}{b_{\sigma}}%
\newcommand{\AlIop}{{\cal A}(\Io)_{+}}%
\newcommand{\AlIom}{{\cal A}(\Io)_{-}}%
\newcommand{\AlJpm}{{\cal A}(\J)_{\pm}}%
\newcommand{\AlIpm}{{\cal A}(\I)_{\pm}}%
\newcommand{\AlIcp}{{\cal A}({\I}^{c})_{+}}%
\newcommand{\AlIcm}{{\cal A}({\I}^{c})_{-}}%
\newcommand{\HJIpot}{H_{\pot\, \J}(\I) }%
\newcommand{\HJLpot}{H_{\pot\, \J}(\LL) }%
\newcommand{\HIpot}{H_{\pot}(\I)}%
\newcommand{\UIpot}{U_{\pot}(\I)}%
\newcommand{\WIpot}{W_{\pot}(\I)}%
\newcommand{\Cpot}{C_{\pot}}%
\newcommand{\HIpots}{H_{\pots}(\I)}%
\newcommand{\UIpots}{U_{\pots}(\I)}%
\newcommand{\WIpots}{W_{\pots}(\I)}%
\newcommand{\HJpot}{H_{\pot}(\J)}%
\newcommand{\HIaJpot}{H_{\pot}(\IaJ)}%
\newcommand{\HLpot}{H_{\pot}(\LL)}%
\newcommand{\HIpotpsi}{H_{\potpsi}(\I)}%
\newcommand{\HKpot}{H_{\pot}(\K)}%
\newcommand{\delpot}{\delta_{\pot}}%
\newcommand{\delpotb}{\delta_{\beta\pot}}%
\newcommand{\delpoto}{\delta_{\poto}}%
\newcommand{\HJIdel}{H_{\delta\,\J}(\I)}%
\newcommand{\HIdel}{H_{\delta}(\I)}%
\newcommand{\HKdel}{HK{\delta}(\K)}%
\newcommand{\UIdel}{U_{\delta}(\I)}%
\newcommand{\potdel}{\pot_{\delta}}%
\newcommand{\potdelI}{\pot_{\delta}(\I)}%
\newcommand{\potdelJ}{\pot_{\delta}(\J)}%
\newcommand{\potdelK}{\pot_{\delta}(\K)}%
\newcommand{\Alana}{\Al_{ent}}%
\newcommand{\Dbeta}{D_{\beta}}%
\newcommand{\Dbetac}{\overline{D_{\beta}}}%
\newcommand{\Dbetao}{\stackrel{\circ}{D}_{\beta} }%
\newcommand{\Doc}{\overline{D_{0}}}%
\newcommand{\Do}{D_{0}}%
\newcommand{\AlIcpm}{{\cal A}_{\pm}({\I}^{c})}%
\newcommand{\Ali}{\Al_{1}}%
\newcommand{\Alii}{\Al_{2}}%
\newcommand{\Alk}{\Al_{k}}%
\newcommand{\All}{\Al_{l}}%
\newcommand{\Alj}{\Al_{j}}%
\newcommand{\AlJi}{{\cal A}({\J}_1)}%
\newcommand{\AlJii}{{\cal A}({\J}_2)}%
\newcommand{\AlJiii}{{\cal A}({\J}_3)}%
\newcommand{\AlLi}{\Al(\LL_{1})}%
\newcommand{\AlLii}{\Al(\LL_{2})}%
\newcommand{\AlLiii}{\Al(\LL_{3})}%
\newcommand{\AlLj}{\Al(\LL_{j})}%
\newcommand{\AlLk}{\Al(\LL_{k})}%
\newcommand{\AlJa} {{\cal A}({\J}_a)}%
\newcommand{\AlJb} {{\cal A}({\J}_b)}%
\newcommand{\AlJip}{{\cal A}({\J}_1)_{+}}%
\newcommand{\AlJiip}{{\cal A}({\J}_2)_{+}}%
\newcommand{\AlJkp}{{\cal A}(\Jk)_{+}}
\newcommand{\AlJlp}{{\cal A}(\Jl)_{+}}
\newcommand{\Jip}{{\J}_1,\{+\}}%
\newcommand{\Jkp}{{\J}_k,\{+\}}%
%
\newcommand{\AlJim}{{\cal A}_{\! -}({\J}_1)}%
\newcommand{\AlJiim}{{\cal A}_{\! -}({\J}_2)}%
\newcommand{\AlJicapii}{\Al(\Jicapii)}%
\newcommand{\AlJacapb}{\Al(\Jacb)}%
\newcommand{\AlIn}{\Al(\In)}%
\newcommand{\AlIk}{\Al(\Ik)}%
\newcommand{\AlIna}{\Al\bigl( \Ina \bigr)}%
\newcommand{\AlIma}{\Al \bigl(\Ima \bigr)}%
\newcommand{\AlImna}{\Al(\Ima \cup \Ina )}%
\newcommand{\Ih}{{\I}_1}%
\newcommand{\If}{{\I}_2}%
\newcommand{\Imi}{{\I}_3}%
\newcommand{\Ii}{{\I}_i}%
\newcommand{\Ij}{{\I}_j}%
\newcommand{\Iij}{{\I}_{i_j}}%
\newcommand{\AlIi}{{\cal A}(\Ii)}%
\newcommand{\AlIj}{{\cal A}(\Ij)}%
\newcommand{\AlIij}{{\cal A}\bigl( \Iij \bigr)}%
%
\newcommand{\AlIh}{{\cal A}({\I}_1)}%
\newcommand{\AlIf}{{\cal A}({\I}_2)}%
\newcommand{\AlImi}{{\cal A}({\I}_3)}%
\newcommand{\Alempty}{\Al(\emptyset)}
\newcommand{\EMNi}{ E^{\mathcal{M}}_{\Ni}}%
\newcommand{\EMNii}{E^{\M}_{\Nii}}%
\newcommand{\EMNibot}{E^{\M}_{\Nibot}}%
\newcommand{\PMNibot}{P^{\M}_{\Nibot}}%
\newcommand{\EMNhf}{\EMNi\EMNii}%
\newcommand{\EMNfh}{\EMNii\EMNi}%
\newcommand{\ENiP}{E^{\Ni}_{\Pa}}%
\newcommand{\ENP}{E^{\N}_{\Pa}}%
\newcommand{\ENiiP}{E^{\Nii}_{\Pa}}%
\newcommand{\EMP}{E^{\M}_{\Pa}}%
\newcommand{\EMQ}{E^{\M}_{\Qa}}%
\newcommand{\EMQc}{E^{\M}_{\Qc}}%
\newcommand{\PMQc}{P^{\M}_{\Qc}}%
\newcommand{\EJJi}{E^{\AlJ}_{\AlJi}}%
\newcommand{\Il}{\I_{l}}%
\newcommand{\Ik}{\I_{k}}%
\newcommand{\Iku}{\I^{k}}%
\newcommand{\Ilu}{\I^{l}}%
\newcommand{\Ikui}{\I^{k}_{i}}%
\newcommand{\Ikuj}{\I^{k}_{j}}%
\newcommand{\Ilui}{\I^{l}_{i}}%
\newcommand{\Iluj}{\I^{l}_{j}}%
\newcommand{\AlIku}{\Al(\Iku)}%
\newcommand{\AlIkui}{\Al(\Ikui)}%
\newcommand{\AlIkuj}{\Al(\Ikuj)}%
\newcommand{\AlIlu}{\Al(\Ilu)}%
\newcommand{\AlIlui}{\Al(\Ilui)}%
\newcommand{\AlIluj}{\Al(\Iluj)}%
\newcommand{\Aku}{A^{k}}%
\newcommand{\Akui}{\Aku_{i}}%
\newcommand{\Akuj}{\Aku_{j}}%
\newcommand{\Akun}{\\Aku_{n}}%
\newcommand{\Ial}{\I_{\alpha}} %
\newcommand{\Ibl}{\I_{\beta}} %
\newcommand{\Ialf}{\I^{1}_{\alpha}} %
\newcommand{\Ibls}{\I^{2}_{\beta}} %
\newcommand{\Jbl}{\J_{\beta}} %
\newcommand{\Jblhat}{\J^{\beta}} %
\newcommand{\Irl}{\I_{\gamma}} %
\newcommand{\Ipbl}{\I^{\prime}_{\beta}} %
\newcommand{\Ip}{\I_{p}}%
\newcommand{\Is}{\I_{s}}%
\newcommand{\It}{\I_{t}}%
\newcommand{\Imn}{\I_{m(n)}}%
\newcommand{\Iln}{\I_{l(n)}}%
\newcommand{\EJ}{E_{\J}}%
\newcommand{\Eempty}{E_{\emptyset}}%
\newcommand{\EI}{E_{\I}}%
\newcommand{\EIal}{E_{\Ial}}%
\newcommand{\EIbl}{E_{\Ibl}}%
\newcommand{\EJbl}{E_{\Jbl}}%
\newcommand{\EJblhat}{E_{\Jblhat}}%
\newcommand{\EIc}{E_{\Ic}}%
\newcommand{\EJc}{E_{\Jc}}%
\newcommand{\EIaJc}{E_{(\I\cap\J)^{c} }}%
\newcommand{\EKc}{E_{\Kc}}%
\newcommand{\EK}{E_{\K}}%
\newcommand{\EL}{E_{\LL}}%
\newcommand{\ko}{k_{0}}%
\newcommand{\lo}{l_{0}}%
\newcommand{\Lko}{\LL_{\ko}}%
\newcommand{\ELko}{E_{\Lko}}%
\newcommand{\Lkep}{\LL_{\kep}}%
\newcommand{\ELkep}{E_{\Lkep}}%
\newcommand{\ELk}{E_{\Lk}}%
\newcommand{\Lk}{{\LL}_{k}}%
\newcommand{\Ln}{{\LL}_{n}}%
\newcommand{\EIJ}{E^{\I}_{\J}}%
\newcommand{\EJI}{E^{\J}_{\I}}%
\newcommand{\ELK}{E^{\LL}_{\K}}%
\newcommand{\EKL}{E^{\K}_{\LL}}%
\newcommand{\EIuJI}{E^{\I \cup \J}_{\I}}%
\newcommand{\EIuJJ}{E^{\I \cup \J}_{\J}}%
\newcommand{\EIIaJ}{E^{\I }_{\I \cap \J}}%
\newcommand{\EJIaJ}{E^{\J}_{\I \cap \J}}%
\newcommand{\EIuJIaJ}{E^{\I\cup \J}_{\I \cap \J}}%
\newcommand{\EJuIJaI}{E^{\J\cup \I}_{\J \cap \I}}%
\newcommand{\EpI}{E^{\prime}_{\I}}%
\newcommand{\EIweak}{{\overline{E}}_{\I}}%
\newcommand{\EJaI}{E_{\J \cap \I} }%
\newcommand{\EIaJ}{E_{\I \cap \J} }%
\newcommand{\EJuI}{E_{\J \cup \I} }%
\newcommand{\EIuJ}{E_{\I \cup \J} }%
\newcommand{\EIn}{E_{\In}}%
\newcommand{\EIl}{E_{\Il}}%
\newcommand{\EIk}{E_{\Ik}}%
\newcommand{\EJn}{E_{\Jn}}%
\newcommand{\EJnm}{E_{\J_{n-1}}}%
\newcommand{\EJl}{E_{\Jl}}%
\newcommand{\EJk}{E_{\Jk}}%
\newcommand{\EIp}{E_{\Ip}}%
\newcommand{\EIs}{E_{\Is}}%
\newcommand{\EIt}{E_{\It}}%
\newcommand{\EIi}{E_{\Ii}}%
\newcommand{\EIj}{E_{\Ij}}%
\newcommand{\EIln}{E_{\Iln}}%
\newcommand{\EImn}{E_{\Imn}}%
\newcommand{\PMN}{P^{\M}_{\N}}%
\newcommand{\PMNi}{P^{\M}_{\Ni}}%
\newcommand{\PMNii}{P^{\M}_{\Nii}}%
\newcommand{\PMPa}{P^{\M}_{\Pa}}%
\newcommand{\aal}{a(\al)}
\newcommand{\alphat}{\alpha_{t}}%
\newcommand{\alphas}{\alpha_{s}}%
\newcommand{\alphaMbt}{\alpha_{-\beta t}}%
\newcommand{\alphaMbs}{\alpha_{-\beta s}}%
\newcommand{\alphaMbst}{\alpha_{-\beta (s+t)}}%
\newcommand{\alphaPbt}{\alpha_{\beta t}}%
\newcommand{\alphaPbs}{\alpha_{\beta s}}%
\newcommand{\alphaPbst}{\alpha_{\beta (s+t)}}%
\newcommand{\alt}{\alpha_{t}}%
\newcommand{\al}{\alpha}%
\newcommand{\alp}{\alpha^{\prime}}%
\newcommand{\alpt}{\alpha_{t}}%
\newcommand{\alps}{\alpha_{s}}%
\newcommand{\alpmbt}{\alpha_{-\beta t}}%
\newcommand{\alpmbs}{\alpha_{-\beta s}}%
\newcommand{\alpmbst}{\alpha_{-\beta (s+t)}}%
\newcommand{\alpbt}{\alpha_{\beta t}}%
\newcommand{\alpbs}{\alpha_{\beta s}}%
\newcommand{\alppbst}{\alpha_{\beta (s+t)}}%
\newcommand{\alIt}{\alpha^{\I}_{t}}%
\newcommand{\alphaI}{\alpha^{\I}}%
\newcommand{\alphaIMbt}{\alpha^{\I}_{-\beta t}}%
\newcommand{\alphaIMbs}{\alpha^{\I}_{-\beta s}}%
\newcommand{\alphaIMbst}{\alpha^{\I}_{-\beta (s+t)}}%
\newcommand{\alphaIPbt}{\alpha^{\I}_{\beta t}}%
\newcommand{\alphaIPbs}{\alpha^{\I}_{\beta s}}%
\newcommand{\alphaIPbst}{\alpha^{\I}_{\beta (s+t)}}%
\newcommand{\state}{\Al_{+, 1 }^{\ast}}%
\newcommand{\posfunctional}{\Al_{+}^{\ast}}%
\newcommand{\functional}{\Al^{\ast}}%
\newcommand{\stateI}{\Al(\I)_{+, 1 }^{\ast}}%
\newcommand{\stateJ}{\Al(\J)_{+, 1 }^{\ast}}%
\newcommand{\stateIJ}{\Al(\I \cup \J)_{+, 1 }^{\ast}}%
\newcommand{\stateIn}{\Al\bigl(\Ina\bigr)_{+, 1 }^{\ast} }%
\newcommand{\stateIm}{\Al\bigl(\Ima\bigr)_{+, 1 }^{\ast}    }%
\newcommand{\stateInIm}{\Al\bigl(\Ina \cup \Ima\bigr)_{+, 1 }^{\ast}    }%
\newcommand{\invstate}{\Al_{+, 1  }^{\ast \, \tau}}%
\newcommand{\invfunctional}{\Al_{+  }^{\ast \, \tau}}%
\newcommand{\evenstate}{\Al^{\ast \, \Theta}_{+, 1 }}%
\newcommand{\evenfunctional}{\Al^{\ast \, \Theta}_{+ }}%
\newcommand{\evenstateI}{\AlI^{\ast \, \Theta}_{+, 1 }}%
\newcommand{\stateforevenpartI}{\{\AlIp\}^{\ast }_{+, 1 }}%
\newcommand{\evenstatecam}{\Al( \Cam \bigr)^{\ast \, \Theta}_{+, 1 }}%
\newcommand{\vpc}{\vp^{c}}%
\newcommand{\Ca}{\C_{a}}%
\newcommand{\Cb}{\C_{b}}%
\newcommand{\Camb}{\C_{a-b}}%
\newcommand{\Cd}{\C_{d}}%
\newcommand{\Calo}{\C_{a}+\lo}%
\newcommand{\Cblo}{\C_{b}-\lo}%
\newcommand{\Ea}{E_{a}}%
\newcommand{\Cone}{\C_{1}}%
\newcommand{\Ctwo}{\C_{2}}%
\newcommand{\Caal}{\C_{a(\al)}}%
\newcommand{\Cn}{\C_{n}}%
\newcommand{\Ck}{\C_{k}}%
\newcommand{\Ckn}{\C_{kn}}%
\newcommand{\Cnak}{\C_{na}+k}%
\newcommand{\Cna}{\C_{na}}%
\newcommand{\Cnpoa}{\C_{(n+1)a}}%
\newcommand{\Cnmoa}{\C_{(n-1)a}}%
%
%
\newcommand{\canb}{\Ca + a n}%
\newcommand{\Can}{\C_{a+ a n}}%
\newcommand{\camb}{\C_{a} +a m}%
\newcommand{\Cam}{\C_{a}+am}%
\newcommand{\Camprime}{\C_{a}+am^{\prime}}%
\newcommand{\Cal}{\C_{a}+al }%
\newcommand{\Caml}{\Cam \cup \Cal}%
\newcommand{\Camn}{\Cam \cup \Can}%
\newcommand{\Calcn}{\C_{\alc}^{n}}
\newcommand{\Cnk}{\C_{n}^{k}}
\newcommand{\Cka}{\C_{ka}}
\newcommand{\Alhfm}{{\cal A}({\J}_1 \cup {\J}_2 \cup {\J}_3)}%
\newcommand{\Alhm}{{\cal A}({\J}_1  \cup {\J}_3)}%
\newcommand{\Alfm}{{\cal A}({\J}_2 \cup {\J}_3)}%
\newcommand{\A}{{\cal A}}%
\newcommand{\Znu}{\Z^{\nu}}%
\newcommand{\Nnu}{\NN^{\nu}}%
\newcommand{\cstar}{{\bf C}^{\ast}}%
\newcommand{\wstar}{{\bf W}^{\ast}}%
\newcommand{\vH}{\I \to \infty}
\newcommand{\vHZ}{\I \to \Znu}
\newcommand{\ItoZ}{\I \nearrow \Znu}
\newcommand{\lvH}{ {\rm{v.H.}}  \lim_{\vH }}
\newcommand{\lvHZ}{{\rm{v.H.}} \lim_{\vHZ}}
\newcommand{\Jnlim}{\Jn \nearrow }
\newcommand{\Jnlimd}{\Jn \searrow }
\newcommand{\Klim}{\K \nearrow }
\newcommand{\Klimd}{\K \searrow }
\newcommand{\Knlim}{\Kn \nearrow }
\newcommand{\Knlimd}{\Kn \searrow }
\newcommand{\IlimZ}{\I \nearrow \Znu}
\newcommand{\JlimZ}{\J \nearrow \Znu}
\newcommand{\Iallim}{\Ial\nearrow }
\newcommand{\IallimZ}{\Ia \nearrow \Znu}
\newcommand{\IallimI}{\Ia \nearrow \I}
\newcommand{\IallimdI}{\Ial\searrow\I}
\newcommand{\Ibllim}{\Ibl\nearrow }
\newcommand{\IbllimZ}{\Ibl \nearrow \Znu}
\newcommand{\IbllimI}{\Ibl \nearrow \I}
\newcommand{\IbllimdI}{\Iblt\searrow\I}
\newcommand{\JbllimI}{\Jbl \nearrow \I}
\newcommand{\JbllimdI}{\Jblhat\searrow\I}
\newcommand{\Jeps}{\J_{\varepsilon}}
\newcommand{\Iolim}{\Io \nearrow }
\newcommand{\lJlim}{\lim_{\JlimZ}}
\newcommand{\lIlim}{\lim_{\IlimZ}}
\newcommand{\JvH}{\J \to \infty}
\newcommand{\Lnlim}{\Ln \nearrow }
\newcommand{\LnlimZ}{\Ln\nearrow\Znu}
\newcommand{\Inlim}{\In\nearrow }
\newcommand{\InlimI}{\In\nearrow\I}
\newcommand{\InlimZ}{\In\nearrow\Znu}
%
\newcommand{\Iklim}{\Ik\nearrow }
\newcommand{\Kklim}{\Kk\nearrow }
\newcommand{\Jklim}{\Jk\nearrow }
\newcommand{\IklimI}{\Ik\nearrow\I}
\newcommand{\JnlimI}{\Jn\nearrow\I}
\newcommand{\JklimI}{\Jk\nearrow\I}
\newcommand{\Lnlimd}{\Ln\searrow }
\newcommand{\LklimZ}{\Lk \nearrow \Znu}
\newcommand{\Iklimd}{\Ik\searrow }
\newcommand{\Jklimd}{\Jk\searrow }
\newcommand{\AlJmi}{\Al(\J\setminus \!\!\{i\})}
\newcommand{\AlZmi}{\Al(\Znu\setminus \!\!\{i\})}
\newcommand{\AlJmip}{\Al_{+}(\J\setminus\! \{i\})}
\newcommand{\AlJmim}{\Al_{-}(\J\setminus \!\{i\})}
\newcommand{\AlJnmi}{\Al(\Jn\setminus \!\{i\})}
\newcommand{\AlJnmip}{\Al_{+}(\Jn\setminus \!\{i\})}
\newcommand{\AlJnmim}{\Al_{-}(\Jn\setminus \!\{i\})}
\newcommand{\AlJkmi}{\Al(\Jk\setminus \!\{i\})}
\newcommand{\AlJkmip}{\Al_{+}(\Jk\setminus \!\{i\})}
\newcommand{\AlJkmim}{\Al_{-}(\Jk\setminus \!\{i\})}
\newcommand{\Lam}{\Lambda}%
\newcommand{\Ia}{{\I}_a}%
\newcommand{\Ala}{\Al(\Ia)}%
\newcommand{\Alb}{\Al(\Ib)}%
\newcommand{\nbf}{{\bf n}} %
\newcommand{\nvhp}{\nbf^{a}_{+}(\I)}%
\newcommand{\nvhm}{\nbf^{a}_{-}(\I)}%
\newcommand{\nvhaIp}{\nbf^{+}_{a}(\I)}%
\newcommand{\nvhaIm}{\nbf^{-}_{a}(\I)}%
\newcommand{\nvhaIap}{\nbf^{+}_{a}(\Ial)}%
\newcommand{\nvhaIam}{\nbf^{-}_{a}(\Ial)}%
\newcommand{\cofnvhaIap}{\nbf^{+}_{a}(\Ial)}%
\newcommand{\cofnvhaIam}{\nbf^{-}_{a}(\Ial)}%
\newcommand{\potI}{\pot(\I)}%
\newcommand{\potJ}{\pot(\J)}%
\newcommand{\potK}{\pot(\K)}%
\newcommand{\potL}{\pot(\LL)}%
\newcommand{\bpot}{\beta \pot}%
\newcommand{\sigx}{{\sigma}_x}%
\newcommand{\sigm}{{\sigma}_m}%
\newcommand{\sign}{{\sigma}_n}%
\newcommand{\sigam}{{\sigma}_{a\cdot m}}%
\newcommand{\sigan}{{\sigma}_{a\cdot n}}%
\newcommand{\B}{{\cal B}}%
\newcommand{\trI}{\Tr_{\I}}%
\newcommand{\ebu}{e^{-\beta \UI}}%
\newcommand{\UCn}{U(\Cn)}%
\newcommand{\UCa}{U(\Ca)}%
\newcommand{\UCkn}{U(\Ckn)}%
\newcommand{\UCka}{U(\Cka)}%
\newcommand{\UCnnm}{U(\Cn+nm)}%
\newcommand{\ebuCn}{e^{-\beta \UCn}}%
\newcommand{\ebuCa}{e^{-\beta U(\Ca)}}%
\newcommand{\Alcirc}{{\cal A}^{\circ}}%
\newcommand{\omegal}{\omega_{1}}%
\newcommand{\omegar}{\omega_{2}}%
\newcommand{\omegacirc}{\omega^{\circ}}%
\newcommand{\AlIandJ}{\AlI \vee \AlJ}%
\newcommand{\Altwo}{{\cal A}(\I \cup \J)}%
\newcommand{\Altwop}{{\cal A}_{+}(\I \cup \J)}%
\newcommand{\Altwom}{{\cal A}_{-}(\I \cup \J)}%
\newcommand{\cip}{c^{\I}_{+}}%
\newcommand{\cjp}{c^{\J}_{+}}%
\newcommand{\Etwocirc}{E^{\Altwo}_{\Alcirc}}%
\newcommand{\la}{\Lambda}%
\newcommand{\vpac}{\vp_a^c}%
\newcommand{\vpnc}{\vp_n^c}%
\newcommand{\vplc}{\vp_l^c}%
\newcommand{\vpkc}{\vp_k^c}%
\newcommand{\vpal}{{\vp_\alpha}}%
\newcommand{\vpIac}{\vp_{\Ia}^c}%
\newcommand{\vpacn}{ \sigma_{\vec n}^* {\vpac}}%
\newcommand{\vpachat}{\widehat {\vpac}}%
\newcommand{\vpnchat}{\widehat {\vpnc}}%
\newcommand{\vpnchatp}{{\widehat {\vpnc}}^{\prime}}%
\newcommand{\vpachatp}{{\widehat {\vpac}}^{\prime}}%
\newcommand{\vplchat}{\widehat {\vplc}}%
\newcommand{\vpCnkc}{\vp_{\Cnk}^c}%
\newcommand{\csign}{{\sigma_{ n}}^{\ast}}
\newcommand{\csigm}{\sigma_m^{\ast} }%
\newcommand{\vpInc}{\vp_{\Ina}^c}%
\newcommand{\vpIc}{\vp_\I^c}%
\newcommand{\vpImc}{\vp_{\Ima}^c}%
\newcommand{\vpImnten}{\vp_{\{\Ima \otimes  \Ina\}}^c}%
\newcommand{\vpImnregion}{\vp_{\{\Ima \cup  \Ina\}}^c}%
\newcommand{\vpacten}{\vp_{\cup_{\!  {\vec m}} \Ima}^c}%
\newcommand{\equiset}{\Lambda_\pot}%
\newcommand{\equisetb}{\Lambda_{\beta \pot}}%
\newcommand{\Alal}{\mathfrak{A}}%
\newcommand{\Imag}{\bf{Im}}%
\newcommand{\vnM}{{\mathfrak{M}}_{\vp}}%
\newcommand{\vNM}{{\mathfrak{M}}}%
\newcommand{\omeh}{\ome^{h}}%
\newcommand{\GNS}{\bigl\{\pivp,{\cal H}_\vp, \mit\Omega_\vp \bigr\}}%
\newcommand{\Hilvp}{{\cal H}_{\vp}}%
\newcommand{\Hiltau}{{\cal H}_{\tau}}%
\newcommand{\HiltauN}{{\cal H}_{\tau}^{\N}}%
\newcommand{\tauN}{{\tau}_{\N}}%
\newcommand{\tauvec}{{\mit\Omega}_\tau}%
\newcommand{\tauvecN}{{\mit\Omega}_\tau^{\N}}%
\newcommand{\tauvecM}{{\mit\Omega}_\tau^{\M}}%
\newcommand{\origv}{\mit\Omega_\varphi}%
\newcommand{\perth}{\mit\Omega_\varphi^h}%
\newcommand{\pertW}{\mit\Omega_\varphi^{\beta \WI }}%
\newcommand{\pertH}{\mit\Omega_\vp ^{\beta \HI } }%
\newcommand{\pertWT}{\mit\Omega_{{\vp \circ \Theta}}^{\beta \WI } }%
\newcommand{\pertsu}{\left[ {\varphi^{\beta U_{\I}}} \right]}%
\newcommand{\pertsuw}{\left[ {\vp^{\beta (\UI+\WI)}} \right]}%
\newcommand{\pertsw}{\left[ {\vp^{\beta \WI } } \right]}%
\newcommand{\pertswbetaone}{\left[ \vp^{ \WI  } \right]}%
\newcommand{\pertsH}{\left[ {\vp^{\beta \HI}} \right]}%
\newcommand{\pertfu}{ {\vp^{\beta \UI } } }%
\newcommand{\pertfuw}{ {\vp^{\beta (\UI+\WI )}} }%
\newcommand{\pertfh}{ {\vp^{\beta (\HI )}} }%
\newcommand{\pertfw}{\varphi^{\beta \WI }}%
\newcommand{\pertspsiu}{\left[ {\psi^{\beta \UI }} \right]}%
\newcommand{\pertvpfh}{ {\vp^{\pivp(\beta\HI )}} }%
\newcommand{\pertvpfw}{\varphi^{ \pivp(\beta\WI) }}%
\newcommand{\hi}{h(\I)}%
\newcommand{\wi}{w(\I)}%
\newcommand{\ui}{u(\I)}%
\newcommand{\bhi}{\beta h(\I)}%
\newcommand{\bwi}{\beta w(\I)}%
\newcommand{\bui}{\beta u(\I)}%
\newcommand{\vphi}{\vp^{\hi}}%
\newcommand{\vpwi}{\vp^{\wi}}%
\newcommand{\vpui}{\vp^{\ui}}%
\newcommand{\vpbhi}{\vp^{\bhi}}%
\newcommand{\vpbwi}{\vp^{\bwi}}%
\newcommand{\vpbui}{\vp^{\bui}}%
\newcommand{\psih}{\psi_{1}}%
\newcommand{\psiItil}{{\tilde{\psi}}_{\I}}%
\newcommand{\psiIh}{\psi_{\I\,1}}%
\newcommand{\sigpsi}{\sigma^{\psi}}%
\newcommand{\sigtpsi}{\sigma_t^{\psi}}%
\newcommand{\sigtpsih}{\sigma_t^{\psih}}%
\newcommand{\hp}{h^{\prime}}%
\newcommand{\hpp}{h^{\prime\prime}}%
\newcommand{\vph}{\vp^{h}}%
\newcommand{\vpu}{\vp^{u}}%
\newcommand{\vpw}{\vp^{w}}%
\newcommand{\hha}{\hat{h}}%
\newcommand{\wha}{\hat{w}}%
\newcommand{\uha}{\hat{u}}%
\newcommand{\vphha}{\vp^{\hha}}%
\newcommand{\vpuha}{\vp^{\uha}}%
\newcommand{\vpwha}{\vp^{\wha}}%
\newcommand{\vphp}{\vp^{\hp}}%
\newcommand{\vphpp}{\vp^{\hpp}}%
\newcommand{\vphn}{[\vp^{h}]}%
\newcommand{\sigvp}{\sigma^{\varphi}}%
\newcommand{\sigtvp}{\sigma_t^{\vp}}%
\newcommand{\sigsvp}{\sigma_s^{\vp}}%
\newcommand{\sigmtvp}{\sigma_{-t}^{\vp}}%
\newcommand{\sigmsvp}{\sigma_{-s}^{\vp}}%
\newcommand{\sigvph}{\sigma^{\vph}}%
\newcommand{\sigtvph}{\sigma_t^{\vph}}%
\newcommand{\sigsvph}{\sigma_s^{\vph}}%
\newcommand{\sigvpw}{\sigma^{\vpw}}%
\newcommand{\sigtvpw}{\sigma_t^{\vpw}}%
\newcommand{\sigsvpw}{\sigma_s^{\vpw}}%
\newcommand{\sigvpu}{\sigma^{\vpu}}%
\newcommand{\sigtvpu}{\sigma_t^{\vpu}}%
\newcommand{\sigsvpu}{\sigma_s^{\vpu}}%
\newcommand{\sigvphha}{\sigma^{\vphha}}%
\newcommand{\sigtvphha}{\sigma_t^{\vphha}}%
\newcommand{\sigsvphha}{\sigma_s^{\vphha}}%
\newcommand{\sigvpwha}{\sigma^{\vpwha}}%
\newcommand{\sigtvpwha}{\sigma_t^{\vpwha}}%
\newcommand{\sigmtvpwha}{\sigma_{-t}^{\vpwha}}%
\newcommand{\sigsvpwha}{\sigma_s^{\vpwha}}%
\newcommand{\sigvpuha}{\sigma^{\vpuha}}%
\newcommand{\sigtvpuha}{\sigma_t^{\vpuha}}%
\newcommand{\sigsvpuha}{\sigma_s^{\vpuha}}%
\newcommand{\sigvpwuhaa}{\sigma^{\vpwuha   }}%
\newcommand{\sigvphi}{\sigma^{\vphi}}%
\newcommand{\sigtvphi}{\sigma_t^{\vphi}}%
\newcommand{\sigsvphi}{\sigma_s^{\vphi}}%
\newcommand{\sigvpwi}{\sigma^{\vpwi}}%
\newcommand{\sigtvpwi}{\sigma_t^{\vpwi}}%
\newcommand{\sigsvpwi}{\sigma_s^{\vpwi}}%
\newcommand{\sigvpui}{\sigma^{\vpui}}%
\newcommand{\sigtvpui}{\sigma_t^{\vpui}}%
\newcommand{\sigsvpui}{\sigma_s^{\vpui}}%
\newcommand{\sigvpb}{\sigma^{\vpb}}%
\newcommand{\sigtvpb}{\sigma_t^{\vpb}}%
\newcommand{\sigsvpb}{\sigma_s^{\vpb}}%
\newcommand{\sigvpbh}{\sigma^{\vpbh}}%
\newcommand{\sigtvpbh}{\sigma_t^{\vpbh}}%
\newcommand{\sigsvpbh}{\sigma_s^{\vpbh}}%
\newcommand{\sigvpbw}{\sigma^{\vpbw}}%
\newcommand{\sigtvpbw}{\sigma_t^{\vpbw}}%
\newcommand{\sigsvpbw}{\sigma_s^{\vpbw}}%
\newcommand{\sigtome}{\sigma_t^{\omega}}%
\newcommand{\sigtomeh}{\sigma_t^{\omeh}}%
\newcommand{\sigstvp}{\sigma_{s+t}^{\vp}}%
\newcommand{\pivp}{\pi_{\vp}}%
\newcommand{\pitau}{\pi_{\tau}}%
\newcommand{\pitauN}{\pi_{\tau}^{\N}}%
\newcommand{\pitauM}{\pi_{\tau}^{\M}}%
\newcommand{\sigspsi}{\sigma_s^{\psi}}%
\newcommand{\sigtpsiI}{\sigma_t^{\psiI}}%
\newcommand{\sigspsiI}{\sigma_s^{\psiI}}%
\newcommand{\sigtpsiItil}{ \sigma_t^{\psiItil} }%
\newcommand{\sigtpsiIc}{\sigma_t^{{\psi}_{\Ic}}}%
\newcommand{\sigtpsiIcomp}{\sigma_t^{{\psi}_{\Znu \setminus \I}}}%
\newcommand{\sigtpsitilde}{\tilde{\sigma_{t}}^{\psi}}%
\newcommand{\pivpAlI}{\pi_{\vp}\bigl(\AlI\bigr)}%
\newcommand{\pivpAlIcvN}{\pi_{\vp}\bigl(\AlIc\bigr)^{\prime \prime}}%
\newcommand{\pivpAlIc}{\pi_{\vp}\bigl( \AlIc \bigr)}%
\newcommand{\ThetaIhat}{\Theta^{\I}}%
\newcommand{\ThetaIchat}{\Theta^{\Ic}}%
\newcommand{\ThetaJihat}{\Theta^{\Ji}}%
\newcommand{\ThetaJaihat}{\Theta^{\Jai}}%
%
\newcommand{\ThetaI}{\Theta_{\I}}%
\newcommand{\ThetaJ}{\Theta_{\J}}%
\newcommand{\Hil}{{\cal H}}%
\newcommand{\Hilh}{{\cal H}_{1}}%
\newcommand{\Hilf}{{\cal H}_{2}}%
\newcommand{\Ome}{\mit\Omega}%
\newcommand{\Omevp}{\Ome_{\vp}}%
\newcommand{\Omevph}{\Ome_{\vp}^{h}}%
\newcommand{\Omeone}{{\mit\Omega}_1}%
\newcommand{\Ometwo}{{\mit\Omega}_2}%
\newcommand{\Omeh}{{\Ome}^{h}}%
\newcommand{\Omeho}{{\Ome}^{h_{1} }}%
\newcommand{\Omeht}{{\Ome}^{h_{2} }}%
\newcommand{\Omehot}{{\Ome}^{h_{1}+h_{2} }}%
\newcommand{\Omehvp}{\mit\Omega_\varphi^h}%
\newcommand{\ho}{h_{1}}%
\newcommand{\htwo}{h_{2}}%
\newcommand{\hotwo}{\ho+\htwo}%
\newcommand{\omeho}{\ome^{\ho}}%
\newcommand{\omeht}{\ome^{\htwo}}%
\newcommand{\omehotwo}{\ome^{\hotwo}}%
\newcommand{\omehotwok}{\ome^{(\hotwo)}}%
\newcommand{\Omef}{\Ome_{1}}%
\newcommand{\Omes}{\Ome_{2}}%
\newcommand{\etaI}{\eta_{\I}}%
\newcommand{\Bll}{{\cal B}_{l}}%
\newcommand{\vpk}{\vp_{k}}%
\newcommand{\vpl}{\vp_{l}}%
\newcommand{\vpJk}{\vp_{\Jk}}%
\newcommand{\vpJj}{\vp_{\Jj}}%
\newcommand{\vpJl}{\vp_{\Jl}}%
\newcommand{\gammatW}{\gamma_{t}^{\WI} }%
\newcommand{\gammat}{\gamma_{t}}%
\newcommand{\omegaI}{\omega_{\I}}%
\newcommand{\omegaJ}{\omega_{\J}}%
\newcommand{\omeI}{\omega_{\I}}%
\newcommand{\omeJ}{\omega_{\J}}%
\newcommand{\omeIJ}{\omega_{\I \cup \J}}%
\newcommand{\omegaIa}{\omega_{\Ia}}%
\newcommand{\omegaIb}{\omega_{\Ib}}%
\newcommand{\vpI}{\vp_{\I}}%
\newcommand{\GNSvpvec}{\mit\Omega_\vp}%
\newcommand{\Io}{\I_{0}}%
\newcommand{\AlIo}{\Al(\Io) }%
\newcommand{\UIo}{U(\Io) }%
\newcommand{\Jo}{\J_{0}}%
\newcommand{\Ko}{\K_{0}}%
\newcommand{\LLo}{\LL_{0}}%
\newcommand{\EIo}{E_{\Io} }%
\newcommand{\EJo}{E_{\Jo} }%
\newcommand{\EKo}{E_{\Ko} }%
\newcommand{\ELLo}{E_{\LLo} }%
\newcommand{\HJoI}{H_{\Jo}(\I)}%
\newcommand{\HIoI}{H_{\Io}(\I)}%
\newcommand{\HKoI}{H_{\Ko}(\I)}%
\newcommand{\HLLoI}{H_{\LLo}(\I)}%
\newcommand{\UI}{U({\I})}%
\newcommand{\HI}{H({\I})}%
\newcommand{\UIn}{U_{n}({\I})}%
\newcommand{\HIn}{H_{n}({\I})}%
\newcommand{\UJn}{U_{n}({\J})}%
\newcommand{\HJn}{H_{n}({\J})}%
\newcommand{\UKn}{U_{n}({\K})}%
\newcommand{\HKn}{H_{n}({\K})}%
\newcommand{\potn}{\pot_{n}}%
\newcommand{\potIn}{\pot_{n}(\I)}%
\newcommand{\potJn}{\pot_{n}(\J)}%
\newcommand{\potKn}{\pot_{n}(\K)}%
\newcommand{\potLLn}{\pot_{n}(\LL)}%
\newcommand{\UIinf}{U_{\infty}({\I})}%
\newcommand{\HIinf}{H_{\infty}({\I})}%
\newcommand{\UKinf}{U_{\infty}({\K})}%
\newcommand{\HKinf}{H_{\infty}({\K})}%
\newcommand{\UJinf}{U_{\infty}({\J})}%
\newcommand{\HJinf}{H_{\infty}({\J})}%
\newcommand{\potinf}{\pot_{\infty}}%
\newcommand{\potIinf}{\potinf(\I) }%
\newcommand{\potJinf}{\potinf(\J) }%
\newcommand{\potKinf}{\potinf(\K) }%
\newcommand{\delinf}{\delta_{\infty}}%
\newcommand{\potnor}{\Vert \pot \Vert}%
\newcommand{\potpsinor}{\Vert \potpsi \Vert}%
\newcommand{\UJI}{U_{\J}({\I})}%
\newcommand{\HJI}{H_{\J}({\I})}%
\newcommand{\HJK}{H_{\J}({\K})}%
\newcommand{\HJL}{H_{\J}({\LL})}%
\newcommand{\HLJ}{H_{\LL}({\J})}%
\newcommand{\HLI}{H_{\LL}({\I})}%
%
\newcommand{\WI}{W(\I)}%
\newcommand{\UJ}{U(\J)}%
\newcommand{\HJ}{H(\J)}%
\newcommand{\WJ}{W(\J)}%
\newcommand{\UK}{U(\K)}%
\newcommand{\HK}{H(\K)}%
\newcommand{\WK}{W(\K)}%
\newcommand{\UL}{U(\LL)}%
\newcommand{\HL}{H(\LL)}%
\newcommand{\WL}{W(\LL)}%
\newcommand{\UIs}{U_{\I}}%
\newcommand{\HIs}{H_{\I}}%
\newcommand{\WIs}{W_{\I}}%
\newcommand{\UJs}{U_{\J}}%
\newcommand{\HJs}{H_{\J}}%
\newcommand{\WJs}{W_{\J}}%
\newcommand{\UKs}{U_{\K}}%
\newcommand{\HKs}{H_{\K}}%
\newcommand{\WKs}{W_{\K}}%
\newcommand{\ULs}{U_{\LL}}%
\newcommand{\HLs}{H_{\LL}}%
\newcommand{\WLs}{W_{\LL}}%
\newcommand{\Hvp}{{\cal H}_\vp}
\newcommand{\modvp}{ \mit\Delta_\vp}
\newcommand{\modvpb}{ \mit\Delta_\vpb}
\newcommand{\modome}{ \mit\Delta_{\Ome}}
\newcommand{\modvpit}{ {\modvp}^{it}}
\newcommand{\modvpmit}{ {\modvp}^{-it}}
\newcommand{\modvpis}{ {\modvp}^{is}}
\newcommand{\modvpmis}{ {\modvp}^{-is}}
\newcommand{\modvpbit}{ {\modvpb}^{it}}
\newcommand{\modvpbmit}{ {\modvpb}^{-it}}
\newcommand{\modomeit}{ {\modome}^{it}}
\newcommand{\modomemit}{ {\modome}^{-it}}
\newcommand{\bdel}{\overline{\delta}}
\newcommand{\del}{\delta}
\newcommand{\bdelta}{\overline{\delta}}
\newcommand{\delcore}{{\delta}_{0}}
\newcommand{\bdelcore}{\overline{\delcore}}
\newcommand{\Dteigi}{\mathfrak{D}(\bdelta)}
\newcommand{\Danal}{{\mathfrak{D}}_{anal}(\bdelta)}
%
%
\newcommand{\SwI}{S(\omegaI)}%
\newcommand{\SIwI}{S_{\I}(\omegaI)}%
\newcommand{\SwIa}{S(\omegaIa)}%
\newcommand{\SwIb}{S(\omegaIb)}%
\newcommand{\SIawIa}{S_{\Ia}(\omegaIa)}%
\newcommand{\SIbwIb}{S_{\Ib}(\omegaIb)}%
\newcommand{\indh}{i(1)}%
\newcommand{\indf}{i(2)}%
\newcommand{\indk}{i(k)}%
\newcommand{\indl}{i(l)} 
\newcommand{\dex}{i(1) i(2) \ldots i(k)}%
\newcommand{\dexcomma}{i(1), i(2), \ldots ,i(k)}%
\newcommand{\Alkdex}{ \Ali, \Alii, \ldots, \Alk}%
\newcommand{\AlLkdex}{ \AlLi, \AlLii, \ldots, \AlLk}%
\newcommand{\AlJkdex}{ \AlJi, \AlJii, \ldots, \AlJk}%
\newcommand{\AlJkpdex}{ \AlJip, \AlJiip, \ldots, \AlJkp}%
\newcommand{\AlNkdexsig}{N, \sigma(N), \ldots, \sigma^{k-1}(N)}%
\newcommand{\AlNkdexshift}{N,\, \shift(N), \cdots,\, \shift^{k-1}(N)}%
\newcommand{\Hwk}{H_{\omega}{(\Alkdex)}}%
\newcommand{\HwLk}{H_{\omega}{(\AlLkdex)}}%
\newcommand{\Hwksig}{H_{\omega}{(\AlNkdexsig)}}%
\newcommand{\HwAlJkdex}{\Hw \bigl( \AlJkdex \bigr)}%
\newcommand{\omelil}{\omega^{l}_{i(l)} }%
\newcommand{\omehatlil}{ {\hat{\omega}}^{l}_{\indl} }%
\newcommand{\hws}{h_{\omega}(\shift)}%
\newcommand{\homeones}{h_{\omeone}(\shift)}%
\newcommand{\hometwos}{h_{\ometwo}(\shift)}%
\newcommand{\hwshift}{h_{\omega}(\shift)}%
\newcommand{\hvpshift}{h_{\vp}(\shift)}%
\newcommand{\hwsn}{h_{\omega, \shift}(N_{n})}%
\newcommand{\hwsN}{h_{\omega, \shift}(N)}%
\newcommand{\sumll}{\sum_{l=1}^{k} \sum_{i(l)}}%
\newcommand{\sumllprim}{\sum_{l=1}^{k} \sum_{i(l)\in \IIlprim}}%
\newcommand{\suml}{\sum_{l=1}^{k}}%
\newcommand{\Hw}{H_{\omega}}%
\newcommand{\omegaJl}{\omega_{\Jl}}%
\newcommand{\omegaJlp}{\omega|_{\AlJlp}}%
\newcommand{\omei}{\omega_{i}}%
\newcommand{\omeihat}{\hat{\omei}}%
\newcommand{\Plil}{P^{l}_{\indl}}
\newcommand{\plil}{{\lambda}^{l}_{\indl}}
\newcommand{\pitougou}{{\lambda}_{i}}
\newcommand{\Phih}{P^{1}_{\indh}}
\newcommand{\Pfif}{P^{2}_{\indf}}
\newcommand{\Pkik}{P^{k}_{\indk}}
\newcommand{\II}{\mathfrak{I}}%
\newcommand{\IIprim}{\II^{\prime}}%
\newcommand{\IIlprim}{\II_{l}^{\prime}}%
\newcommand{\Pri}{P_{i}}%
\newcommand{\AlJmIprim}{\Al^{\prime}(\I \subset \J)}%
\newcommand{\AlJmI}{\Al(\J \setminus \I)}
\newcommand{\Stil}{\widetilde{S}}
\newcommand{\StilI}{\Stil_{\I}}
\newcommand{\Stilcar}{\widetilde{S^{car}}}
\newcommand{\StilIcar}{\widehat{S^{car}_{\I} }}
\newcommand{\vpJ}{\vp_{\J}}
\newcommand{\vpJmIprim}{\vp|_{\AlJmIprim}}
\newcommand{\vpip}{\varpi^{\prime}}
\newcommand{\vpipSOTO}{\varpi^{\prime}|{\AlJmIprim}}
\newcommand{\rhovpip}{\rhohat_{\vpip}}
\newcommand{\rhovpipSOTO}{\rhohat_{\vpipSOTO}}
\newcommand{\tauI}{\tau_{\I}}
\newcommand{\Ftil}{\widetilde{F}}
\newcommand{\FtilI}{\widetilde{F_{\I}} }
\newcommand{\AlIcommut}{\AlI^{\prime}}%
\newcommand{\AlIcommutp}{(\AlI^{\prime})_{+}}%
\newcommand{\AlIcommutm}{(\AlI^{\prime})_{-}}%
\newcommand{\AlIcommutpm}{(\AlI^{\prime})_{\pm}}%
\newcommand{\AlJcommut}{\AlJ^{\prime}}%
\newcommand{\AlIpcommut}{\bigl(\AlI_{+}\bigr)^{\prime}}%
\newcommand{\AlInpcommut}{\bigl(\AlIn_{+}\bigr)^{\prime}}%
\newcommand{\AlImcommut}{\bigl(\AlI_{-}\bigr)^{\prime}}%
\newcommand{\AlIpmcommut}{\bigl(\AlI_{\pm}\bigr)^{\prime}}%
\newcommand{\AlIprime}{\Al(\I)^{\prime}}%
\newcommand{\AlIoprime}{\Al(\Io)^{\prime}}%
\newcommand{\AlInprime}{\Al(\In)^{\prime}}%
\newcommand{\AlIkprime}{\Al(\Ik)^{\prime}}%
\newcommand{\AlKkprime}{\Al(\Kk)^{\prime}}%
\newcommand{\Imc}{{\I_m}^{c}}%
\newcommand{\Inc}{{\In}^{c}}%
\newcommand{\AlImc}{\Al(\{\I_{m} \}^{c})}%
\newcommand{\AlInc}{\Al(\{\In\}^{c})}%
\newcommand{\bn}{b_{n}}%
\newcommand{\bkk}{b_{k}}%
\newcommand{\bkdouble}{b_{kk}}%
\newcommand{\psiprim}{\psi^{\prime}}%
\newcommand{\psiIprim}{\psi^{\prime}_{\I}}%
\newcommand{\phiprim}{\phi^{\prime}}%
\newcommand{\phiIprim}{\phi^{\prime}_{\I}}%
\newcommand{\psiIpBu}{\tauI \otimes \phiIprim}%
\newcommand{\namadef}{\!\!\!\!\! \!\!\!\!\! &&\Hwk  \nonumber  \\
&\equiv&
\!\!\!\!\!\!\!\!\!\!\!\! \sup_{  {\sum \omega_{\dex}}  =\omega}
\biggl[   
 \sum_{\dexcomma} \!\!\!\!\!\!
\eta \Bigl( \omega_{\dex}\bigl(\identitybf\bigr) \Bigr)  
-\sum_{l=1}^{k} \sum_{\indl} \eta 
\Bigl( \omelil \bigl(\identitybf\bigr) \Bigr)  \nonum \\
%
 &+&
\suml S\bigl(\omega | \All \bigr)- 
\sumll \omelil(\identitybf) S\bigl( \omehatlil | \All \bigr)
\;\biggl]
}
\newcommand{\namadefsec}{\!\!\!\!\! \!\!\!\!\! &&\Hwk  \nonumber  \\
&\equiv&
\label{eq:namadefsec}
\!\!\! \sup
\biggl[   
 \sum_{\dexcomma} \!\!\!\!\!\!
\eta \Bigl( \omega_{\dex}\bigl(\identitybf\bigr) \Bigr)  
-\sum_{l=1}^{k} \sum_{\indl} \eta 
\Bigl( \omelil \bigl(\identitybf\bigr) \Bigr)  \nonum \\
%
 &+&
\suml S\bigl(\omega | \All \bigr)- 
\sumll \omelil(\identitybf) S\bigl( \omehatlil | \All \bigr)
\;\biggl]
}
\newcommand{\Classicalpart}{ \sum_{\dexcomma} \!\!\!\!\!\!
\eta \Bigl( \omega_{\dex}\bigl(\identitybf\bigr) \Bigr)  
-\sum_{l=1}^{k} \sum_{\indl} \eta 
\Bigl( \omelil \bigl(\identitybf\bigr) \Bigr)}%
\newcommand{\Defectpart}{\sumll 
\omelil(\identitybf) S\bigl( \omehatlil | \All \bigr)}%
\newcommand{\Corepart}{\suml S\bigl(\omega | \All \bigr)}%
\newcommand{\htwsig}{ht_{\omega}(\sigma)}
\newcommand{\htwshift}{ht_{\omega}(\shift)}
\newcommand{\vi}{v_{i}}%
\newcommand{\vj}{v_{j}}%
\newcommand{\vl}{v_{l}}%
\newcommand{\vI}{v_{\I}}%
\newcommand{\vJ}{v_{\J}}%
\newcommand{\vK}{v_{\K}}%
\newcommand{\vKn}{v_{\Kn}}%
\newcommand{\vKk}{v_{\Kk}}%
\newcommand{\vIn}{v_{\In}}%
\newcommand{\uij}{u_{ij}}%
\newcommand{\uji}{u_{ji}}%
\newcommand{\uki}{u_{ki}}%
\newcommand{\ujk}{u_{jk}}%
\newcommand{\ulm}{u_{lm}}%
\newcommand{\ull}{u_{ll}}%
\newcommand{\uml}{u_{ml}}%
\newcommand{\nrmI}{2^{-|\I|}}%
\newcommand{\upq}{u_{pq}}%
\newcommand{\uqp}{u_{qp}}%
\newcommand{\upp}{u_{pp}}%
\newcommand{\uqq}{u_{qq}}%
\newcommand{\upl}{u_{pl}}%
\newcommand{\uoo}{u_{11}}%
\newcommand{\uot}{u_{12}}%
\newcommand{\uto}{u_{21}}%
\newcommand{\utt}{u_{22}}%
\newcommand{\uii}{u_{ii}}%
\newcommand{\ujj}{u_{jj}}%
\newcommand{\ujl}{u_{jl}}%
\newcommand{\ukl}{u_{kl}}%
\newcommand{\ulk}{u_{lk}}%
\newcommand{\ukm}{u_{km}}%
\newcommand{\umk}{u_{mk}}%
\newcommand{\umm}{u_{mm}}%
\newcommand{\uab}{u_{\alpha\beta}}%
\newcommand{\uai}{u_{\alpha i}}%
\newcommand{\uia}{u_{i\alpha}}%
\newcommand{\uja}{u_{j\alpha}}%
\newcommand{\uaj}{u_{\alpha j}}%
\newcommand{\uma}{u_{m\alpha}}%
\newcommand{\uam}{u_{\alpha m}}%
\newcommand{\ujb}{u_{j\beta}}%
\newcommand{\ubj}{u_{\beta j}}%
\newcommand{\uib}{u_{i\beta}}%
\newcommand{\ubi}{u_{\beta i}}%
\newcommand{\umb}{u_{m\beta}}%
\newcommand{\ubm}{u_{\beta m}}%
\newcommand{\uil}{u_{il}}%
\newcommand{\uli}{u_{li}}%
\newcommand{\uijI}{u_{ij}(\I)}%
\newcommand{\ukiI}{u_{ki}(\I)}%
\newcommand{\ujkI}{u_{jk}(\I)}%
\newcommand{\ulmI}{u_{lm}(\I)}%
\newcommand{\uooI}{u_{11}(\I)}%
\newcommand{\uotI}{u_{12}(\I)}%
\newcommand{\utoI}{u_{21}(\I)}%
\newcommand{\uttI}{u_{22}(\I)}%
\newcommand{\uiiI}{u_{ii}(\I)}%
\newcommand{\ujjI}{u_{jj}(\I)}%
\newcommand{\ujlI}{u_{jl}(\I)}%
\newcommand{\umlI}{u_{ml}(\I)}%
\newcommand{\uklI}{u_{kl}(\I)}%
\newcommand{\ulkI}{u_{lk}(\I)}%
\newcommand{\ukmI}{u_{km}(\I)}%
\newcommand{\umkI}{u_{mk}(\I)}%
\newcommand{\uabI}{u_{\alpha\beta}(\I)}%
\newcommand{\uaiI}{u_{\alpha i}(\I)}%
\newcommand{\uiaI}{u_{i\alpha}(\I)}%
\newcommand{\ujaI}{u_{j\alpha}(\I)}%
\newcommand{\uajI}{u_{\alpha j}(\I)}%
\newcommand{\umaI}{u_{m\alpha}(\I)}%
\newcommand{\uamI}{u_{\alpha m}(\I)}%
\newcommand{\ujbI}{u_{j\beta}(\I)}%
\newcommand{\ubjI}{u_{\beta j}(\I)}%
\newcommand{\uibI}{u_{i\beta}(\I)}%
\newcommand{\ubiI}{u_{\beta i}(\I)}%
\newcommand{\umbI}{u_{m\beta}(\I)}%
\newcommand{\ubmI}{u_{\beta m}(\I)}%
\newcommand{\uooi}{u_{11}(i)}%
\newcommand{\uoti}{u_{12}(i)}%
\newcommand{\utoi}{u_{21}(i)}%
\newcommand{\utti}{u_{22}(i)}%
\newcommand{\aij}{a_{ij}}%
\newcommand{\aijp}{a_{ij+}}%
\newcommand{\aijm}{a_{ij-}}%
\newcommand{\alm}{a_{lm}}%
\newcommand{\almp}{a_{lm+}}%
\newcommand{\almm}{a_{lm-}}%
\newcommand{\akl}{a_{kl}}%
\newcommand{\apq}{a_{pq}}%
\newcommand{\aklp}{a_{kl+}}%
\newcommand{\aklm}{a_{kl-}}%
\newcommand{\aklpm}{a_{kl\pm}}%
\newcommand{\aoo}{a_{11}}%
\newcommand{\aot}{a_{12}}%
\newcommand{\ato}{a_{21}}%
\newcommand{\att}{a_{22}}%
\newcommand{\aoop}{a_{11+}}%
\newcommand{\aotp}{a_{12+}}%
\newcommand{\atopp}{a_{21+}}%
\newcommand{\attp}{a_{22+}}%
\newcommand{\aoom}{a_{11-}}%
\newcommand{\aotm}{a_{12-}}%
\newcommand{\atom}{a_{21-}}%
\newcommand{\attm}{a_{22-}}%
\newcommand{\bij}{b_{ij}}%
\newcommand{\bijp}{b_{ij+}}%
\newcommand{\bijm}{b_{ij-}}%
\newcommand{\blm}{b_{lm}}%
\newcommand{\blmp}{b_{lm+}}%
\newcommand{\blmm}{b_{lm-}}%
\newcommand{\bk}{b_{\K}}%
\newcommand{\bkp}{b_{\K}^{\prime}}%
\newcommand{\bkpp}{b_{\K}^{\prime\prime}}%
\newcommand{\bkl}{b_{kl}}%
\newcommand{\bklh}{b_{kl 1}}%
\newcommand{\bklf}{b_{kl 2}}%
\newcommand{\bklp}{b_{kl+}}%
\newcommand{\bklm}{b_{kl-}}%
\newcommand{\bklpm}{b_{kl\pm}}%
\newcommand{\boo}{b_{11}}%
\newcommand{\bonetwo}{b_{12}}
\newcommand{\bto}{b_{21}}%
\newcommand{\btt}{b_{22}}%
\newcommand{\boop}{b_{11+}}%
\newcommand{\botp}{b_{12+}}%
\newcommand{\btopp}{b_{21+}}%
\newcommand{\bttp}{b_{22+}}%
\newcommand{\boom}{b_{11-}}%
\newcommand{\botm}{b_{12-}}%
\newcommand{\btom}{b_{21-}}%
\newcommand{\bttm}{b_{22-}}%
\newcommand{\Done}{D_{1}}%
\newcommand{\Di}{D_{i}}%
\newcommand{\Dpi}{D^{\prime}_{i}}%
\newcommand{\Dp}{D^{\prime}}%
\newcommand{\DN}{D_{N}}%
\newcommand{\Dpone}{D^{\prime}_{1}}%
\newcommand{\DpN}{D^{\prime}_{N}}%
\newcommand{\Ap}{A_{+}}%
\newcommand{\Ami}{A_{-}}%
\newcommand{\Apm}{A_{\pm}}%
\newcommand{\Br}{B_{r}}%
\newcommand{\Brn}{B_{r}(n)}%
\newcommand{\Brnz}{B_{r}^{\Znu}( n ) }%
\newcommand{\Broz}{B_{r}^{\Znu}(0 ) }%
\newcommand{\Bro}{B_{r}( 0 ) }%
\newcommand{\znu}{\Znu}%
\newcommand{\surf}{{\rm{surf}}_{r}}%
\newcommand{\deltap}{\delta_{\pot}}%
\newcommand{\deltapp}{\delta_{{\pot}^{\prime}}}%
\newcommand{\altb}{(\alt,\,\beta)}%
\newcommand{\Aast}{A^{\ast}}%
\newcommand{\AI}{A(\I)}%
\newcommand{\AJ}{A(\J)}%
\newcommand{\AK}{A(\K)}%
\newcommand{\Ddel}{D({\delta})}%
\newcommand{\Ddelal}{D({\delal})}%
\newcommand{\vepsi}{\varepsilon}%
\newcommand{\leps}{l_\varepsilon}%
\newcommand{\avepsi}{a_{\varepsilon}}%
\newcommand{\an}{a_{n}}%
\newcommand{\xii}{x_{i}}%
\newcommand{\xj}{x_{j}}%
\newcommand{\yi}{y_{i}}%
\newcommand{\yj}{y_{j}}%
\newcommand{\sumi}{\sum_{i}}%
\newcommand{\sumj}{\sum_{j}}%
\newcommand{\sumk}{\sum_{k}}%
\newcommand{\delal}{\delta_{\alpha}}%
\newcommand{\notni}{{\not\ni}}%
\newcommand{\deltaij}{\delta_{ij}}%
\newcommand{\AlIl}{\Al(\I_{l})}%
\newcommand{\Domdel}{D(\delta)}%
\newcommand{\Domdelal}{D(\delal)}%
\newcommand{\HoI}{H^{0}_{\I}}%
\newcommand{\hij}{h_{ij}}%
\newcommand{\hji}{h_{ji}}%
\newcommand{\hii}{h_{ii}}%
\newcommand{\hia}{h_{i\alpha}}%
\newcommand{\hbj}{h_{\beta j}}%
\newcommand{\delh}{\delta^{h}}%
\newcommand{\alth}{\alpha_t^{h}}%
\newcommand{\altone}{\alpha_{t_{1}} }%
\newcommand{\altm}{\alpha_{t_{m}} }%
\newcommand{\altk}{\alpha_{t_{k}} }%
\newcommand{\alsh}{\alpha_s^{h}}%
\newcommand{\als}{\alpha_s}%
\newcommand{\uht}{u_t^{h}}%
\newcommand{\uhs}{u_s^{h}}%
\newcommand{\uhst}{u_{s+t}^{h}}%
\newcommand{\vpvec}{\GNSvpvec}%
\newcommand{\vpvech}{\vpvec^{h}}%
\newcommand{\vpvechha}{\vpvec^{\hha}}%
\newcommand{\vpvecuha}{\vpvec^{\uha}}%
\newcommand{\vpvecwha}{\vpvec^{\wha}}%
\newcommand{\npot}{\bigl\Vert \pot \bigr\Vert }%
\newcommand{\normkuu}{\bigl\Vert  \bigr\Vert_{\scriptstyle{\PBI}}}%
\newcommand{\nni}{n_{i}}%
\newcommand{\nnI}{n_{|\I|}}%
\newcommand{\nno}{n_{1}}%
\newcommand{\nnim}{n_{i-1}}%
\newcommand{\Ennmc}{E_{\{\nno,\ldots, \nnim  \}^{c}}   }%
\newcommand{\limn}{\lim_{n}}%
\newcommand{\limal}{\lim_{\alpha}}%
\newcommand{\vpccaorig}{\vp^{c}_{\Ca}}
\newcommand{\vpccm}{\vp^{c}_{\Cam}}%
\newcommand{\vpccmmink}{\vp^{c}_{\Ca^{m-k} }}%
\newcommand{\vpccn}{\vp^{c}_{\Can}}%
\newcommand{\vpccl}{\vp^{c}_{\Cal}}%
\newcommand{\vpccml}{\vpc_{ \{ \Cam \cup \,\Cal  \}}    }%
\newcommand{\vpccmn}{\vpc_{ \{ \Cam \cup \Can  \}}    }%
\newcommand{\Rs}{R_{s}}%
\newcommand{\Rss}{R_{ {s^{\scriptscriptstyle{\prime}}} } }%
\newcommand{\omers}{\ome_{ {\scriptscriptstyle{\Rs}}  }}%
\newcommand{\omerss}{\ome_{ {\scriptscriptstyle{\Rss}}  }}%
\newcommand{\Somers}{S(\omers)}%
\newcommand{\Somerss}{S(\omerss)}%
\newcommand{\Spsivp}{S(\psi,\, \vp)}%
\newcommand{\eps}{\varepsilon}%
\newcommand{\epso}{\varepsilon_{1}}%
\newcommand{\epst}{\varepsilon_{2}}%
\newcommand{\surfr}{{\rm{surf}}_{r}}%
\newcommand{\surfd}{{\rm{surf}}_{d}}%
\newcommand{\Op}{O^{\prime}}%
\newcommand{\Oap}{O_{\alpha}^{\prime}}%
\newcommand{\Oa}{O_{\alpha}}%
\newcommand{\WnI}{W_{n}(\I)}%
\newcommand{\Hn}{H(\{n\})}%
\newcommand{\ppot}{p(\pot)}%
\newcommand{\ppotpsi}{p(\potpsi)}%
\newcommand{\Uca}{U(\Ca)}%
\newcommand{\pbpot}{p(\beta\pot)}%
\newcommand{\Ppot}{P(\pot)}%
\newcommand{\Ppotpsi}{P(\potpsi)}%
\newcommand{\Pbpot}{P(\beta\pot)}%
\newcommand{\Pbpotp}{P(\beta{\pot}^{\prime})}%
\newcommand{\Pbpotr}{P(\betar\potr)}%
\newcommand{\finf}{f_{\infty}}%
\newcommand{\alc}{\alpha_{\circ}}%
\newcommand{\blc}{\beta_{\circ}}%
\newcommand{\rlc}{\gamma_{\circ}}%
\newcommand{\alpr}{\alpha^{\prime}}%
\newcommand{\bpr}{\beta^{\prime}}%
\newcommand{\alone}{\alpha_{1}}%
\newcommand{\HIal}{H(\Ial)}%
\newcommand{\UIal}{U(\Ial)}%
\newcommand{\HIbl}{H(\Ibl)}%
\newcommand{\UIbl}{U(\Ibl)}%
\newcommand{\Diaa}{D_{i}^{(a,\al)}}%
\newcommand{\Dpaa}{D^{\prime\,(a,\al)}}%
\newcommand{\Daa}{D^{(a,\al)}}%
\newcommand{\HDiaa}{H(\Diaa)}%
\newcommand{\UDiaa}{U(\Diaa)}%
\newcommand{\WDiaa}{W(\Diaa)}%
\newcommand{\nam}{\nvhaIam}%
\newcommand{\nap}{\nvhaIap}%
\newcommand{\ao}{a_{0}}%
\newcommand{\alo}{\alpha_{0}}%
\newcommand{\aloa}{\alo(a)}%
\newcommand{\AhJ}{\Ah(\J)}
\newcommand{\potr}{\pot_{\gamma}}%
\newcommand{\potrm}{\pot_{\gamma(\mu)}}%
\newcommand{\vpr}{\vp_{\gamma}}%
\newcommand{\betar}{\beta_{\gamma}}%
\newcommand{\epot}{e_{\pot}}%
\newcommand{\epots}{e_{\pots}}%
\newcommand{\epotr}{e_{\potr}}%
\newcommand{\epo}{e_{\pot}(\ome)}%
\newcommand{\epor}{e_{\pot}(\omer)}%
\newcommand{\epotvp}{e_{\pot}(\vp)}%
\newcommand{\omer}{\omega_{\gamma}}%
\newcommand{\gammae}{\gamma_{\eps}}%
\newcommand{\vpic}{\vp_{\I}^{c}}%
\newcommand{\rhovpic}{\rhohat_{\vpic}}%
\newcommand{\rhon}{\rho_{n}}%
\newcommand{\Alcm}{\Al(\Cam)}%
\newcommand{\Alcn}{\Al(\Can)}%
\newcommand{\Alcl}{\Al(\Cal)}%
\newcommand{\Alcml}{\Al(\Caml)}%
\newcommand{\Alcmn}{\Al(\Camn)}%
\newcommand{\svpnchat}{s(\vpnchat)}%
\newcommand{\evpnchat}{e_{\pot}(\vpnchat)}%
\newcommand{\vpcCn}{\vp^{c}_{\Cn}}%
\newcommand{\vpcCa}{\vp^{c}_{\Ca}}%
\newcommand{\mo}{m_{1}}%
\newcommand{\mj}{m_{j}}%
\newcommand{\mi}{m_{i}}%
\newcommand{\mnu}{m_{\nu}}%
\newcommand{\mpr}{m^{\prime}}
\newcommand{\mpro}{\mpr_{1}}%
\newcommand{\mpri}{\mpr_{i}}%
\newcommand{\mprj}{\mpr_{j}}%
\newcommand{\mprnu}{\mpr_{\nu}}%
\newcommand{\vpbw}{\vp^{\beta \WI}}%
\newcommand{\vpbwstate}{\left[\vp^{\beta \WI} \right]}%
\newcommand{\deps}{d_{\varepsilon}}%
\newcommand{\Cnone}{\Cn^{1}}%
\newcommand{\Cntwo}{\Cn^{2}}%
\newcommand{\vppot}{\vp_{\pot}}%
\newcommand{\vppsi}{\vp_{\potpsi}}%
\newcommand{\Bdpot}{B_{d}(\pot)}%
\newcommand{\accpot}{\Gamma({\pot})}%
\newcommand{\potpsia}{\potpsi_{\alpha}}%
\newcommand{\psia}{\potpsi_{\alpha}}%
\newcommand{\vppsia}{\vp_{\psia}}%
\newcommand{\ad}{a^{\prime}}%
\newcommand{\add}{a^{\prime \prime}}%
\newcommand{\bd}{b^{\prime}}%
\newcommand{\bdd}{b^{\prime \prime}}%
\newcommand{\co}{c_{1}}%
\newcommand{\ct}{c_{2}}%
\newcommand{\aone}{a_{1}}%
\newcommand{\at}{a_{2}}%
\newcommand{\bo}{b_{1}}%
\newcommand{\bt}{b_{2}}%
\newcommand{\kep}{k_{\varepsilon}}%
\newcommand{\nep}{n_{\varepsilon}}%
\newcommand{\alep}{\alpha_{\varepsilon}}%
\newcommand{\blep}{\beta_{\varepsilon}}%
\newcommand{\ELn}{E_{\Ln}}%
\newcommand{\Eic}{E_{\{i\}^{c}}}%
\newcommand{\no}{n_{0}}%
\newcommand{\Lno}{\LL_{\no}}%
\newcommand{\uioo}{u_{11}^{(i)}}%
\newcommand{\uiot}{u_{12}^{(i)}}%
\newcommand{\uito}{u_{21}^{(i)}}%
\newcommand{\uitt}{u_{22}^{(i)}}%
\newcommand{\uiaa}{u_{\alpha\alpha}^{(i)}}%
\newcommand{\uiab}{u_{\alpha\beta}^{(i)}}%
\newcommand{\uiba}{u_{\beta\alpha}^{(i)}}%
\newcommand{\uibb}{u_{\beta\beta}^{(i)}}%
\newcommand{\kn}{k_{n} }%
\newcommand{\lnn}{l_{n} }%
\newcommand{\upkln}{u^{\prime(n)}_{\kn \lnn} }%
\newcommand{\upaan}{u^{\prime(n)}_{\alpha \alpha} }%
\newcommand{\upabn}{u^{\prime(n)}_{\alpha \beta} }%
\newcommand{\upban}{u^{\prime(n)}_{\beta  \alpha} }%
\newcommand{\upbbn}{u^{\prime(n)}_{\beta  \beta} }%
\newcommand{\uaan}{u^{(n)}_{\alpha \alpha} }%
\newcommand{\uabn}{u^{(n)}_{\alpha \beta} }%
\newcommand{\uban}{u^{(n)}_{\beta  \alpha} }%
\newcommand{\ubbn}{u^{(n)}_{\beta  \beta} }%
%
\newcommand{\kj}{k_{j}}
\newcommand{\lj}{l_{j}}
\newcommand{\upjkl}{u^{\prime({i_j})}_{\kj \lj} }%
\newcommand{\upjaa}{u^{\prime({i_j})}_{\alpha \alpha} }%
\newcommand{\upjab}{u^{\prime({i_j})}_{\alpha \beta} }%
\newcommand{\upjba}{u^{\prime({i_j})}_{\beta  \alpha} }%
\newcommand{\upjbb}{u^{\prime({i_j})}_{\beta  \beta} }%
\newcommand{\ujaa}{u^{({i_j})}_{\alpha \alpha} }%
\newcommand{\ujab}{u^{({i_j})}_{\alpha \beta} }%
\newcommand{\ujba}{u^{({i_j})}_{\beta  \alpha} }%
\newcommand{\ujbb}{u^{({i_j})}_{\beta  \beta} }%
\newcommand{\sigkl}{\sigma(k,l) }%
\newcommand{\ukp}{u_{kp} }%
\newcommand{\ukq}{u_{kq} }%
\newcommand{\ukk}{u_{kk} }%
\newcommand{\uql}{u_{ql} }%
\newcommand{\ulq}{u_{lq} }%
\newcommand{\blp}{b_{lp} }%
\newcommand{\blq}{b_{lq} }%
\newcommand{\bqk}{b_{qk} }%
\newcommand{\bll}{b_{ll} }%
\newcommand{\Ue}{U_{0} }%
\newcommand{\Uodd}{U_{1} }%
\newcommand{\be}{b_{0} }%
\newcommand{\bodd}{b_{1} }%
\newcommand{\aincr}{a_{i_{n}}^{\ast}}%
\newcommand{\ain}{a_{i_n}}%
\newcommand{\nk}{{n}_{k}}%
\newcommand{\mk}{{m}_{k}}%
\newcommand{\ml}{{m}_{l}}%
\newcommand{\beps}{b_{\varepsilon}}%
\newcommand{\neps}{n_{\varepsilon}}%
\newcommand{\Lleps}{\LL_{\leps}}%
\newcommand{\AlLleps}{\Al\bigl(\Lleps\bigr)}%
\newcommand{\bnf}{b_{n}^{0}}%
\newcommand{\bmkf}{b_{\mk}^{0}}%
\newcommand{\bksec}{b(k)}%
\newcommand{\bnkf}{b_{n}^{0}}%
\newcommand{\bns}{b_{n}^{1}}%
\newcommand{\Ineps}{\I_{\neps}}%
\newcommand{\Iml}{\I_{\ml}}%
\newcommand{\AlIml}{\Al(\Iml )}%
\newcommand{\Dpsi}{D_{\potpsi}}%
\newcommand{\Dpsio}{D_{\potpsio}}%
\newcommand{\Dpsit}{D_{\potpsit}}%
\newcommand{\Dpsipm}{D^{\pm}_{\potpsi}}%
\newcommand{\Dpsip}{D^{+}_{\potpsi}}%
\newcommand{\Dpsim}{D^{-}_{\potpsi}}%
\newcommand{\Dpsiopm}{D^{\pm}_{\potpsio}}%
\newcommand{\Dpsitpm}{D^{\pm}_{\potpsit}}%
\newcommand{\alprim}{\alpha^{\prime}}%
\newcommand{\laI}{l(a,\,\I)}
\newcommand{\laJ}{l(a,\,\J)}
\newcommand{\laIm}{l(a,\,\I+m)}
\newcommand{\Hset}{\mathbf{H}}
\newcommand{\HsetI}{{\mathbf{H}}_{\shift}}
\newcommand{\HIaJ}{H(\I \cap \J)}%
\newcommand{\lab}{l(a,\,b)}
\newcommand{\laCb}{l(a,\,\Cb)}
\newcommand{\laCbmk}{l(a,\,\Cb-k)}
\newcommand{\diam}{{\rm{diam}}}
%
\newcommand{\vpi}{\vp_{i}}%
\newcommand{\vpn}{\vp_{n}}%
\newcommand{\vpj}{\vp_{j}}%
\newcommand{\vpij}{\vp_{i_j}}%
\newcommand{\vpkuij}{\vpku_{i_j}}%
\newcommand{\vrho}{\varrho}%
\newcommand{\vrhoi}{\vrho_{i}}%
\newcommand{\vrhoj}{\vrho_{j}}%
\newcommand{\vrhoij}{\vrho_{i_j}}%
\newcommand{\vrhok}{\vrho_{k}}%
\newcommand{\vrhon}{\vrho_{n}}%
\newcommand{\vrhoh}{\vrho_{1}}%
\newcommand{\alvp}{\alpha_{\varphi}}%
\newcommand{\alvpone}{\alpha_{\vp_{1}}}%
\newcommand{\alvptwo}{\alpha_{\vp_{2}}}%
\newcommand{\alpot}{\alpha_{\pot}}%
\newcommand{\shat}{\hat{s}}
\newcommand{\vpku}{\vp^{k}}%
\newcommand{\vplu}{\vp^{l}}%
\newcommand{\vpcirc}{\vp_{\circ}}
\newcommand{\Ckaum}{\Cka^{m}}
\newcommand{\EIfara}{E^{(1)}_{\I} }%
\newcommand{\EIcfara}{E^{(1)}_{\Ic} }%
\newcommand{\EIsara}{E^{(2)}_{\I} }%
\newcommand{\EIplu}{E_{\I(+)} }%
\newcommand{\EJplu}{E_{\J(+)} }%
\newcommand{\EJaiplu}{E_{\Jai(+)} }%
\newcommand{\EJiplu}{E_{\Ji(+)} }%
\newcommand{\EJkplu}{E_{\Jk(+)} }%
\newcommand{\EJiJkplu}{E_{\Ji,\cdots,\Jk(+)}}
\newcommand{\EJiJkpluome}{E^{\ome}_{\Ji,\cdots,\Jk(+)}}
\begin{abstract}
 We study equilibrium statistical mechanics of Fermion lattice systems
 which require a different treatment compared with spin lattice systems
 due to the non-commutativity of local algebras for  disjoint regions.

Our major result is the 
equivalence of the KMS 
condition and the variational principle with a minimal assumption
 for the dynamics and without any explicit 
 assumption  on the potential.
Its  proof applies to  spin lattice systems as well, yielding 
a vast improvement over known results. 

All formulations are  in terms of a $\cstar$-dynamical systems
 for the 
 Fermion (CAR) algebra $\Al$ 
 with all or a part of the following assumptions
:\\
\ (I) The interaction is even, namely, 
 the dynamics $\alt$ commutes with 
the even-oddness automorphism $\Theta$. (Automatically 
 satisfied when (IV) is assumed.)\\
\ (II) The domain of the generator $\delal$ of $\alt$ contains
the set $\Alinfty$ of all strictly local elements of $\Al$.\\
\ (III) The set  $\Alinfty$ is the core of 
$\delal$.\\
\ (IV) The dynamics $\alt$ commutes with lattice translation automorphism
 group  $\tau$
 of $\Al$.

A major technical tool  is the conditional expectation from $\Al$
  onto its $\cstar$-subalgebras $\AlI$ for any subset $\I$ 
of the lattice,
 which induces a system of commuting squares.
This technique overcomes the lack  of tensor product structures 
for Fermion systems and even simplifies many known arguments 
  for spin lattice systems.
  
In particular, this tool is used for obtaining 
 the isomorphism between the real vector space of all $\ast$-derivations
  with their domain $\Alinfty$, commuting with $\Theta$, 
 and that of all $\Theta$-even standard potentials which satisfy
 a specific norm convergence condition for the one point
 interaction energy.
 This makes it possible to associate a unique standard potential
 to every dynamics satisfying (I) and (II).
The convergence condition for the potential is a consequence 
 of its definition in terms of the $\ast$-derivation and not an additional 
 assumption.

If translation invariance is imposed on $\ast$-derivations and 
potentials, then the isomorphism is kept and the space of translation
 covariant standard 
 potentials becomes
 a separable Banach space with respect to the norm of the one point interaction energy.
This is a crucial basis for an application of convex analysis 
 to the equivalence proof in the major  result.

Everything goes in parallel for spin lattice systems 
without the evenness assumption (I).
\end{abstract}
%
%
%
%
\tableofcontents
\section{Introduction}
\label{sec:INTRO}
 We investigate the equilibrium 
statistical mechanics of Fermion lattice systems.
While equilibrium statistical mechanics of spin lattice systems has
   been well studied (see e.g. \cite{BRA2}, \cite{ISR}
 and \cite{SIMON}), 
 there is a crucial difference between 
 spin and Fermion  cases. Namely, local algebras for disjoint
 regions commute elementwise for spin lattice systems,  but 
 do not commute for Fermion lattice systems.
 Due to this  difference, the known formulations
 and proof in the case of spin lattice systems do
 not necessarily go over to the case of Fermion lattice systems
and that is the motivation for  this investigation. 
An example of a Fermion lattice system is the well-studied 
Hubbard model, to which our results apply.

It turned out that, in the matter of the equivalence of the 
KMS condition  and the variational principle (i.e. the minimum free energy) for translation invariant states,
 we obtain its proof without any explicit assumption 
 on the potential except for the condition that 
 it  is the  standard potential
corresponding  to  a translation invariant
 even dynamics, a minimal condition for a proper formulation
 of the problem.
 Without any change in the methods of proof,
 this strong result holds 
 for spin lattice systems as well\ --\ a vast 
 improvement over known results for spin lattice systems
 and a solution of a problem posed by Bratteli and Robinson 
(Remark after Theorem 6.2.42. \cite{BRA2}).
In addition to this major result, we hope that 
the present work supplies a general mathematical foundation 
for equilibrium statistical mechanics of 
 Fermion lattice systems, which was lacking so far.

There are two distinctive features of our approach.
 One feature is the central role of the time
 derivative (i.e. the generator of the dynamics).
 On one hand, this enables us to deal with all 
 types of potentials without any explicit conditions 
 on their long range or many body behavior, as long as the first 
 time derivative of strictly localized operators 
 can be defined.
On the other hand, the existence of the dynamics for a given potential 
 is separated from the problems treated here and we can bypass that existence 
 problem via Assumption (III) below.

Another feature is the use of 
conditional expectations instead 
 of the tensor product structure traditionally used for 
spin lattice systems. 
They provide not only a substitute tool 
(for the tensor product structure), which is applicable 
 for both spin and Fermion lattice systems, but also 
 a method of estimates which does not use the norm of individual
 potentials, for which we do not impose any explicit condition.

The main subject of our paper is the characterization of equilibrium states
 in terms of the KMS condition and the variational 
 principle, 
which have an entirely different appearance but are shown to be 
 equivalent. They refer to canonical ensembles 
 in the infinite volume limit.
 However, they also refer to grand canonical ensembles if the 
 dynamics is modified by gauge transformations with respect to
 Fermion numbers \cite{CHEMICAL}. 
Namely, in the language of potentials, we may add a one-body
potential, which consists of the particle number operator(s)
times c-number chemical potential(s), and then the canonical
ensemble for the so-modified potential is the grand canonical
ensemble for the original potential, so that the grand
canonical ensemble can be studied as a canonical ensemble for
a modified potential, which is in the scope of our theory.

For the sake of notational simplicity, our presentation 
 is for the case of one Fermion at each lattice site.
Our results and proofs hold without any essential change for more general 
 case where a finite numbers of Fermions and finite spins coexist
 at each lattice site. The even-oddness in that case
 refers to the total Fermion number. For example, 
 for Hubbard model, there are two Fermions at each 
 lattice site, representing the two components 
 of a spin $1/2$ Fermion.

Our starting point is a $\cstar$-dynamical system $(\Al,\,\alt)$,
 where $\Al$
 is the $\cstar$-algebra of Fermion creation and annihilation operators
 on lattice sites of  $\Znu$ with local subalgebras $\AlI$
 for finite subsets  $\I \subset \Znu$
 and $\alt$ is a given strongly continuous 
one-parameter group of $*$-automorphisms
 of $\Al$.

Since the normal starting point in statistical mechanics is
a potential, a digression on our formulation and strategy
starting from a given dynamics may be appropriate at this point.
The KMS condition, which is formulated in terms of the dynamics,
is one of two main components of our equivalence result. On the
other hand, the variational principle, which is formulated in
terms of the potential, is the other main component. Therefore
both dynamics and potential are indispensable for our main
results and their mutual relation is of at most importance.

The key equation for that relation is the following formula.
For any operator $A$ localized in a finite subset $\I$ of the
lattice, its time derivative is given by
\begin{eqnarray*}
\frac{{\text{d}}}{{\text{d}t}}\alpha _t(A)=\alpha _t(i[\HI,\,A])
\end{eqnarray*}
where $\HI$ is described as a sum of potentials $\potJ$,
based on a finite subset $\J$ of the lattice, the sum being over
all $\J$ except those $\J$ for which $\potJ$ commutes with
any $A$ localized in $\I$, thus $\HI$ depending on $\I$.

The problem of construction of $\alpha _t$ from a given
class of potentials is not a straight-forward task and
has been studied by many people. As a result, a large
number of results are known for quantum spin lattice systems
(see e.g. \cite{BRA2}) and most of them can be applied to Fermion
lattice systems. There are also some specific analyses for
Fermion lattice systems (see e.g. \cite{MATSUI96}).

In parallel, the equivalence of the KMS condition and the
variational principle for translation invariant states has
been proved for a wide class of potentials for quantum spin
lattice systems. The same proof also works 
 for Fermion lattice systems in most cases; for example
this is the case for finite range potentials (see e.g.
page 113 of \cite{MATSUI98qpc}).

While these results cover a wide range of explicit models, it
seems difficult to decide exactly which class of potentials
determine a dynamics and to show the equivalence in question
in most general cases (which is not explicitly known) from the
potential point of view.

In the present work, we do not intend to make any contribution
to the problem of either construction of a dynamics from a
potential, or giving a complete criterion for potentials, which
give rise to a unique dynamics. (Thus we do not directly
contribute to the study of explicit models.)

On the contrary, we avoid these difficult problems by assuming
that the dynamics is already given (since this is needed in any
case for the KMS condition) and prove the equivalence result in
question under minimal (general) assumptions on the dynamics,
explained immediately below.

Note that we do not make any explicit assumptions about the
existence of a potential for a given dynamics nor about its
property (such as the absolute convergence of the sum defining
$\HI$ in terms of the potential).

For any given dynamics, for which all finitely localized
operators have the time derivative at $t=0$ (Assumption (II)
below) and which is lattice translation invariant
(Assumption (IV) below), we show the existence of a
corresponding potential, of which $\HI$ is a sum (as in
usual formulation) convergent in a well-defined sense.
\ \\

We now explain our assumptions and interconnection of
dynamics with potentials in more detail. The following
two assumptions make it possible to associate a potential
to any given dynamics satisfying them.\\
 \ \\
\quad (I) The dynamics is even. In other words, 
$\alt\, \Theta=\Theta \,\alt$
 for any $t \in \R$,
  where $\Theta$ is an involutive automorphism of $\Al$,
 multiplying $-1$ on all creation and annihilation operators.\\
 \ \\
\quad (II) The domain $\Domdelal$ of the generator $\delal$ of $\alt$ 
includes 
 $\Alinfty$, the union of all $\AlI$ for all finite subsets
 $\I$
 of the  lattice.
\ \\

It should be noted that Assumption (I) follows 
 from Assumption (IV) below. 
(See Proposition \ref{pro:shift-theta-alt}.)

We denote  by $\DB$ the  set of all $*$-derivations with $\Alinfty$
 as their domain  and their values in $\Al$, commuting with $\Theta$
 $($on $\Alinfty)$. Then the generator $\delal$ of our 
 $\alt$, when restricted to $\Alinfty$, belongs to $\DB$.

It is shown that $\DB$ is in one-to-one correspondence 
 with the set   $\PB$ of standard even potentials, which 
are functionals $\potI$ of all finite subsets $\I$
 of the lattice with values in the self-adjoint $\Theta$-even 
 part of the local algebra $\AlI$, satisfying 
our standardness condition and a topological
 convergence condition (Theorem\,\ref{thm:DB-PB}).

The topological convergence condition 
($(\pot$-$\rm{e})$ in Definition$\,$
\ref{df:STANDARDPOT}) 
is required in 
order that the potential is associated with a 
$\ast$-derivation on $\Alinfty$
 and refers to the convergence of the interaction energy operator 
 for every
 finite subset $\I$
\begin{eqnarray*}
\HI&=&
 \sum_{\K} \bigl\{ \potK;\ \K \cap \I \ne \emptyset \bigr\},
\end{eqnarray*}
  where a finite sum is first taken over $\K$ contained in a finite 
subset $\J$ and the limit of $\J$ tending
 to the whole lattice is to converge in the norm topology 
 of $\Al$.
(If this condition is satisfied for every  one-point set
 $\I=\{n \}\ (n\in \Znu)$, 
 then it is satisfied for 
 all finite subsets $\I$.) Note the difference from 
 conventional topological conditions, such as summability
  of $\Vert\potI \Vert$ over all $\I$ containing a point $n$,
   which are assumed for the sake of mathematical convenience. 
 
 For $\pot \in \PB$, internal energy $\UI$  and surface
 energy $\WI$ are also given  in terms of $\pot$
by  the conventional formulae for every finite $\I$.

The connection of the derivation $\del$
 and the corresponding potential $\pot$
 is given by
  \begin{eqnarray*}
\delta A= i[\HI,\,A] \quad \bigl(A \in \AlI \bigr).
\end{eqnarray*}
Due to the $\Theta$-evenness assumption (I), the replacement
 of $\HI$ by $\HK$ with $\K \supset \I$
 gives 
 the same $\del$ on $\AlI$, 
 a necessary condition for consistency.
 
The standardness 
($(\pot$-$\rm{d})$ in Definition$\,$\ref{df:STANDARDPOT})
 is formulated in terms of conditional expectations and 
 picks up a unique potential for each $\del\in\DB$. 
 Without the standardness condition, there are many different 
 potentials (called equivalent potentials)
 which yield exactly the same $\del$ through 
 the above formulae. Through the one-to-one correspondence
  between $\del(\in \DB)$ and $\pot(\in \PB)$,  
 any dynamics $\alt$ satisfying our standing assumptions (I)
 and (II) is associated with a unique 
 standard potential $\pot \in \PB$. 
 This is a crucial point of our formulation, leading to
 our major result.

When we  want to derive a statement involving $\alt$
 from a condition involving the potential $\pot$, 
we need the following assumption, guaranteeing
 the unique determination of $\alt$ from 
 the given $\pot$:\\
 \ \\
\quad (III) $\Alinfty$ is  the core of  the
 generator $\delal$ of the dynamics $\alt$.\\

For the discussion of variational principle, we need 
 the translation invariance assumption for the dynamics:\\
\ \\
\quad (IV) $\alt \shiftk=\shiftk \alt$,
 where $\shiftk$, $k \in \Znu$, is the automorphism group of $\Al$
 representing the lattice translations.\\

The above Assumptions (I) - (IV) are the only assumptions needed
for our theory below. On the other hand, if a potential $\pot$
(say, in the class $\PB$) is first given for any model,
it is a hard problem in general to show that the corresponding
derivation $\delpot \in \DB$ is given
by some dynamics satisfying Assumptions (II) and (III), or
equivalently that the closure of $\delpot$ is a generator
of a dynamics (i.e. it can be exponentiated to a one-parameter
group of automorphisms of $\Al$).
\ \\

We now present our main theorem after the explanation about
the variational principle and its ingredients.
The set $\PBI$
  of all translation covariant potentials in $\PB$
 forms a Banach space (Proposition\,\ref{pro:BANACH}) 
 with respect to 
  the norm
\begin{eqnarray*}
\potnor \equiv \Vert \Hn \Vert,
\end{eqnarray*} 
 which is independent of the lattice point $n$.
 The finite range potentials  are shown to be dense
 in $\PBI$ with respect to this norm
 and to imply 
 separability of $\PBI$ 
  (Theorem$\,$\ref{thm:separable} and 
  Corollary$\,$\ref{cor:separable}).

In terms of this norm,  
we obtain  the energy estimate 
\begin{eqnarray*}
\Vert\UI \Vert\le \Vert \HI \Vert \le \npot \cdot |\I|,
\end{eqnarray*}
 where $|\I|$ is the cardinality of $\I$ 
 (Lemma\,\ref{lem:EE}).
Then the conventional estimate
for $\WI$ follows. 
These estimates are used to show the existence 
 of the thermodynamic functionals, such as 
pressure $\Ppot$ and mean energy $\epo$.
All these estimates are carried out by the technique of conditional
 expectations without using the norm of the individual 
 $\potI$.

 For any state $\ome$ of $\Al$, its local entropy 
 $S_{\AlI}(\ome)=S(\ome|_{\AlI})$ is given as usual
by the von Neumann entropy $S(\cdot)$.
 Due to the non-commutativity of local algebras for  disjoint regions,
 not  all known properties  of  entropy 
 for spin lattice systems hold for our Fermion case
 \cite{MORIYAentangle}.
However, the strong subadditivity of entropy 
(SSA) for Fermion systems holds.
Then the existence of the mean entropy $s(\ome)$ 
for any translation invariant state $\ome$ for 
 Fermion lattice systems follows by a known method
 of  spin lattice systems.

The variational principle refers to the following equation for a translation
 invariant state $\vp$ of $\Al$ for a given translation
 covariant potential $\pot(\in \PBI)$
 and $\beta \in \R$:
\begin{eqnarray}
\label{eq:INTVP}
 \Pbpot= s(\vp)- \beta \epot(\vp) 
\end{eqnarray}

Our major result can be formulated as the following two theorems.
\ \\
\\{\bf{Theorem\ A.}} {\it{Under Assumptions}} 
(II) {\it{and}} (IV) {\it{for the dynamics $\alt$,
 any translation invariant state, which satisfies 
 the KMS condition for $\alt$ at the inverse temperature $\beta$,
 is a solution of the equation (\ref{eq:INTVP}), 
 where $\pot$ is the unique standard potential corresponding to $\alt$.}}\\
\ \\
{\bf{Theorem\ B.}} {\it{Under Assumptions}} 
(II), (III) {\it{and}} (IV) {\it{for the dynamics 
$\alt$,
any solution $\vp$ of  (\ref{eq:INTVP})
satisfies the KMS condition for $\alt$ at $\beta$.}}
\ \\
\begin{rem}
These two theorems hold also for spin lattice systems.
\end{rem}
\ 
\par We now present an over-all picture of the proof of our
main results above.
The proof of Theorem A and Theorem B
 will be carried out through the
 following steps:\\
 (1) KMS condition $\Rightarrow$ Gibbs condition.\\
(2) Gibbs condition $\Rightarrow$ Variational principle.\\
(3) Variational principle 
$\Rightarrow$ dKMS condition on $\Alinfty$.\\
(4) dKMS condition on $\Alinfty$ $\Rightarrow$ dKMS condition on $\Ddelal$.\\
(5) dKMS condition on $\Ddelal$ $\Rightarrow$ KMS condition.

Assumptions (I) and  (II)  are used  throughout  (1)-(5).
Assumption (IV) is used for the formulation of the variational 
 principle and necessarily 
  for (2) and (3). It is also used to derive Assumption (I), which
      is not included in the  premise of Theorems A and B.

Assumption (III) 
 is used only for (4).

The differential KMS (abbreviated as dKMS) condition 
in (4) and (5) refers to a known condition,
 which is entirely described in terms of the generator 
$\delal$ of $\alt$ and 
without use of $\alt$ (Definition$\,$\ref{df:DKMS}). This condition on the full  domain $\Ddelal$ of the generator $\delal$ of $\alt$ is known to be equivalent
 to the KMS condition (which is Step (5)).
The differential KMS condition for our purpose
 is the condition for the restriction 
 of $\delal$ to $\Alinfty$. Thus we need to show 
 Step (4)
 using the additional assumption (III) on  $\alt$.

For  Steps (1) and (2), we follow the 
proof for  spin lattice systems
 in principle.
However, the Gibbs condition for 
 Fermion lattice systems requires a careful
 definition. We define the Gibbs condition for a state $\vp$ as 
  the requirement that the local algebra $\AlI$
 is in the centralizer of the perturbed functional 
 $\vp^{\beta \HI}$, 
  which is obtained from
 $\vp$
 by a perturbation $\beta \HI$, 
for each finite subset $\I$ of the lattice
 (Definition$\,$\ref{def:Gibbs} and Lemma$\,$\ref{lem:EGIBBS}).
When $\AlI$ and $\AlIc$ commute (as in the case of 
spin lattice systems),
this condition reduces to the product type characterization
 which was introduced and  called the Gibbs 
 condition  by Araki and Ion 
  for quantum spin lattice systems 
\cite{ARAKIION}. With our definition of the Gibbs condition,
we have been able to prove Steps (1) and (2).

The product type characterization mentioned above 
is the condition 
that 
$\vp^{\beta \HI}$
 is  the  product of the tracial  state of  $\AlI$ and its 
 restriction to the complement algebra
 $\AlIc$. In  the present case of  
 Fermion lattice systems,
we  show that a Gibbs state satisfies this condition 
 if and 
 only if it is an even state of $\Al$ 
 (Proposition$\,$\ref{pro:evenGibbsProduct}).

The same kind of formulation and 
result are valid for a perturbation $\beta \WI$.

For Step (3)  as well as 
for the proof of the variational equality
\begin{eqnarray}
\label{eq:INTVE1}
 \Pbpot=\sup_{\omega \in   \invstate} 
\Bigl\{  s(\omega)- \beta \epot(\omega)      \Bigr\},
\end{eqnarray} 
 which is crucial for the variational principle, 
 we need a product state of local Gibbs state.
For this purpose, we have a technical result about the existence of 
a joint extension from  states of local algebras for disjoint
 subsets of the lattice to a state of the algebra for their union,
 which holds if the individual 
 states are even possibly except one (Theorem$\,$\ref{thm:EXT}). 

As an aside, the converse of Step (1) is 
shown under Assumptions (I), (II) and (III)
 (Theorem$\,$\ref{thm:GtoK}).

A major tool of our analysis is the $\cstar$-algebra
 conditional expectation 
$\EI: \Al \mapsto \AlI$ with respect to the unique tracial
 state $\tau$ of $\Al$.
 Its existence is shown not only for finite subsets
 but for all subsets $\I$ of the lattice
 (Theorem$\,$\ref{thm:EIEJ}). 
 Based on the product property of $\tau$
for subalgebras $\AlI$ and $\AlJ$ for disjoint $\I$
 and $\J$, 
we obtain the following commuting square of $\cstar$-subalgebras
 (Theorem\,\ref{thm:CARsquare}) for 
 Fermion systems. (It holds trivially for 
 spin systems.)
\begin{eqnarray*}
 \begin{CD}
\AlIuJ @>\EI>> \AlI\\
@V{\EJ}VV @VV{\EIaJ}V \\
\AlJ @>>\EIaJ>  \AlIaJ.
\end{CD}
\end{eqnarray*}
This serves as a replacement for the tensor-product structure 
in  traditional arguments
 for spin lattice systems.

As by-products, we obtain a few useful results on the CAR
 algebra: The even-odd automorphism $\Theta$
 is shown to be outer for any infinite CAR algebra
 (Corollary$\,$\ref{cor:ThetaOut}) and formulae
 for commutants of $\AlI$ and $\AlIp$ in $\Al$
 for finite and infinite $\I$ are obtained
 (Theorem $\,$\ref{thm:SOUGOUI} and 
 Theorem $\,$\ref{thm:SOUGOUIp}).

Some more results contained in this paper are as follows.

We show the validity of the variational
 equality (\ref{eq:INTVE1}) 
 when the Connes-Narnhofer-Thirring entropy $\hws$ 
 with respect to the group of lattice translation automorphisms 
 $\shift$ is used 
 in  place of the mean entropy $s(\ome)$ 
 (Theorem$\,$\ref{thm:VarCNT}). Note that 
our system $(\Al,\,\tau)$, where $\shift$ 
denotes the group of lattice translation automorphisms,  does not
 belong to the class 
of $\cstar$-systems 
considered in \cite{NESHSTOvariation}, 
being a non-abelian system. 

We define general potentials 
as those which satisfy all conditions for those in $\PB$
except for the standardness. They include 
all potentials satisfying
the following condition: 
\begin{eqnarray}
\label{eq:INTROEquivPot7}
\sum_{\I \ni n}\Vert \potI \Vert<\infty
\end{eqnarray}
for every lattice point $n$.
For each general potential, 
the corresponding $\HI$
 and $\del$ are defined and 
there is a unique standard potential in $\PB$
 with the same $\del$ as a given general 
 potential as described earlier.

Restricting our attention to those general potentials
satisfying (\ref{eq:INTROEquivPot7}) 
(a condition which is introduced also 
 in some discussion of spin lattice systems), 
 we are able to
 show by a straightforward argument that 
 the set of solutions of variational 
 principle for a general 
 translation covariant potential satisfying 
(\ref{eq:INTROEquivPot7})
 coincide with those for the equivalent standard potential  
(which is automatically translation covariant)
(Remark 1 to Proposition$\,$\ref{pro:EquivPot}),
 although the pressure and the mean energy 
  may be different between the two potentials.

\section{Conditional Expectations}
\label{sec:CE}
\subsection{Basic  Properties}
\label{subsec:CEbasic}
The following proposition is  well-known (see,
 e.g., Proposition 2.36, Chapter V
 \cite{TAKESAKI1}).
\begin{pro}  
\label{pro:CE1}
    Let $\M$ be a von Neumann algebra  with  a 
    faithful normal tracial state $\tau$ 
        and $\N$ be its von Neumann subalgebra.
 Then there exists a unique conditional expectation
\begin{eqnarray*}
 \EMN :
 a \in \M  \to \EMN(a) \in \N 
\end{eqnarray*}
  satisfying 
\begin{eqnarray}
\label{eq:CE2}
\tau(ab)=\tau\bigl( \EMN(a)b \bigr)
\end{eqnarray}
for any $b \in \Ne$.
  \end{pro}
\ \\
\begin{rem}
A conditional expectation $\EMN$ is linear, positive, unital, and
 satisfies 
\begin{eqnarray}
\label{eq:CE3}
\EMN(ab) =\EMN(a)b, \quad \EMN(ba)=b \EMN(a), 
\end{eqnarray}
for any $a \in \M$ and $b \in \Ne$, and
\begin{eqnarray}
\label{eq:CE4}
\Vert \EMN \Vert =1.
\end{eqnarray}
\end{rem}
\ \\

We shall obtain a $\cstar$-version of this proposition
 for the Fermion algebra in $\S$$\,$\ref{sec:FLS},
 where $\M$ and $\N$ are $\cstar$-algebras with a unique tracial 
state $\tau$.
The main step of its proof is the existence of $\EMN(a) \in \N$
 for every $a \in \M$ satisfying (\ref{eq:CE2}).
Once it is established, the map $\EMN$ is a conditional expectation
 by  standard argument, which we formulate for the sake of 
 completeness as follows.
\begin{lem}
\label{lem:CEunique}
Let $\M$ be a unital $\cstar$-algebra with a faithful 
tracial state $\tau$
 and $\N$ be its subalgebra containing the identity 
of $\M$. 
Suppose that for every $a \in \M$ there exists an element $\EMN(a)$ of $\N$
 satisfying (\ref{eq:CE2}).
Then the map $\EMN$ from $\M$ to $\N$ is the unique conditional 
 expectation from $\M$ to $\N$ with respect to $\tau$,
 possessing the following properties$:$\\
\ $(${\rm{1}}$)$  $\EMN$ is linear, positive and unital map from 
 $\M$ onto $\N$.\\
\ $(${\rm{2}}$)$  For any $a \in \M$ and $b \in \Ne$,
\[ \EMN(ab) =\EMN(a)b, \ \EMN(ba)=b \EMN(a).\]
\ $(${\rm{3}}$)$  $\EMN$ is a projection of norm $1$.
\end{lem}
\proof
\ First we prove the uniqueness of $\EMN(a) \in \N$ satisfying 
(\ref{eq:CE2}) for a given $a \in \M$.
Let $\ad$ and $\add$ in $\N$ satisfy (\ref{eq:CE2}), namely,

\begin{eqnarray*}
\tau(ab)=\tau(\ad b)=\tau(\add b)
\end{eqnarray*}
 for all $b \in \N$.
Then 
\begin{eqnarray*}
\tau\bigl(b(\ad-\add) \bigr)=0.
\end{eqnarray*}
By taking $b=(\ad-\add)^{\ast}$ and using the faithfulness of $\tau$,
 we obtain $\ad-\add=0$,
hence the uniqueness of $\EMN(a) \in \N$ 
for each $a \in \M$.

Except for the positivity, (1) and (2) can be shown in the same pattern
 as follows.
Let $a=\co\aone+\ct \at$ where $\aone, \at \in \M$
 and $\co, \ct \in \Com$.
 Then for any $b \in \N$,
\begin{eqnarray*}
\tau(ab)&=&\co\tau(\aone b)+\ct\tau(\at b)=\co\tau\bigl(\EMN(\aone)b\bigr)+
\ct\tau\bigl(\EMN(\at)b\bigr)\\
&=&\tau\bigl(  \bigl\{\co \EMN(\aone)+\ct\EMN(\at)\bigr\}b  \bigr).
\end{eqnarray*}
Since $\co \EMN(\aone)+\ct\EMN(\at) \in \N$, 
the uniqueness already shown implies 
\begin{eqnarray*}
\co \EMN(\aone)+\ct\EMN(\at)=\EMN(a).
\end{eqnarray*}
Therefore, $\EMN$ is linear.

In the same way, for any $a \in \M$ and $b \in \N$,
\begin{eqnarray*}
\tau(ab\bd )=\tau\bigl( \EMN(a)b \bd \bigr)
\end{eqnarray*}
holds 
for all $\bd \in \N$ and hence 
 \begin{eqnarray*}
  \EMN(ab)=\EMN(a)b. 
\end{eqnarray*}
Also
\begin{eqnarray*}
\tau(b a\bd )&=&\tau( a\bd b)=\tau\bigl( \EMN(a)\bd b \bigr)\\
&=&\tau(b \EMN(a) \bd )
\end{eqnarray*}
implies 
\begin{eqnarray*}
\EMN(ba)=b\EMN(a).
\end{eqnarray*}
If $a \in \N$, then the identity $\tau(ab)=\tau\bigl( \EMN(a)b   \bigr)$ with 
$b \in \N$ and the uniqueness result imply 
\begin{eqnarray*}
\EMN(a)=a.
\end{eqnarray*}
Therefore $\EMN$ is a map onto $\N$.
By taking $a=\identitybf(\in \N)$, we have 
\begin{eqnarray*}
\EMN(\identitybf)=\identitybf.
\end{eqnarray*}
Hence  $\EMN$ is unital.

Since $\EMN(a)\in \N$ for any $a \in \M$, we have 
 $\EMN\bigl(\EMN(a)\bigr)=\EMN(a)$.
Therefore $\EMN$ is a projection.

To show the positivity of the map $\EMN$,
 we consider the GNS triplet for the tracial state $\tauN$
 of $\N$ (which is the restriction of $\tau$ to $\N$)
 consisting of a Hilbert space $\HiltauN$,
 a representation $\pitauN$ of $\N$
 on $\HiltauN$ and a unit  vector $\tauvecN \in \HiltauN$,
giving rise to  the state $\tauN(A)
 =\tau(A)=(\tauvecN,\,\pitauN(A)\tauvecN)$
 for $A \in \N$.

If $a \in \M$ and $a \ge 0$, then for $b \in \N$
\begin{eqnarray*}
\bigr( \pitauN(b)\tauvecN,\,\pitauN(\EMN(a)) \pitauN(b)\tauvecN \bigl)
&=&\tauN\bigl( b^{\ast}\EMNa b   \bigr)\\
&=&\tauN\bigl( \EMNa b  b^{\ast} \bigr)=\tau(ab b^{\ast})=
\tau(b^{\ast} a b)\ge 0.
\end{eqnarray*}
Since $\pitauN(b)\tauvecN $, $b \in \N$ is dense in $\HiltauN$,
 we obtain  
\begin{eqnarray*}
 \pitauN\bigl(\EMN(a)\bigr) \ge 0.
\end{eqnarray*}
Since $\pitauN$ is faithful, 
\begin{eqnarray*}
\EMN(a)\ge 0,
\end{eqnarray*}
 and the positivity
 of $\EMN$ is shown.

For any $a \in \M$, the faithfulness of $\pitauN$ implies     
\begin{eqnarray}
\label{eq:EMNconti}
&&\Vert \EMNa \Vert \nonum \\
&=&\bigl\Vert \pitauN \bigl(\EMNa \bigr)\bigr\Vert \nonum \\
&=&\sup_{\bo,\bt \in \N} \Bigl\{ \bigl|\bigl(
\pitauN(\bo)\tauvecN,\, \bigl\{ \pitauN(\EMNa)\bigr\}\pitauN(\bt)\tauvecN \bigr)\bigl|
\ ;\ 
 \Vert \pitauN(\bo)\tauvecN\Vert\le 1,\  
\Vert \pitau(\bt)\tauvecN\Vert\le 1 \bigl)\Bigr|\Bigr\} \nonum \\
&=&\sup_{\bo,\bt \in \N} \Bigl\{ \bigl|\bigl(
\tau\bigl(\bo^{\ast}\EMN(a)\bt   \bigr)|
\ ;\  \tau(\bo^{\ast}\bo) \le 1,\  
\tau(\bt^{\ast}\bt) \le 1 \Bigr\} \nonum \\
&=&\sup_{\bo,\bt \in \N} \Bigl\{ \bigl|\bigl(
\tau\bigl(\bo^{\ast} a \bt   \bigr)|
\ ;\  \tau(\bo^{\ast}\bo) \le 1,\  
\tau(\bt^{\ast}\bt) \le 1 \Bigr\} \nonum \\
&=&\sup_{\bo,\bt \in \N} \Bigl\{ \bigl|\bigl(
\pitauM(\bo)\tauvecM,\,  \pitauM(a) \pitauM(\bt)\tauvecM \bigr)\bigl|
\ ;\ 
 \Vert \pitauM(\bo)\tauvecM\Vert\le 1,\  
\Vert \pitauM(\bt)\tauvecM\Vert\le 1 \bigl)\Bigr|\Bigr\} \nonum \\
&\le&\Vert \pitauM ( a ) \Vert
=\Vert a \Vert,
\end{eqnarray}
 where we have used the cyclicity of $\pitauN(\N)$ for $\HiltauN$
 for the second equality,
\begin{eqnarray*}
\tau\bigl(\bo^{\ast}\EMNa \bt \bigr)=
\tau\bigl(\EMNa \bt \bo^{\ast}\bigr)=
\tau\bigl(a \bt \bo^{\ast}\bigr)=\tau\bigl(\bo^{\ast} a \bt \bigr).
\end{eqnarray*}
 for the fourth equality, 
and the same computation backwards 
 replacing $\N$ by $\M$ for the fifth  equality. 
Due to $\EMN(\identitybf)=\identitybf$ and (\ref{eq:EMNconti}),
 we have 
\begin{eqnarray*}
\Vert \EMN \Vert =1.
\end{eqnarray*}
We have completed the proof.\proofend
\subsection{Geometrical Lemma}
Let us consider
 finite type {{\bf I}} factors (i.e., full matrix algebras)
   $\M$ and $\N$ such that 
$\M \supset \N$.
We have   
the isomorphisms $\M \simeq \N  \otimes {\N}_1$, 
$\N \simeq \N \otimes {\mathbf {1}}$, and 
$\tau={\tau}_{\N} \otimes {\tau}_{\Ni}$
 where
$\Ni \equiv \M \cap {\N}^{\prime}$ is  a finite type {\bf{I}} factor.

A conditional expectation satisfying (\ref{eq:CE2}) is given by
 the slice map:
\begin{eqnarray}
\label{eq:slice}
 \EMN(bb_1)=\tau(b_1)b \quad(b\in\N, b_1 \in {\N}_1).
\end{eqnarray} 

We give this $\EMN$ 
 a  geometrical picture which we find 
useful.   
We introduce  the following 
inner product on $\M$:
\begin{eqnarray*}
<a,\: b> \equiv \tau(a^{\ast}b), \quad (a,b \in \M).
\end{eqnarray*}
$\M$ is then a  (finite-dimensional) Hilbert space 
with this inner product. 
 Let $\PMN$ be the orthogonal projection onto  the subspace $\N$ of $\M$.
 We show that $\PMN$ is the same as  $\EMN$ 
as a map $\M \mapsto \N$.
\begin{lem} 
\label{lem:CEGEO}
With the notation  above,  
 \begin{eqnarray}
\label{eq:EPid}
 \PMN a = \EMN(a).
 \end{eqnarray}
for any $a \in \M$.
\end{lem}
\proof
\ Any $a \in \M$ can be decomposed as $a=\PMN a + a^{\prime}$
 where $a^{\prime} \in \N^{\bot}$.
 For any $b \in \N$, we have $b^{\ast} \in \N$ and hence
 \begin{eqnarray*}
 \tau(ab)&=&<b^{\ast},\ a>=<b^{\ast},\ \PMN a>
+<b^{\ast},\ a^{\prime}> \\
         &=&<b^{\ast},\ \PMN a>=\tau \bigl((\PMN a)  b\bigr).
 \end{eqnarray*}
 Since $\PMN a \in \N$, it follows from Proposition \ref{pro:CE1} that
 \begin{eqnarray*}
 \PMN a = \EMN(a).
 \end{eqnarray*}
 \proofend \\

\subsection{Commuting Square}
\label{subsec:SQUARE}
We introduce  the following  equivalent conditions for a commuting square.
(See e.g. \cite{COMMUTINGSQUARE}.)
\begin{pro}
\label{pro:square}
Let $\M, \Ni, \Nii$ and  
$\Pa$ be finite type {{\bf I}} factors satisfying
\begin{eqnarray*}
\M \supset \Ni \supset \Pa,\ \ \M \supset \Nii \supset \Pa.
\end{eqnarray*}

  Then the following conditions are equivalent$:$\\
 $(${\it{1}}$)$  $\EMNi \vert_{\Nii}=\ENiiP$\\
  $(${\it{2}}$)$   $\EMNii \vert_{\Ni}=\ENiP$\\
$(${\it{3}}$)$ $\Pa=\Ni \cap \Nii$ and $\EMNi \EMNii=\EMNii \EMNi$\\
$(${\it{4}}$)$ $\EMNi\EMNii=\EMP$ \\
 $(${\it{5}}$)$     $\EMNii\EMNi=\EMP$.
\end{pro}
\proof
 
$(${\it{1}}$)$ $\Longleftrightarrow$  $(${\it{4}}$)$:\\
Assume (${\it{1}}$). Let $a \in \M$ and $b \in \Pa$.
 By the assumption, we have 
\begin{eqnarray*}
\EMNi\bigl( \EMNii(a) \bigr)=
 \ENiiP \bigl( \EMNii(a) \bigr)
\in \Pa. 
\end{eqnarray*}
due to  $\EMNii(a) \in \Nii$.
 On the other hand, 
\begin{gather*}
\begin{split}
\tau \left(\EMNi \bigl(\EMNii(a)\bigr) b \right)
&=\tau\left(  \EMNii(a)b  \right)
 \quad  (\text{due to} \ b \in (\Pa \subset) \Ni ) \nonum \\
&=\tau \left(  ab \right) 
 \quad  (\text{due to} \ b \in (\Pa \subset)\Nii).
\end{split}
\end{gather*}
Hence $\EMNi\bigl( \EMNii(a)\bigr)=\EMP(a)$ and so 
 $\EMNi \EMNii=\EMP$.

The converse is obvious: for $a \in \Nii$,  $(${\it{4}}$)$
 implies 
\begin{eqnarray*}
\ENiiP(a)=\EMP(a)=\EMNi\EMNii(a)=\EMNi(a)
\end{eqnarray*}
 and hence $(${\it{1}}$)$.

 $(${\it{2}}$)$ $\Longleftrightarrow$  $(${\it{5}}$)$ :\\
Exactly the same proof as above, with $\Ni$ and $\Nii$ interchanged.

 $(${\it{4}}$)$ $\Longleftrightarrow$  $(${\it{3}}$)$:\\
Assume  $(${\it{4}}$)$. By Lemma$\,$\ref{lem:CEGEO},
$(${\it{4}}$)$ implies 
\begin{eqnarray*}
\PMNi\PMNii=\PMPa.
\end{eqnarray*}
Taking adjoints, we obtain
\begin{eqnarray*}
\PMNii\PMNi=\PMPa.
\end{eqnarray*}
This implies
\begin{eqnarray*}
\EMNii\EMNi=\EMP=\EMNi\EMNii,
\end{eqnarray*}
 the last equality being due to $(${\it{4}}$)$.

Due to $\Ni \supset \Pa$ and $\Nii \supset \Pa$,
 we have $\Pa \subset \Ni \cap \Nii$. 
If  $b \in \Ni \cap \Nii$,
then 
\begin{eqnarray*}
b=\EMNi\EMNii(b)= \EMP(b) \in \Pa
\end{eqnarray*} 
by $(${\it{4}}$)$.
Hence  $\Pa=\Ni\cap \Nii$.
This completes  the proof of  
$(${\it{4}}$)$ $\Longrightarrow$  $(${\it{3}}$)$.

Assume $(${\it{3}}$)$.
 For  any $a \in \M$, $(${\it{3}}$)$
 implies 
$\EMNi \bigl(\EMNii(a)\bigr)=\EMNii\bigl(\EMNi(a)\bigr)\in 
 \Ni \cap \Nii=\Pa$ because the range of $\EMNi$
 is $\Ni$ and the range of $\EMNii$
 is $\Nii$.
 For any $b \in \Pa$ and  $a \in \M$,
\begin{eqnarray*}
\tau \left(\EMNi \bigl(\EMNii(a)\bigr) b \right)=
\tau \left( \EMNii(a) b \right)=\tau(ab).
\end{eqnarray*}
Hence $\EMNi \bigl(\EMNii(a)\bigr)=\EMP(a)$.
This implies $(${\it{4}}$)$.

 $(${\it{5}}$)$ $\Longleftrightarrow$  $(${\it{3}}$)$:\\ 
Exactly the same proof as above, with $\Ni$ and $\Nii$ interchanged.
\proofend
\section{Entropy and Relative Entropy}
\label{sec:ENT}
\subsection{Definitions}
\label{subsec:DEFent}
We introduce some definitions and related 
lemmas needed for formulation
of the main result of this section.
\begin{lem}
\label{lem:ADJUSTED}%
Let $\M$ be a finite type {{\bf I}} factor. \\
{\rm (i)} Let $\vp$ be a positive linear functional on $\M$.
Then there exists a unique $\rhovp \in \M_{+}$
(called  adjusted density matrix) satisfying
\begin{eqnarray*}
\vp(a)=\tau(\rhovp a)
\end{eqnarray*}
for all $a\in \M$.\\
{\rm (ii)}
 Let $\N$ be a subfactor of  $\M$  and
$\vpN$ be the restriction of $\vp$ to $\N$.
Then 
\begin{eqnarray*}
\rhodensityN=\EMN(\rhodensity)
\end{eqnarray*}
\end{lem}
\proof
\ {\mbox (i)} is well-known.\\
(ii) For $b \in \N$,  
$\vpN(b)=\vp(b)=\tau(\rhodensity b)=\tau\bigl( \EMN(\rhodensity)b \bigr).$
Since $\EMN(\rhodensity) \in {\N}_+$, we have
\begin{eqnarray*}
\rhodensityN=\EMN(\rhodensity).
\end{eqnarray*}
\proofend \\
\begin{rem} 
The above definition of density matrix is given
 in terms of the tracial state in contrast to 
the standard definition 
 using the matrix trace $\Tr$.
 Hence we 
  use the word `adjusted'.
\end{rem}

\begin{df}
\label{df:adjustedENT}
Let $\rhodensity$ be the adjusted density matrix of a positive linear
  functional $\vp$ of a finite type {{\bf I}} factor.
Then
\begin{eqnarray*}
\widehat{S}(\vp) \equiv
 -\vp(\log \rhodensity)  
\end{eqnarray*}
 is called the adjusted entropy of $\vp$.
\end{df}
\begin{rem}
The adjusted density matrix and the adjusted entropy 
 for a type 
${{\bf I}_{n}}$ factor $\M$ with the dimension $\Tr(\id)=n$ 
are related to the usual ones by the following relations:
  \begin{eqnarray}
  \rhodensity=n \rhod, \ 
 \widehat{S}(\vp)=S(\vp)-\vp(\mathbf{1}) \log n.
 \label{eq:normalization}
  \end{eqnarray} 
\end{rem}
\ \\

The range of the values of  entropy is given 
 by the following well-known  lemma.
\begin{lem}
\label{lem:ENTrange}
If $\M$ is a type 
${{\bf I}_{n}}$ factor and $\vp$ is a state of $\M$, then
  \begin{eqnarray}
\label{eq:ENTlogn}
 0 \le S(\vp) \le \log n.
  \end{eqnarray} 
The equality $S(\vp)=0$ holds if and only if $\vp$
 is a pure state of $\M$.
The equality $S(\vp)=\log n$ holds if and only if $\vp$
 is the  tracial state $\tau$ of $\M$.
\end{lem}
\begin{df}
\label{df:RENT}
The relative entropy  of $\varrho$ and $\sigma$ in ${\M}_+$ as well as
 that of positive linear functionals  $\vp$  and $\psi$ are defined by 
 \begin{eqnarray}
  S(\sigma,\ \varrho)& =&\tau\bigl(\varrho 
      (\log \varrho - \log \sigma)\bigr) \label{eq:rel1}\\
 S(\psi,\ \vp)& =& 
 \vp(\log \rhodensity - \log \rhodensitypsi) 
(= \tau  \Bigl(\rhodensity \log \rhodensity - \rhodensity\log \rhodensitypsi
 \Bigr) )
\label{eq:rel2}.
 \end{eqnarray}
\end{df} 
\ \\
\begin{rem}
$S(\psi, \ \vp)$ remains the same if $\rhodensity$ and $\rhodensitypsi$
 are replaced by 
 the density matrices  $\rho_{\vp}$ and $\rho_{\psi}$
 with respect to $\Tr$.
 The right-hand sides of (\ref{eq:rel1}) and (\ref{eq:rel2}) are 
well-defined
 when $\varrho$, $\sigma$, $\rhodensity$ and $\rhodensitypsi$ are regular.
 Otherwise, one may define them as the limit of regular cases,
 for example by taking the limit $\varepsilon \to 0$ for
 $(1-\varepsilon)\vp+\varepsilon \tau$, $(1-\varepsilon)\psi+
 \varepsilon \tau$ for (\ref{eq:rel2}), and similarly for (\ref{eq:rel1}).
 The value of $S(\psi, \vp)$ is 
 real or $+\infty$ for 
positive linear functionals  $\vp$  and $\psi$.
\end{rem}
\ \\
The following lemma is also well-known.
\begin{lem}
Let $\vp$ and $\psi$ be states. Then
 $\Spsivp$  
 is non-negative. 
It vanishes if and only if $\vp=\psi$.
\end{lem}

\begin{rem}
We  note that there are different 
notations  for the relative entropy and 
 that we  adopt that 
of Araki \cite{ARAKI76rims} 
 and Kosaki \cite{KOSAKI86}. 
In comparison with our notation, the order of two states is 
reversed in that of Umegaki \cite{UME62}, while
 both the order of states and the sign are  reversed  
 in that of Bratteli and
Robinson  \cite{BRA2}.
\end{rem}
\subsection{Monotone Property}
\label{subsec:MONO}
Under any conditional expectation $E$
 and under restriction to any subalgebra, 
the relative entropy is known to
 be non-increasing:
\begin{eqnarray}
\label{eq:MONOE}
S(\psi \circ E,\;  \vp \circ E) &\leq& S(\psi, \vp),\\
\label{eq:monoN}
S(\psiN,\; \vpN) &\leq& S(\psi, \ \vp).   
\end{eqnarray}
(For example, (\ref{eq:monoN}) 
 is Theorem 4.1(iv) of \cite{KOSAKI86}.
(\ref{eq:MONOE}) follows from Theorem 4.1(v) of \cite{KOSAKI86},
 because $E$ is a Schwarz map \cite{TOMIYAMA}.)

 When we want to exhibit the dependence of entropy on $\M$
more explicitly, we  use the notation $\sm$ and $\smh$
instead of $S$ and $\Shat$.
The relation between the entropy and the relative entropy 
for a state $\vp$ 
is given by 
\begin{eqnarray*}
\Shat(\vp)=-S(\tau, \ \vp)=S(\vp)-S(\tau).
\end{eqnarray*} 
Note that $ S(\tau)=\log n $ for a type 
${{\bf I}_{n}}$ factor  $\M$.

We identify $\M$
 with  $\N \otimes (\Nex)$ and use the notation 
 $\vpN \otimes {\tau}_{\Nex}$.
We also  identify 
 $A \in \N \subset \M$ with $A \otimes \mathbf{1} \in 
 \N \otimes (\Nex)$.

\begin{lem} 
\label{lem:SN-SM}
Let $\M \supset \N$ be finite type {{\bf I}} factors, and 
$\vp$ be a state on $\M$. Then
\begin{eqnarray}
\label{eq:SN-SM}                    
\snh(\vpN)-\smh(\vp)=\sm(\vpN \otimes \tau_{\Nex}, \vp)
                    =\sm(\vp \circ \EMN, \vp). 
\end{eqnarray}
\end{lem}
\proof
\ If $\vp$ is a faithful state, we show the above identity
by a straight-forward calculation. 
If $\vp$ is not faithful, we add 
$\varepsilon \cdot \tau$ to $(1-\varepsilon)\vp$ and  
then take  the limit $\varepsilon \to 0$.
\proofend\\
\begin{rem}
$\Shat$ in the above Lemma cannot be replaced by $S$. 
\end{rem}
\subsection{Strong Subadditivity }
\label{subsec:SSAgen}
If the system under consideration enjoys 
the commuting square property 
with respect to a  tracial state,   
the strong subadditivity property 
for the adjusted entropy $\Shat$ holds 
(see Theorem 12 in \cite{PETZ}).
\begin{thm}
\label{thm:SSAgen}
Let $\M, \Ni, \Nii$ and 
 $\Pa$ be finite type {{\bf I}} factors satisfying
 one of the equivalent conditions of Proposition \ref{pro:square}. 
Let $\psi$ be a state on $\M$.
 Then
\begin{eqnarray}
\label{eq:SSAgen}
\Shat(\psi)-\Shat(\psiNi)-\Shat(\psiNii)+\Shat(\psiP) \leq 0.
\end{eqnarray}  
\end{thm}
\proof
 \ By (\ref{eq:SN-SM}) and (\ref{eq:MONOE}) 
\begin{eqnarray*}
\sniih(\psiNii)-\smh(\psi)=\sm(\psi \circ \EMNii, \ \psi)
 \geq \sm(\psi \circ \EMNii \circ \EMNi, \psi \circ \EMNi)
 \end{eqnarray*}
By the  assumption,  
$\EMNii \EMNi = \EMNi \EMNii =\EMP$.
 Hence,
 \begin{eqnarray*}
\sm(\psi \circ \EMNii \circ \EMNi, \ \psi \circ \EMNi)
&=&\sm(\psiP \otimes \tau_{\Pex},\  \psiNi \otimes \tau_{\Niex})\nonumber \\
&=&\sn(\psiP \otimes \tau_{\PexNi},\ \psiNi)\nonumber \\
&=&\sph(\psiP)-\snih(\psiNi),
\end{eqnarray*}
where the second equality is due to 
$\tau_{\Pex}= \tau_{\PexNi} \otimes \tau_{\Niex}$
 and the last equality due to 
 (\ref{eq:SN-SM}). Therefore we obtain (\ref{eq:SSAgen}).
\proofend\\
\section{Fermion Lattice Systems}
\label{sec:FLS}
\subsection{Fermion Algebra} 
\label{subsec:FAl}
We introduce Fermion lattice systems
where there exists one  spinless 
Fermion at  each lattice site  and  they interact with each other.
 The  restriction to spinless particle (i.e.,
 one degree of freedom for each site)
  is  just a matter of simplification
 of notation.
All 
 results and their proofs in the present work
 go over  
 to the case of an arbitrary (constant) finite number of degrees
 of freedom
 at each lattice site without any essential alteration. 

The lattice we consider is 
  $\nu$-dimensional lattice $\Znu$\,($\nu \in \NN$,
an arbitrary positive integer).
\begin{df}
\label{df:CARalg}
The Fermion $\cstar$-algebra $\Al$
 is a unital $\cstar$ algebra 
 satisfying the following conditions
 and generated by elements in $(${\rm{1-1}}$)$$:$\\
$(${\rm{1-1}}$)$  For each lattice site $i \in \Znu$, there are 
 elements  $\ai$ and $\aicr$ of  $\Al$ called annihilation
 and creation operators, respectively,
 where $\aicr$ is the adjoint of $\ai$. \\
$(${\rm{1-2}}$)$ The CAR(canonical anticommutation relations)
 are satisfied for any $i,j \in \Znu$:
\begin{eqnarray}
\label{eq:CAR1}
\{ \aicr, \aj \}&=&\delta_{i,j}\, \identitybf \nonumber \\
\{ \aicr, \ajcr \}&=&\{ \ai, \aj \}=0.
\end{eqnarray}
Here $\{A, B\}=AB+BA$ (anticommutator), 
 $\delta_{i,j}=1$ for $i = j$, and
  $\delta_{i,j}=0$ for  $i\neq j$.\\
$(${\rm{1-3}}$)$ Let $\Alinfty$ be the $\ast$-algebra generated by 
 all $\ai$ and $\aicr$ ($i\in \Znu$), namely
 the (algebraic) linear span of their monomials
 $A_{1}\cdots A_{n}$ where $A_{k}$ is $\aik$ or $\aikcr$,
 $\ik \in \Znu$.
\ \\
$(${\rm{2}}$)$  For each  subset $\I$  of $\Znu$,
 the $\cstar$-subalgebra of $\Al$ generated by   
 $\ai$,  $\aicr$, $i  \in \I$,  is denoted
 by $\AlI$. 
 If the cardinality $|\I|$ of the set $\I$ is finite,
 then $\AlI$ is referred to as a local algebra or more specifically
 the local algebra for  $\I$.
 For the empty set $\emptyset$,
 we define $\Alempty =\Com \identitybf$.
\end{df}
\ \\
{\it{Remark 1.}}
$\Alinfty$ is dense in $\Al$.
\ \\
{\it{Remark 2.}}
For  finite  $\I$,
$\AlI$ is known to be isomorphic to  
  the tensor product of $\vert \I \vert$
  copies of the full $2\times2$ matrix algebra 
  ${\mathrm M}_{2}(\Com)$ and hence isomorphic to 
 ${\mathrm M}_{2^{|\I|}}(\Com)$.
Then
 \begin{eqnarray*}
\Alinfty=\bigcup_{|\I| < \infty }\AlI
\end{eqnarray*}
has the unique $\cstar$-norm.
 $\Al$ together with its individual elements 
$\{\ai,\,\aicr| i \in \Znu\}$
 is uniquely defined up to isomorphism and 
 is isomorphic to the UHF-algebra
  ${{\overline{\otimes}}_{i \in \Znu} {\mathrm M}_2(\Com)}$,
  where the bar denotes the norm completion. 
$\Al$ has the unique tracial state $\tau$
 as the extension of the unique tracial state of $\AlI$, 
 $|\I|<\infty$.
\ \\
\ \\
{\it{Remark 3.}}
Since $\aicr$'s  and $\ai$'s
 anti-commute among different indices, $\aicr$
 and  $\ai$ with a specific $i$ can be brought together
  at any spot in a monomial, 
  with a possible sign change\,(without changing the 
ordering among themselves),
   and this can be done for each $i$. Therefore, the monomials 
   of the form 
\begin{eqnarray}
\label{eq:order}
\Aih \cdots \Aik
\end{eqnarray}
 together with $\identitybf$
 have a dense linear span in $\AlI$,
 where   
 the indices 
$\ih, \cdots, \ik  \in  \I$ are distinct and  
$\Aialpha$ is one of 
$\acrial$, $\aial$, $\acrial\aial$, $\aial \acrial$.

 \begin{df}
\label{df:THETA}
$\ThetaIhat$ denotes a unique automorphism of $\Al$
 satisfying
 \begin{eqnarray}
\label{eq:Theta1}
\ThetaIhat(\ai)&=&-\ai, 
\quad\ThetaIhat(\aicr)=-\aicr, \quad  ( i \in \I),\\
\ThetaIhat(\ai)&=&\ai, \quad\ThetaIhat(\aicr)=\aicr, \quad ( i \ne \I).\nonum
\end{eqnarray}
In particular, we denote 
$\Theta=\Theta^{\Znu}$.

The  even  and odd parts of $\Al$ are defined as
  \begin{eqnarray}
\label{eq:EO}
 \Alp \equiv \{a \in \A \; \bigl| \;   \Theta(a)=a  \},\quad  
 \Alminus \equiv \{a \in \A  \; \bigl| \;  \Theta(a)=-a  \}. 
 \end{eqnarray}
\end{df}
\noindent {\it{Remark 1.}}
Such  $\Theta$ exists and
 is unique because (\ref{eq:Theta1})
 preserves CAR. It obviously satisfies 
\begin{eqnarray}
\label{eq:Theta2}
\Theta^{2}={\mbox{id}}.
\end{eqnarray}
\\
\ \\
{\it{Remark 2.}}
 For any $a \in \AlI$, 
\begin{eqnarray}
\label{eq:EO1}
  a=a_{+} +a_{-},\ \ a_{\pm}\equiv \frac{1}{2}
  \bigl(a \pm \Theta(a)   \bigr) 
\end{eqnarray}
  gives the (unique) splitting of $a$ into a sum of 
  $a_{+} \in \AlIp  $ and $a_{-} \in \AlIm$, where
  the  even and odd parts of $\AlI$
 are denoted  
   by $\AlIp$ and $\AlIm$. \\
\ \\
{\it{Remark 3.}}
 For any $a \in \Alminus$, we have
  \begin{eqnarray}
 \tau(a)=\tau\bigl(\Theta(a) \bigr)=-\tau(a)=0.
 \label{eq:oddkill} 
  \end{eqnarray} 
\begin{lem}
\label{lem:IJrel}
Let $\I$ and $\J$ be mutually disjoint and 
$\asig \in \AlIsig$, $\bsig \in \AlJsig$  where
 $\sigma=+$ or $-$.  
Then 
\begin{eqnarray}
\label{eq:epsCAR}
\asig b_{\sigma^{\prime}}=\epsilon(\sigma, \sigma^{\prime})
 b_{\sigma^{\prime}} \asig,
\end{eqnarray}
 where 
\begin{eqnarray*}
\epsilon(\sigma, \sigma^{\prime})&=&-1\ \ \mbox{if}\  
\sigma=\sigma^{\prime}=- \\ &=&+1\ \ \mbox{otherwise}.
\end{eqnarray*}
\end{lem}
\proof
\ Since $\AlI$ is generated by $\ai$
  and $\aicr$, $i\in \I$, polynomials
 $p$ of $\ai$ and $\aicr$, $i\in \I$,
 are dense in $\AlI$. 
For any $\varepsilon>0$ and a given $a_{\sigma}$, 
$\sigma=+$ or $-$.
   there exists a polynomial $p$, i.e. a linear
   combination $p$ of monomials of $a_i$ and $a_i^{\ast}$,
   $i\in \I$, satisfying 
$\Vert a_{\sigma}-p\Vert <{\varepsilon}$.
   Since 
$E_{\sigma}\equiv (1/2)({\mbox{id}} + {\sigma}{\Theta})$ satisfies
   $E_{\sigma}a_{\sigma}=a_{\sigma}$ and 
$\Vert E_{\sigma} \Vert =1$, we
   have 
\begin{eqnarray*}
\Vert E_{\sigma}(a_{\sigma}-p)\Vert 
=\Vert a_{\sigma}-p_{\sigma}\Vert
   <\varepsilon
\end{eqnarray*}
 where 
$p_{\sigma}=E_{\sigma}p$. Since $E_{\sigma}$
   selects even or odd monomials (annihilating others)
   according as $\sigma$ is + or -, $p_{\sigma}$ is a linear
   combination of even or odd monomials of $a_i$ and 
$a_i^{\ast}$,
   $i\in \I$. Similarly there exits a linear combination
   $q_{\sigma^\prime}$ of even or odd monomials of 
$a_j$ and $a_j^{\ast}$,
   $j\in \J$, satisfying 
$
\Vert b_{{\sigma}^{\prime}}-q_{{\sigma}^{\prime} }
\Vert<\varepsilon.
$
Since the graded commutation relation (\ref{eq:epsCAR})
   holds for $p_{\sigma}$ and $q_{{\sigma}^{\prime}}$, it holds for
   $a_{\sigma}$ and $b_{{\sigma}^{\prime}}$.
\proofend
\begin{df}
\label{df:shiftk}
$(${\rm{1}}$)$  For each $k \in \Znu$, 
$\shiftk$  denotes a unique  automorphism 
 of $\Al$ satisfying 
\begin{eqnarray}
\label{eq:shiftk}
\shiftk(a_i^{\ast})=a_{i+k}^{\ast},\quad
 \shiftk(a_i)=a_{i+k}, \quad (i \in \Znu).
\end{eqnarray}
$(${\rm{2}}$)$  For a state $\vp$ of $\Al$, the adjoint action of $\shiftk$
 is defined by
\begin{eqnarray}
\label{eq:dualk}
\dshiftk\vp(A)=\vp\bigl( \shiftk(A) \bigr),\quad (A \in \Al).
\end{eqnarray}
\end{df}

\begin{rem}
 The automorphism $\shiftk$ 
 represents the lattice translation by the amount $k \in \Znu$.
 The map $k \in \Znu \mapsto \shiftk$
 is a group of automorphisms:\\ 
\begin{eqnarray*}
\shiftk \shiftl =\shiftkl,\quad  (k,l \in \Znu).
\end{eqnarray*}
The subalgebras transform covariantly under this group:
\begin{eqnarray}
 \shiftk \bigl(\AlI \bigr)= \Al(\I+k),
\end{eqnarray}
 where $\I+k=\{i+k;\ i\in \I\}$
 for any subset of $\I$ of $\Znu$ and any $k \in \Znu$.
\end{rem}
\begin{df}
\label{df:STATEzoku}
The sets of 
 all states  and all positive linear functionals of $\Al$ 
 are denoted by
  $\state$  and $\posfunctional$;
the sets of all  
 $\Theta$ invariant  
and all $\shift$ invariant ones  by
 $\evenstate$, $\evenfunctional$ 
 and $\invstate$,
 $\invfunctional$,\, respectively.
For any  subset $\I$ of $\Znu$, 
the set of all states of $\AlI$
is denoted by $\stateI$; the set of all 
$\Theta$ invariant ones by $\evenstateI$.
\end{df}
\ \\
{{\it Remark 1.}}
Any  translation  invariant state is 
automatically even
(see, e.g., 
  Example 5.2.21 of  \cite{BRA2}):
\begin{eqnarray}
\label{eq:TRAeven}
\invstate \subset \evenstate.
\end{eqnarray}
\ \\
{{\it Remark 2.}}
For each subset $\I$ of $\Znu$,
 we can consider
 the set of all states  $\stateforevenpartI$ on the 
 even subalgebra 
$\AlIp$.
There exists an obvious one-to-one correspondence 
between $\evenstateI$ and $\stateforevenpartI$ due to (\ref{eq:oddkill})
 by the restriction and  the unique $\Theta$ invariant extension.
\subsection{Product Property of the Tracial State}
\label{subsec:PROPRO}
The following proposition provides a basis for the present section.
%
\begin{pro}
\label{pro:PROPRO}
If  $\Ji$ and $\Jii$ are  disjoint, 
then
\begin{eqnarray}
\label{eq:PROPRO}
\tau(ab)=\tau(a) \tau(b)
\end{eqnarray}
for arbitrary  $a \in \AlJi$ and  $b \in \AlJii$, 
\end{pro}
\proof
\ It is enough to  prove the formula when $a$ and $b$
are monomials
 of the form 
(\ref{eq:order}).
Let $a=\Ai \aprim$, where $i \in \Ji$, 
$\aprim \in \AlJisetmini$
is a monomial of the form (\ref{eq:order}) and $\Ai$ is one of 
$\aicr$, $\ai$, $\aicr \ai$, $\ai \aicr$.
   We will now show
 \begin{eqnarray}
\tau(ab)=\tau(\Ai)\tau(\aprim b).
\label{eq:starthree} 
 \end{eqnarray}  
If $\aprim b$ is a $\Theta$-odd monomial, then $\tau(\aprim b)=0$ by 
(\ref{eq:oddkill}).
 If $\Ai$ is $\Theta$-even, then $ab$ is odd and  $\tau(ab)=0$, 
  implying (\ref{eq:starthree}).
If $\Ai$ is odd, then $\Ai(\aprim b)=-(\aprim b)\Ai$.
Hence 
\begin{eqnarray*}
 \tau(ab)&=&\tau \bigl( \Ai(\aprim b) \bigr)=-\tau \bigl( 
  (\aprim b) \Ai \bigr)=-\tau \bigl(\Ai (\aprim b) \bigr) \\
  &=&0,
 \end{eqnarray*}
 where the third  equality  
 is due to the tracial property of $\tau$.
 So (\ref{eq:starthree}) holds in either case.

 If $\aprim b$ is even and $\Ai$ is odd, then
 $\tau(\Ai)=0$ because $\Ai$ is odd and $\tau(ab)=0$ because
 $ab=\Ai(\aprim b)$ is odd. Again (\ref{eq:starthree}) holds.

 Finally, if $\aprim b$ is even and $\Ai=\aicr \ai$, then 
 $\aicr$ commutes with $\aprim b$ due to CAR and hence  
\begin{eqnarray*}
 \tau(ab)&=&\tau \bigl( (\aicr \ai)(\aprim b)  \bigr)
 =\tau \bigl(\ai (\aprim b) \aicr  \bigr) \\
 &=&\tau \bigl( \ai \aicr (\aprim b)  \bigr)
  \quad ({\mbox {due to}}\  [\aicr,\; \aprim b]=0)\\
  &=& \frac{1}{2} \tau \bigl( (\aicr \ai+\ai \aicr)(\aprim b)  \bigr)
    =\frac{1}{2} \tau(\aprim b).
 \end{eqnarray*}  
 The same formula for $\aprim b=1$ yields $\tau(\Ai)=\frac{1}{2}$
 and hence
\begin{eqnarray*}
 \tau(ab)=\tau(\Ai)\tau(\aprim b).
\end{eqnarray*}  
 If $\aprim b$ is even and $\Ai=\ai \aicr$, the above formula holds
 in the same way.
We have now proved (\ref{eq:starthree}) for all cases.

Let $a$ be now given by (\ref{eq:order}). 
By using (\ref{eq:starthree}) for $i_{1},\ i_{2}, \cdots,\ i_{k}$
 successively, we obtain 
 \begin{eqnarray*}
 \tau(ab)=\tau(\Aih) \cdots \tau(\Aik) \tau(b).
 \end{eqnarray*} 
 The same equality for $b=1$ yields
 \begin{eqnarray*}
 \tau(a)=\tau(\Aih) \cdots \tau(\Aik).
 \end{eqnarray*} 
 Hence we have
 \begin{eqnarray*}
 \tau(ab) =\tau(a) \tau(b).
 \end{eqnarray*} 
 This completes the proof. \proofend\\ 
We may say that the tracial state $\tau$ is a `product' state
although $\AlJi$ and $\AlJii$ do not commute.
We will show in the next subsections that
this product   property of the tracial state 
implies   the 
 commuting square property for the  conditional expectations.
\subsection{Conditional Expectations for Fermion Algebras}
\label{subsec:CECAR}
We prove the $\cstar$-algebraic version of Proposition$\,$\ref{pro:CE1}
 for the Fermion algebra $\Al$ and its subalgebras.
 We note that $\AlI$ is  not a von Neumann algebra unless 
 $\I$ is a finite subset of $\Znu$.
 Hence Proposition$\,$\ref{pro:CE1}
is not directly applicable to the Fermion algebra.
\begin{thm}  
\label{thm:EIEJ}
For any  subset  $\I$ of $\Znu$, 
 there exists a conditional expectation
\begin{eqnarray}
\label{eq:EIEJ0}
\EI: \ a \in \Al \mapsto \EI(a)\in \AlI
\end{eqnarray}
uniquely determined by $\EI(a) \in \AlI$ and 
\begin{eqnarray}
\label{eq:EIEJ1}
\tau (ab)=\tau \bigl(\EI(a)b \bigl) \quad (b \in \AlI).
\end{eqnarray}

For any second subset $\J$ of $\Znu$,
\begin{eqnarray}
\label{eq:EIEJ2}
\EI(a) \in \AlIaJ
\end{eqnarray}
 for any $a \in \AlJ$, and  
\begin{eqnarray}
\label{eq:EIEJ3}
\EI \EJ=\EJ \EI=\EIaJ.
\end{eqnarray}
\end{thm}
\proof
\ The $\cstar$-subalgebra of $\Al$
 generated by $\AlI$ and $\AlIcp$
 is isomorphic to their tensor product and will 
 be denoted as $\AlI\otimes\AlIcp$.
 Let 
 \begin{eqnarray}
\label{eq:EIfaraEQ}
\EIfara \equiv\frac{1}{2}\bigl( {\mbox{id}}+\ThetaIchat \bigr).
\end{eqnarray} 
It maps $\Al$ onto $\AlI\otimes\AlIcp$. Since 
 \begin{eqnarray*}
\tau\bigl(\ThetaIchat(a) b \bigr)=
\tau\bigl(\ThetaIchat(a b) \bigr)=
\tau(ab)
\end{eqnarray*} 
 for all $a\in \Al$ and $b \in \AlI\otimes\AlIcp$,
  $\EIfara$ satisfies (\ref{eq:EIEJ1}).

Since $\tau$ is a product state for the tensor product
 $\AlI\otimes\AlIcp$, there exists a conditional 
 expectation $\EIsara$ from $\AlI\otimes\AlIcp$
 onto $\AlI$ satisfying (\ref{eq:EIEJ1}), 
 characterized by 
$\EIsara(cd)=\tau(d)c$ for $c\in \AlI$
 and $d\in \AlIcp$ and called a slice map.
 Therefore 
 \begin{eqnarray}
\EI= \EIsara \EIfara
\end{eqnarray} 
 is a map from $\Al$ onto $\AlI$ satisfying (\ref{eq:EIEJ1}).
 By Lemma$\,$\ref{lem:CEunique}, it is a unique conditional 
expectation from $\Al$ onto $\AlI$ satisfying (\ref{eq:EIEJ1}).

 To show (\ref{eq:EIEJ2}), note that 
 $\AlJ$ is generated by $\AlJaI$ and $\AlJaIc$, namely,
 the linear span of products $ab$
 with $a \in \AlJaI$ and $b\in \AlJaIc$ is dense in 
 $\AlJ$. Due to the linearity of $\EI$ and 
 $\Vert\EI \Vert=1$, it is enough to show (\ref{eq:EIEJ2}) 
  for such products. 
We have $\EIfara(b)\in \AlIcp$ and hence
 \begin{eqnarray*}
\EI(ab)=\EIsara\bigl( a\EIfara(b)\bigr)
=a\tau\bigl(\EIfara(b)\bigr)=
a \tau(b)\in \AlJaI,
\end{eqnarray*} 
 which proves (\ref{eq:EIEJ2}).

For any  $a \in \Al$, $\EJ(a) \in \AlJ$
 and hence  $\EI\bigl( \EJ(a) \bigr)\in \AlIaJ$. 
For $b \in \AlIaJ$, (\ref{eq:EIEJ1}) implies 
\begin{eqnarray*}
\tau \bigl(\EI \bigr(\EJ(a)\bigr) b \bigr)=\tau\bl(\EJ(a)b\br)=\tau(ab),
\end{eqnarray*} 
 where the first equality is due to $b \in \AlI$, while
  the second  equality is due to $b \in \AlJ$.
This equality and $\EI (\EJ(a))\in \AlIaJ$
 imply 
\begin{eqnarray*}
\EIaJ(a)=\EI \bigl(\EJ(a)\bigr)
\end{eqnarray*} 
by the uniqueness result.
By interchanging $\I$ and $\J$,
we obtain 
\begin{eqnarray*}
\EI \EJ=\EJ \EI=\EIaJ,
\end{eqnarray*} 
 which proves the last  statement (\ref{eq:EIEJ3}).
\proofend 
\ \\
{\it{Remark 1.}} For spin lattice systems,
 the conditional expectation $\EI$ can be obtained simply as a 
 slice map with respect to the tracial state $\tau$.
 When spins and Fermions coexist at each lattice site, 
 $\EI$ can be obtained in exactly the same way as 
Theorem$\,$\ref{thm:EIEJ} (by including spin operators
 in the even part $\AlIp$), provided that the degree of freedom at each 
 lattice site is finite (i.e. $\AlI$ is a finite factor of type I
 for any finite $\I$). In all these cases, the results of our paper 
 are valid as they are proved by the use of conditional 
 expectations $\EI$.

\ \\
{\it{Remark 2.}} Theorem$\,$\ref{thm:EIEJ} can be shown 
 by a more elementary (lengthy) method by giving $\EI$
 explicitly for a finite $\I$
 and then giving $\EJ$ for an infinite $\J$ as a limit 
 of $\EIn$ for an increasing sequence of finite subsets
 $\In$ of $\Znu$ tending to $\J$.
 Proof presented above is by a suggestion of a referee.

 \begin{cor}
\label{pro:ETheta}
For each subset  $\I$  of $\Znu$, 
\begin{eqnarray}
\label{eq:ETheta}
\EI \Theta =\Theta \EI.
\end{eqnarray}
\end{cor}
\proof 
\ For any $a \in \Al$ and $b \in \AlI$, 
\begin{eqnarray*}
\tau\bigl( \EI\bigl(\Theta(a)\bigr)b\bigr)
&=&\tau\bigl( \Theta(a) b\bigr) \nonum \\
&=&\tau\bigl( \Theta \{ \Theta(a) b \}\bigr)=
\tau\bigl(a \Theta(b) \bigr) \\
&=&\tau\bigl( \EI(a) \Theta(b)   \bigr)=
\tau\bigl( \Theta\{ \EI(a) \Theta(b) \}  \bigr)\\
&=&\tau\bigl( \Theta\bigl(\EI(a)\bigr) b  \bigr).
\end{eqnarray*}
Since $\AlI$ is invariant under $\Theta$ as a set, 
 we have $\Theta \bigl(\EI(a)\bigr)= 
\EI \bigl(\Theta(a)\bigr)$ due to the uniqueness 
 of $\EI$ in  the preceding theorem.
\proofend

We now show a continuous dependence
 of $\EI$ on the subsets $\I$ of $\Znu$. 
We use the following  notation for various limits of subsets 
of $\Znu$.
If $\{\Ial\}$ is a monotone (not necessarily strictly) increasing or 
  decreasing 
net of subsets converging
 to a subset $\I$ of $\Znu$, we
 write 
$\Ial \nearrow \I$ or  $\Ial \searrow \I$.
For these cases, $\I=\cup_{\al}\Ial$ or  $\I=\cap_{\al}\Ial$, 
respectively. We use  $\Ial \to \I$
 for the standard convergence of a net  $\Ial$
 to  $\I$ (i.e., $\limsup_{\alpha}\Ial= \liminf_{\alpha}\Ial=\I$). 
By $\J \nearrow \Znu$ (which 
is written without any  index), 
 we mean a net of all finite subsets 
tending to $\Znu$ 
with the set inclusion as its partial ordering.
(In the same way, we use 
$\J \nearrow \I$.) 
In this case, $\J$ itself serves as the net index and  it is a
 monotone increasing net.
 Later in $\S$ \ref{sec:THERLIM} and 
$\S$ \ref{sec:ENTROPY}, we use a 
more restrictive notion of a van Hove net $\{\Ial\}$ tending to 
$\Znu$ or to `$\infty$' (see Appendix for detailed explanation).
 \begin{lem}
\label{lem:EIlim}
Let $\{\Ial\}$ be an increasing net of (finite or infinite)
 subsets of $\I$ such that their 
union is $\I$.
 For any $a \in \Al$, 
\begin{eqnarray}
\label{eq:EIlim-inc}
\lim_{\alpha} \EIal(a)=\EI(a).
\end{eqnarray}
As a special case $\I=\Znu$,  
\begin{eqnarray}
\label{eq:EZlim}
\lim_{\Ial \nearrow \Znu}\EIal(a)=a.
\end{eqnarray}
\end{lem}
\proof
  \ Since polynomials of $\ai$ and $\aicr$, $i\in \I$, 
 are dense in $\AlI$, there exists a finite subset 
 $\Jn$ of $\I$ and $\an\in \AlJn$ such that
\begin{eqnarray*}
\Vert \EI(a)-\an \Vert < \frac{1}{n}.
\end{eqnarray*}

Because $\Jn$ is a finite subset of $\I$
 and $\cup_{\al}\Ial=\I$, there exists a finite number of 
$\Ial$, say, 
 $\I_{\alpha(1)},\cdots \I_{\alpha(k)}$,
 such that 
 $\cup_{l=1}^{k}\I_{\alpha(l)} \supset \Jn$.
Since $\Ial$ is a net, there exists an index 
$\alpha_{n}> \alpha(1), \cdots,\alpha(k)$.
 Since $\Ial$ is increasing, 
$\I_{{\alpha_n}} \supset \I_{\alpha(1)}\cup \cdots 
  \I_{\alpha(k)} \supset \Jn$.

  For any $\alpha \ge \alpha_{n}$,
 $\Ial \supset \Jn$ and so $\EIal(\an)=\an$.
Hence by $\I \supset \Ial$, we have 
\begin{eqnarray*}  
\Vert \EIal(a)-\an \Vert 
=\Vert \EIal \bigl(\EI(a)-\an \bigr)\Vert
\le  \Vert \EI(a)-\an \Vert
< \frac{1}{n}
\end{eqnarray*}
due to $\Vert \EIal\Vert \le 1$. Thus 
\begin{eqnarray*}  
\Vert \EIal(a)-\EI(a) \Vert \le 
\Vert \EIal(a)-\an \Vert
+ \Vert \EI(a)-\an \Vert
< \frac{2}{n},
\end{eqnarray*}
for all $\alpha \ge \alpha_{n}$, 
which proves the assertion (\ref{eq:EIlim-inc}).
\proofend
\begin{lem}
\label{lem:EIlimd}
Let $\{\Ial\}$ be a decreasing net of (finite or infinite)
 subsets of $\Znu$ such that their 
intersection  is $\I$.
 For any $a \in \Al$, 
\begin{eqnarray}
\label{eq:EIlimd}
\lim_{\alpha}\EIal(a)=\EI(a).
\end{eqnarray}
\end{lem}
\proof\ 
Let $\Lk$ be a monotone increasing sequence of finite subsets of $\Znu$
 such that their union is $\Znu$. 
For any $\varepsilon>0$, there exists
 $\kep$ such that 
\begin{eqnarray*}
\Vert a -\ELk(a)\Vert< \varepsilon
\end{eqnarray*}
for all $k \ge \kep$ by Lemma$\;$\ref{lem:EIlim}.
Hence 
\begin{align}
\label{eq:EIlimd2}
\Vert\EI(a)-E_{\I\cap\Lk}(a)\Vert&=
\Vert \EI \bigl(a-E_{\Lk}(a) \bigr)  \Vert<\varepsilon, \\
\label{eq:EIlimd2-2}
\Vert\EIal(a)-E_{\Ial\cap\Lk}(a)\Vert&=
\Vert \EIal\bigl(a-E_{\Lk}(a) \bigr)  \Vert<\varepsilon
\end{align}
for all $k \ge \kep$ and all $\alpha$ due to 
$\Vert\EI \Vert\le 1$ and 
$\Vert\EIal\Vert\le 1$. 

Since $\IallimdI$, we have 
$(\Ial\cap\Lk )\searrow (\I\cap \Lk)$.
Since $\Lkep$ is 
a finite set, there exists $\alep$ 
 such that 
 $\Ial\cap\Lkep=\I\cap \Lkep$
 and hence $E_{\Ial \cap \Lkep}=E_{\I\cap\Lkep}$ 
 for all $\al\ge \alep$.
Therefore, 
we obtain
\begin{eqnarray*}
\Vert \EIal(a)-\EI(a) \Vert 
&\le& \Vert \EIal(a)-E_{\Ial\cap\Lkep}(a) \Vert +
\Vert E_{\Ial\cap\Lkep }(a)-\EI(a)  \Vert \nonum \\
&=& \Vert \EIal(a)-E_{\Ial\cap\Lkep}(a) \Vert +
\Vert E_{\I \cap\Lkep }(a)-\EI(a)  \Vert 
< 2\varepsilon
\end{eqnarray*}
 for all $\al \ge \alep$, where the first term is estimated by 
(\ref{eq:EIlimd2-2}), and 
the second by 
(\ref{eq:EIlimd2}).
Hence we obtain
\begin{eqnarray*}
\lim_{\al}\EIal(a)=\EI(a).
\end{eqnarray*}
\proofend 
%
%
\begin{thm}
\label{thm:EIlim}
If a net $\{\Ial\}$ converges to $\I$, then
\begin{eqnarray}
\label{eq:EIlim}
\lim_{\alpha}\EIal(a)=\EI(a).
\end{eqnarray}
for all $a \in \Al$,.
\end{thm}
\proof\ By definition, $\Ia\to\I$ means
\begin{eqnarray*}
\I=\cap_{\beta}\big(\cup_{\alpha\ge \beta}\Ial\bigr)=
\cup_{\beta}\big(\cap_{\alpha\ge \beta}\Ial\bigr).
\end{eqnarray*}
Set
\begin{eqnarray*}
\Jblhat\equiv\cup_{\alpha\ge \beta}\Ial,\quad
\Jbl\equiv\cap_{\alpha\ge \beta}\Ial.
\end{eqnarray*}
Then $\JbllimdI$ and $\JbllimI$.
By Lemmas$\,$\ref{lem:EIlimd} and \ref{lem:EIlim},
there exists a $\blep$ for any given $\varepsilon>0$
such that for all $\beta\ge\blep$
\begin{eqnarray*}
\Vert\EJblhat(a)-\EI(a)\Vert<\varepsilon,\quad
\Vert\EJbl(a)-\EI(a)\Vert<\varepsilon.
\end{eqnarray*}
Hence
 \begin{eqnarray*}
\Vert\EJblhat(a)-\EJbl(a)\Vert<2\varepsilon.
\end{eqnarray*}
Since $\Jblhat\supset\Ibl\supset \Jbl$, we have
$\EIbl\EJblhat=\EIbl$, $\EIbl\EJbl=\EJbl$ and 
 \begin{eqnarray*}
\Vert\EIbl(a)-\EJbl(a)\Vert
=\Vert\EIbl \bigl(\EJblhat(a)-\EJbl(a)\bigr)\Vert
<2\varepsilon.
\end{eqnarray*}
Therefore
\begin{eqnarray*}
\Vert\EIbl(a)-\EI(a)\Vert<3\varepsilon
\end{eqnarray*}
for all $\beta\ge\blep$.
This proves (\ref{eq:EIlim}).\proofend

The following corollary follows immediately 
from the  results obtained in this subsection.
\begin{cor}
\label{cor:Iintsec}
For any countable family $\{\In\}$ of subsets of  $\Znu$,
\begin{eqnarray}
\label{eq:Iintsec}
\cap_{n=1}^{\infty}\AlIn=\Al\left(\cap_{n=1}^{\infty}\In\right). 
\end{eqnarray}
\end{cor}
 \proof\ 
Let $\Jn\equiv{\displaystyle{\cap_{k=1}^{n}}} \Ik$ and 
$\I\equiv \displaystyle{\cap_{n=1}^{\infty}}\In$.
Then $\Jnlimd \I$.
By (\ref{eq:EIEJ3}), $\EJnm\EIn=\EJn$ and hence
 $\EJn=\prod_{k=1}^{n}\EIk$.
 On one hand, $\Jn \subset \Ik$ for $k=1,\ldots,n$, and hence 
$\AlJn \subset \cap_{k=1}^{n}\AlIk$.
On the other hand, $a \in \cap_{k=1}^{n}\AlIk$ satisfies 
$\EIk(a)=a$ for all $k=1,\ldots,n$
 and hence $\EJn(a)=a \in \AlJn$. Therefore
\begin{eqnarray*}
\AlJn=\cap_{k=1}^{n}\AlIk.
\end{eqnarray*}
Since $\Jn\supset \I$, we have $\AlJn \supset \AlI$ and hence
\begin{eqnarray*}
\cap_{n=1}^{\infty} \AlIn =\cap_{n=1}^{\infty} \AlJn \supset \AlI.
\end{eqnarray*}
For $a \in \cap_{n=1}^{\infty} \AlJn$, $\EJn(a)=a$ for any $n$.
Since $\lim_{n}\EJn(a)=\EI(a)$ by Lemma$\;$\ref{lem:EIlimd},
 we have $a=\EI(a) \in \AlI$. Now we obtain the desired conclusion
\begin{eqnarray*}
\cap_{n=1}^{\infty}\AlIn=\AlI.
\end{eqnarray*}
\proofend
\subsection{Commuting Squares for Fermion Algebras}
\label{subsec:CARsquare}
In the following theorem, 
we show that  
 any  two subsets  $\I$ and $\J$ of $\Znu$
are associated with  a  commuting square of the conditional expectations 
with  respect to the tracial state $\tau$.
For $\K \subset \LL \subset \Znu$,
denote the restriction of $\EK$ to $\AlL$ by $\ELK$.
Then it is a conditional expectation
 from $\AlL$ to $\AlK$ 
 with respect to the tracial state.
\begin{thm}
\label{thm:CARsquare}
For any subsets $\I$ and $\J$ of $\Znu$, the following 
 subalgebras of $\Al$ form a  
 commuting square$:$ \\
\setlength{\unitlength}{0.7mm}
\begin{picture}(100,48)(-30,0)
\put(0,16){\makebox(20,17)[r]{$\AlIuJ$}}
\put(45,1){\makebox(10,10)[c]{ $\AlJ$ }}
\put(45,37){\makebox(10,10)[c]{$\AlI$ }}
\put(80,16){\makebox(20,17)[l]{$\AlIaJ,$}}
\put(23,24){\vector(1,0){54}}
\put(23,22){\vector(3,-2){22}}
\put(23,26){\vector(3,2){21}}
\put(55,7){\vector(3,2){22}}
\put(55,40){\vector(3,-2){22}}.
\end{picture} \\
 Here the arrow from $\AlL$ to $\AlK$
 represents the conditional expectation $\ELK$.
\end{thm}
\proof
\ It follows from (\ref{eq:EIEJ3}) that 
\begin{eqnarray*}
\EIIaJ \EIuJI 
=\EIuJIaJ =\EJIaJ \EIuJJ,
\end{eqnarray*}
  which shows the assertion. \proofend 
\subsection{Commutants of Subalgebras}
\label{subsec:COMMUAlI}
We are going to determine the commutants of subalgebras of $\Al$.
\begin{lem}
\label{lem:IfpCOMM}
For a finite $\I$,
\begin{eqnarray}
\label{eq:IfpCOMM}
\AlIpcommut\cap\Al =\AlIc + \vI \,\AlIc,
\end{eqnarray}
 where $\vI$ is a self-adjoint unitary in $\AlIp$
 given by
\begin{eqnarray}
\label{eq:vIEQ}
 \vI \equiv \prod_{i \in \I}\vi,\quad \vi \equiv \aicr\ai-\ai\aicr. 
\end{eqnarray}
 and implementing $\ThetaIhat$ on $\Al$.
\end{lem}
\proof \ By CAR, 
\begin{eqnarray*}
\aicr \vi=-\aicr,\  \ai \vi=\ai,\ 
\vi \aicr=\aicr,\  \vi \ai=-\ai.
\end{eqnarray*}
Thus  $\vi$ anticommutes with $\ai$ and $\aicr$.
 If $j \ne i$, $\vi$ commutes with $\aj$ and $\ajcr$
 due to $\vi \in \Alisite_{+}$.
Therefore  for any $a \in \AlI$,
 we have    
\begin{eqnarray}
\label{eq:advIin}
   ({\mbox {Ad}} \vI) a \equiv \vI a \vI^{\ast}=\Theta(a),
\end{eqnarray}
 or equivalently,
\begin{eqnarray}
\label{eq:advIin1}
 \vI a =\Theta(a)\vI.
\end{eqnarray}
 For any $a \in \AlIc$,
\begin{eqnarray}
\label{eq:advIc}
\vI a =a \vI.
\end{eqnarray}
Due to $\vI^{\ast}=\vI=\vI^{2}$, 
 $\vI$ is a self-adjoint unitary implementing $\ThetaIhat$
 on $\Al$.

Since $\vI\in \AlIp$ implements $\ThetaIhat$,
 $\AlIpcommut$ is contained in the fixed point subalgebra 
 $\Al^{\ThetaIhat}$.
In terms of  
$\EIcfara=\frac{1}{2}\bigl( {\mbox{id}}+\ThetaIhat \bigr)$,
we have 
\begin{eqnarray*}
\AlIpcommut\subset \Al^{\ThetaIhat}=\EIcfara(\Al)=
\AlIp \otimes\AlIc. 
\end{eqnarray*}
Since $\AlIc$ is in $\AlIpcommut$, we have 
\begin{eqnarray}
\label{eq:CENTER1}
\AlIpcommut= {\cal{Z}} \bigl(\AlIp\bigr) \otimes\AlIc 
\end{eqnarray}
where ${\cal{Z}} \bigl(\AlIp\bigr)$ is the center of $\AlIp$.
Since $\AlIp=\{\vI \}^{\prime}\cap \AlI$, $\vI$
 is a self-adjoint unitary in $\AlI$
 and $\AlI$ is a full matrix algebra for a finite $\I$,
 we have  
\begin{eqnarray}
\label{eq:CENTER2}
{\cal{Z}} \bigl(\AlIp\bigr) = \Com \identitybf +\Com \vI. 
\end{eqnarray}
By (\ref{eq:CENTER1}) and (\ref{eq:CENTER2}),
 we obtain (\ref{eq:IfpCOMM}).
\proofend
\begin{lem}
\label{lem:IfCOMM}
For a finite $\I$,
\begin{eqnarray}
\label{eq:IfCOMM}
\AlIcommut\cap\Al =\AlIcp + \vI \,\AlIcm,
\end{eqnarray}
\end{lem}
\proof\ By  Lemma$\,$\ref{lem:IfpCOMM} and 
$\AlIcommut\subset \AlIpcommut$,
 any element $a\in \AlIcommut$ is of the form
 \begin{eqnarray*}
a=\ah+\vI\af, \quad \ah,\af \in \AlIc. 
\end{eqnarray*}
Take any unitary $u \in \AlIm$
 (e.g., $u=\ai+\aicr$, $i\in \I$).
Then we have 
\begin{eqnarray*}
a&=&\frac{1}{2}(a+ua u^{\ast})
=\frac{1}{2}(\ah+u\ah u^{\ast})
+\frac{1}{2}\vI(\af-u\af u^{\ast})\\
&=&(\ah)_{+}+\vI(\af)_{-}
\end{eqnarray*}
 due to $u\vI=-\vI u$, where 
\begin{eqnarray*}
(\ah)_{+}=
\frac{1}{2}\bigl(\ah+\Theta(\ah) \bigr) \in \AlIcp,\quad 
(\af)_{-}=\frac{1}{2} \bigl(\af-\Theta(\af) \bigr)\in \AlIcm.
\end{eqnarray*}
Hence 
\begin{eqnarray*}
\AlIcommut \subset \AlIcp + \vI \,\AlIcm.
\end{eqnarray*}
The inverse inclusion follows from (\ref{eq:advIin1})
 and Lemma$\,$\ref{lem:IJrel}.
Hence  (\ref{eq:IfCOMM}) holds.
\proofend
\begin{lem}
\label{lem:IinfCOMM}
For an infinite $\I$, 
\begin{eqnarray}
\label{eq:IinfCOMM}
\AlIprime\cap \Al=\AlIcp.
\end{eqnarray}
\end{lem}
\proof \ 
It is clear that elements of $\AlIcp$ and $\AlI$ commute.
Hence it is enough to prove $\AlIprime\cap \Al \subset \AlIcp$.

Let $a \in \AlIprime\cap \Al$. Then 
\begin{eqnarray*}
a_{\pm}=\frac{1}{2}\bigl(a\pm\Theta(a)\bigr) \in \AlIprime\cap \Al
\end{eqnarray*}
 because $\Theta\bigl(\AlI\bigr)=\AlI$.
 For any finite subset $\K$ of $\I$,
 $a_{\pm}\in \AlKprimepm$. 
Hence by Lemma$\;$\ref{lem:IfCOMM},  
\begin{eqnarray*}
a_{+} \in \AlKcp.
\end{eqnarray*}
Consider an 
increasing sequence of finite subsets  $\Knlim \I$.
We apply Corollary$\,$\ref{cor:Iintsec}
 to $(\Kn)^{c}\searrow \Ic$,
and  obtain 
\begin{eqnarray}
\label{eq:IinfCOMM3}
a_{+} \in \cap_{n=1}^{\infty} \AlKncp=\AlIcp.
\end{eqnarray}
We now prove $a_{-}=0$, which yields the desired conclusion 
due to $a=a_{+}+a_{-}$ and (\ref{eq:IinfCOMM3}). 
For a monotone increasing sequence of finite subsets $\Ln$ of $\Znu$
 such that $\LnlimZ$, we have $\lim_{n}\ELn(a_{-})=a_{-}$
 and hence there exists $\nep$ 
  for any given $\varepsilon>0$ such that
\begin{eqnarray}
\label{eq:IinfCOMM4}
\Vert \ELn(a_{-})-a_{-}\Vert < \varepsilon
\end{eqnarray}
 for $n \ge\nep$.
 For any $k$, we set
$\Kk \equiv \I \cap \Lk (\subset \I)$. 
Then  $a_{-} \in \AlKkprime$
 and by  Lemma$\,$\ref{lem:IfCOMM} we have
\begin{eqnarray*}
a_{-}=\vKk \bkk
\end{eqnarray*}
 for some $\bkk\in \AlKkcm$.
 For any $i \in \Kk$,
\begin{eqnarray}
\label{eq:IinfCOMM5}
\Eic(a_{-})=\tau(\vi)v_{(\Kk\setminus\{i\})}\bkk=0.
\end{eqnarray}
Now take an  $\no \ge \nep$.
Since $\Kklim \I$ and $\I$ is an infinite set
 while any $\Lno$ is a finite set, there exists
 a number $k$ such that $\Kk$ contains a point $i$ of $\Znu$
 such that $i \notin \Lno$.
Then $\Lno \subset \{i\}^{c}$. It follows from  
(\ref{eq:IinfCOMM5}) that  
\begin{eqnarray*}
E_{\Lno}(a_{-})=E_{\Lno} \Eic(a_{-})=0.
\end{eqnarray*}
This and (\ref{eq:IinfCOMM4}) imply
\begin{eqnarray*}
\Vert a_{-}\Vert < \varepsilon.
\end{eqnarray*}
Since $\varepsilon$ is arbitrary, we obtain $a_{-}=0$.
\proofend
\ \\
Combining Lemma$\,$\ref{lem:IfCOMM} and
 Lemma$\,$\ref{lem:IinfCOMM}, we obtain
\begin{thm}
\label{thm:SOUGOUI}
{\rm{(1)}}  For a finite $\I$, 
\begin{eqnarray*}
\AlIcommut\cap\Al=\AlIcp + \vI \,\AlIcm,
\end{eqnarray*}
where $\vI$ is given by (\ref{eq:vIEQ}).\\
{\rm{(2)}}  For an infinite $\I$, 
\begin{eqnarray*}
\AlIprime \cap \Al=\AlIcp.
\end{eqnarray*}
\end{thm}
As a preparation for the remaining case
 (the commutant of $\AlIp$ for infinite $\I$), 
we present the following technical 
 Lemma for the sake of completeness. We define 
\begin{eqnarray}
\uioo\equiv \aicr \ai, 
\ \uiot \equiv \aicr, \ \uito\equiv  \ai, \ \uitt\equiv \ai \aicr.
\end{eqnarray}
\begin{lem}
\label{lem:preEICOMM}
 Let $\I=(i_{1},\cdots,i_{|\I|})$ be a finite subset of $\Znu$.
Put
\begin{eqnarray}
\upjaa \equiv \ujaa  \ \,{\text{for}}\,\  \alpha=1,2, \quad
 \upjab \equiv \ujab v_{\{i_{1},\cdots,i_{j-1}\}} 
 \ \,{\text{for}}\,\ \alpha\ne\beta.
\end{eqnarray}
Define 
\begin{eqnarray}
\ukl\equiv \prod_{j=1}^{|\I|} \upjkl,
\end{eqnarray}
   where 
 $\kn$ and $\lnn$ are either 1 or 2, respectively,
$k=(k_{1},\cdots,k_{|\I|})$ and 
 $l=(l_{1},\cdots,l_{|\I|})$. Then the following holds.
\\
\ $(${\rm{1}}$)$ The set of all $\ukl$ form a
 self-adjoint system of  matrix units of $\AlI$.\\
\ $(${\rm{2}}$)$  Let $\sigkl$ be the number of $n$ such that $\kn \ne \lnn$.
Then 
\begin{eqnarray}
\label{eq:ukl2}
\Theta(\ukl)=(-1)^{\sigkl}\ukl.
\end{eqnarray}
 \ $(${\rm{3}}$)$  
Any $a \in \Al$ has a unique expansion
\begin{eqnarray}
\label{eq:aklexp}
a=\sum_{k,l}\ukl \akl
\end{eqnarray}
with $\akl \in \AlIc$ and   $\akl$ is uniquely given by 
\begin{eqnarray}
\label{eq:aklexpunique}
\akl=2^{|\I|} \EIc(\ulk a).
\end{eqnarray}
 \end{lem}
\proof \
(1) By using (\ref{eq:CAR1}) for the case of $i=j$,
$\bigl\{\uiab\bigr\}_{\alpha\beta} (\alpha,\beta=1,2)$
 satisfies the relations
\begin{eqnarray}
\label{eq:preEICOMM1}
\bigl( \uiab \bigr)^{\ast}=\uiba,\quad  
\uiab u^{(i)}_{\alpr\bpr}=\delta_{\beta \alpr}
u^{(i)}_{\alpha \bpr},
\quad  \sum_{\alpha}\uiaa=1,
\end{eqnarray}
for a self-adjoint system of matrix units.
 Since $v_{\{i_{1},\cdots,i_{j-1}\}}$ is a self-adjoint unitary 
commuting with $a_{\ij}$ and $a^{\ast}_{\ij}$, the same 
computation shows that  
$\bigl\{\upjab\bigr\}_{\alpha\beta}$ $(\alpha,\beta=1,2)$
 satisfies the same relations. 

Since $v_{\{i_{1},\cdots,i_{j-1}\}}$  anticommutes with 
$\aik$ and $\aikcr$ for $k<j$ and commutes with 
them for $k \ge j$, 
 $\bigl\{\upjab\bigr\}_{\alpha\beta}$ commutes with each other 
 for different $j$.

Since they generate all $\Al(\{\ik \})$ recursively for $k=1,\cdots,n$,
 they form a self-adjoint 
 system of  matrix units of $\AlI$.

(2) $\Theta(\uiaa)=\uiaa$, $\Theta(\uiab)=-\uiab$ for $\alpha\ne \beta$, 
 and $\Theta(v_{\{i_{1},\cdots,i_{j-1} \}} )=
v_{\{i_{1},\cdots,i_{j-1}\}} $ imply (\ref{eq:ukl2}).

(3) For a full matrix algebra $\AlI$ contained in 
 a $\cstar$-algebra
 $\Al$, the following expansion of any $a \in \Al$
 in term of a self-adjoint system of a matrix units
  $\{ \ukl \}$ of $\AlI$ is well-known.
\begin{eqnarray}
a&=&\sum_{k,l}\ukl \bkl, \nonum\\
\bkl&=&\sum_{m}\umk a  \ulm \in \AlIcommut.
\end{eqnarray}

By Lemma$\,$ \ref{lem:IfCOMM}, there are $\bklh$
 and $\bklf$ in $\AlIc$
 satisfying 
\begin{eqnarray}
\bkl=\bklh+\vI \bklf.
\end{eqnarray}
By direct computation, 
 $\ukl\vI=\pm \ukl$ where the sign depends on 
$k$ and $l$.
 Thus we have the expansion (\ref{eq:aklexp})
 with $\akl=\bklh \pm \bklf\in \AlIc$.

The coefficient $\akl \in \AlIc$
is uniquely determined 
by the following computation and given by  (\ref{eq:aklexpunique}). 
\begin{eqnarray*}
\EIc(\ulk a)&=&\EIc (\sum_{l^{\prime}} u_{ll^{\prime}} a_{kl^{\prime}} )\\
&=&\sum_{l^{\prime}}\EIc(u_{ll^{\prime}}) a_{kl^{\prime}}=
 \sum_{l^{\prime}}\tau(u_{ll^{\prime}} ) a_{kl^{\prime}}=2^{-|\I|}\akl.
\end{eqnarray*}
Here we have used the following relation:
\begin{eqnarray*}
\tau(\ukl)&=&\tau(\ukm \uml)=\tau(\uml \ukm)\nonum \\
&=&\delta_{kl}\tau(\umm)=\delta_{kl}2^{-|\I|}\tau(\sum_{m}\umm)=2^{-|\I|}
 \delta_{kl}.
\end{eqnarray*}
\proofend 
\begin{thm}
\label{thm:SOUGOUIp}
{\rm{(1)}} For a finite $\I$, 
\begin{eqnarray}
\AlIpcommut\cap\Al=\AlIc+\vI \AlIc,
\end{eqnarray}
where $\vI$ is given by (\ref{eq:vIEQ}).\\
{\rm{(2)}} For an infinite $\I$, 
\begin{eqnarray}
\AlIpcommut\cap\Al=\AlIc.
\end{eqnarray}
\end{thm}
\proof
\ (1) is given by Lemma$\,$\ref{lem:IfpCOMM}.

To prove (2), we consider an infinite $\I$.
 Clearly $\AlIpcommut\cap\Al \supset \AlIc$ due to (\ref{eq:epsCAR}).
 Hence it is enough to prove that
 any $b \in \AlIpcommut\cap\Al$ belongs to $\AlIc$

Let $\{\Ln\}$ be an increasing sequence of finite subsets of $\Znu$
 such that their union is $\Znu$. 
Set $\In \equiv \Ln\cap \I$. Then $\InlimI$.

For any $\varepsilon>0$, there exist a positive integer 
 $\leps$ and an element $\beps$ of $\AlLleps$
 satisfying 
\begin{eqnarray*}
\Vert b-\beps \Vert<\varepsilon.
\end{eqnarray*}
For any $n$, $b \in \AlInpcommut$ due to $\In \subset \I$
 and $b \in \AlIpcommut$.
The conclusion of (1) implies 
\begin{eqnarray}
\label{eq:EIinfCOMM2}
b=\bnf+\vIn \bns,
\end{eqnarray}
 where $\bnf,\;\bns \in \AlInc$.

Since $\InlimI$ and $\I$ is infinite, there exists an $\neps$
 such that $\Ineps$ contains a point $i$
 which does not belong to $\Lleps$.
 Then $i \in \In$ for all $n \ge \neps$.
Due to $\beps \in \AlLleps$ and $\{i\}^{c}\supset \Lleps$, 
\begin{eqnarray}
\label{eq:EIinfCOMM3}
\Eic(\beps)=\beps.
\end{eqnarray}
Since $\bnf,\;\bns \in \AlInc \subset \Alic$
 for all $n \ge \neps$, we have
\begin{eqnarray}
\Eic(\bnf)&=&\bnf \nonum \\ 
\Eic(\vIn \bns)&=&\tau(\vi)v_{{\In\setminus \{i \}}} \bns=0.
\end{eqnarray}
This implies
\begin{eqnarray}
\label{eq:EIinfCOMM5}
\Eic(b)=
\Eic(\bnf)+\Eic(\vIn \bns)=\bnf.
\end{eqnarray}
It follows from (\ref{eq:EIinfCOMM3}) and (\ref{eq:EIinfCOMM5})
\begin{eqnarray*}
\Vert \beps-\bnf \Vert=\Vert \Eic(\beps)-\Eic(b) \Vert
\le \Vert\beps -b \Vert<\varepsilon.
\end{eqnarray*}
Therefore,
\begin{eqnarray}
\label{eq:EIinfCOMM10}
\Vert b-\bnf \Vert \le \Vert b-\beps \Vert+
 \Vert \beps-\bnf \Vert <2 \varepsilon
\end{eqnarray}
 for all $n \ge \neps$.
 Hence 
\begin{eqnarray*}
b=\lim_{n}\bnf.
\end{eqnarray*}
For any fixed $m \in \NN$,  $\bnf \in \AlInc \subset \AlImc$
 for all $n \ge m$ due to $\In \supset \I_m$.
Thus $b \in \AlImc$ for any $m$.
By Corollary $\,$\ref{cor:Iintsec},
\begin{eqnarray*}
b \in \cap_{m}\AlImc=
 \Al\bigl(\cap_{m} (\Imc) \bigr)=
\Al\bigl( \bigl\{\cup_{m}\I_m \bigr\}^{c} \bigr)=\AlIc.
\end{eqnarray*}
\proofend 
As a by-product, we obtain the following.
\begin{cor}
\label{cor:ThetaOut}
For any infinite  $\I$, 
the restriction of $\Theta$ to $\AlI$ is outer.
\end{cor}
\proof\ We denote the restriction of $\Theta$
 by the same letter.
For any infinite subsets $\I$ and $\J$, 
$(\AlI,\,\Theta)$
 is isomorphic to $(\AlJ,\,\Theta)$ as a pair of $\cstar$-algebra
  and its  automorphism
   through any bijective  map between $\I$ and $\J$.
   Therefore it is enough to show the assertion
 for  a proper  infinite subset $\I$ 
 of $\Znu$.
    
Suppose that $u$ is a unitary element in $\AlI$
     such that 
 \begin{eqnarray*}
u^{\ast}au=\Theta(a),
\end{eqnarray*}
for all $a \in \AlI$. 
Substituting $u$ into $a$, we have $\Theta(u)=u$.
Let $b \in \AlIcm$ and $b \ne 0$.
Then $ub \in \Alminus$.
By (\ref{eq:epsCAR})
\begin{eqnarray*}
ba=\Theta(a)b.
\end{eqnarray*}
Hence $ub \in \AlIcommut$.
Therefore $ub \in \AlIcommutm$, which implies 
$ub=0$ by Lemma$\,$\ref{lem:IinfCOMM}.
This implies
\begin{eqnarray*}
b=u^{\ast}(ub)=0,
\end{eqnarray*}
 a contradiction.
\proofend
%
%
\section{Dynamics}
\label{sec:DYNAMICS}
\subsection{Assumptions}
\label{subsec:ASS}%
 We  consider a one-parameter group of $*$-automorphisms
 $\alt$ of the Fermion algebra $\Al$.
 Throughout this work, 
  $\alt$ is  assumed to be strongly continuous, that is, 
$t\in \R  \mapsto  \alt(A) \in \Al$  is norm continuous for each $A  \in \Al$.
 In order to associate a potential to the dynamics $\alt$
(see $\S$$\;$\ref{subsec:POT} for details), 
 we need the following two assumptions 
on $\alt$ and its  generator $\delal$ 
  with  the  domain $\Ddelal$
: \\
\ \\
\quad (I) $\alt\, \Theta=\Theta \,\alt  \quad 
{\mbox{for all}}\ t \in \R$. \\
\ \\
\quad (II) $\Alinfty$ is in the domain of $\delal$,
 namely, $\Alinfty \subset \Ddelal$. \\

The assumption (I) of $\Theta$-even dynamics comes from 
 two sources.
On the physical side, the generator 
 of the time translation $\alt$ should be 
 $i=\sqrt{-1}$ times the commutator  with
 the energy operator which is a physical observable and hence 
$\Theta$-even.

On the technical side, the potential to be introduced below
 has to commute with a fixed local element of $\Al$
 when the support region of the potential is far away
 in order that the expression for the action of the generator on that   
 local element converges and makes sense.

 For $\alt$ to be uniquely specified by the associated potential 
to be introduced in $\S$$\;$\ref{subsec:POT},
 we need the following assumption:\\
\ \\
\quad (III) $\Alinfty$ is  the core of  $\delal$, namely, if 
 $\del$ denotes the restriction of $\delal$ to $\Alinfty$,
  its closure $\overline{\del}$ is $\delal$. \\

The assumption (III) will  be used 
 to  derive  a conclusion involving
  $\alt$ such as the KMS  condition
 from other conditions involving the associated potential such as the Gibbs condition and the variational principle.
 
Later, when we discuss translation invariant equilibrium states,
 we will add the assumption  of translation invariance:\\
\ \\
\quad (IV) $\alt \, \shiftk=\shiftk \,\alt 
\quad {\mbox{for any}}\ t \in \R,\; k \in \Znu$.\\

Later in Proposition \ref{pro:shift-theta-alt}, it will be shown that
         Assumption (IV) implies Assumption (I).

By Assumptions (I) and (II), the restriction 
$\del$ of $\delal$ to $\Alinfty$ satisfies 
\begin{eqnarray}
\label{eq:Thetadel}
\delta \Theta(A)=\Theta(\delta A)
\end{eqnarray}
 for any $A \in \Alinfty$.
In the rest of this section, we deal with an arbitrary 
$\ast$-derivation $\del$ with the domain $\Alinfty$ commuting with $\Theta$ 
 (eq.(\ref{eq:Thetadel})) irrespective of whether 
 it comes from a dynamics $\alt$ or not.
 Of course, we can use the results about such a general
  $\del$ for the restriction of $\delal$
   to $\Alinfty$.

\subsection{Local Hamiltonians}
\label{subsec:LH}
 Since $\AlI$ is a finite type {\bf{I}}  
   factor for each finite subset $\I$ of $\Znu\!\!$, 
 there exists a self-adjoint element $\HoI \in \Al$
 satisfying 
\begin{eqnarray}
\label{eq:SAKAI1}%
\delta A=i[\HoI,\,A ]
\end{eqnarray}
 for any $A \in \AlI$ where $\del$ is any $\ast$-derivation
  with its domain $\Alinfty$ and values in $\Al$
   (i.e., $\del$ is a linear map from $\Alinfty$ into $\Al$
    satisfying $\del(AB)=(\del A)B+A(\del B)$
     and $\del(A^{\ast})=(\del A)^{\ast}$).
   Although this is well-known (see, e.g., \cite{SAKAI76}),
 we include its  proof for the sake of completeness.
\begin{lem}
\label{lem:SAKAI}
Let $\{\uij \}$ be a self-adjoint 
system of matrix units of $\AlI$.
Define 
\begin{eqnarray*}
\hij \equiv  \sum_{l}\uli \delta \ujl 
 -\deltaij \nrmI \sum_{l} \sum_{m}
  \ulm \delta \uml.
\end{eqnarray*}
Then $\hij \in \AlIcommut$. Define 
\begin{eqnarray*}
iH \equiv \sum_{i,j}\uij \hij.
\end{eqnarray*}
 It satisfies $H^{\ast}=H$ and  
\begin{eqnarray*}
[iH,\,A]=\delta A 
\end{eqnarray*}
 for $A \in \AlI$.
 Furthermore, 
\begin{eqnarray}
\label{eq:SAKAIlem1}
\EIc(H)=0.
\end{eqnarray}
\end{lem}
\proof
\ 
(1) We first prove  $\hij \in \AlIcommut$.
 If $i \ne j$, 
\begin{eqnarray*}
[\hij,\,\uab]&=&\sum_{l} \uli (\delta \ujl) \uab
 -\uai \delta \ujb  \nonumber \\
&=&\sum_{l} \uli  \bigl( \delta(\ujl \uab) -\ujl \delta \uab \bigr)
 -\uai \delta \ujb \nonumber \\
&=& \uai \delta \ujb
 -\uai \delta \ujb=0.
\end{eqnarray*}
If $i=j$,
\begin{eqnarray*}
[\hii,\,\uab]&=&\sum_{l} \uli (\delta \uil) \uab
 -\uai \delta \uib 
 \nonumber \\
 \ &\ &- \sum_{l}\sum_{m} \nrmI
 \Bigl\{   
 \ulm (\delta \uml) \uab-\uab \ulm \delta \uml
\Bigr\} \nonumber \\
&=& 
  \sum_{l} \uli \bigl( \delta(\uil\uab)-\uil \delta \uab \bigr)
 -\uai \delta \uib 
  \nonumber \\
 \ &\ &- \sum_{l}\sum_{m}\nrmI
 \Bigl\{   
 \ulm \bigr( \delta(\uml\uab)-\uml \delta\uab \bigl) 
-\uab \ulm \delta \uml
\Bigr\} \nonumber \\
&=&\uai \delta \uib-\delta \uab -\uai \delta \uib \nonumber \\
 \ &\ &- \sum_{m} \nrmI \uam \delta \umb+
\nrmI (2^{|\I|}\identitybf) \delta \uab
+\nrmI \sum_{m} \uam \delta \umb \nonumber\\
  &=&0.
\end{eqnarray*}
(2) We prove $[iH,\,\uab]=\delta \uab$, 
 which yields 
$[iH,\,A]=\delta A$
 for any $A \in \AlI$ by linearity.
\begin{eqnarray*}
[iH,\,\uab]&=&\sum_{i,j}[\uij,\,\uab]\hij=
\sum_{i}\uib \hia -\sum_{j} \uaj \hbj \nonumber \\
&=& \sum_{i}\uii \delta \uab -\sum_{m} \nrmI
\uam \delta \umb  \nonumber \\
&\ &- \uab \sum_{j} \delta \ujj+\sum_{m} \nrmI
\uam \delta \umb  \nonumber \\
&=& \delta \uab -\uab   \delta\bigl( \sum_{j}\ujj \bigr)   \nonumber \\
&=& \delta \uab -\uab   \delta  \identitybf   \nonumber \\
&=&\delta \uab,
\end{eqnarray*}
 where   we have used 
  $\hij \in \AlIcommut$ for the first equality. \\
(3) Next we prove $H^{\ast}=H$ or $iH+(iH)^{\ast}=0$.
 By using $\uij^{\ast}=\uji$ and $(\delta a)^{\ast}=\delta a^{\ast}$,
 we obtain 
\begin{eqnarray*}
iH+(iH)^{\ast}&=&\sum \uij(\hij+\hji^{\ast}), \\
\hij+\hji^{\ast}&=&\sum_{l}
 \Bigl\{   
 \uli  \delta \ujl + (\delta \uli) \ujl
\Bigr\} -\delta_{ij}\nrmI \sum_{l} \sum_{m}
 \Bigl\{   
 \ulm \delta \uml + (\delta \ulm) \uml
\Bigr\} \\
&=&\sum_{l}
\delta(\uli\ujl)-\delta_{ij}\nrmI
\sum_{l}\sum_{m}
 \delta(\ulm\uml)\\
&=&\delta_{ij}\del\Bigl(\sum_{l}\ull\Bigr)-\delta_{ij}
\delta\Bigl(\sum_{l}\ull\Bigr)=0.
\end{eqnarray*}
Hence $iH+(iH)^{\ast}=0$.\\
(4) We prove the last statement.
Note that 
$
\tau(\uij)=\nrmI \delta_{ij}.$ Hence
\begin{eqnarray*}
i \EIc (H)&=&\nrmI \sum_{i}\hii 
=
\sum_{i}  \Bigl\{  \sum_{l}\uli \delta \uil
 -\nrmI \sum_{l} \sum_{m}
  \ulm \delta \uml \Bigr\}   \nonumber \\
&=&0.
\end{eqnarray*}
\proofend

We denote this $H$  by $\HoI$. 
\begin{lem}
\label{lem:SAKAIeven}
If $\del$ is a $\ast$-derivation
with domain $\Alinfty$ and values in $\Al$ commuting with $\Theta$, 
then there exists a self-adjoint element $\HI \in \Alp$
 satisfying 
\begin{eqnarray*}
\delta A=i[\HI,\,A ]
\end{eqnarray*}
 for all  $A \in \AlI$
 and 
\begin{eqnarray*}
\EIc(\HI)=0.
\end{eqnarray*}
\end{lem}
\proof
\ Due to commutativity of $\del$ and $\Theta$
and $\Theta^{2}=\identitybf$, we have
\begin{eqnarray*}
\delta A&=&\Theta \bigl( \delta \Theta(A)  \bigr)
=\Theta  \bigl( i[\HoI,\,\Theta(A)] \bigr) \nonumber \\
&=&i[\Theta(\HoI),\,A]
\end{eqnarray*}
for any $A \in \AlI$.
Set 
\begin{eqnarray}
\label{eq:SAKAIeven3}
\HI \equiv (\HoI)_{+}=\frac{1}{2} 
\bigl( \HoI + \Theta(\HoI)  \bigr) \;(\in \Alp).
\end{eqnarray}
Then we have $\HI^{\ast}=\HI$ and 
\begin{eqnarray*}
\delta A=i[\HI,\,A] \quad (A \in \AlI).
\end{eqnarray*}
Since $\EIc(\HoI)=0$, it follows from (\ref{eq:SAKAIeven3})
 and (\ref{eq:ETheta})
 that
\begin{eqnarray*}
\EIc(\HI)=0.
\end{eqnarray*}
\proofend

  The local Hamiltoinian operator $\HI$
 obtained in the above lemma has the  following properties:\\  
\quad($H$-i) $\HI^{\ast}=\HI \in \Al$. \\
 \quad($H$-ii) $\Theta\bigl(\HI\bigr)=\HI$   \ $\bigl($ i.e. $\HI \in \Alp \bigr)$. \\
\quad($H$-iii) $\delta A=i[\HI,\,A]$  \ $(A \in \AlI)$. \\
\quad($H$-iv) $\EIc\bigl( \HI \bigr)=0.$\\
\ \\
\begin{rem}
The property $(H$-${\rm{iv}})$ implies 
\begin{eqnarray}
\label{eq:tauHI}
\tau\bigl(\HI \bigr)=\tau\bigl( \EIc (\HI)\bigr)=0.
\end{eqnarray}
\end{rem}
\begin{lem}
\label{lem:HIfromDEL}%
$\HI$ satisfying $(H$-${\rm{ii}})$--$(H$-${\rm{iv}})$ 
is uniquely determined by 
$\delta$.
\end{lem}
\proof
\ If $\HI$ and $\HI^{\prime}$ satisfy ($H$-ii)-($H$-iv), then 
$\Delta=\HI-\HI^{\prime}$
 satisfies 
  $[\Delta,\,A]=0$
 for all $A \in \AlI$ due to ($H$-iii).
By  Lemma$\,$\ref{lem:IfCOMM} and ($H$-ii) for $\Delta$,
\begin{eqnarray*}
\Delta \in \AlIcommut \cap \Alp=\AlIcp.
\end{eqnarray*}
 Hence ($H$-iv) implies 
\begin{eqnarray*}
\Delta=\EIc(\Delta)=\EIc\bigl(\HI \bigr)-\EIc\bigl(\HI^{\prime}\bigr)=0.
\end{eqnarray*}
Therefore $\HI$ satisfying 
($H$-ii)-($H$-iv) is unique. 
\proofend

We call $\HI$  the standard Hamiltonian for the region 
$\I$.
\ \\
\begin{rem}
For the empty set $\emptyset$, $H(\emptyset)=0$
 by ($H$-iv).

Under the conditions ($H$-ii)-($H$-iv), 
 the property $\HI^{\ast}=\HI$ of ($H$-i)
 and the property $(\del A)^{\ast}=\del A^{\ast}  (A \in \AlI)$
 for $\del$ are equivalent, because of the following reason.
 If $\HI^{\ast}=\HI$, then 
$(\del A)^{\ast}=\del A^{\ast}$ immediately follows from 
  ($H$-iii). If $(\del A)^{\ast}=\del A^{\ast}$, then 
$\HI^{\ast}$ satisfies  ($H$-iii) along with 
 ($H$-ii) and ($H$-iv). Hence $\HI^{\ast}=\HI$ by the uniqueness result 
 Lemma$\,$
 \ref{lem:HIfromDEL}.
\end{rem}
\begin{lem}
\label{lem:HIHJ}
If $\I \subset \J$ is 
 a pair of finite subsets,
then 
\begin{eqnarray}
\label{eq:HIHJ}
\HI=\HJ-\EIc\bigl(\HJ  \bigr).
\end{eqnarray}
\end{lem}
\proof
\ $\HJ$ satisfies ($H$-ii) and ($H$-iii) for the region  $\I(\subset \J)$. Furthermore, $\EIc\bigl(\HJ\bigr) \in \AlIcp$ due to ($H$-ii)
 for $\HJ$ and hence it commutes with $A \in \AlI$.
Therefore 
$\HJ-\EIc\bigl(\HJ  \bigr)$ satisfies ($H$-ii)-($H$-iv) 
for the region  $\I$.
 By the uniqueness (Lemma$\,$\ref{lem:HIfromDEL}), we obtain 
$\HI=\HJ-\EIc\bigl(\HJ  \bigr)$.
\proofend

 We  give the number ($H$-v) to the condition above: \\
($H$-v) $\HI=\HJ-\EIc\bigl(\HJ  \bigr)$ for any finite subsets 
 $\I \subset \J$ of $\Znu$.\\
\ \\
The proof above has shown that ($H$-v) is derived
 from ($H$-ii)-($H$-iv).

So far we have derived the properties 
($H$-i), ($H$-ii), ($H$-iv) and ($H$-v)
 for the family $\bigl\{ \HI \bigr\}$ from its definition 
in terms of $\del$ through the relation
 ($H$-iii).

In the converse direction, any family of 
 an element $\HI \in \Al$ for each finite subset $\I$
 of $\Znu$ defines a derivation $\del$ on $\Alinfty$
 by ($H$-iii).

This definition requires a consistency:
 if $A \in \AlI$ and $A \in \AlJ$, 
 we have a definition of $\del(A)$ by $\HI$ and $\HJ$.
The proof that they are the same is given as follows.
First we note that $A \in \AlI \cap \AlJ=\AlIaJ$.
Thus it is enough  to show 
 \begin{eqnarray}
\label{eq:consistent}%
[\HI,\,A]=[\HK,\,A]
\end{eqnarray}
 for any  $\K \subset \I$ and $A \in \AlK$, because,
using this identity for the pair $\I \supset K=\I\cap \J$
 and $\J \supset \K$; we obtain $[\HI,\,A]=[\HJ,\,A]$
 for any $A \in \AlIaJ$.
 
Since $\EKc\bigl(\HI\bigr)$ is $\Theta$-even 
by ($H$-ii) and (\ref{eq:ETheta}), 
$\EKc\bigl(\HI \bigr)$ is in $\AlKcp$ and commutes with 
$A \in \AlK$. By ($H$-v), 
\begin{eqnarray*}
\HK=\HI-\EKc\bigl(\HI \bigr)
\end{eqnarray*}         
which leads to 
 the consistency equation (\ref{eq:consistent}).

$\del$ defined by ($H$-iii)
 is a $\ast$-derivation with domain $\Alinfty$ 
 due to  ($H$-i), 
and commutes
 with $\Theta$ by ($H$-ii).

We have not used ($H$-iv) in this argument, but  have imposed 
it  on $\HI$ to obtain the uniqueness of $\HI$ for a given $\delta$. 
 Namely, by 
 Lemma$\,$\ref{lem:SAKAIeven} and 
Lemma$\,$\ref{lem:HIfromDEL}, the correspondence of $\del$
 and $\HI$ is bijective, for which the condition 
($H$-iv) is used.

Summarizing  the argument so far,  we have obtained  
Theorem $\,$\ref{thm:DEL-HI} stated below 
 after introduction of  two definitions.
\begin{df}
\label{df:DB}
The real vector space  of all $\ast$-derivations 
with their definition domain $\Alinfty$ and 
 commuting with $\Theta$ (on $\Alinfty$) is denoted by $\DB$.
 \end{df}
\begin{rem}
Under Assumptions (I) and (II), the restriction $\del$
 of the generator $\delal$ of $\alt$ belongs to
 $\DB$ 
\end{rem}
\begin{df}
\label{df:HI}
 The real vector space of functions $\HI$ of  finite subsets $\I$
 satisfying  
 the following four conditions
 is denoted by $\Hset$ and 
its element $H$ is called a local Hamiltonian. \\
 \quad $(H$-${\rm{i}})$ $\HI^{\ast}=\HI  \in \Al$, \\
 \quad $(H$-${\rm{ii}})$ 
$\Theta\bigl(\HI\bigr)=\HI$   \ $($i.e. $\HI \in \Alp)$ \\
\quad $(H$-${\rm{iv}})$ $\EIc\bigl( \HI \bigr)=0$,\\
 \quad $(H$-${\rm{v}})$  $\HI=\HJ-\EIc\bigl(\HJ  \bigr)$ 
for any finite subsets 
 $\I \subset \J$ of $\Znu$.
 \end{df}
\begin{thm}
\label{thm:DEL-HI}
The 
 following relation between $H \in \Hset$ and $\del \in \DB$
 gives a bijective, real linear map from  $\Hset$ to $\DB$.\\
 \quad $(H$-${\rm{iii}})$\ 
  $\delta A=i[\HI,\,A]$\ $(A \in \AlI)$. 
\end{thm}
%
\begin{rem}
The value $\del A$ of the derivation $\del \in \DB$
 for $A \in \Alinfty$ is in general not in $\Alinfty$.
\end{rem}
\subsection{Internal Energy}
\label{subsec:InternalE}
For a finite subset $\I$ of $\Znu$, set 
\begin{eqnarray}
\label{eq:POTUI}
\UI \equiv \EI  \bigl( \HI \bigr)\  \bigl(\in \AlI\bigr)
\end{eqnarray}
 and call it the internal energy for the region $\I$.
Due to $H(\emptyset)=0$, $U(\emptyset)=0$.

 Due to the property (\ref{eq:tauHI}),
\begin{eqnarray*}
\EI\EIc\bigl( (\HJ) \bigr)=\tau\bigl((\HJ)\bigr)=0.
\end{eqnarray*}
By $(H$-{\rm{v}}$)$,
 we obtain 
 for $\I \subset \J$
\begin{eqnarray}
\label{eq:POTUI1.5}
\UI&=&\EI \HI= \EI \left(  \bigl\{ \HJ-\EIc\bigl(\HJ\bigr) \bigr\} 
\right)\nonum \\
&=&\EI \HJ=\EI\EJ \HJ=\EI\UJ.
\end{eqnarray}
 Furthermore, for any finite subset $\I$ and any subset $\J$
 of $\Znu$, 
 we have 
\begin{eqnarray}
\label{eq:POTUI2}
\EJ \bigl( \UI\bigr)=\EJ \EI \bigl( \UI\bigr)=
\EJaI \bigl( \UI\bigr)=U(\I\cap \J),
\end{eqnarray}
 where the last equality is due to (\ref{eq:POTUI1.5}).
Due to (\ref{eq:tauHI}),
\begin{eqnarray}
\label{eq:tauUI}
\tau \bigl( \UI\bigr)=\tau\bigl(\EI(\HI) \bigr)=\tau\bigl( \HI \bigr)=0.
\end{eqnarray}
Let us denote 
\begin{eqnarray}
\label{eq:HJIdef}
\HJI\equiv \EJ \bigl( \HI\bigr).
\end{eqnarray}
\begin{lem}
\label{lem:HJI}
$(\rm{1})$ For any  pair of finite subsets $\I$ and $\J$,
  \begin{eqnarray}
\label{eq:HJI1}
 \HJI=U(\J) -U(\Ic \cap \J).
\end{eqnarray}
$(\rm{2})$ For any finite subset $\I$, 
 \begin{eqnarray}
\label{eq:HIlim}%
 \HI=\lim_{\JlimZ} 
 \bigl( \UJ-U(\Ic\cap\J) \bigr). 
\end{eqnarray}
\end{lem}
\proof\ (1): By applying  $(H$-${\rm{v}})$
 for pairs $\I\supset \IaJ$ and $\J\supset \IaJ$, we obtain
\begin{eqnarray*}
 \HIaJ&=&\HI -\EIaJc \bigl(\HI \bigr),\\
   \HIaJ&=&\HJ -\EIaJc \bigl(\HJ \bigr).
\end{eqnarray*}
Therefore 
 \begin{eqnarray*}
 \HI =\HJ-\EIaJc \bigl(\HJ-\HI \bigr).
\end{eqnarray*}
By applying $\EJ$ to this equation, we obtain
 \begin{eqnarray*}
 \HJI =\UJ-\EJ \EIaJc \bigl(\HJ-\HI \bigr).
\end{eqnarray*}
Since
 \begin{eqnarray*}
 \J\cap (\I \cap \J)^{c} =\J \cap (\Ic \cup \Jc)=(\J \cap \Ic)\cup (\J \cap \Jc)=\J \cap \Ic,
\end{eqnarray*}
we obtain 
\begin{eqnarray*}
 \EJ E_{ (\I \cap \J)^{c}  } =
 E_{\J\cap{(\I \cap \J)^{c}  }}=E_{\J \cap \Ic}=\EJ \EIc=\EIc \EJ.
\end{eqnarray*}
 Since 
  $\EIc\bigl( \HI \bigr)=0$ by  $(H$-${\rm{iv}})$, we have
 \begin{eqnarray*}
 \EJ E_{(\IaJ)^{c}} \bigl( \HJ-\HI\bigr)=\EIc \EJ\bigl( \HJ\bigr)
 =\EIc \bigl( \UJ \bigr). 
\end{eqnarray*}
Thus 
 \begin{eqnarray*}
 \HJI =\UJ-\EIc \bigl( \UJ \bigr).
\end{eqnarray*}
By this and (\ref{eq:POTUI2}), 
we arrive at (\ref{eq:HJI1}).

(2): By (\ref{eq:EZlim}),
 we have 
\begin{eqnarray}
\label{eq:HJIlim}
\HI=\lim_{\JlimZ} \HJI.
\end{eqnarray}
 This and (\ref{eq:HJI1}) imply the desired (\ref{eq:HIlim}).
\proofend
\subsection{Potential}
\label{subsec:POT}
We introduce  the potential $\{\potI\}$ in terms of $\{\HI\}$
 and derive its characterizing properties. 
 As a consequence, we establish the one-to-one 
correspondence
 between $\{\potI\}$ and  $\{\HI\}$.
\begin{lem}
\label{lem:POTfrHI}   
 For a given
$\{\HI\} \in \Hset$ and the corresponding $\{ \UI\}$, 
there exists one and only one family
 of $\Bigl\{ \potI \in \Al 
;\ {\rm{finite}} \ \I  \subset \Znu \Bigr\}$
 satisfying the following conditions $:$ \\
  $(${\it{1}}$)$ $\potI \in \AlI$. \\%
$(${\it{2}}$)$ $\potI^{\ast}=\potI$, 
$\Theta\bigl(\potI \bigr)=\potI$, $\pot(\emptyset)=0$. \\
 $(${\it{3}}$)$ $\EJ \bigl( \potI\bigr)=0$ if $\J \subset \I$
   and $\J \ne \I$. \\
  $(${\it{4}}$)$  $\UI=\sum_{\K \subset \I} \potK$. \\
 $(${\it{5}}$)$  
$\HI=\lJlim 
\sum_{\K} \bigl\{ \potK;\ \K \cap \I \ne \emptyset,\ \K \subset \J\  \bigr\}.$
\end{lem}
\proof
\ We show this lemma in several steps.\\
\underline{Step 1.} Existence of $\pot$ satisfying 
$(${\it{1}}$)$ and  
$(${\it{4}}$)$ 
for all finite $\I$.

 The following expression for 
$\potI$ in terms of $\UK$, $\K\subset \I$ satisfies $(${\it{1}}$)$ 
 and  $(${\it{4}}$)$ 
  for all $\I$ and hence the existence.
\begin{eqnarray}
\label{eq:POTfrUI2}
\potI=\sum_{\K \subset \I}(-1)^{\vert\I\vert-\vert\K\vert} \UK.
\end{eqnarray}
In fact, substituting this expression into $\sum_{\J\subset \I}\potJ$,
 we obtain 
\begin{eqnarray}
\label{eq:POTfrUI3}
\sum_{\J \subset \I}\sum_{\K \subset \J}
(-1)^{|\J|-|\K|}\UK =\sum_{\K \subset \I} \alpha(\K)\UK, \nonum\\
\alpha(\K)=\sum_{\J: \K \subset \J \subset \I}(-1)^{|\J|-|\K|}=
\sum_{m=|\K|}^{|\I|}(-1)^{m-|\K|} \beta_{m},
\end{eqnarray}
where $\beta_{m}$
 is the number of distinct $\J$ satisfying
\begin{eqnarray*}
\K \subset \J \subset \I,\quad |\J|=m.
\end{eqnarray*}
This is the number of way for choosing $m-|\K|$
 elements (for $\J \setminus \K$)
 out of $\I\setminus \K$, which is ${|\I|-|\K|}\choose{m-|\K|}$.
Putting $l=m-|\K|$, $n=|\I|-|\K|$,
we obtain
\begin{eqnarray*}
\alpha(\K)=\sum_{l=0}^{n}(-1)^{l} {n\choose l}=(1-1)^{n}=0
\end{eqnarray*}
for all $\K \ne \I$ (then $n \ge 1$), while we have $\alpha(\I)=1$.
 Hence  $(${\it{4}}$)$ is satisfied by $\potI$
 given as (\ref{eq:POTfrUI2}) for all  $\I$.\\
\underline{Step 2.} Uniqueness  of $\pot$ satisfying  
 $(${\it{4}}$)$. 

The relation $(${\it{4}}$)$ implies 
\begin{eqnarray}
\label{eq:POTfrUI4}
\potI=\UI-\sum_{\K \subset \I, \K \ne \I} \potK
\end{eqnarray}
 which obviously  determines $\potI$ uniquely
 for a given $\{\UI\}$ 
by the  mathematical induction on $|\I|=m$
 starting from $\pot(\emptyset)=U(\emptyset)=0$.\\
\underline{Step 3.} Property $(${\it{2}}$)$.

We already obtain $\pot(\emptyset)=0$. 
Since $\UI^{\ast}=\UI$ and $\Theta \bigl(\UI\bigr)=\UI$,
 $\potI$  defined by (\ref{eq:POTfrUI2})
 as a real linear combination of $\UK$, $\K \subset \I$
satisfies $(${\it{2}}$)$.\\
\underline{Step 4.} Property $(${\it{3}}$)$.

 We note that $(${\it{3}}$)$ is equivalent to the 
following condition:\\
\begin{eqnarray}
\label{eq:3prime}
\EJ \bigl( \potI\bigr)=0, 
\quad {\mbox{for}}\  \J\not\supset\I,
\end{eqnarray}
because $\EJ \bigl( \potI\bigr)=\EJ \EI \bigl( \potI\bigr)=
\EJaI\bigl( \potI\bigr)$ 
 by Theorem$\,$\ref{thm:EIEJ}, $\J\cap \I \subset \I$, and 
$\J\cap \I\ne \I$ if and only if $\J \not\supset\I$.
On the other hand, $\EJ\bigl(\potI \bigr)=\potI$
 if $\J \supset \I$ due to $\potI \in \AlI \subset \AlJ$.

We now prove   $(${\it{3}}$)$ by  the mathematical induction on $|\I|=m$.
 For $m=1$, the only $\J$ satisfying $\J \subset \I$
 and $\J \ne \I$ is $\J=\emptyset$ for which $\potJ=0$.
 Then $\potI=\UI $  and 
\begin{eqnarray*}
\EJ\bigl( \potI \bigr)=\tau\bigl( \potI \bigr)\identitybf=\tau\bigl(\UI\bigr)=0
\end{eqnarray*}
due to (\ref{eq:tauUI}).
Suppose $(${\it{3}}$)$ holds for $|\I|<m$. We consider 
$\I$ with $|\I|=m$.
We apply $\EJ$ (for $\J\subset \I$, $\J \ne \I$)
 on both sides of (\ref{eq:POTfrUI4}).
 All $\K$ in the summation on the right-hand side satisfy
 $|\K|<m$ due to $\K \subset \I$ and $\K \ne \I$.
 Hence the inductive assumption is applicable to $\potK$
 on the right-hand side.
 If $\K \not\subset\J$, we have $\EJ(\potK)=0$
  by  (\ref{eq:3prime}).
 If $\K \subset \J$, we have $\EJ(\potK)=\potK$.
 Therefore, by using $\EJ\UI=\UJ$ (due to $\J \subset \I$),
 we obtain 
 \begin{eqnarray*}
\EJ\potI&=&\EJ\UI-\sum_{\K\subset \I, \K \ne \I} \EJ \potK \nonumber \\
&=&\UJ-\sum_{\K\subset \J} \potK =0.
\end{eqnarray*}
 This  proves  $(${\it{3}}$)$. \\
\underline{Step 5.} Property $(${\it{5}}$)$.

For a finite subset $\J$ and  $\I \subset \J$,
$\HJI$
is written  in terms of $\pot$
 by  (\ref{eq:HJI1})  and  $(${\it{4}}$)$ as
\begin{eqnarray}
\label{eq:HJIsum}
\HJI( =\EJ\bigl(\HI\bigr) )=
 \sum_{\K} \bigl\{ \potK;\ \K \cap \I \ne \emptyset,\ \K \subset \J \bigr\}.
\end{eqnarray}
Due to (\ref{eq:HJIlim}),  $\pot$ satisfies $(${\it{5}}$)$.
\proofend 

We collect useful formulae for 
$U$ and $H$ in terms of  $\pot$ which have been obtained above:
\begin{eqnarray}
\label{eq:colUI}
\UI&=& \sum_{\K \subset \I} \potK,\\
\label{eq:colHJI}
\HJI&=&
 \sum_{\K} \bigl\{ \potK;\ \K \cap \I \ne \emptyset,\ \K \subset \J \bigr\},\\
\label{eq:colHI}
\HI&=& \lJlim \left(
 \sum_{\K} 
 \bigl\{ \potK;\ \K \cap \I \ne \emptyset,\ \K \subset \J \bigr\}
\right)
\bigl(=\lJlim \HJI\bigr).
\end{eqnarray}
\begin{df}
\label{df:STANDARDPOT}
 A function 
$\pot$ of finite subsets $\I$
 of $\Znu$ with the value $\potI$
 in $\Al$ is called a standard potential 
if it satisfies the following conditions $:$\\
 $(\pot$-$\rm{a})$ $\potI\in \AlI$, $\pot(\emptyset)=0$. \\
  $(\pot$-$\rm{b})$  $\potI^{\ast}=\potI$.\\
$(\pot$-$\rm{c})$  $\Theta\bigl( \potI \bigr)=\potI$.\\
 $(\pot$-$\rm{d})$  $\EJ \bigl( \potI\bigr)=0$ 
if $\J \subset \I$ and $\J \ne \I$. \\
$(\pot$-$\rm{e})$ For each fixed 
finite subset $\I$ of $\Znu$,
the net 
\begin{eqnarray*}
\HJI=
\sum_{\K} \bigl\{ \potK;\ \K \cap \I \ne \emptyset,\ \K \subset \J \bigr\},
\end{eqnarray*}
 is a Cauchy net in the norm topology
 of $\Al$ for $\JlimZ$.
 The  index set for the net  is  the set of all finite 
 subsets $\J$ of $\Znu$,
 partially ordered by the set inclusion.
\end{df}
\begin{rem}
\label{rem:STANDARD}
 $(\pot$-$\rm{d})$  
is equivalent to the following condition:  \\ 
$(\pot$-$\rm{d})^{\prime}$ 
   $\EJ \bigl( \potI\bigr)=0$ 
unless  $\I \subset \J$, \\
because 
$
\EJ\bigl( \potI\bigr)=\EJ \EI\bigl( \potI\bigr)=\EJaI
\bigl( \potI\bigr).$
\end{rem}
\begin{df}
\label{df:PB}
The real vector space of all standard potentials is denoted by $\PB$. 
\end{df}

\begin{rem}
\label{rem:PB}
$\PB$ is a real vector space as a function space,
 where the linear operation is defined by
\begin{eqnarray}
\label{eq:linearST}
(c \pot +d \potpsi)(\I)=c \potI+ d\potpsi(I), 
\quad c,d \in \R,\quad  \pot,\potpsi \in \PB.
\end{eqnarray}
\end{rem}

We show the  one-to-one correspondence of $\pot \in \PB$ and 
$H \in \Hset$. 
\begin{thm}
\label{thm:PB-Hset}
The equations (\ref{eq:colHJI})
 and (\ref{eq:colHI}) for $\pot \in \PB$ and $H\in \Hset$
 give a bijective, real linear map from $\PB$ to $\Hset$.
\end{thm}
\proof 
\ First note that  $(${\it{4}}$)$ of Lemma$\,$\ref{lem:POTfrHI}
 is satisfied for $\UI=\EI\bigl( \HI\bigr)$
 due to $(\pot$-$\rm{d})$, 
 if (\ref{eq:colHJI})
 and (\ref{eq:colHI}) are satisfied. 
 By  Lemma$\,$\ref{lem:POTfrHI}, there exists a unique 
 $\pot \in \PB$ satisfying (\ref{eq:colHJI})
 and (\ref{eq:colHI})
 for any given $H \in \Hset$.
 
 The map is evidently linear. The only remaining task is to prove 
 the property ($H$-i),  ($H$-ii), ($H$-iv) and ($H$-v)
 for the $\HI$ given by (\ref{eq:colHJI})
 and (\ref{eq:colHI}),  on the basis of 
  $(\pot$-$\rm{a})$-$(\pot$-$\rm{e})$.
($H$-i), ($H$-ii) and ($H$-iv) follow from 
$(\pot$-$\rm{b})$, $(\pot$-$\rm{c})$ and $(\pot$-$\rm{d})^{\prime}$,
 respectively.

 To show   ($H$-v), let $\LL$ be a finite subset containing 
  $\J \supset \I$. Then 
\begin{eqnarray*}
\HLJ-\HLI&=&\sum_{\K} \bigl\{ \potK;\ \K \cap \J \ne \emptyset,
 \K \cap \I =\emptyset,
\ \K \subset \LL \bigr\}\nonum \\
&=&\EIc \left( \sum_{\K} \bigl\{ \potK;\ \K \cap \J \ne \emptyset,
\ \K \subset \LL \bigr\} \right) \nonum \\
&=&\EIc \bigl( \HLJ \bigr)
\end{eqnarray*}
due to (\ref{eq:colHJI}), $(\pot$-$\rm{a})$
 and $(\pot$-$\rm{d})^{\prime}$.
 By taking limit $\LL \nearrow \Znu$, we obtain 
\begin{eqnarray*}
\HJ-\HI&=&\EIc \bigl( \HJ \bigr),
\end{eqnarray*}
 where the convergence is 
due to $(\pot$-$\rm{e})$ and $\Vert \EIc\Vert =1$.
\proofend

\begin{rem} 
We will use later the real linearity of the above map: 
\begin{eqnarray}
\label{eq:linearH}
H_{c\pot +d \potpsi }(\I)&=&c H_{\pot }(\I)+ d H_{ \potpsi }(\I),
 \quad c,d \in \R,\quad  \pot,\potpsi \in \PB,  \\
\label{eq:linearU}
U_{c\pot +d \potpsi }(\I)&=&c U_{\pot }(\I)+ d U_{ \potpsi }(\I),
 \quad c,d \in \R,\quad  \pot,\potpsi \in \PB,
\end{eqnarray}
 where $H_{\pot}(\I)$ and $U_{\pot}(\I)$
 denote $\HI$ and $\UI$ corresponding to $\pot \in \PB$.
\end{rem}

\begin{thm}
\label{thm:DB-PB}
The following relation  between
 $\pot \in \PB$ and 
$\delpot \in \DB$
 gives a bijective, real linear map
 from  $\PB$ to  $\DB$.
 \begin{eqnarray}
\label{eq:DB-PB1}
\delpot A&=& i[\HI,\,A] \quad \bigl(A \in \AlI \bigr), \\
\label{eq:DB-PB2}
\HI&=&\lim_{\JlimZ}
 \sum_{\K} \bigl\{ \potK;\ \K \cap \I \ne \emptyset,\ \K \subset \J \bigr\}.
\end{eqnarray}
\end{thm}  
\proof 
This is a consequence of 
Theorem$\,$\ref{thm:DEL-HI} and Theorem$\,$\ref{thm:PB-Hset}.
\proofend

\ \\
{\it{Remark 1.}} The technique using the conditional
 expectations for associating a unique standard potential with a
 a given $\ast$-derivation has been developed for quantum spin
 lattice systems by one of the authors
\cite{ARAKI78springer650}.
 The corresponding formalism for classical lattice systems 
 is developed   in \cite{ARAKIiwanami}.
Also see \cite{ISR} where $\EI$ for the quantum 
spin case is called a partial trace.

\ \\
{\it{Remark 2.}} We note that $\PB$ is a Fr\'echet 
 space with respect to a countable family of seminorms 
$\bigl\{ \Vert H(\{i\}) \Vert\bigr\} $, $i \in \Znu$.

\subsection{General Potential}%
\label{subsec:GP}
 If the  function 
\begin{eqnarray}
\pot: \I\in \{ {\mbox{finite subsets of}}\  \Znu\}
 \longmapsto \potI
\end{eqnarray}
satisfies ($\pot$-a), ($\pot$-b), ($\pot$-c) and 
($\pot$-e), we 
call it  a general potential.

By ($\pot$-e), we define $\HI$
 by (\ref{eq:colHI}) and (\ref{eq:colHJI}).
Then, for any finite subsets $\K \supset \I$, 
\begin{eqnarray}
\label{eq:POTG1}
\HK-\HI
=\lim_{\JlimZ} \sum_{\LL}
\left\{\potL;\ \LL\cap \K \ne \emptyset,\ \LL\cap \I=\emptyset,\ 
\LL\subset \J \right\}
\end{eqnarray}
 due to ($\pot$-e).
Therefore,  we can define $\delpot$ with the domain $\Alinfty$ by
\begin{eqnarray}
\label{eq:POTG2}
\deltap A=i[\HI,\,A] \quad {\mbox {for}}\ A \in \AlI,
\end{eqnarray}
 which is a consistent definition due to (\ref{eq:POTG1}) by  
 essentially  the same argument as the one 
leading to (\ref{eq:consistent}).
The properties  ($\pot$-a), ($\pot$-b), ($\pot$-c), 
and ($\pot$-e)
 imply that $\deltap \in \DB$.
 Two general potentials $\pot$ and $\pot^{\prime}$
 are said to  be equivalent if $\deltap=\deltapp$.
 It follows from Theorem$\,$\ref{thm:DB-PB} that there is a unique standard
potential which is equivalent to any given general potential
 defined above.
 The equivalence is discussed, e.g., 
in \cite{ISR} and \cite{SIMON} with the name of physical equivalence.
 We will consider the consequence of equivalence 
 for a specific class  of general potentials 
 in $\S$\,\ref{sec:DISCUSSION}.
 %
\section{KMS  Condition}
\label{sec:KMSC}
\subsection{KMS Condition}
\label{subsec:DefKMS}%
We recall the definition of the KMS condition for 
a given dynamics $\alt$
of $\Al$
(see e.g. \cite{BRA2}).
\begin{df}
\label{df:KMS}
A state $\vp$ of $\Al$ is called  an $\alt$-KMS state
 at the inverse temperature $\beta \in \R$ or
  $(\alt,\,\beta)$-KMS  state (or more simply KMS state)
  if it satisfies  one of the following 
two equivalent conditions $:$\\
\ $(${\rm{A}}$)$ Let $\Dbeta$ be the strip region 
\begin{eqnarray*}
\Dbeta&=&\Bigl\{z \in \C;\  0 \le {\Imag}z \le \beta    \Bigr\}
 \quad {\mbox{if}}\ \beta \ge 0, \nonumber \\ 
 &=&\Bigl\{z \in \C;\  \beta\le {\Imag}z \le  0    \Bigr\}
 \quad {\mbox{if}}\ \beta < 0, \nonumber \\ 
\end{eqnarray*} 
 in the complex plane $\C$
 and $\Dbetao$ be its interior. 

For every $A$ and $B$ in $\Al$, there exists a function $F(z)$
 of $z \in \Dbeta$ (depending on $A$ and $B$) such that \\
\quad $(${\rm{1}}$)$\ $F(z)$ is analytic in $\Dbetao$,\\
\quad $(${\rm{2}}$)$\ $F(z)$ is continuous and bounded on  $\Dbeta$,\\
\quad $(${\rm{3}}$)$\ For all real $t \in \R$,  
\begin{eqnarray*} 
F(t)=\vp \bigl(A \alpha_{t}(B) \bigr),\ \  
F(t+i \beta)=\vp \bigl(\alpha_{t}(B)A \bigr).
\end{eqnarray*} 
\ \\
\ $(${\rm{B}}$)$ Let  $\Alana$ be  the set of all $B \in \Al$
 for which $\alt(B)$ has an analytic extension 
 to $\Al$-valued entire function  $\alpha_{z}(B)$
 as a function of $z \in \Com$. 
For  $\A \in \Al$ and $B \in \Alana$,
\begin{eqnarray*}
\vp\bigl(A\al_{i\beta}(B)\bigr)=\vp(BA).
\end{eqnarray*} 
\end{df}

\begin{rem}
In (A), the condition (1) is empty if $\beta=0$.
The boundedness in (2) can be omitted 
(see, e.g., proposition 5.3.7
 in \cite{BRA2}). $\Alana$ is known to be dense in $\Al$.
\end{rem}
For a state $\varphi $  on  $\A $,     
let 
$\Bigl\{{\cal H}_\varphi, \pi_\varphi,  
\mit\Omega_\varphi \Bigr\}$ 
denote its GNS triplet, namely, 
$\pivp$ is a (GNS) representation of $\Al$ on the 
Hilbert space $\Hilvp$, and 
$\mit\Omega_\varphi$ is a 
cyclic unit vector in $\Hilvp$, representing $\vp$ 
as the vector state. 
If $\vp$ is an $(\alt$, $\beta)$-KMS  state, 
 then  $\Omega_\varphi$ is separating for the generated von Neumann
 algebra
$\vnM \equiv \pi_\varphi(\A)^{\prime \prime}$.
Let $\mit\Delta_\varphi$  and $\sigtvp$
 be  the modular operator
and  modular automorphisms for $\Omega_\varphi$ and  $\vp$,
  respectively, 
 \cite{TAKESAKIspringer128}.

The KMS condition implies   that 
\begin{eqnarray} 
\label{eq:KMSscale}
\sigtvp \bigl(\pivp(A)\bigr)
=\pivp\bigl(\alpha_{-\beta t}(A) \bigr),\quad  A \in \Al.
\end{eqnarray}    

It is  a result of Takesaki  \cite{TAKESAKIspringer128}
 that the KMS  condition of  a one-parameter automorphism group
 of a von Neumann algebra with respect to a cyclic vector
  implies the separating property of  
 the vector, and the modular automorphism  group of the  von Neumann algebra
  with respect to the  cyclic and separating vector is characterized 
 by the KMS condition at $\beta=-1$ with respect to 
the state given by that vector.

For the sake of brevity in stating an assumption later,
 we use the following terminology. 
\begin{df}
\label{def:MODULAR}
A state $\vp$ is said to be modular if 
 $\Omega_\varphi$ is separating for 
 $\pivp(\A)^{\prime \prime}$.
\end{df}

\subsection{Differential KMS Condition}
\label{subsec:dKMS}%
It is convenient to introduce 
 the following  condition in terms of the generator $\delal$ of 
the dynamics $\alt$, equivalent to the KMS condition
 with respect to $\alt$.
\begin{df}
\label{df:DKMS}
Let $\del$ be a  $\ast$-derivation of $\Al$
 with its domain $\Ddel$.
A state $\vp$ is  said to satisfy the differential 
$(\del,\,\beta)$-KMS
 condition (or briefly, $(\del,\,\beta)$-dKMS condition)
 if the following two conditions are  satisfied

$(${\rm{C-1}}$)$\ 
$\vp\bigl(A^{\ast} \del A \bigr)$ is pure imaginary for all 
  $A \in \Ddel$.

%
  $(${\rm{C-2}}$)$ $-i \beta \vp\bigl(A^{\ast}\del A \bigr)
      \geq  S\bigl(\vp(A \Aast),\,\vp(\Aast A)  \bigr)$ 
    for all $A \in \Ddel$ where the function $S(x,\,y)$
 is given for $x \geq 0,\,y \geq0$
 by:
\begin{eqnarray*}
S(x,\,y)&=&y\log y -y \log x  \quad {\mbox{if}}\ x>0,\ y>0,\nonumber \\
S(x,\,y)&=&+ \infty  \quad {\mbox{if}}\ x=0,\ y>0,\nonumber \\
S(x,\,y)&=&0  \quad {\mbox{if}}\ x \geq 0,\ y=0.
\end{eqnarray*}
\end{df}
 We use the following known result 
 (see, e.g., Theorem$\,$5.3.15
 in \cite{BRA2}).
\begin{thm}
\label{thm:RoepstroffSewell}
Let $\delal$ be a generator of $\alt$, namely, 
$e^{t \delal }=\alt$.
Then the $(\delal,\,\beta)$-dKMS condition and 
the $(\alt,\,\beta)$-KMS condition are equivalent.
\end{thm}
\begin{rem}
 The function $S(x,\,y)$  is the relative entropy for  
 linear functionals of one-dimensional $\ast$-algebra.
The order of the arguments $x, y$ in our notation is 
opposite  to that of the definition in  \cite{UME62}.
(Both the order of the argument and the sign are opposite
 to those in  \cite{BRA2}.)
 Our definition here is in accordance with our definition 
 of the relative entropy previously given.
\end{rem}
\begin{lem}
\label{lem:Sxy}
$S(x,\,y)$ is convex and lower semi-continuous in $x,\, y$.
\end{lem}
\proof 
 \ A convenient expression for $S(x,\,y)$
 is 
\begin{eqnarray}
\label{eq:KOSAKI1}%
S(x,\,y)=\sup_{n}\sup_{s(t)}
   \Bigl\{y\log n- \int_{\frac{1}{n}}^{\infty}
\Bigl( ys(t)^{2}+t^{-1}x \bigl\{1-s(t)\bigr\}^{2} \Bigr)  \frac{dt}{t} \Bigr\},
\end{eqnarray}
 where $s(t)$ varies over the linear span of characteristic functions
 of finite intervals in $[0,\,+\infty)$.
 The equality is immediate for $x=0$, $y>0$
 as well as for $x \geq 0$, $y=0$.
 For $x>0$, $y>0$, 
(\ref{eq:KOSAKI1}) follows from identities for $\lambda=x/y$.
\begin{eqnarray*}
y(\log y-\log x)&=&\sup_{n} \Bigl\{-y\log \Bigl(\frac{x}{y}+
 \frac{1}{n} \Bigr) \Bigr\} \nonumber \\
&=&\sup_{n} \Bigl\{ y\log n-\int_{\frac{1}{n}}^{\infty}
y \,\frac{\lam}{t+\lam} \frac{dt}{t}\Bigr\},  \nonum \\
-y \,\frac{\lam}{t+\lam}&=&\sup_{s \in \R}
\Bigl\{-\Bigl( ys^{2}+xt^{-1}(1-s)^{2} \Bigr)\Bigr\}. 
\end{eqnarray*}

From the expression above, $S(x,\,y)$
  is seen to be convex and lower semi-continuous in $(x,\,y)$
 because it is a supremum of homogeneous linear
 functions of $(x,\,y)$.

(The variational expression (\ref{eq:KOSAKI1}) for 
 general von Neumann algebras is established by Kosaki
\cite{KOSAKI86}.
 This expression indicates manifestly some 
 basic properties of relative entropy for the general case.) 
\proofend
\begin{lem}
\label{lem:stableC}
The conditions {\rm{(C-1)}} and 
 {\rm{(C-2)}} are stable under the simultaneous limit of 
 $A$ and $\del A$ in norm topology and $\vp$ in the weak$\ast$
 topology as well as under the convex combination of states $\vp$.
\end{lem}
\proof\  
Let $A_{n}, A \in \Ddel$,  
 $\Vert A_{n} - A\Vert \to 0$, 
$\Vert \del A_{n}- \del A\Vert \to 0$,
 $\bigl| \vp_{n}(B)-\vp(B)\bigr| \to 0$ 
  for every $B \in \Al$.
Then 
\begin{eqnarray*} 
&& \bigl| \vp_{n}\bigl(A_{n}^{\ast}\delta A_{n} \bigr) 
-   \vp\bigl(A^{\ast}\delta A \bigr) \bigr| \nonum \\
&\le& \bigl| \vp_{n}\bigl(A_{n}^{\ast}\delta A_{n}-A\delta A \bigr) \bigr|
+ \bigl| \vp_{n}\bigl(A \delta A\bigr) 
-\vp\bigl( A \delta A \bigr) \bigr|,
\end{eqnarray*}
which converges  to $0$ as $n \to \infty$.
Therefore, the condition (C-1) holds for 
  $\vp$ and $A$ if it holds for $\vp_n$ and $A_n$.

Similarly, 
\begin{eqnarray*} 
\vp_{n}\bigl(A_{n}  A_{n}^{\ast} \bigr)\to \vp(A A^{\ast}),\  
\vp_{n}\bigl(A_{n}^{\ast} A_{n} \bigr)\to \vp(A^{\ast}A), 
\end{eqnarray*}
as $n \to \infty$.
By the lower semi-continuity of $S(x,\,y)$ in $(x,\,y)$, 
we then obtain  
\begin{eqnarray*}
 S\bigl(\vp(A \Aast),\,\vp(\Aast A)  \bigr)
 \le \liminf_{n} 
  S\bigl(\vp( A_{n} A_{n}^{\ast}),\,\vp( A_{n}^{\ast}  A_{n} )  \bigr).
\end{eqnarray*}
Hence we obtain the condition (C-2) for $\vp$ and $A$ 
if it holds for $\vp_n$ and $A_n$.
Since $\vp\bigl(A^{\ast}\del A \bigr)$ is affine in $\vp$
 while  $S\bigl(\vp(A \Aast),\,\vp(\Aast A)  \bigr)$
is convex in $\vp$, the conditions 
(C-1)  and (C-2) are  stable under the convex combination of $\vp$.
\proofend
\begin{cor}
\label{cor:DKMS-KMS}
Let $\alt$ be a one-parameter group of $*$-automorphisms of $\Al$
 satisfying  the  conditions  
{\rm{(II)}}  and {\rm{(III)}}. Let $\delal$ be
 the generator of $\alt$.
Then  a state $\vp$ is an $(\alt,\,\beta)$-KMS state if and only if 
 it is a $(\del,\,\beta)$-dKMS state, where
$\del$ denotes  the restriction of $\delal$ 
 to $\Alinfty$.
\end{cor}
\proof\ 
The restriction $\del$ of $\delal$ to $\Alinfty$ makes sense due to
 the assumption (II). 
By Theorem\,\ref{thm:RoepstroffSewell},
 it suffices to prove that the dKMS condition for $\del$
 implies the same for $\delal$.
By Assumption (III), there exists 
 a sequence $A_n \in \Alinfty$
 for any given $A \in \Ddelal$ such that 
$\Vert A_n -A \Vert \to 0$, $\Vert \del A_n -\delal A \Vert \to 0$.
 Hence  the conditions (C-1) and (C-2) for $\delta$ imply the same for 
 $\delal$ due to Lemma$\,${\ref{lem:stableC}}.
\proofend
\section{Gibbs Condition}
\label{sec:GIBBS}
In this section, we define the Gibbs condition.
We first recall the notion of perturbation 
of dynamics and states.
\subsection{Inner Perturbation}
\label{subsec:INNER}
Consider a given dynamics 
$\alt$ of $\Al$ with its generator $\delta$ on  
 the domain $D(\delta)$.
 For each $h=h^{\ast} \in \Al$, there exists
 the unique perturbed  dynamics $\alth$ of $\Al$
 with its generator $\delh$ given by
\begin{eqnarray}
\label{eq:INN0}
\delh(A)\equiv\del(A)+i[h,\,A] \quad (A \in \Ddel)
\end{eqnarray}
 on the same domain as the generator $\delta$ of $\alt$.
This $\alth(A)$ is explicitly given by
\begin{eqnarray}
\label{eq:INN1}
\alth(A)=\uht\alt(A) (\uht)^{\ast}
\end{eqnarray}
 where 
\begin{eqnarray}
\label{eq:INN2}  
  \uht \equiv \identitybf+
\sum_{m=1}^\infty  i^{m}
  \int_0^{t}\!\!dt_1 
  \int_0^{t_1}\!\!\!dt_2
  \cdots \!\!
  \int_0^{t_{m-1}} \!\!\!\!\!\!\!dt_m
\,\altm(h) 
 \cdots
\altone(h).
\end{eqnarray}
This  is unitary and satisfies the following cocycle equation:
 \begin{eqnarray*}
\uhs\als(\uht)=\uhst
\end{eqnarray*}
 The same statements hold for a von Neumann algebra
  $\vNM$ and its one 
 parameter group of $\ast$-automorphisms $\alt$; 
the $t$-continuity of $\alt$ for each fixed $x \in \vNM$ 
 in  the  strong  operator topology of $\vNM$
 is to be assumed.

 Let $\Ome$ be a cyclic and separating vector for $\vNM$.
Let $\modome$ be the  modular operator 
 for $\Ome$ and $\sigtome$ be the corresponding modular automorphism
 group
\begin{eqnarray*}
\sigtome(x)=\modomeit x \modomemit,
\end{eqnarray*}
 where $\ome$ indicates the positive linear functional
 \begin{eqnarray*}
\ome(x)=(\Ome,\,x\Ome), \quad (x \in \vNM).
\end{eqnarray*}
For $h=h^{\ast}\in \vNM$, the perturbed vector $\Omeh$
 is given by
\begin{eqnarray}
\label{eq:Omeh1}  
 &{\:}& \Omeh \nonumber  \\  
 &\equiv&\!\!\!\!\!
\sum_{m=0}^\infty  
  \int_0^{\frac12}\!\!dt_1 
  \int_0^{t_1}\!\!\!dt_2
  \cdots \!\!
  \int_0^{t_{m-1}} \!\!\!\!\!\!\!dt_m\,
  \mit\Delta_\varphi^{t_m}  \pi_\varphi(h)
  \mit\Delta_\varphi^{t_{m-1}-t_m}\pi_\varphi(h)
  \cdots
  \mit\Delta_\varphi^{t_1    -t_2}\pi_\varphi(h)
\Ome \nonumber \\
&=&\!\!\!\!
{\mbox{Exp}}_r\left(\int_0^{\frac{1}{2}}\; ; 
\mit\Delta_\varphi^t \pi_\varphi(h)
\mit\Delta_\varphi^{-t}dt\right) \Ome,
\end{eqnarray}
 where the sum is known to converge 
absolutely (\cite{ARAKI73relative hamiltonian}).
 The notation ${\mbox{Exp}}_r$
is taken from  \cite{ARAKI73expansional}. 

 The positive linear  functional  $\omeh$ on $\vNM$  
is defined  by 
\begin{eqnarray}
\omeh(x)\equiv
  \left(\Omeh,\,x \Omeh\right) \quad (x \in \vNM).
\end{eqnarray}
 The vector $\Omeh$  defined above  is cyclic and separating for $\vNM$.
 Its modular automorphism group $\sigtomeh$ of $\vNM$
 coincides with  $(\sigtome)^{h}$,  i.e. the perturbed dynamics of 
$(\sigtome,\,\vNM)$
 by $h$.
$\Omeh$ is  in the natural positive cone of $(\Ome,\,\vNM)$
 (see, e.g., \cite{TAKESAKI1} and \cite{BRA2})
 for any self-adjoint element $h \in \vNM$ 
  and satisfies
\begin{eqnarray}
(\Omeho)^{\htwo}=\Omehot
\end{eqnarray} 
 for any self-adjoint elements $\ho, \htwo  \in \vNM$. 
 We have
\begin{eqnarray}
\label{eq:pert1}
 (\omeho)^{\htwo} = \ome^{\hotwo}, \quad 
 \sigma_{t}^{\{\omehotwok\}} \bigl(
= \bigl(\sigma_{t}^{\ome}\bigr)^{(\ho+\htwo)} \bigr)
=  \bigl\{(\sigma_{t}^{\ome})^{\ho} {\bigl\}}^{\htwo},
\end{eqnarray}
 where $\bigl\{(\sigma_{t}^{\ome})^{\ho} {\bigl\}}^{\htwo}$
 indicates the dynamics which is given by 
  the successive  perturbations  first by $\ho$ and 
 then by $\htwo$.
 We denote the normalization of $\omeh$ by $[\omeh]$: 
\begin{eqnarray}
\label{eq:pertstate}
[\omeh]= \omeh(\identitybf)^{-1}\omeh=
\ome^{(h-\{\log\omeh(\identitybf)\}\identitybf )}.
\end{eqnarray}
We use the following  estimates (Theorem 2 of 
\cite{ARAKI73GP})
and a formula 
(e.g. (3.5) of \cite{ARAKI76CNRS} and Theorem 3.10 
of \cite{ARAKI77rims}) later.
\begin{eqnarray}
\label{eq:pertstate2}
\Vert \Omeh \Vert \le \exp \frac{1}{2}\Vert h \Vert,
 \quad \log \omeh(\identitybf)\le \Vert h \Vert.
\end{eqnarray}
\begin{eqnarray}
\label{eq:pertstate3}
S(\vph,\,\vp)=-\vp(h). 
\end{eqnarray}
\subsection{Surface Energy}%
\label{subsec:SURFACE}
 Let us consider $\pot \in \PB$.
For any finite subset $\I$ of $\Znu$, we define
 \begin{eqnarray}
\WI \equiv \HI-\UI.
\end{eqnarray} 
 By (\ref{eq:colUI}), (\ref{eq:colHJI}) and (\ref{eq:colHI}),
 the expression for $\WI$ 
  in terms of the potential is  given as follows. 
\begin{eqnarray}
\label{eq:WIterm}
  \WI &=& 
\sum_{\K} 
\bigl\{ 
{\pot(\K) ; \K \cap \I \ne  \emptyset,\; 
                              \K \cap \Ic \ne \emptyset} 
\bigr\} \\
\Bigl( &=& \lim_{\JlimZ} 
\bigl( \sum_{\K} \bigl\{ 
{\pot(\K) ; \K \cap \I \ne  \emptyset, \;
                              \K \cap \Ic \ne \emptyset,\ \K\subset \J}
\bigr\} \bigr)  \Bigr).\nonum
\end{eqnarray} 
 $\WI$ is 
the sum of all (interaction) potentials 
between the  inside and the outside of $\I$ by definition, and 
 will be called the surface energy. \\
\subsection{Gibbs Condition}%
\label{subsec:DefGibbs}%
 We are now in a position to introduce our Gibbs condition 
 for a state $\vp$ of $\Al$ for a given $\del \in \DB$.
 We use the following notation in its definition below.
 As in $\S$$\,$\ref{subsec:DefKMS},
$\Bigl\{{\cal H}_\varphi, \pi_\varphi, 
\mit\Omega_\varphi \Bigr\}$ 
is the  GNS triplet for $\vp$.
  The normal extension of $\vp$ to the weak closure 
$\vnM(= \pi_\varphi(\A)^{\prime \prime})$ is denoted by
  the same letter $\vp$:
\begin{eqnarray*}
\vp(x)&=&(\vpvec,\,x \vpvec) \quad(x \in \vnM), \nonumber \\
\vp\bigl(\pivp(a)\bigr)&=& \vp(a) \quad (a  \in \Al).  
\end{eqnarray*}
 Let $\potI$, $\HI$, $\UI$ and $\WI$  
  be those uniquely associated with $\del$.
  The following operators will be used for perturbations of dynamics
 and states
\begin{gather}
\hha=\pivp\bigl(\beta\HI \bigr),\; 
\uha=\pivp\bigl(\beta\UI \bigr), \; 
\wha=\pivp\bigl(\beta\WI \bigr).
\end{gather}
\begin{df}
\label{def:Gibbs}
For $\del \in \DB$, 
a state $\vp$ of $\Al$ is said to satisfy 
the $(\del,\, \beta)$-Gibbs condition, or alternatively 
the $(\pot,\, \beta)$-Gibbs condition, if
 the following two conditions 
are  satisfied. \\
\ {\rm{(D-1)}} $\vp$  is a modular state. 
{\rm{(}}See Definition \ref{def:MODULAR}.{\rm{)}}\\
\ \\
\ {\rm{(D-2)}}  For each finite subset $\I$ of $\Znu$,
 $\sigtvpwha$ 
 satisfies  
\begin{eqnarray*}
\sigtvpwha\bigl( \pivp(A)\bigr)=
 \pivp \Bigl( e^{-i \beta \UI t} A  e^{i \beta \UI t}
  \Bigr)
\end{eqnarray*}
 for all $A \in \AlI$.
\end{df}

The condition (D-2) is equivalent to the following condition (D-2)$^{\prime}$
 as shown in the subsequent  Lemma and hence 
 we may define the $(\del,\, \beta)$-Gibbs condition
 by (D-1) and (D-2$)^{\prime}$. \\
\ \\
\ (D-2)$^{\prime}$
 {\it{For each finite subset $\I$ of $\Znu$ and $A \in \AlI$,
 $\pivp(A)$ is  $\sigtvphha$-invariant, namely,  
$\pivp\bigl( \AlI \bigr)$ is in the centralizer of the positive linear
 functional  $\vphha$.}}
\ \\
\begin{lem}
\label{lem:EGIBBS}
The conditions {\rm{(D-2)}} and   {\rm{(D-2)}}$^\prime$ are equivalent. 
\end{lem}
\proof
\ First  assume (D-2).
 Since $\hha=\wha+\uha$, we have $\vp^{\hha}={(\vpwha)}^{\uha}$
 and hence 
\begin{eqnarray*}
\sigma_{t}^{\vphha} &=& 
\Bigl\{(\sigma_{t}^{\vp})^{\wha} {\Bigl\}}^{\uha}\nonumber \\
&=& \Bigl(\sigma_{t}^{\vpwha}\Bigr)^{\uha}
\end{eqnarray*}
Since 
$e^{-i \beta \UI t} \,\UI\,  e^{i \beta \UI t}=\UI$, 
   $\pivp\bigl(\UI \bigr)$  is invariant 
 under $\sigtvpwha$ by (D-2).
Then  unitary cocycle bridging $\sigtvpwha$ and $\sigtvphha$
 becomes $e^{i \uha t}$. 
Hence
 \begin{eqnarray*}
\sigtvph= {\mbox {Ad}}(e^{i \uha t}) \circ \sigtvpwha.
\end{eqnarray*}
Therefore, for  $\pivp\bigl( A \bigr)$,  $A \in \AlI$,
 we have 
\begin{eqnarray*}
\sigtvph(  \pivp\bigl( A \bigr) )&=& 
e^{i \uha t} \sigtvpw(\pivp\bigl( A \bigr))  
e^{-i \uha t} \nonumber \\
&=&  \pivp \bigl( {\mbox {Ad}}(e^{i \beta \UI t})
\circ   {\mbox {Ad}}(e^{-i \beta  \UI t}) 
 \circ    A \bigr) \nonumber \\
&=& \pivp\bigl(   A \bigr).
\end{eqnarray*} 
 Thus (D-2)$^{\prime}$ is satisfied.

 We show the converse. Assume  (D-2)$^{\prime}$.
 Since $\wha=\hha-\uha$, $\sigtvpwha$ is the perturbed dynamics of 
 $\sigtvphha$ by $-\uha$.
 Since $u \in \AlI$ is 
 $\sigtvphha$-invariant (being in the centralizer),
 the corresponding unitary cocycle is $e^{-i \uha t}$.
 Hence, for 
 $\pivp\bigl( A \bigr)$, $A \in \AlI$,
 we have
\begin{eqnarray*}
\sigtvpwha\bigl( \pivp\bigl( A \bigr)     \bigr)&=&
e^{-i \uha t} \sigtvphha( \pivp\bigl( A \bigr) ) e^{+i \uha t} \nonumber \\
&=&e^{-i \beta\pivp\bigl(\UI \bigr)t}
\pivp\bigl( A \bigr)  e^{i \beta\pivp\bigl(\UI \bigr)t} \nonumber \\
 &=&
 \pivp\Bigl(e^{-i \beta \UI t} A  
e^{i \beta \UI t} \Bigr),
 \end{eqnarray*} 
 and  (D-2) is derived.
\proofend\\
We introduce the local Gibbs state.
\begin{df}
\label{def:LocalGibbs}
For finite $\I$, the local Gibbs state 
 of $\AlI$ (or local Gibbs state for $\I$) with respect 
 to $(\del,\,\beta)$ is given by
\begin{eqnarray}
\label{eq:vpIcEQ}
\vpIc(A)\equiv \frac{\tau (e^{-\beta \UI} A)}
{\tau(e^{-\beta \UI})}, \quad  A \in \AlI.
\end{eqnarray}
\end{df}
\begin{cor}
\label{cor:LocalGibbs}
If $\vp$ satisfies the $(\del,\,\beta)$-Gibbs condition,
 then the restriction of $\vphha$ to $\AlI$
 is $\vphha(\identitybf)$ times the tracial state $\tau$
 and that of $\vpwha$ is $\vpwha(\identitybf)$ 
times the local Gibbs state
 $\vpIc$ given by (\ref{eq:vpIcEQ}).
\end{cor}
\proof \ 
Since $\vphha$ has the tracial property 
 for $\AlI$ by (D-2)$^{\prime}$, 
its restriction to $\AlI$ must be 
$\vphha(\identitybf)$ times the unique tracial state $\tau$.

Since the inner automorphism group 
\begin{eqnarray}
\alIt \equiv  {\mbox {Ad}}(e^{-i \beta  \UI t}) 
\end{eqnarray}  
leaves $\AlI$ invariant and has the same action on $\AlI$
 as the modular automorphism of $\vpwha|_{\AlI}$
 (the restriction of $\vpwha$ to $\AlI$), 
 $\vpwha|_{\AlI}$ satisfies 
$(\alIt,\,-1)$ KMS condition and hence must be 
$\vpwha(\identitybf)$ times 
the unique KMS 
 state given by the local Gibbs state $\vpIc$.
\proofend
%
\subsection{Equivalence to  KMS  Condition}
\label{subsec:KtoG}
\begin{thm}
\label{thm:KtoG}
Let $\alt$ be dynamics
of $\Al$ satisfying  
 conditions  ${\rm{(I)}}$ and ${\rm{(II)}}$
 and $\del$ be the restriction of its generator $\delal$ 
 to $\Alinfty$.
Then any $(\alt,\,\beta)$-KMS state $\vp$ of $\Al$ 
satisfies $(\del,\,\beta)$-Gibbs condition.
\end{thm}
\proof
\ As already indicated, it is known that the KMS condition implies
 (D-1).
 It remains to  show (D-2).
 We have
\begin{eqnarray*}
(d/ds) \left(   
\sigsvpwha(x)-\sigsvp(x)
\right)_{s=0} = i  
\left[\wha,\, x   \right],
\end{eqnarray*}
 for $x \in \vnM$.
By the group property of the automorphisms,
\begin{eqnarray*}
(d/dt) \sigtvpwha(x)
=\sigtvpwha
\left\{ (d/ds)\sigsvpwha(x) \Bigl|_{s=0}  \right\}
\end{eqnarray*}
for $x$ in the domain of the generator  of
 $\sigtvpwha$.
For the same $x$,  
 we have
\begin{eqnarray*}
(d/dt) \sigtvpwha(x)
=\sigtvpwha
\left\{ (d/ds)\sigsvp(x) \Bigl|_{s=0}+ i[\wha,\,x]  \right\}.
\end{eqnarray*}
The KMS condition implies that 
\begin{eqnarray*} 
\sigsvp \bigl(\pivp(A)\bigr)
=\pivp\bigl(\alpha_{-\beta s}(A) \bigr),\quad  A \in \Al.
\end{eqnarray*}    
Therefore, if  $A \in \Al$ is in the domain of the generator of $\alt$,
 we have
\begin{eqnarray*}
(d/dt) \sigtvpwha\Bigl(\pivp(A) \Bigr)
=\sigtvpwha
\left\{ (d/ds)\Bigl( \pivp\bigl\{\alpha_{-\beta s}(A)
\bigr\}\Bigr) \Bigl|_{s=0}\right\} +  \sigtvpwha \Bigl(
\pivp\bigl\{[i \beta \WI,\,A]\bigr\} \Bigr).
\end{eqnarray*}
 Now we take $A \in \AlI$. By ($H$-iii), 
\begin{eqnarray*}
(d/dt) \sigtvpwha\Bigl(\pivp(A) \Bigr)
&=&\sigtvpwha
\Bigl(  -i \beta  \pivp\bigl\{ [\HI,\,A] \bigr\}  \Bigr) 
+\sigtvpwha \Bigl(
i\beta \pivp \bigl\{ [\WI,\,A] \bigr\}\Bigr) \nonumber \\
&=&-i \beta \sigtvpwha\Bigl(\pivp \bigl\{ [\UI,\,A] \bigr\} \Bigr).
\end{eqnarray*}
For $A \in \AlI$, 
  $e^{i \beta \UI t} A  e^{-i \beta \UI t} \in \AlI$,
 and  we have 
\begin{eqnarray*}
\ &&(d/dt) \sigtvpwha
\Bigl(\pivp\bigl\{e^{i \beta \UI t} A e^{-i \beta \UI t}  \bigr\} \Bigr)
\nonumber \\
 &=&\sigtvpwha
\left\{ (d/ds)\sigsvpwha
\Bigr( \pivp \bigl\{ e^{i \beta \UI(t+s) }Ae^{-i \beta \UI(t+s)} \bigr\} \Bigr)
 \Bigl|_{s=0}  \right\} \nonumber \\
&=&
 \sigtvpwha\!\left( -i \beta
\pivp\bigl\{ [\UI,\, e^{i \beta \UI t}Ae^{-i \beta \UI t} ] \bigr\}
 + \pivp \Bigl\{ d/ds 
\bigl(  e^{i \beta \UI (t+s)} A e^{-i \beta \UI (t+s)} \bigr)
\Bigl|_{s=0}  \Bigr\}   \right) \nonumber \\
&=&0.
\end{eqnarray*}
This implies that 
\begin{eqnarray*}
\sigtvpwha
\Bigl(\pivp\bigl\{e^{i \beta \UI t} A e^{-i \beta \UI t}  \bigr\} \Bigr)
\end{eqnarray*}
 is a constant function of $t$ and hence equals to its value 
at $t=0$, which is $\pivp(A)$.
Thus 
 \begin{eqnarray*}
 \sigmtvpwha\bigl(\pivp(A)\bigr)=
\pivp\bigl( e^{i \beta \UI t} A e^{-i \beta \UI t}  \bigr)
\end{eqnarray*}
 and (D-2) is shown.
 \proofend \\

To show the converse, we need  the  assumption (III) 
 for the dynamics $\alt$.  
\begin{thm}
\label{thm:GtoK}
Let $\alt$ be a dynamics of $\Al$
 satisfying   
 the conditions ${\rm{(I)}}$,  ${\rm{(II)}}$  and ${\rm{(III)}}$. 
Let  $\del$ be the restriction of its generator $\delal$ 
to $\Alinfty$.
Then any  $(\del,\,\beta)$-Gibbs state $\vp$ of $\Al$
 satisfies 
 $(\alt,\,\beta)$-KMS condition.
\end{thm}
\proof
\ We use (D-2)$^{\prime}$.
 It says that
\begin{eqnarray*}
(d/dt) \sigtvphha\bigl( \pivp(A) \bigr)=0
\end{eqnarray*}
 for all $A \in \AlI$.
By the group property of the automorphism, 
\begin{eqnarray*}
(d/dt) \sigtvp(x)
=\sigtvp
\left\{ (d/ds)\sigsvp(x) \Bigl|_{s=0}  \right\}.
\end{eqnarray*}
For any $A \in \Alinfty$, there exists a finite subset $\I$
such that $A \in \AlI$.
Since  $\vp=(\vphha)^{-\hha}$,  we have 
\begin{eqnarray}
\label{eq:KtoG3}%
(d/dt) \sigtvp\bigr( \pivp(A) \bigr)
&=&\sigtvp
\left\{ (d/ds)\sigsvphha \bigl(\pivp(A)\bigr)
 \Bigl|_{s=0}    -[i\hha,\,\pivp(A)]   \right\} \nonumber \\
&=&\sigtvp\bigl(-i \beta \pivp([\HI,\, A]) \bigr) \nonum \\
&=& -\beta \sigtvp\bigl(\pivp (\delta A ) \bigr).
\end{eqnarray}
 We note that  for any $A \in \Al$
\begin{eqnarray*} 
\sigtvp \bigl( \pivp(A) \bigr)&=&\modvpit \pivp(A) \modvpmit, \quad
\modvp \GNSvpvec =\GNSvpvec.
\end{eqnarray*}
By applying (\ref{eq:KtoG3}) on $\vpvec$ and setting $t=0$, we conclude
 that $\pivp(A) \vpvec$ is in the domain of $\log \modvp$ and
 \begin{eqnarray}
\label{eq:KtoG4}
i (\log \modvp) \pivp(A) \vpvec=-\beta \pivp\bigl(\delta(A)\bigr) \vpvec
\end{eqnarray}
 for all $A \in \Alinfty$.

By Assumption (III), 
for every  $A \in \Ddelal$,
  there exists a sequence $\{A_{n}\}$, $A_{n} \in \Alinfty$
   such that $\{A_n\}$  and $\{  \delta A_n (=\delal A_n  )  \}$
converge to $A$ and $ \delal A (=\bar{\delta}A )$, respectively,
 in the norm topology of $\Al$.
 Since $\log \modvp$  is a (self-adjoint) closed operator, 
 $\pivp(A) \vpvec$  must be in the domain of $\log \modvp$
 and (\ref{eq:KtoG4}) holds for any $A \in \Ddelal$. 

For $A \in \Ddelal$ and $t \in \R$, we set  
\begin{eqnarray*}
\xi_{t}\equiv
 \sigtvp \Bigl( \pivp \bigl\{ \alpbt(A)\bigr\} \Bigr) \vpvec
=\modvpit \pivp(\alpbt(A)) \vpvec.
 \end{eqnarray*}
 For $A \in \Ddelal$, $\alt(A)$ is in $\Ddelal$ for any $t \in \R$.
Therefore, we can substitute $\alpbt(A)$ into $A$ 
of (\ref{eq:KtoG4}) and obtain
 \begin{eqnarray*}
(d/dt) \xi_{t}&=& 
\modvpit
\left\{ (d/ds)\modvpis \pivp \bigl\{ \alpbt(A) \bigl\} \vpvec
 \Bigl|_{s=0}           \right\} 
+ \modvpit\Bigl( (d/dt) \pivp \bigl\{ \alpbt(A) \bigl\} \vpvec  \Bigr)
\nonumber \\
&=& \modvpit \left\{-\beta 
\pivp \bigl\{ \delta(\alpbt(A)) \bigl\} \vpvec  
+ 
\pivp \bigl\{ \beta \delta(\alpbt(A)) \bigl\} \vpvec  
  \right\} \nonum \\
&=&0.
 \end{eqnarray*}
Therefore, we have $\xi_{t}=\xi_{0}$  and 
\begin{eqnarray*}
 \sigtvp \Bigl( \pivp \bigl\{ \alpbt(A)\bigr\} \Bigr) \vpvec
= \pivp(A) \vpvec.
\end{eqnarray*}
Since $\vpvec$ is separating for $\vnM$, we obtain 
\begin{eqnarray*}
 \sigtvp \Bigl( \pivp \bigl\{ \alpbt(A)\bigr\} \Bigr) 
= \pivp(A).
\end{eqnarray*}
This implies 
\begin{eqnarray*}
 \pivp \bigl\{ \alpbt(A)\bigr\}  
= \sigmtvp \bigl( \pivp(A) \bigr).
\end{eqnarray*}
Since  $\Ddelal(\supset \Alinfty)$ is norm dense in  $\Al$, we have
\begin{eqnarray*}
  \pivp \bigl\{ \alpmbt(A)\bigr\}  
= \sigtvp \bigl( \pivp(A) \bigr),
\end{eqnarray*}
 for every $A \in \Al$.

 Since $\vp$  satisfies $(\sigtvp,\,-1)$-KMS condition as a state 
 of  $\vnM$,
 we obtain the  $(\alt,\,\beta)$-KMS condition for $\vp$.
\proofend 
\subsection{Product Form of the Gibbs Condition}
\label{subsec:GibbsProduct}
In the case of quantum spin lattice systems, 
for any  region $\I \subset \Znu$, $\Al= \AlI \otimes \AlIc$.
 In this situation, the Gibbs condition implies that
 $\vpwha(=\pertvpfw)$ is a product of the local 
  Gibbs state of $\AlI$
 and  its restriction to $\AlIc$,
  or equivalently $\vphha(=\pertvpfh)$ is a  product  of
   the tracial state of $\AlI$ and its restriction to $\AlIc$
 for any finite region $\I$ \cite{ARAKIION}.
  
  However,
  this product property for $\vpwha$  and  $\vphha$
 for  the present Fermion case 
 does not seem to be automatic in general.
   We  show  that such a product property holds  
   if and only if the Gibbs state $\vp$  is $\Theta$-even, 
 where the product property refers to the validity of the formula
\begin{eqnarray}
\label{eq:ProductGibbs0}
\psi(AB)=\psi(A)\psi(B)/\psi(\identitybf),\quad A\in \AlI, \ B \in \AlIc
\end{eqnarray}
 for $\psi=\vphha$ and for $\psi=\vpwha$
\begin{pro}
\label{pro:evenGibbsProduct}
Assume the conditions {\rm{(I)}} and {\rm{(II)}}
 for the dynamics.
Let $\I$ be a non-empty finite subset of $\Znu$.
If  $\vp$ satisfies  the Gibbs condition,
then $\pertvpfw$ has  the  product property 
(\ref{eq:ProductGibbs0})
if and only if $\vp$ is $\Theta$-even. 
The same is true for $\pertvpfh$.
\end{pro}
\proof
 \ First assume that $\vp$ is even.
It follows from the Gibbs condition that
$\AlI$ is in the centralizer of $\vphha$ and the restriction
of $\vphha$ to $\AlI$ is tracial.
We will  show
\begin{eqnarray} 
\label{eq:PRODUCT1}
\vphha\bigl( [\Ah,\,\Af]B \bigr)=0
\end{eqnarray} 
for any $\Ah,\,\Af \in \AlI$ and any $B \in \AlIc$.
It  is enough to show this for all combinations of 
even and  odd $\Ah$, $\Af$ and  $B$ because
 the general case follows from these 
 cases by linearity. 

Since $\Ah$ and $\Af$ are in the centralizer of 
 $\vphha$, we have
\begin{eqnarray*} 
\vphha(\Ah\Af B)=\vphha(\Af B\Ah), \quad \vphha(\Af\Ah B)
=\vphha(\Ah B\Af).
\end{eqnarray*}
If one, or more of $\Ah$, $\Af$, $B$ is even, then 
$B\Ah=\Ah B$ or $B \Af=\Af B$ holds. Hence (\ref{eq:PRODUCT1})
 follows for this case.

 The remaining case is when  $\Ah$, $\Af$, $B$ are all odd.
 We now  show that 
   $\vphha$ is even so that (\ref{eq:PRODUCT1}) holds 
 in this case.

 Since $\vp$ is assumed to be even at this part of proof,
 $\Theta$ leaves $\vp$ invariant and hence there exists an involutive unitary 
 $\Unit$ on the GNS representation space $\Hilvp$ of $\vp$,
 satisfying  
\begin{eqnarray}
 \label{eq:PRODUCT2}
\Unit \pivp (A){\Unit}^{\ast}&=&\pivp \bigl( \Theta(A) \bigr),
 \quad (A \in \Al), \\
 \label{eq:PRODUCT3}
\Unit \GNSvpvec&=&\GNSvpvec.
\end{eqnarray}
 Since $\HI$ is even by assumption, it follows 
 from the commutativity of $\Unit$ with $\modvp$ \cite{TAKESAKIspringer128}
 and  the above equations (\ref{eq:PRODUCT2}), (\ref{eq:PRODUCT3}) 
  that the perturbed vector $\vpvechha$ 
 is $\Unit$ invariant. Therefore  $\vphha$ is even,
 since it is the  vector functional by $\vpvechha$. 
Hence    $\vphha$ vanishes on every odd element and  
(\ref{eq:PRODUCT1}) is satisfied 
if $\Ah$, $\Af$ and $B$ are all odd. 
Now $(\ref{eq:PRODUCT1})$ is proved for all the cases.

Since $\AlI$ is a $2^{\vert \I \vert} \times 2^{\vert \I \vert}$
  full matrix algebra,
  any element $A \in \AlI$ can be written as 
 \begin{eqnarray*}
 A=\tau(A)+\sum_{j}[A_{j1},\, A_{j2}]
 \end{eqnarray*} 
 for some $A_{j1}, A_{j2} \in \AlI$. Hence $(\ref{eq:PRODUCT1})$ implies
\begin{eqnarray} 
\label{eq:PRODUCT4}
\vphha(AB)=\tau(A)\vphha(B)
\end{eqnarray}
 for any $A \in \AlI$ and  $B \in \AlIc$. 
This means that 
 $\vphha$ has a form of  the product of $\tau$  of  $\AlI$
  and its restriction to
 $\AlIc$. 
 
 Since $\UI$ is in the centralizer of $\vphha$,
  we have
\begin{eqnarray*} 
\vpwha=\{\vphha\}^{-\uha}=\vphha\cdot e^{-\uha}.
\end{eqnarray*}
Hence,  for any $A \in \AlI$ and  $B \in \AlIc$,
\begin{eqnarray*}
\vpwha(AB)
=\tau(e^{-\uha})\vpIc(A) \vphha(B).
\end{eqnarray*}
By setting $A=\identitybf$, we have 
\begin{eqnarray*}
\vpwha(B)= \tau(e^{-\uha})    \vphha(B).
\end{eqnarray*}
Therefore
\begin{eqnarray}
\label{eq:PRODUCT5} 
\vpwha(AB)=\vpIc(A)\vpwha(B).
\end{eqnarray}
Hence  we have the desired product property of $\vpwha$.

We  now prove the converse, starting from the 
assumption that $\vphha$ has  a product form (\ref{eq:ProductGibbs0}).

We note that
\begin{eqnarray*}
\tau(\ai\aicr)=\tau(\aicr\ai)
=\tau\Bigl( \frac{1}{2}(\ai\aicr+\aicr\ai) \Bigr)
=\tau\Bigl( \frac{1}{2} \identitybf \Bigr)
=\frac{1}{2} 
\end{eqnarray*}
 due to CAR. On the other hand, $\ai$ anticommutes with 
any odd element $B$ in 
$\AlIc$
 and hence
 \begin{eqnarray}
 \label{eq:PRODUCT7}
\vphha(\ai \aicr B)=\vphha(\aicr B \ai)=-\vphha(\aicr \ai B), 
 \end{eqnarray} 
 where the first equality follows  because $\ai$ is in the centralizer 
  of $\vphha$ due to the Gibbs condition.
  By the product form assumption,
\begin{eqnarray*} 
\vphha(AB)=\vphha(A)\vphha(B)/\vphha(\identitybf)
\end{eqnarray*}
   for $A \in \AlI$ and  $B \in \AlIc$.
Since  $A$ is in the centralizer, 
 $\vphha(A)/\vphha(\identitybf)=\tau(A)$
 for the unique tracial state $\tau$ of $\AlI$.
Hence
 \begin{eqnarray}
 \label{eq:PRODUCT8}
\vphha(\ai \aicr B) &=&\tau(\ai \aicr)\vphha(B) =\frac{1}{2}\vphha(B),
\nonumber \\
\vphha(\aicr \ai B) &=&\tau(\aicr \ai)\vphha(B) =\frac{1}{2}\vphha(B).
 \end{eqnarray} 
 From (\ref{eq:PRODUCT7}) and (\ref{eq:PRODUCT8}), we obtain
 \begin{eqnarray}
 \vphha(B)=0
 \end{eqnarray} 
 for any $B \in \AlIcm$. 
Since $\Al_{-}=\AlIp \AlIcm+\AlIm \AlIcp$
 for a finite $\I$, $\vphha$ vanishes  on odd elements of $\Al$.
We conclude  that 
 $\vphha$ is  even.
 This implies that
 $\vp$
 is also even by the same argument as in the first part of this  proof
 due to $\vp=\{\vphha\}^{-\hha}$.
 \proofend  \\

\begin{rem}
By the above Proposition, we have already shown that 
 if a Gibbs state $\vp$ satisfies the condition that 
$\pertvpfw$ has the  product property (\ref{eq:ProductGibbs0}) 
for the pair  ($\AlI$, $\AlIc$) 
for  one  non-empty  finite  $\I$, then $\vp$ has
 this   product property 
  for every  finite subset $\I$.
\end{rem}
\ \\

In connection with Proposition$\,$\ref{pro:evenGibbsProduct}, if  $\AlIc$   
 is replaced by the commutant algebra $\AlIprime$
 in the product property (\ref{eq:ProductGibbs0}), 
then $\vpwha$ is a product of the local Gibbs state
of $\AlI$ and its restriction to $\AlIprime$ 
 for every finite region $\I$ 
 irrespective of whether $\vp$ is even or not as is shown 
in the following
 corollary.
 This situation is much  the same  as in quantum  spin 
lattice systems.
\begin{cor}   
\label{cor:SpinGibbs}
Assume the conditions {\rm{(I)}} and {\rm{(II)}}
 for the dynamics.
Let $\vp$ be a modular state. The state $\vp$ satisfies
the  Gibbs condition if and only if 
the perturbed functional $\vpwha$ is a 
product of the local Gibbs state $\vpIc$ of $\AlI$
 and its restriction to $\AlIprime$ 
 for every finite  $\I$. 
\end{cor}
\proof\ 
 For a finite $\I$, 
$\AlI$ is a full matrix algebra and hence
$\Al$ is an (algebraic) tensor product 
 of $\AlI$ and $\AlIprime$. 

If $\vpwha$ has the product property described  above,
 then the GNS representation of $\Al$
 associated with $\vpwha$ is the tensor product of those
 for $(\AlI,\,\vpIc)$ and $(\AlIprime,\,\psi)$ where 
 $\psi=\vpwha|_{\AlIprime}$.
Therefore the product of the modular automorphisms 
 for these two pairs satisfies the KMS condition 
(with $\beta=-1$) for $(\Al,\,\vpwha)$ and must be the modular 
 operator for  $(\Al,\,\vpwha)$.
In particular, the restriction of the modular automorphisms of 
   $(\Al,\,\vpwha)$ to $\AlI$
coincides with the modular automorphisms 
 $
\alIt \bigl( ={\mbox {Ad}}(e^{-i \beta  \UI t}) \bigr)
$
 for $(\AlI,\,\vpIc)$. 
 Hence the Gibbs condition is satisfied. 

Conversely, assume that the  Gibbs condition is satisfied 
 for $\vp$.
By the elementwise commutativity of  $\AlI$ and $\AlIprime$,
 we can show directly (\ref{eq:PRODUCT1}) in  
Proposition$\,$\ref{pro:evenGibbsProduct}  in this case 
for any $\Ah, \Af \in \AlI$ and 
$B \in \AlIprime$ skipping  the  previous discussion
about even and odd elements.
 The    argument
 showing  (\ref{eq:PRODUCT4}) and (\ref{eq:PRODUCT5}) are  
still valid
  after  we replace $\AlIc$   by $\AlIprime$. 
\proofend
%
%
\section{Translation Invariant Dynamics}
\label{sec:TID}
\subsection{Translation Invariance and Covariance}%
\label{subsec:TI}%
From now on, 
 we need  the following assumption  for
the dynamics $\alt$ for the most part of our theory. \\
\ \\
\quad (IV) $\alt\, \shiftk=\shiftk \,\alt$
for all $t \in \R$ and  $k \in \Znu$.
\ \\

If (IV) holds, $\alt$ 
 is said to be translation invariant.
 This assumption implies our earlier assumption (I)
 due to the following Proposition, which we owe to a referee.
\begin{pro}
\label{pro:shift-theta-alt}
Any automorphism $\alt$ commuting with the lattice 
 translation $\shiftk$, $k\in \Znu$, must commute with $\Theta$.
\end{pro}
For its proof, we need the following Lemma.
\begin{lem}
\label{lem:ASYMP}
An element $x\in \Al$ is $\Theta$-even if and only if 
 the following asymptotically central property holds.
\begin{eqnarray}
\label{eq:ASYMPEQ}
\lim_{k\to \infty}\Vert [\shiftk(x),\, y] \Vert =0
\end{eqnarray}
 for all $y\in \Al$.
\end{lem}
\proof\ 
 If $x\in (\Alinfty)_{+}$ and $y\in \Alinfty$, then 
 $[\shiftk(x),\,y]=0$ for sufficiently large $k$.
 By the density of $(\Alinfty)_{+}$ in $\Alp$
 and $\Alinfty$ in $\Al$, we obtain (\ref{eq:ASYMPEQ})
 for $x\in \Alp$ and $y\in \Al$.

In the converse direction, consider a general
 $x\in \Al$ and define 
$x_{\pm}=1/2\bigl(x\pm \Theta(x)\bigr)\in \Al_{\pm}$.
 Due to the validity of (\ref{eq:ASYMPEQ})
 for $x_{+}$, which is just shown, we have  
\begin{eqnarray*}
\lim_{k\to \infty}\Vert [\shiftk(x),\, y] \Vert =
\lim_{k\to \infty}\Vert [\shiftk(x_{-}),\, y] \Vert.
\end{eqnarray*}
Take a unitary $y\in \Alm$ (e.g., $\ai+\aicr$).
Then 
\begin{eqnarray*}
\Vert [\shiftk(x_{-}),\, y] \Vert=
2\Vert \shiftk(x_{-}) y \Vert=2\Vert x_{-}\Vert.
\end{eqnarray*}
Hence (\ref{eq:ASYMPEQ}) for $x$ implies $x_{-}=0$, 
namely $x\in \Alp$. \proofend
\ \\
\underline{{\it{Proof of Proposition\ref{pro:shift-theta-alt} }} }:\\
 Due to $\shiftk \al=\al \shiftk$, we have 
\begin{eqnarray*}
\bigl\Vert [\shiftk\bigl( \al(x) \bigr),\, \al(y)] \bigr\Vert=
\bigl\Vert \alpha \bigl\{ [\shiftk(x),\, y] \bigr\} \bigr\Vert=
\bigl\Vert  [\shiftk(x),\, y]  \bigl\Vert.
\end{eqnarray*}
Hence $\al(x)\in \Alp$ if and only if $x\in \Alp$
 by Lemma$\,$\ref{lem:ASYMP}.
Let
\begin{eqnarray}
E_{+}\equiv\frac{1}{2}({\mbox{id}}+ \Theta).
\end{eqnarray}
It is the conditional expectation  from $\Al$ onto 
 $\Alp$, characterized by $E_{+}(x)\in \Alp$ for all $x\in \Al$
 and $\tau(xy)=\tau\bigl(E_{+}(x)y \bigr)$ for all 
 $x\in \Al$ and $y\in \Alp$.
Then $\al\bigl(\al^{-1}(y)\bigr)=y\in \Alp$
 implies $\al^{-1}(y)\in \Alp$ and 
\begin{eqnarray*}
\tau\bigl(E_{+}(\al(x) )y \bigr)&=&
\tau\bigl( \al(x) y \bigr)=
\tau\bigl( \al (x \al^{-1}(y) \bigr)=\tau\bigl( x \al^{-1}(y) \bigr)\\
&=& \tau\bigl(E_{+}(x) \al^{-1}(y) \bigr)=
\tau\bigl(\al^{-1} \bigl\{\al (E_{+}(x)) y\bigr\} \bigr)=
\tau\bigl(\al (E_{+}(x)) y \bigr),
\end{eqnarray*}
 where we have used $\al^{-1}(y)\in \Alp$ in the fourth equality.
 Since  $E_{+}(\al(x))\in \Alp$ and 
  $\al(E_{+}(x))\in \Alp$ (due to $E_{+}(x) \in \Alp $), 
 we have $ E_{+}(\al(x)) =  \al(E_{+}(x))$.
 Therefore $E_{+} \al=\al E_{+}$
 and $\al$ commutes with $\Theta$. \proofend
\ \\
\begin{rem}
A referee pointed out the following approach 
(which we have not adopted).
Under assumption IV,
 any $\alt|_{\Alp}$-KMS state of $\Alp$ has a unique even 
extension to an $\alt$-KMS state of $\Al$ 
 (e.g. by \cite{CHEMICAL}).
 This allows one to  reduce the analysis of KMS states
 to the case of asymptotically abelian system 
 due to (\ref{eq:ASYMPEQ}).
\end{rem}

The dynamics $\alt$ is translation invariant if and only if 
its generator $\alt$ 
 commutes with every $\shiftk$ $(k \in \Znu)$. 
 (This statement includes
 the $\shiftk$-invariance of the domain of the generator.)

The corresponding standard potential (which exists under the assumptions
 (I) and (II)) satisfies the following 
 translation covariance condition:\\
\ \\
\ ($\pot$-f) $\shiftk \potI =\pot(\I+k)$, 
 for all finite subsets $\I$ of $\Znu$ and all $k \in \Znu$.
\ \\

Such a potential will be  said to be  translation covariant.
 
We consider the set $\PBI$ of  all translation covariant potentials 
  in  $\PB$. Namely, $\PBI$ is defined to be the set of 
all $\pot$ satisfying all conditions of 
Definition$\,$\ref{df:STANDARDPOT}, i.e., 
($\pot$-a,b,c,d,e) and the translation covariance ($\pot$-f).

 We make $\PBI$ a real vector space 
 as a function space on the set of finite subsets of $\Znu$
by the linear
 operation given in  (\ref{eq:linearST}).

In the same way, we define $\HsetI$
 to be the subspace  of $\Hset$
 such that each element $H$ satisfies the 
 following translation covariance condition:\\
\ $(H$-$\rm{vi})$ $\shiftk(\HI)=H(\I+k)$ for all $k \in \Znu$.

We denote the set of all translation invariant
 derivations in $\DB$ by $\DBI$.
Namely, $\DBI$ is the set of all $*$-derivations with $\Alinfty$
 as their domain, commuting with $\Theta$ and also with
 $\tau$.

 From Theorems$\,$\ref{thm:DEL-HI}, \ref{thm:PB-Hset} and 
 \ref{thm:DB-PB},  the  following corollaries obviously follow.
\begin{cor}
The relation 
$(H$-${\rm{iii}})$  
(as given in $\S$$\,$\ref{subsec:LH}) between 
$H \in \HsetI$ and $\del \in \DBI$
 gives a bijective, real linear map from  $\HsetI$ to $\DBI$.\\
\end{cor}
\begin{cor}
The equations (\ref{eq:colHJI})
 and (\ref{eq:colHI}) for $\pot \in \PBI$ and $H\in \HsetI$
 give a bijective, real linear map from $\PBI$ to $\HsetI$.
\end{cor}
\begin{cor}
\label{cor:DB-PB}
The equations  (\ref{eq:DB-PB1}) and (\ref{eq:DB-PB2})
 between
 $\pot \in \PBI$ and 
$\delpot \in \DBI$
 gives a bijective, real linear map
 from  $\PBI$ to  $\DBI$.
\end{cor}  

 For $\pot \in \PBI$, we define 
\begin{eqnarray*}
\npot \equiv \bigl\Vert H(\{ n\}) \bigr\Vert 
\end{eqnarray*}
 which is independent of $n \in \Znu$
 due to the translation covariance  of $\pot$. It defines
a norm on $\PBI$.
 We show that this norm makes $\PBI$ a Banach space, 
 after giving the following energy estimates.
\begin{lem}
\label{lem:EE}%
For $\pot \in \PBI$, the following estimate hold $\rm{:}$
\begin{eqnarray}
\label{eq:EE1}%
\Vert \UI \Vert \le\Vert \HI \Vert \le \npot \cdot |\I|,
\end{eqnarray}
In particular, if $\npot=0$, 
$H=U=\pot=0$ (as functions of finite subsets $\I$ of $\Znu$). 
\end{lem}
\proof
\ For $\I=\emptyset$, both sides of the above inequalities are $0$.

For $\I=\{\nno, \ldots, \nnI\}$, 
 we obtain  
\begin{eqnarray*}
\HI&=&\lJlim
\sum_{\K}\, 
\Bigl\{ \potK;\ \K \cap \I \ne \emptyset,\ \K \subset \J \Bigr\} \nonum \\
&=&
 \lJlim \sum_{i=1}^{|\I|}
\sum_{\K}\, 
\Bigl\{ \potK;\  \K \ni \nni, \, \K  \not \ni \nno,\ldots, \nnim,\,  
\K \subset \J \Bigr\} \nonum \\
&=&\lJlim \sum_{i=1}^{|\I|}
 \Ennmc  \sum_{\K}\, 
\Bigl\{ \potK;\  \K \ni \nni,\,   
\K \subset \J \Bigr\} \nonum \\
 &=& \sum_{i=1}^{|\I|}
 \Ennmc  
 H\bigl(\{\nni\}\bigr),
\end{eqnarray*}
 where the third equality comes from the following identities
\begin{eqnarray*}
\Ennmc \potK= \left\{ 
  \begin{array}{rl}
  0\ & \mbox{if}\  \{\nno,\ldots,\nnim \} \cap \K  \ne \emptyset,
  \; \mbox{i.e.}\; \{\nno,\ldots,\nnim   \}^{c} \not \supset \K \\
 \potK& \mbox{if}\   \nno,\ldots,\nnim  \not \in \K,
  \;\mbox{i.e.}\; \{\nno,\ldots,\nnim   \}^{c}  \supset \K,
\end{array} \right. 
\end{eqnarray*}
and the interchange of $\lJlim$ and $\Ennmc$ in the fourth equality
 is allowed due to 
$\Vert \Ennmc \Vert =1$.

The following estimate follows:
\begin{eqnarray}
\label{eq:EE55}%
\Vert \HI \Vert 
 &\le& \sum_{i=1}^{|\I|}
 \Vert \Ennmc  
 H\bigl(\{\nni\}\bigr)\Vert \nonum\\
  &\le& \sum_{i=1}^{|\I|}
 \Vert   
 H\bigl(\{\nni\}\bigr)\Vert =|\I| \cdot  \npot. 
\end{eqnarray}
Since $\UI=\EI \bigl(\HI\bigr)$ and $\Vert \EI \Vert=1$, we obtain
\begin{eqnarray*}
\Vert \UI \Vert \le \Vert \HI \Vert  \le \npot \cdot |\I|.
\end{eqnarray*}
If $\potnor=0$, then $\HI=\UI=0$ for all $\I$ by this estimate
 and hence $\potI=0$ by (\ref{eq:POTfrUI2}).
\proofend

The following estimate will be used later.
\begin{lem}
\label{lem:UIUJ}
For disjoint finite subsets $\I$ and $\J$ of $\Znu$,
\begin{eqnarray}
\label{eq:UIUJ}
\Vert U(\I \cup \J)-\UI \Vert \le \potnor \cdot |\J|.
\end{eqnarray}
\end{lem}
\proof
\ Due to $\I \cap \J =\emptyset$, 
\begin{eqnarray*}
U(\I \cup \J)-\UI=
\Bigl\{ \potK;\  \K \cap \J \ne \emptyset,\ \K\subset \I\cup\J \Bigr\}.
\end{eqnarray*}
 Therefore, we have
\begin{eqnarray*}
U(\I \cup \J)-\UI=E_{\I\cup\J}\HJ,
\end{eqnarray*}
because  $\HJ$ is the sum of $\potK$ for all $\K$
satisfying $\K \cap \J \ne \emptyset$,  and $E_{\I\cup\J}$
 annihilates all $\potK$ for which 
 $\K$ is not contained in $\I \cup \J$   while  it retains
 $\potK$ unchanged if  $\K$ is  contained in $\I \cup \J$. 
 Hence 
\begin{eqnarray*}
\Vert U(\I \cup \J)-\UI\Vert=\Vert \EIuJ \HJ\Vert
\le  \Vert \HJ \Vert \le \potnor \cdot |\J|.
\end{eqnarray*}
\proofend
\begin{pro}
\label{pro:BANACH}
$\PBI$ is a real Banach space with respect to the norm 
$\npot = \Vert H(\{ n\}) \Vert$.
\end{pro}
\proof
\ $\PBI$ is a normed space 
 with respect to $\npot $,
 because 
\begin{eqnarray*}
\bigl\Vert \pot_{1}+\pot_{2} \bigr\Vert
&=&\Vert H_{\pot_{1}+\pot_{2}}
\bigl(\{n \}\bigr)  \Vert \nonum \\
&=&\Vert H_{\pot_{1}}\bigl(\{n \}\bigr)  +  H_{\pot_{2}}\bigl(\{n \}\bigr)  
\Vert \nonum \\
&\le& 
\Vert H_{\pot_{1}}\bigl(\{n \}\bigr)\Vert  +  
 \Vert H_{\pot_{2}}\bigl(\{n \}\bigr)  \Vert  \nonum \\
&=&  \bigl\Vert \pot_{1}\bigr\Vert
 +\bigl\Vert \pot_{2} \bigr\Vert
\end{eqnarray*}
\begin{eqnarray*}
\bigl\Vert c \pot \bigr\Vert
&=& \Vert c H_{\pot}\bigl(\{n \}\bigr) \Vert \nonum \\
&=&
|c| \Vert  H_{\pot}\bigl(\{n \}\bigr) \Vert=|c| \potnor,
\end{eqnarray*}
for $\pot_{1}, \pot_{2}, \pot \in \PBI$, and $c \in \R$, due to the linear dependence of $H_{\pot}$ on $\pot$
 and because $\npot=0$ implies 
$\potI=0$ for all $\I$ due to Lemma$\,$\ref{lem:EE}
 and (\ref{eq:POTfrUI2}).
 
We now  show its  completeness.
Suppose $\bigl\{\potn \bigr\}$ is a Cauchy sequence 
in  $\PBI$
 with respect to the norm 
$\Vert\cdot\Vert$.
 Let us denote the corresponding $\HI$ and $\UI$ for $\potn$
 by $\HIn$ and $\UIn$, respectively.
The linear dependence of $\HI$ on $\pot$ and 
Lemma$\,$\ref{lem:EE} imply that 
  $\bigl\{\HIn \bigr\}$ is a 
Cauchy sequence in  $\Al$ with respect to the $\cstar$-norm.
 Since $\Al$ is a $\cstar$-algebra, 
  $\bigl\{\HIn \bigr\}$ has a unique limit in $\Al$, 
which will be denoted by $\HIinf$.

Since $\UI=\EI \bigl(\HI\bigr)$ with $\Vert \EI \Vert=1$, 
$\{ \UIn\}$ is also a Cauchy sequence in  $\Al$, 
has a unique 
 limit $\UIinf$,
and  $\UIinf=\EI\bigl(\HIinf\bigr)$.

For each  finite subset $\I$ of $\Znu$,
 $\{\potIn\}$ also converges to the potential 
 $\potinf(\I)$ for $\UIinf$
in the $\cstar$-norm
  because 
   $\potI$ is a finite linear combination  of  $\UJ$, 
$\J \subset \I$ due to (\ref{eq:POTfrUI2}), 
and $\{\UJn\}$ converges to $\UJinf$
 in the $\cstar$-norm for every such $\J$. 
 For any finite subsets $\I$, $\J$ of $\Znu$,
we obtain
\begin{eqnarray*}
&\ &\sum_{\K} 
\bigl\{ \potKinf;\ \K \cap \I \ne \emptyset,\ \K \subset \J \bigr\} \nonum \\
&=& \sum_{\K} \limn
 \bigl\{ \potKn;\ \K \cap \I \ne \emptyset,\ \K \subset \J \bigr\} %
= \limn \sum_{\K} 
 \bigl\{ \potKn;\ \K \cap \I \ne \emptyset,\ \K \subset \J \bigr\} \nonum \\
&=&\limn \EJ\bigl(\HIn \bigr) %
= \EJ \Bigl( \limn \HIn  \Bigr) \nonum \\
&=& \EJ  \bigl(\HIinf\bigr),
\end{eqnarray*}
 where the third equality is due to (\ref{eq:HJIsum}).
Hence, by (\ref{eq:EZlim}) we have 
\begin{eqnarray*}
\lJlim  \left(\sum_{\K} 
\bigl\{ \potKinf;\ \K \cap \I \ne \emptyset,\ \K \subset \J \bigr\} \right)
=\lJlim \EJ\bigl(\HIinf \bigr)=\HIinf.
\end{eqnarray*}
 Thus  $\potinf$ satisfies the condition 
($\pot$-e) in the definition
 of  $\PBI$.
 The other conditions ($\pot$-a), ($\pot$-b), 
($\pot$-c), ($\pot$-d), and ($\pot$-f) are satisfied
 since each $\potn$ satisfies them and $\limn\potIn=\potinf(\I)$
 for every finite subset $\I$ of $\Znu$.
In conclusion, we have  $\potinf \in \PBI$.

Finally,  we have
\begin{eqnarray*}
\limn\Vert \potn -\potinf\Vert=
\limn\Vert H_{n}(\{0\}) -H_{\infty}(\{0\})\Vert
=0.
\end{eqnarray*}
  We have now shown the completeness of $\PBI$.
\proofend\\
%
%
%
\subsection{Finite Range Potentials}
\label{subsec:Frange}
\begin{df}
\label{df:Frange}
$(${\rm{1}}$)$ A potential $\pot \in \PBI$ is said to be  of a finite range
 if there exists an $r\ge 0$ such that $\potI=0$
whenever 
\begin{eqnarray}
\diam(\I)=\max \bigl\{ |i-j|;\ i,j\in \I \bigr\}>r.
\end{eqnarray}
The infimum of such  $r$ is called the range of $\pot$.

$(${\rm{2}}$)$ The subspace of $\PB$ consisting of all potentials
 $\pot \in \PB$   of a finite range
 is denoted  by $\PBf$.
Furthermore, we denote
\begin{eqnarray}
\PBIf \equiv \PBf\cap \PBI.
\end{eqnarray}
\end{df}

For  $a \in \NN$,
$\Ca$  denotes
 the following cube in $\Znu$ 
\begin{eqnarray}
\label{eq:CaEQ}
 \Ca  \equiv \{x \in \Znu\ ;\  0 \le x_{i} \le a-1, 
 \ i=1, \cdots,\nu  \}.
\end{eqnarray}
We introduce the following averaged conditional 
expectation.
\begin{eqnarray}
\label{eq:EaEQ}
 \Ea  \equiv \frac{1}{|\Ca|}\sum_{i \in \Ca}E_{\Ca-i},
\end{eqnarray}
 where $|\Ca|=a^{\nu}$ is the number of  lattice points in $\Ca$,
 called the volume of $\Ca$.
(The sum in the above equation 
 is over all translates of $\Ca$ 
 which contain the origin $0 \in \Znu$.)

 For any finite subset $\I\subset \Znu$, 
 $\laI$ denotes the number of translates of $\Ca$ 
 containing $\I$. 
By definition, for any $m \in \Znu$,
\begin{eqnarray}
\label{eq:laImEQ}
 \laI=\laIm
\end{eqnarray}
 We need the following lemma in this subsection and later.
\begin{lem}
\label{lem:laI-Ca}
For a finite  $\I$,
\begin{eqnarray}
\label{eq:laIEQ}
\lim_{a\to \infty}\frac{\laI}{|\Ca|}=1
\end{eqnarray}
\end{lem}
\proof\  Let $d\in \NN$ be fixed such that 
there exists  a translate $\Cd+k\; (k \in \Znu)$ of $\Cd$  containing  $\I$.
For $a>d$, a translate of $\Ca$ contains $\I$
 if it contains $\Cd+k$.
 Hence $\laI$ is bigger than the number of translates of $\Ca$
 which contains $\Cd$, which is $(a-d+1)^{\nu}$.
Hence 
\begin{eqnarray*}
1\ge \frac{\laI}{|\Ca|}\ge \frac{(a-d+1)^{\nu}}{|\Ca|}=
\left( 1-\frac{(d-1)}{a}\right)^{\nu}\to1 \ (a\to \infty).
\end{eqnarray*}
This shows (\ref{eq:laIEQ}).
\proofend 

In order to prove that the subspace $\PBIf$ is dense in $\PBI$,
 we need the following  Lemma.
\begin{lem}
\label{lem:Ealim}
For any $A \in \Al$,
\begin{eqnarray}
\label{eq:Ealim}
\lim_{a\to \infty}\Ea(A)=A.
\end{eqnarray}
\end{lem}
\proof\ Since $\Alinfty$ is dense in $\Al$, there exists  
 $\Aeps\in \Alinfty$ for any $\varepsilon>0$ such that 
\begin{eqnarray}
\label{eq:Ealim2}
\Vert \Aeps -A\Vert< \varepsilon.
\end{eqnarray}
Let $\Aeps\in \Al(\I_{\varepsilon})$
 for a finite $\I_{\varepsilon}$.
Then there exists a sufficiently large positive integer 
$b$  such that 
 a translate of $\Cb$, say $\Cb-k$,  contains both  $0$ (the origin 
 of $\Znu$)
  and  $\I_{\varepsilon}$.
%
%
 If a translate $\Ca-i$ of $\Ca$
 contains  $\Cb-k$, then 
$E_{\Ca-i}(\Aeps)=\Aeps$ 
 because 
$\Ca-i \supset \Cb-k\supset \I_{\varepsilon}$
and $\Aeps\in \Al(\I_{\varepsilon})$.
Such $i$ belongs to $\Ca$
due to  $0 \in \Cb-k \subset \Ca-i$.
The number of translates 
 $\Ca-i$ of $\Ca$ which contains
$\Cb-k$ is equal to $\laCb$ (the number
of translates of $\Ca$
 which contains $\Cb$). Therefore,  we obtain
\begin{eqnarray*}
&&\Vert \Aeps-\Ea(\Aeps) \Vert \nonum\\
&=&\left\Vert 
 \left(1-\frac{\laCb}{|\Ca|} \right)\Aeps
-\frac{1}{|\Ca|}\sum \bigl\{E_{\Ca-i}(\Aeps);\ i \in \Ca,\; \Ca-i 
\not\supset
\Cb-k \bigr\}
\right\Vert.
\end{eqnarray*}
Hence, by using $\Vert E_{\Ca-i}(\Aeps) \Vert \le \Vert \Aeps \Vert$
 due to $\Vert E_{\Ca-i}\Vert=1$,
 we obtain 
\begin{eqnarray*}
 \Vert \Aeps-\Ea(\Aeps) \Vert
&\le &
 \left( \Bigl\{ 1-\frac{\laCb}{|\Ca|} \Bigr\} 
+ \frac{1}{|\Ca|}\bigl\{ |\Ca|-\laCb \bigr\} \right)
\Vert \Aeps \Vert \nonum \\
&=&2  \left(1-\frac{\laCb}{|\Ca|} \right) \Vert \Aeps \Vert.
\end{eqnarray*}
 By Lemma$\,$\ref{lem:laI-Ca}
\begin{eqnarray*}
\lim_{a \to \infty}\frac{\laCb}{|\Ca|} =1.
\end{eqnarray*}
Hence,  there exists $n_{\varepsilon} \in \NN$ such that for 
$a\ge n_{\varepsilon}$, 
\begin{eqnarray}
\label{eq:Aeps2}
 \Vert \Aeps-\Ea(\Aeps) \Vert
<\varepsilon.
\end{eqnarray}
Hence, for $a \ge n_{\varepsilon}$,
\begin{eqnarray*}
 \Vert A-\Ea(A) \Vert&\le& 
\Vert A-\Aeps \Vert+
\Vert \Aeps -\Ea(\Aeps)\Vert+
\Vert \Ea(\Aeps -A) \Vert\\
&<&3 \varepsilon
\end{eqnarray*}
 by (\ref{eq:Ealim2}), (\ref{eq:Aeps2}) and $\Vert \Ea \Vert=1$.
 \proofend

\begin{thm}
\label{thm:separable}
$\PBIf$ is dense in $\PBI$.
\end{thm}
\proof\ Let $\pot \in \PBI$.
 For any finite $\I \subset \Znu$ containing the origin
 $0$ of $\Znu$, 
\begin{eqnarray}
\Ea\bigl( \potI \bigr)=\frac{\laI}{|\Ca|}\potI,
\end{eqnarray}
because  $E_{\Ca-i}\bigl( \potI \bigr)=\potI$
 if $\Ca-i$ contains $\I$
 while $E_{\Ca-i}\bigl( \potI \bigr)=0$
 if $\Ca-i$ does not contain $\I$ due to $(\pot$-$\rm{d})$.
 Note that all translates of $\Ca$ which contains $\I$
 appear in the sum (\ref{eq:EaEQ})
 since  $\I$ is assumed to contain $0$. 

We now consider the following potential
\begin{eqnarray}
\potaI=\frac{\laI}{|\Ca|}\potI.
\end{eqnarray}
Due to  $\pot \in \PBI$, 
  $(\pot$-$\rm{a})$,  $(\pot$-$\rm{b})$, $(\pot$-$\rm{c})$ 
 and  $(\pot$-$\rm{d})$ for $\pota$ follow automatically.
Since $\pot\in \PBI$ is translation covariant and $\laI$
 is translation invariant under translation of $\I$
by  (\ref{eq:laImEQ}), $\pota$ satisfies the translation covariance
 $(\pot$-$\rm{f})$.
 $\pota$ is of a finite range because  
  there is no translates of $\Ca$ containing 
 $\I$ if $\diam(\I)>\sqrt{\nu} (a-1)$ and hence 
$\laI=0$ for such $\I$ and $a(\in \NN)$.
Hence  $(\pot$-$\rm{e})$ is automatically satisfied.
 Therefore we conclude that $\pota \in \PBIf$.
We compute 
\begin{eqnarray*}
\Ea\bigl( H_{\pot}(\{0\})\bigr)&=&\sum_{\J \ni 0}\frac{1}{|\Ca|}
 \sum_{i\in \Ca}E_{\Ca-i}\bigl( \potJ \bigr) \nonum \\
&=&\sum_{\J \ni 0}\frac{\laJ}{|\Ca|}
  \potJ = H_{\pota}(\{0\}),
\end{eqnarray*}
where we have used 
 $E_{\Ca-i}\bigl( \potJ \bigr)=\potJ $ for $\Ca-i\supset\J$
 and  $E_{\Ca-i}\bigl( \potJ \bigr)=0 $ for $\Ca-i\not\supset\J$
 due to $(\pot$-$\rm{d})$.
(Note that if a translate $\Ca-i$ contains $\J$,
 then $i \in \Ca$ due to $0 \in \J$
 and hence the number of $i\in \Ca$, for which $\Ca-i\supset\J$,
 is $\laJ$.) 

 By Lemma$\,$\ref{lem:Ealim}, we obtain
\begin{eqnarray*}
\lim_{a\to \infty}\Vert \pot -\pota \Vert&=&
\lim_{a\to \infty}\Vert H_\pot(\{0\}) -H_{\pota}(\{0\}) \Vert \nonum \\
&=&\lim_{a\to \infty}\Vert H_\pot(\{0\}) -\Ea \bigl( H_\pot(\{0\}) \bigr) 
\Vert=0.
\end{eqnarray*}
 This  completes  the proof. \proofend

\begin{cor}
\label{cor:separable}
$\PBI$ is a separable Banach space.
\end{cor}
\proof\  For each  $n \in \N$, the set of all $\pot \in \PBIf$
 with its range not exceeding $n$ is a finite dimensional 
  subspace of $\PBI$, because
 such $\pot$ is determined by $\potI$
 for a finite number of $\I$ containing the origin 
 and satisfying $\diam(\I) \le n$,
 and so has a dense countable subset. Taking union over $n \in \NN$, 
 we have a  countable dense subset of   $\PBIf$.
By Theorem$\,$\ref{thm:separable}, 
the same countable subset is dense in 
$\PBI$. We have now shown that $\PBI$ is  separable.
\proofend 
%
 %
%
\section{Thermodynamic Limit}%
\label{sec:THERLIM}
The van Hove limits of the densities (volume average) of extensive quantities are usually called 
 thermodynamic limits.
 We now provide their existence theorems.
The same proof as the case of spin lattice systems
 (see, e.g., \cite{BRA2}, \cite{ISR} and \cite{SIMON}) 
is applicable to the present Fermion lattice case.
We, however, present slightly simplified proof by using methods
 different from those of the known proof.
First we derive a surface energy estimate which we will find 
 useful and  crucial in the argument of the present section.
\subsection{Surface Energy Estimate}
\begin{lem}
\label{lem:SKILL}%
For $\pot \in \PBI$,
\begin{eqnarray}
\label{eq:SKILL}%
\lvH \frac{\Vert \WI \Vert }{|\I|}=0
\end{eqnarray}
\end{lem}
\proof
\ Let $\{\Ial\}$ be an arbitrary van Hove net  
 of $\Znu$.
For $n \in \Znu$ and a finite subset $\I$ of $\Znu$, let 
\begin{eqnarray*}
\WnI &\equiv&\lJlim 
 \sum_{\K}\, 
\Bigl\{ \potK;\ \K \ni n,\ \K \cap \I^{c} \ne \emptyset,\ \K \subset \J \Bigr\} \nonum \\
&=&\lJlim \Bigl(H_{\J}(\{n\})-\EI\Bigl\{ H_{\J}(\{n\})\Bigr\} \Bigr) \nonum \\
&=&\Hn -\EI\bigl\{ H(\{n\})\bigr\}.
\end{eqnarray*}
Let $\Brnz$ be the intersection of $\Brn$
(the ball with its center $n$ and  radius $r$)
 and $\Znu$.
 If  $n \in \I$ and $n \not\in \surfr(\I)$, then $\Brnz \subset \I$
 and hence
\begin{eqnarray*}
\EI\Bigl( H_{\Brnz} (\{n\})\Bigr)=H_{\Brnz}(\{n\}).
\end{eqnarray*}
Therefore, 
\begin{eqnarray*}
\WnI =\Hn-H_{\Brnz}(\{n\}) 
-\EI \Bigl\{\Hn -H_{\Brnz}(\{n\})\Bigr\}.
\end{eqnarray*}
From this, we obtain
\begin{eqnarray*}
\Vert \WnI \Vert \le 2 \Vert \Hn-H_{\Brnz}(\{n\}) \Vert.
\end{eqnarray*}
By (\ref{eq:colHI}), 
 for given $\eps >0$, we can take sufficiently large $r>0$
 (hence sufficiently large $\Bro$ )
 satisfying 
\begin{eqnarray*}
\Vert H(\{ 0\})-H_{\Broz}(\{0\})\Vert <\frac{\eps}{4}.
\end{eqnarray*}
By the translation covariance assumption on $\pot$, we have
\begin{eqnarray*}
\Vert H(\{ n\})-H_{\Brn}(\{n\})\Vert 
&=& \left\Vert \tau_{n} \Bigl\{ 
H(\{ 0\})-H_{\Broz} (\{0\}) \Bigr\} \right\Vert  \nonum \\
&=& \Vert H(\{ 0\})-H_{\Broz}(\{0\})\Vert \nonum \\
&<&\frac{\eps}{4}.
\end{eqnarray*}
Hence
\begin{eqnarray}
\label{eq:sd8}
\Vert  \WnI \Vert \le \frac{\eps}{2},
\end{eqnarray}
if $n \in \I$ and $n \not\in \surfr(\I)$.

For $\I=\bigl\{n_{1},\ldots, n_{|\I|} \bigr\}$,
 we have
\begin{eqnarray}
\label{eq:sd9}
\WI=\sum_{i=1}^{|\I|} E_{\{n_{1},\ldots,n_{i-1}\}^{c}}
W_{n_{i}}(\I)
\end{eqnarray}
 and hence 
\begin{eqnarray}
\label{eq:sd10}
\Vert \WI \Vert \le \sum_{i=1}^{|\I|} \Vert W_{n_{i}}(\I) \Vert.
\end{eqnarray}
For $n=n_{i} \not\in\surfr(\I)$, we use the  estimate 
 (\ref{eq:sd8}) for $\Vert \WnI \Vert$.
For $n=n_i \in \surfr(\I)$, we use
\begin{eqnarray*}
\Vert \WnI \Vert=\Vert \Hn - \EI\bigl( \Hn\bigr)\Vert
\le 2\Vert \Hn \Vert=2 \potnor.
\end{eqnarray*}
Then
\begin{eqnarray}
\label{eq:sd12}
\Vert \WI \Vert \le \frac{\eps}{2}\cdot |\I|+2\potnor \cdot
 |\surfr(\I)|.
\end{eqnarray}
Since $\{\Ial\}$ is a van Hove net,
 there exists $\alpha_{\eps}$ such that, for 
$ \alpha \ge \alpha_{\eps}$, 
\begin{eqnarray*}
\frac{|\surfr(\Ial)|}{|\Ial|}
<\frac{\eps}{4\Vert \pot \Vert }.
\end{eqnarray*}
 For such $\alpha$, we obtain
\begin{eqnarray*}
\frac{\Vert W(\Ial) \Vert}{|\Ial|} <\eps,
\end{eqnarray*}
 which completes the proof.
\proofend\\
\begin{lem}
\label{lem:UHBUNK}
Let $\{\Ial\}$
 be a  van Hove net of 
 $\Znu$. For each $\Ial$ and $a \in \NN$,
 take a set of mutually disjoint $\nam$
 translates $\Diaa$ of $\Ca$ which are all packed in $\Ial$.
  For any $\eps>0$,  take an $\ao \in \NN$
 such that $\Vert W(\Ca)\Vert< |\Ca| \,\eps/2$ for all $a>\ao$.
For any such $a$, there exists an 
$\alo(a)$ such that, for $\alpha>\alo(a)$,
\begin{eqnarray}
\label{eq:BUNK0}
\Vert \HIal -\sum_{i=1}^{\nam}\UDiaa\Vert &<&\nvhaIam |\Ca| \eps,\\
\label{eq:BUNK1}
\Vert \UIal -\sum_{i=1}^{\nam}\UDiaa\Vert &<&\nvhaIam |\Ca| \eps,
\end{eqnarray}
 and 
\begin{eqnarray}
\label{eq:BUNK2}
1 \ge  \frac{\nam |\Ca |}{ |\Ial|} \ge 1-\frac{\eps}{\potnor}.
\end{eqnarray}
\end{lem}
\proof
\ Before we start the proof, we note that the existence of $\ao$
 is guaranteed
 by Lemma$\,$\ref{lem:SKILL}.
Let us set
\begin{eqnarray*}
\Daa\equiv \bigcup_{i=1}^{\nam}\Diaa,\quad
\Dpaa \equiv \Ial \setminus \Daa.
\end{eqnarray*}
Obviously
\begin{eqnarray*}
|\Dpaa| \le \Bigl(\nap-\nam \Bigr) |\Ca|, 
\end{eqnarray*}
and 
\begin{eqnarray*}
\nap|\Ca| \ge |\Ial| \ge \nam |\Ca|.
\end{eqnarray*}
From this, we obtain
\begin{eqnarray}
\label{eq:BUNK5}
 1 &\ge& \frac{|\Ial|}{\nap|\Ca|} \ge \frac{\nam}{\nap},  \nonum \\
 1 &\ge& \frac{\nam|\Ca|}{|\Ial|} \ge \frac{\nam}{\nap}.
\end{eqnarray}
On the other hand, 
\begin{eqnarray*}
&&\HIal-\sum_{i=1}^{\nam}U(\Diaa ) \nonum \\
&=&\sum_{i=1}^{\nam} 
E_{ \{\Daa_{1} \cup \cdots  \Daa_{i-1}\}^{c} }
 \Bigl(  W(\Diaa) \Bigr)
 +E_{\{\Daa\}^{c}}\Bigl( H(\Dpaa)\Bigr).
\end{eqnarray*}
Therefore,
\begin{eqnarray}
\label{eq:BUNK8}
\Vert\HIal-\sum_{i=1}^{\nam}U(\Diaa) \Vert &\le&
\sum_{i=1}^{\nam} \Vert W(\Diaa) \Vert+\Vert H(\Dpaa)\Vert \nonum \\
&\le&  \nam |\Ca| \cdot\frac{\eps}{2}+ \potnor |\Dpaa|,
\end{eqnarray}
 where in the second inequality
 the assumption
  $\Vert W(\Ca)\Vert< |\Ca| \,\eps/2$ 
 together with the translation covariance of $\pot$ are
used for $\Vert W(\Diaa) \Vert$, and  Lemma$\,$\ref{lem:EE} is used for  
$\Vert H(\Dpaa)\Vert$.
Due to condition ({\it{1}}) for the van Hove limit,
 there exists $\alo(a)$  for given $\epso>0$ such that,
 for $\al \ge \alo(a)$,
 \begin{eqnarray}
\label{eq:BUNK10}
0 \le  1-\frac{\nam}{\nap} <\epso.
\end{eqnarray}
If $\epso <1$, then
\begin{eqnarray*}
\nap &<&\frac{1}{1-\epso}\nam, \nonum \\
|\Dpaa| &\le& \nap \epso |\Ca| < \frac{\epso}{1-\epso}\nam |\Ca|.
\end{eqnarray*}
Now we choose $\epso$ which satisfies 
\begin{eqnarray}
\label{eq:BUNK13}
\frac{2\epso}{1-\epso} \potnor < \eps,\  {\mbox{and}} \
 (0<)\epso<1.
\end{eqnarray}
Then   from (\ref{eq:BUNK8}) and (\ref{eq:BUNK13}),
  we have
\begin{eqnarray*}
\Vert\HIal-\sum_{i=1}^{\nam}U(\Diaa) \Vert &\le&
  \nam |\Ca | \cdot\frac{\eps}{2}+ \potnor \frac{\epso}{1-\epso}
\nam|\Ca|  \nonum \\
&=& \nam|\Ca| 
\Bigl( \frac{\eps}{2}+\frac{\epso}{1-\epso}\potnor\Bigr) \nonum \\
&<&\nam |\Ca| \eps.
\end{eqnarray*}
We also have
\begin{eqnarray*}
\Vert\UIal-\sum_{i=1}^{\nam}U(\Diaa) \Vert &=&
\Vert E_{\Ial}\Bigl( H(\Ial)-\sum_{i=1}^{\nam}U(\Diaa) \Bigr)
 \Vert \nonum \\
&<&\nam |\Ca| \eps.
\end{eqnarray*}
Due to (\ref{eq:BUNK13}),
\begin{eqnarray*}
\epso< \frac{\eps}{\potnor}
\end{eqnarray*}
By (\ref{eq:BUNK5}), (\ref{eq:BUNK10}) and this inequality,
 we obtain
\begin{eqnarray*}
1 \ge  \frac{\nam |\Ca|}{ |\Ial|} \ge 1-\frac{\eps}{\potnor}.
\end{eqnarray*}
 \proofend
\subsection{Pressure}%
\label{subsec:PRESS}
\begin{thm}
\label{thm:PRESS}
Assume $\pot \in \PBI$.\\
\ $($\it{1}$)$ The following limit exists$:$
\begin{eqnarray}
\label{eq:PRESS1}
p(\pot)&\equiv&\lvH \frac{1}{|\I|} \log \tau \bigl(e^{-\HI} \bigr)  \nonum \\
&=&\lvH \frac{1}{|\I|} \log \tau \bigl(e^{-\UI} \bigr).
\end{eqnarray}
\ $($\it{2}$)$ $\ppot$
 is a convex functional of $\pot$ satisfying the following 
 continuity property$:$
\begin{eqnarray}
\label{eq:pressconti}
\Bigl| \ppot  -\ppotpsi \Bigr| \le \Vert \pot -\potpsi \Vert.
\end{eqnarray}
\end{thm}
\proof
\ We first prove ({\it{1}}) in  four  steps.\\
\underline{Step 1.} 
 We need  the following 
 well-known matrix inequality: 
\begin{eqnarray}
\label{eq:PRESSconti}
| \log \tau(e^{-A})-\log \tau(e^{-B})| \le \Vert A-B \Vert,
\end{eqnarray}
 for $A, B \in \Alinfty$.
This follows from the following computation: 
\begin{eqnarray*}
&&| \log \tau(e^{-A})-\log\tau (e^{-B})| 
= \Bigl| \int_0^1 \frac{d}{d\lam} \Bigl\{
\log \tau \bigl( e^{-\lam A-(1-\lam)B}    \bigr) 
    \Bigr\} d\lam \Bigr| \nonum \\
&=&\Bigl| \int_0^1 
\frac{  \tau \bigl( e^{-\lam A-(1-\lam)B } \cdot(B-A) \bigr)   } 
{        \tau\bigl(  e^{-\lam A-(1-\lam)B} \bigr)     } d \lam  
\Bigr| \le \Vert A-B \Vert,
\end{eqnarray*}
 where we have used  the fact that 
 ${\tau(e^{c}x)}/{\tau(e^{c})}$ for $c=c^{\ast} \in \Alinfty$
 is a state function of $x \in \Al$ and hence bounded 
 by $\Vert x\Vert$.
Setting $B=0$ and noting $\log \tau (e^{-B})=0$ for $B=0$, we have
\begin{eqnarray}
\label{eq:PRESSabove}
| \log \tau(e^{-A})|  \le \Vert  A \Vert.
\end{eqnarray}
\underline{Step 2.}  
We  use the notation in the preceding Lemma.
Because $U(\Diaa)$ with distinct $i$'s 
mutually commute due to the disjointness of $\Diaa$,
(\ref{eq:colUI}), (\ref{eq:epsCAR})  and 
($\pot$-c), we have  
\begin{eqnarray}
\label{eq:PRESS7}
&& \log \tau \Bigl(e^{\{-\sum_{i=1}^{\nam}U(\Diaa)  \}} \Bigr)
= \log \tau \Bigl( \prod_{i=1}^{\nam} e^{-U(\Diaa)}   \Bigr) \nonum \\
&=& \log \prod_{i=1}^{\nam} \tau \bigl( e^{-U(\Diaa)}\bigr)=
\sum_{i=1}^{\nam} \log \tau \bigl(e^{-U(\Diaa)} \bigr) \nonum\\
&=&\nam \log \tau \bigl( e^{-U(\Ca)}  \bigr), 
\end{eqnarray}
 where the second equality is due to the product property 
(\ref{eq:PROPRO}) of the tracial state, and the last equality follows from
 the translation covariance ($\pot$-f).  
By (\ref{eq:PRESSabove}), (\ref{eq:EE1}) and  
(\ref{eq:BUNK2}), we have
\begin{eqnarray}
\label{eq:PRESS8}
\left | \frac{\nam}{|\Ial |} \log \tau \bigl(e^{-\Uca} \bigr)-
 \frac{1}{|\Ca |}\log \tau \bigl(e^{-\Uca} \bigr)   \right|
<\eps.
\end{eqnarray}
\underline{Step 3.}
By (\ref{eq:PRESS7}), (\ref{eq:PRESS8}), 
(\ref{eq:PRESSconti}), 
 (\ref{eq:BUNK0})
  and  (\ref{eq:BUNK2}), 
\begin{eqnarray}
\label{eq:PRESS11}
&&\Bigl| \frac{1}{|\Ial |} \log \tau \bigl(e^{-\HIal} \bigr)-
 \frac{1}{|\Ca| }\log \tau \bigl(e^{-\Uca } \bigr)   \Bigr| \nonum \\
&=& 
\Bigl| \frac{1}{|\Ial |} \log \tau \bigl(e^{-\HIal} \bigr)-
\frac{1}{|\Ial|}
\log \tau \Bigl(e^{\{-\sum_{i=1}^{\nam}U(\Diaa)  \}} \Bigr)
+ \Bigl( \frac{\nam}{|\Ial |}-\frac{1}{|\Ca| } \Bigr)
\log \tau \bigl(e^{-\Uca } \bigr)   \Bigr| \nonum \\
&<&2 \eps
\end{eqnarray}
 for  any $\al>\alo(a)$.
Hence for  any $\al, \beta>\alo(a)$,
we have
\begin{eqnarray*}
\Bigl| \frac{1}{|\Ial |} \log \tau \bigl(e^{-\HIal} \bigr)
-\frac{1}{|\Ibl |} \log \tau \bigl(e^{-\HIbl} \bigr)
\Bigr|
<4 \eps.
\end{eqnarray*}
Therefore,   $\frac{1}{|\Ial |} \log \tau \bigl(e^{-\HIal} \bigr)$
 is a Cauchy net in $\R$  and has the (van Hove) limit.\\
\underline{Step 4.} Due to 
\begin{eqnarray*}
\lvH \frac{\Vert \HI -\UI \Vert}{|\I|}
= \lvH \frac{\Vert \WI \Vert}{|\I|}=0
\end{eqnarray*}
 and 
\begin{eqnarray*}
| \log \tau \bigl(e^{-\HI} \bigr)
- \log \tau \bigl(e^{-\UI} \bigr) | \le \Vert \HI -\UI \Vert,
\end{eqnarray*}
 the convergence  of $\frac{1}{|\Ial |} \log \tau \bigl(e^{-\HIal} \bigr)$
 implies that of  $\frac{1}{|\Ial |} \log \tau \bigl(e^{-\UIal} \bigr)$
   to the same value.

Now we prove ({\it{2}}).
Since 
$\HIpot$ is linear in $\pot$, we have the convexity of 
 $\log \tau \bigl(e^{-\HIpot} \bigr)$ in $\pot$
 due to the well-known convexity of  the function:
 \begin{eqnarray*}
 \lam \mapsto \log \tau(e^{(A+\lam B)})\quad
 {\text{for}}\  A=A^{\ast} \text{and}\  B=B^{\ast}.
 \end{eqnarray*}
 Hence the convexity of $\ppot$ follows.
By (\ref{eq:PRESSconti}), the linearity of $\HIpot$ in $\pot$
 and (\ref{eq:EE1}), we obtain
\begin{eqnarray*}
&&\Bigl| \frac{1}{|\I|}\log \tau \bigl(e^{-\HIpot} \bigr)-
 \frac{1}{|\I|} \log \tau \bigl(e^{-\HIpotpsi} \bigr) \Bigr| \nonum \\
&\le& \frac{1}{|\I|} \Vert \HIpot -\HIpotpsi  \Vert
= \frac{1}{|\I|} \Vert H_{\pot-\potpsi}(\I)  \Vert \nonum \\
&\le& \Vert \pot -\potpsi\Vert
\end{eqnarray*}
 for any finite $\I$.
 Hence   (\ref{eq:pressconti}) 
 follows.
\proofend

The pressure functional  $\Ppot$ of  $\pot \in \PBI$
 is conventionally defined by using the matrix 
trace in contrast to $\ppot$ in the preceding theorem 
defined in terms of the tracial state:
\begin{eqnarray}
\label{eq:PRESSDEF}
\Ppot\equiv \lvH \frac{1}{|\I|} \log \trI \bigl(e^{-\HI} \bigr)
\Bigl(=\lvH \frac{1}{|\I|} \log \trI  \bigl(e^{-\UI} \bigr) \Bigr),
\end{eqnarray}
 where $\trI$ denotes the matrix trace on $\AlI$ 
and hence $\trI=2^{|\I|} \tau$.
Therefore, for any $\pot  \in \PBI$,
\begin{eqnarray}
\label{eq:TRACEpress0}
\Ppot=\ppot + \log 2.
\end{eqnarray}
Due to the preceding theorem, we have obviously
\begin{cor}
\label{cor:TRACEpress}
Assume $\pot \in \PBI$. \\
\ {\rm{(1)}} The following limit exists$:$
\begin{eqnarray}
\label{eq:TRACEpress1}
P(\pot)&\equiv&\lvH \frac{1}{|\I|} \log \trI \bigl(e^{-\HI} \bigr)  \nonum \\
&=&\lvH \frac{1}{|\I|} \log \trI \bigl(e^{-\UI} \bigr).
\end{eqnarray}
\ {\rm{(2)}} $\Ppot$
 is a convex functional of $\pot$ satisfying the following 
 continuity property$:$
\begin{eqnarray}
\label{eq:TRACEpress2}
\Bigl| \Ppot  -\Ppotpsi \Bigr| \le \Vert \pot -\potpsi \Vert.
\end{eqnarray}
\end{cor}

\begin{rem}
We have 
\begin{eqnarray}
p(0)=0, \quad |\ppot|\le \potnor
\end{eqnarray}
 which do not hold for $\Ppot$.
\end{rem}
\subsection{Mean Energy}%
\label{subsec:ME}%
%
%
\begin{thm}
\label{thm:ME}
For  $\pot \in \PBI$ and a translation 
 invariant state $\ome$ of $\Al$, the following limit exists$:$
\begin{eqnarray}
\label{eq:ME1}
\epo &\equiv& \lvH \frac{1}{|\I|} \ome \bigl(\HI  \bigr) \nonum \\
&=& \lvH \frac{1}{|\I|} \ome \bigl(\UI  \bigr).
 \end{eqnarray} 
The mean energy $\epo$ so obtained 
is linear in $\pot$, affine in $\ome$, bounded by $\potnor$,
 and  weak$\ast$ continuous in $\ome$$:$
\begin{eqnarray}
\label{eq:MEprop1}
 e_{c\pot+d{\mit \Psi}  }(\omega)&=&
  c e_{\pot}(\omega) +d e_{\mit\Psi}(\omega)\quad 
  (c,\  d \in \R), \\
\label{eq:MEprop2}
e_{\pot}( \lam \omeone   +   (1-\lam)\ometwo)&=&
  \lam e_{\pot}( \omeone)  +(1-\lambda) e_{\pot} (\ometwo)\ 
 (0 \le\lam\le 1),  \\
\label{eq:MEprop3}
|\epo| &\le& \potnor, \\
\label{eq:MEprop4}
 \lim_{\gamma} e_{\pot}(\omer)&=&\epo,
\end{eqnarray}
 where $\pot$ and ${\mit \Psi}$ are in $\PBI$,
 $\ome, \omeone, \ometwo$ and $\omer$ are 
 in $\invstate$, and 
 $\{\omer\}$ is a net converging to $\ome$
in the weak\,$\ast$ topology.
\end{thm}
\proof
\ By the  argument leading to (\ref{eq:PRESS11})
  in  Theorem$\,$\ref{thm:PRESS},
 there exists $a \in \NN$ and $\aloa$ 
 for any given $\eps>0$
such that 
  for all $\al>\aloa$  
\begin{eqnarray}
\label{eq:ME3}
\left| \frac{1}{|\Ial|} \ome \bigl( \HIal\bigr)-\frac{1}{|\Ca |}
\ome \bigl( \Uca\bigr) \right|<2 \eps,
\end{eqnarray}
 where we can take the same  $a \in \NN$ and $\aloa$  
 uniformly in $\ome \in \state$. 
This estimate implies that
$\left\{ \frac{1}{|\Ial|} \ome \bigl( \HIal\bigr) \right\}_{\al}$
 is a Cauchy net in $\R$ and hence  converges.

Since $\ome\bigl( \HI \bigr)$ is linear in $\pot$ and affine in $\ome$,
 so is $\epo$. Due to (\ref{eq:EE1}), 
 we obtain $|\epo| \le \potnor$.

Finally we show the continuity in $\ome$.  Let $\{ \omer \}_{\gamma}$
 be a net of states converging to $\ome$ in the weak$*$ topology.
 For any $\eps>0$, we fix $a \in \NN$ satisfying 
 (\ref{eq:ME3}) for all $\al>\aloa$ and for all states $\ome$.
From the weak$*$ convergence of 
 $\{ \omer \}_{\gamma}$
  to $\ome$, 
   there exixts $\gammae$ such that for all $\gamma \ge \gammae$
\begin{eqnarray*}
\frac{1}{|\Ca |}
\Bigl| \ome \bigl( \Uca \bigr)-
\omer \bigl( \Uca\bigr) \Bigr|< \eps.
\end{eqnarray*}
Thus we have 
\begin{eqnarray*}
\left| \frac{1}{|\Ial|} \ome \bigl( \HIal\bigr)
-\frac{1}{|\Ial |} \omer \bigl(\HIal  \bigr) \right| <5 \eps,
\end{eqnarray*}
 for all $\al>\aloa$.
 By taking the van Hove limit, 
 we obtain 
\begin{eqnarray*}
 |\epo-\epor|<5 \eps
\end{eqnarray*}
 for all $\gamma \ge \gammae$.
 Hence $\epo$ is continuous in $\ome$  relative to
 the weak$*$ topology.
\proofend
\section{Entropy for Fermion Systems}
\label{sec:ENTROPY}
\subsection{SSA for Fermion Systems}
We first show the SSA property of entropy for the Fermion case,
 which is a  consequence of the  results
on the conditional expectations in $\S$$\,$\ref{sec:ENT}
 and $\S$$\,$\ref{sec:FLS}.
\begin{thm}
\label{thm:SSACAR}%
For finite subsets $\I$ and $\J$ of $\Znu$,
the strong subadditivity (SSA) of $\Shat$ holds for 
  any state $\psi$ of $\Al$\,$:$
\begin{eqnarray}
\label{eq:SSAhatCAR}
\Shat(\opsiIuJ)-\Shat(\opsiI)-\Shat(\opsiJ)+\Shat(\opsiIaJ) \leq 0,
\end{eqnarray}
 where $\opsiK$ denotes the restriction of $\psi$ to $\AlK$.
$\Shat$ in this inequality
can be replaced by $S$\,$:$
\begin{eqnarray}
\label{eq:SSACAR}
S(\opsiIuJ)-S(\opsiI)-S(\opsiJ)+S(\opsiIaJ) \leq 0.
\end{eqnarray}
\end{thm}
\proof
\ The SSA of $\Shat$ follows from Theorem$\,$\ref{thm:SSAgen}
 and Theorem$\,$\ref{thm:CARsquare}.
By (\ref{eq:normalization}) and
\begin{eqnarray*}
\log 2^{\vert \IuJ \vert}
-\log 2^{\vert \I \vert}
-\log 2^{\vert  \J \vert}
+\log 2^{\vert \IaJ \vert} =0,
\end{eqnarray*}
the SSA of $\Shat$ implies that of $S$. 
\proofend \\
{\it{Remark 1.}} The strong subadditivity can be rewritten as 
\begin{eqnarray}
\label{eq:SSA-3}
S(\psi_{123})-S(\psi_{13})-S(\psi_{23})+S(\psi_{3}) \leq 0,
\end{eqnarray}
 for any disjoint subsets $\Ih$, $\If$ and $\Imi$ of $\Znu$,
 where $\psi_{123}$ denotes the restriction of $\psi$
 to $\Al(\Ih \cup \If \cup \Imi)$, and so on.     \\
\ \\
{\it{Remark 2.}} The SSA for Fermion systems above 
does not seem to follow from those for the 
tensor product systems 
(\cite{LIEBRUSKAI73}, \cite{LIEBRUSKAI73lett}) 
in any obvious way.\\
\ \\
{\it{Remark 3.}}
 Note that the SSA for  Fermion systems holds
 whether 
  the state $\psi$ is $\Theta$-even or not. 
For two disjoint finite regions $\I$ and $\J$,
 the so-called  
 triangle inequality of entropy
\begin{eqnarray*}
|S(\opsiI)-S(\opsiJ)| \le S(\opsiIuJ)
\end{eqnarray*}
is known to hold for quantum spin lattice systems 
\cite{ARAKILIEB}. However, it can fail  
for Fermion lattice systems  
 when $\psi$ breaks $\Theta$-evenness (see 
 a concrete example in \cite{MORIYAentangle}).\\

The following is a special case of 
Theorem\,\ref{thm:SSACAR} when $\I \cap \J=\emptyset$.
\begin{cor}
\label{cor:SACAR}%
For disjoint finite subsets $\I$ and $\J$,
the following subadditivity holds.
\begin{eqnarray}
\label{eq:SAhatCAR}
\Shat(\opsiIuJ)\leq \Shat(\opsiI)+\Shat(\opsiJ),\\
\label{eq:SACAR}
S(\opsiIuJ)\leq S(\opsiI)+S(\opsiJ).
\end{eqnarray}
\end{cor}
\subsection{Mean Entropy}
 We now show the existence of mean entropy 
(von Neumann entropy density)
 for translation invariant states of $\Al$.

 For $s=(s_1, \ldots, s_{\nu}) \in \Nnu$,
 we define $\Rs$ as the following box region
 with edges of length $s_i-1$ containing $s_i$ points of $\Znu$
 and with the volume $|\Rs|=\prod_{i=1}^{\nu}s_i$.
 \begin{eqnarray}
 \Rs  \equiv \{x \in \Znu\ ;\  0 \le x_{i} \le s_i-1, \ i=1, \ldots,\nu    \}.
\end{eqnarray}
\begin{thm}
\label{thm:MENT}
Let $\omega$
 be a translation invariant state.
The van Hove limit 
\begin{eqnarray}
\label{eq:vanHove-ent}
s(\omega) \equiv \lvH
 \frac{1}{{\vert \I \vert}}
 {S(\omegaI)}
\end{eqnarray} 
exists and is given as the following infimum
\begin{eqnarray}
\label{eq:rect-ent}
s(\omega)=\inf_{s \in \Nnu} \frac{1}{\vert \Rs \vert} 
S(\omers).
\end{eqnarray}
The mean entropy functional 
 \begin{eqnarray}
\label{eq:meanentvalue}
\omega  \mapsto  s(\omega) \in [0,\log 2] 
\end{eqnarray}
defined  
 on the set $\invstate$ of translation invariant states
is affine and 
upper semicontinuous  with respect to the weak\,$*$ topology.
\end{thm} 
\proof\ 
The  SSA property of von Neumann 
entropy  proved in Theorem$\,$\ref{thm:SSACAR}
 is sufficient for the same proof of this Theorem
 as in the case of quantum spin lattice systems.
(See e.g. Proposition 6.2.38
  of \cite{BRA2}.)
\proofend

The following results about 
Lipschitz continuity
 of bounded affine functions on a state space and, in particular,
 of entropy density are known.
\begin{pro}
\label{pro:ISR108}
A bounded affine function $f$
on $\invstate$ satisfies 
\begin{eqnarray}
\label{eq:ISR108}
|f(\omeone)-f(\ometwo)|\le (M/2)\Vert \omeone-\ometwo\Vert
\end{eqnarray}
for any $\omeone,\ometwo \in \invstate$,
 where 
\begin{eqnarray*}
M\equiv 
 \sup \bigl\{ |f(\omeone)-f(\ometwo)|
\; ;\ \omeone,\ometwo\in \invstate \bigr\}.
\end{eqnarray*}
\end{pro}
%
\begin{cor}
\label{cor:ISRcont}
The mean entropy $s(\ome)$ satisfies
 \begin{eqnarray}
\label{eq:ISRcont}
|s(\omeone)-s(\ometwo)|
\le \frac{1}{2} (\log 2) \Vert\omeone-\ometwo\Vert
\end{eqnarray}
for any $\omeone, \ometwo \in \invstate$.
\end{cor}
Proposition$\,$\ref{pro:ISR108} is the first equation
 on page 108 of \cite{ISR} and Corollary$\,$\ref{cor:ISRcont}
 is Corollary IV.4.3 on the same page of \cite{ISR}. 
The inequality (\ref{eq:ISRcont})
 without $\frac{1}{2}$ factor is obtained in 
\cite{FANNES73}.
The coefficient $\frac{1}{2}\log 2$ 
is best possible,
 the equality being reached by $\ome_1=\tau$
 and any pure translation invariant state $\ome_2$ with 
 vanishing 
 mean entropy $s(\ome_2)=0$, 
 in which case $\Vert\ome_{1}-\ome_{2}\Vert=2$
   because $\pi_{\tau}$
   (type II) and $\pi_{\ome_{2}}$ 
(type I) are disjoint.
 An example of such an $\ome_2$ is given by Theorem$\,$\ref{thm:EXT}
as a `product state extension' of $\Theta$-even pure states 
$\vpi$ of $\Aliten$ ($i\in \Znu$) 
 satisfying the covariance 
 condition $\dshiftk\vpi=\vp_{i+k}$ for all $k\in \Znu$.

We  define mean entropy $\shat(\ome)$ for 
$\ome \in \invstate$ by using trace
$\tau$ instead of matrix 
trace $\trI$ for each finite $\I$: 
\begin{eqnarray}
\label{eq:shatEQ}
\hat{s}(\omega) \equiv \lvH
 \frac{1}{{\vert \I \vert}}
 {\Shat(\omegaI)}.
\end{eqnarray}
It is obviously related to $s(\ome)$ by
\begin{eqnarray}
\label{eq:shatlog2}
s(\ome)=\shat(\ome)+ \log 2,
\end{eqnarray}
for any $\ome \in \invstate$. 
%
\subsection{Entropy Inequalities for  Translation Invariant States}
\label{subsec:KAY}
In addition to Theorem$\,$\ref{thm:MENT},
the SSA property of von Neumann 
entropy plays an essential 
role in the derivation of some basic entropy 
inequalities for the present Fermion lattice systems in the 
same way as for quantum spin lattice systems. 
The following two consequences are about  
monotone properties of entropy as a function on 
the set of box regions of the lattice;
the first one is a monotone decreasing property of the 
finite-volume entropy density and the second one 
is a monotone increasing property of the entropy.
\begin{thm}
\label{thm:KAY}%
Let $\ome$ be a translation invariant state on $\A$ and
let $\Rs$ and  $\Rss$ be finite boxes of $\Znu$
such that $\Rs \subset \Rss$. 
Then 
\begin{eqnarray}
\label{eq:TANCHOmaen}
\frac{1}{\vert \Rs \vert} \Somers &\geq&
\frac{1}{\vert \Rss \vert} \Somerss,\\
\label{eq:TANCHOent}
 S(\omers) &\le& S(\omerss).
\end{eqnarray}
\end{thm}
This theorem follows from \cite{KAYS}, where
(\ref{eq:TANCHOmaen}) and (\ref{eq:TANCHOent}) are 
derived from the following  properties without any
 other input.\\
 \ $\bullet$ Positivity and finiteness of the entropy of every 
 local region,\\
 \ $\bullet$ Strong subadditivity. \\
 \ $\bullet$ Shift invariance.

In \cite{BAUM}, sufficient conditions are given for a sequence of regions    
of more general shape than boxes 
 which guarantee a monotone decreasing property 
of the form (\ref{eq:TANCHOmaen})
  for any translation invariant state $\ome$.
This result also applies to our Fermion lattice systems.

%
%
%
%
\section{Variational Principle}%
We first prove the existence of a (unique) product 
state extension of given states in any (finite or infinite) 
number of mutually disjoint regions under the condition 
that all given states except for at most one are $\Theta$-even.

This result is a crucial tool to overcome possible 
difficulties which originate in the non-commutativity
 of Fermion systems in connection with the proof 
 of variational equality in this section and in
 the equivalence proof of the variational principle with 
the KMS condition in the next section.
\subsection{Extension of Even States}
For each $\I$, $\AlI$ is invariant under $\Theta$
 and hence the restriction of $\Theta$  to $\AlI$ 
is an automorphism of $\AlI$ and will be denoted by 
the same symbol $\Theta$.
%
We need the following lemma.
\begin{lem}
\label{lem:EvenDensity}
Let $\I$ be a finite subset of $\Znu$.
Let $\vp$ be a state of $\AlI$
and $\vrho \in \AlI$
 be its adjusted density matrix $:$
\begin{eqnarray*}
\vp(A)=\tau(\vrho A)=\tau(A \vrho), \quad (A \in \AlI).
\end{eqnarray*}
Then $\vp$ is an  even state if and only if 
$\vrho$ is $\Theta$-even.
\end{lem}
\proof\ 
 Since the tracial state $\tau$ is invariant under any 
 automorphism, we obtain
\begin{eqnarray*}
\vp(A)&=&\vp\bigl( \Theta(A) \bigr)=
 \tau \bigl(\vrho \Theta(A) \bigr)=
  \tau \bigl( \Theta \bigl\{ \vrho \Theta(A) \bigr\} \bigr)\\
&=& \tau \bigl( \Theta(\vrho) A \bigr)
\end{eqnarray*}
if $\vp$ is even.
By the uniqueness of the density matrix, we have
$\Theta(\vrho)=\vrho$.

By the same computation, 
$\vp \bigl( \Theta(A)\bigr)=\vp(A)$ for every $A \in \AlI$
 if $\Theta(\vrho)=\vrho$.
\proofend 
\begin{thm}
\label{thm:EXT}
Let $\{\Ii \}$ be a (finite or infinite) family of mutually 
disjoint subsets of $\Znu$ and $\vpi$ be a state of $\AlIi$ for each $i$.
Let  $\I =\bigcup_{i} \Ii$.
Then there exists a state $\vp$ of $\AlI$ satisfying
\begin{eqnarray}
\label{eq:EXT1}
\vp(\Aih \cdots \Ain)=\prod_{j=1}^{n} \vpij(\Aij)
\end{eqnarray}
 for any  set $(\ih, \cdots, i_{n})$ of distinct 
indices and for any $\Aij \in \AlIij$ if all states  $\vpi$ 
 except for at most one  are $\Theta$-even. 
When such  $\vp$ exists, it is unique.
\end{thm}
\proof\ (Case$\,$1) A finite family of finite subsets $\{\Ii\}$,
 $i=1,\cdots,n$. 

For each $i$, let $\vrhoi$
 be the density matrix of $\vpi$:
\begin{eqnarray*}
\vpi(A)=\tau(\vrhoi A)=\tau(A \vrhoi), \quad (A \in \AlIi), \\
\vrhoi \in \AlIi,\ \vrhoi\ge 0,\  \ \vrhoi(\identitybf)=1.
\end{eqnarray*}
If $\vpi$ is $\Theta$-even, 
 then $\vrhoi$ is $\Theta$-even, namely,
\begin{eqnarray*}
\vrhoi \in \AlIip.
\end{eqnarray*}
If all states $\vpi$ except for one is even,
all $\vrhoi$ except for one belong to $\AlIip$. 
Thus  each $\vrhoi$ commutes with any $\vrhoj$.
 The product
\begin{eqnarray}
\label{eq:EXT4}
\vrho=\vrhon\cdots \vrho_{1}
\end{eqnarray}
is a product of mutually commuting non-negative hermitian 
operators and hence it is positive.
 Define
\begin{eqnarray}
\label{eq:EXT5}
\vp(A)\equiv \tau(\vrho A),\quad A \in \AlI.
\end{eqnarray}
By the product property of $\tau$ (\ref{eq:PROPRO}), we have
\begin{eqnarray*}
\vp(\Ah\cdots \An)&=&\tau(\vrho\Ah\cdots \An)=
\tau(\vrho_{n-1}\cdots \vrhoh
\Ah\cdots A_{n-1}  \An \vrhon)\\
&=& \tau(\vrho_{n-1}\cdots \vrhoh
\Ah\cdots A_{n-1}) \tau(\An \vrhon)\\
&=&
 \tau(\vrho_{n-1}\cdots \vrhoh
\Ah\cdots A_{n-1}) \vpn(\An ).
\end{eqnarray*}
Using this recursively, we obtain
\begin{eqnarray*}
\vp(\Ah\cdots \An)
=\prod_{i=1}^{n}\vpi(\Ai).
\end{eqnarray*}
This also shows $\vp(\identitybf)=1$.
 Hence the existence is proved for  Case$\,$1.

Since the monomials of the form (\ref{eq:order})
 with all indices in $\I$ are total in $\AlI$,
 the uniqueness of a state $\vp$ of $\AlI$ satisfying 
the product property (\ref{eq:EXT1}) follows.

\ (Case$\,$2) A general family  $\{\Ii\}$.

Let $\{\Lk\}$ be an increasing sequence of finite subsets of $\Znu$
 such that their union is $\Znu$. 
Set $\Ikui \equiv \Ii \cap \Lk$
and $\Iku \equiv \I \cap \Lk$ for each $k$.
 For each $k$, only a finite number  (which will be denoted 
 by $n(k)$) of $\Ikui$ are non-empty
 and all of them are finite subsets of $\Znu$.
 Note that the restriction of an even state $\vpi$
 to $\AlIkui$ is even.
 Hence we can apply the result for Case$\,$1
 to $\{\Ikui\}$. We obtain a unique product state $\vpku$ of $\AlIku$
 satisfying
\begin{eqnarray}
\vpku (\Aih \cdots A_{i_{n(k)}})=\prod_{j=1}^{n(k)} \vpkuij(\Aij),
 \quad \Aij \in \Al(\I^k_{i_j}).
\end{eqnarray}
By the uniqueness already proved, the restriction of 
$\vpku$ to $\AlIlu$
 for $l<k$ coincides with $\vplu$.
 There exists a state $\vpcirc$
 of the $\ast$-algebra $\cup_{k}\AlIku$
 defined by  
\begin{eqnarray*}
\vpcirc(A)=\vpku(A)
\end{eqnarray*}
 for $A \in \AlIku$.
Since $\cup_{k}\Iku=\I$, $\cup_{k}\AlIku$ is dense in $\AlI$.
Then there exists a unique continuous extension $\vp$
 of $\vpcirc$ to $\AlI$
 and $\vp$ is a state of $\AlI$.

Take an arbitrary index  $n$.
Let 
\begin{eqnarray*}
 A=\Ah \cdots A_{n}, \quad \Ai\in \AlIi.
\end{eqnarray*}
 Set $\Akui \equiv \ELk(\Ai) \in \AlIkui$.
 Since $\LklimZ$,
\begin{eqnarray*}
\Ai&=&\lim_{k }\Akui,\nonum \\
A&=&\lim_{k } (A^{k}_{1}\cdots A^{k}_{n}).
\end{eqnarray*}
Hence 
\begin{eqnarray*}
\vp(A)&=&\lim_{k}\vp( A^{k}_{1}\cdots A^{k}_{n}  ) \nonum \\
 &=&\lim_{k}\vpku ( A^{k}_{1}\cdots A^{k}_{n}  )
=\lim_{k}\prod_{i=1}^{n}\vp_{i}(A^{k}_{i}) \nonum \\
&=&\prod_{i=1}^{n}\vp_{i}(\Ai).
\end{eqnarray*}
Thus $\vp$ satisfies the product property (\ref{eq:EXT1}).

The uniqueness of $\vp$ is proved in the same way as Case$\,$1.
\proofend 
\ \\
{\it{Remark 1.}} 
This result is given in Theorem 5.4. of Power's Thesis
\cite{POWERS}.
\\
\ \\
{\it{Remark 2.}} 
The unique product state extension $\vp$ is even if and only if  
all $\vpi$ are even. \\
\ \\
{\it{Remark 3.}} 
The condition 
that all $\vpi$ except for at most one are $\Theta$-even can be shown to be necessary for 
 the existence of the product  state extension $\vp$ 
 satisfying  (\ref{eq:EXT1}) \cite{A.M.EXT}.
%
%
\begin{lem}
\label{lem:EXTentropy}
Let $\{\Ii \}$ be a finite
 family of mutually disjoint finite subsets 
 of $\Znu$.
 Let  $\vpi$ be a state of $\AlIi$
 for each $i$ and 
 all   $\vpi$  be $\Theta$-even with at most one exception.
Let $\vp$ be their product state extension 
 given by Theorem$\,$\ref{thm:EXT}.
Then
\begin{eqnarray}
\label{eq:EXTentropy}
S(\vp)=\sum_{i}S(\vpi),\quad \Shat(\vp)=\sum_{i}\Shat(\vpi).
\end{eqnarray}
\end{lem}
\proof\ 
This follows from the computation using the density matrix
 (\ref{eq:EXT4}).
\begin{eqnarray}
\Shat(\vp)=-\vp(\log \vrho)=-\sum_{i}\vp(\log \vrhoi)=
-\sum_{i}\vpi(\log \vrhoi)=\sum_{i}\Shat(\vrhoi).
\end{eqnarray}
Here the mutual commutativity of $\vrhoi$ is  used. 
Due to $|\I|=\sum_{i}|\Ii|$, we can replace $\Shat$ by $S$.
\proofend
\subsection{Variational Inequality}%
We have already quoted 
 the positivity of relative entropy:
 \begin{eqnarray}
\label{eq:VARI1}
 \Spsivp = 
\tau  \Bigl(\rhodensity \log \rhodensity - \rhodensity\log \rhodensitypsi
 \Bigr ) \ge 0,
\end{eqnarray}  
 where the equality holds if and only if 
$\vp=\psi$.

Recall our  notation  (\ref{eq:vpIcEQ})
 for the local Gibbs state $\vpIc$ of $\AlI$
with respect to $(\pot,\,\beta)$.
Let $\ome$ be a state of $\Al$.
 Substituting $\psi=\vpic$ and $\vp=\omeI$
 into (\ref{eq:VARI1}), we obtain
\begin{eqnarray}
\label{eq:VARI3}
S(\vpic,\,\omeI)=
-\Shat(\omeI)+\beta \ome(\UI)+ \log \tau \bigl( \ebu\bigr)
  \ge 0.
\end{eqnarray}
 Now we assume that $\ome$ is translation invariant.
 By dividing the above inequality by $|\I|$ and then taking the van Hove limit
 $\vH$, we obtain the following variational inequality
\begin{eqnarray}
\label{eq:VARI5}
\pbpot \ge \hat{s}(\ome)-\beta \epo,
\end{eqnarray}
where $\hat{s}(\ome)$ is  given by (\ref{eq:shatEQ}). 
Equivalently, we have 
\begin{eqnarray}
\label{eq:VARINE}
\Pbpot \ge s(\ome)-\beta \epo.
\end{eqnarray}
%
%
%
\subsection{Variational Equality}%
\label{subsec:VAREQ}
The variational inequality in the preceding 
 subsection is now strengthened to the following 
 variational equality.
\begin{thm}
\label{thm:VAREQ}
Let $\pot \in \PBI$. Then 
\begin{eqnarray}
\label{eq:VAREQ}
 \Pbpot=\sup_{\omega \in   \invstate} 
\Bigl\{  s(\omega)- \beta e_{\pot}(\omega)   \Bigr\},
\end{eqnarray} 
where $\Pbpot$, $s(\omega)$ and $e_{\pot}$ denote
the pressure, mean entropy and mean energy, respectively, and
$\invstate$ denotes the set of all translation invariant states
 of $\Al$. 
\end{thm}
\proof\ 
The proof below will be carried out in the same way 
as for classical or quantum lattice systems (\cite{RUELLE67} or    
 e.g., Theorem III.4.5 in \cite{SIMON}), with a help 
of the product state extension provided by Theorem$\,$\ref{thm:EXT}.

By the variational inequality (\ref{eq:VARINE}), 
we only have to find a sequence $\{ \rho_{n} \}$
of translation invariant states of $\Al$ satisfying 
\begin{eqnarray}
 \bigr\{s(\rho_{n})- \beta e_{\pot}(\rho_{n})\bigr\}
 \to \Pbpot \quad(n\to\infty).   
\label{eq:VAREQ2}
\end{eqnarray}  

For this purpose, we interrupt the proof and show 
the following lemma about mean entropy and mean energy of 
periodic states. It corresponds to Theorem$\,$\ref{thm:MENT}
 and Theorem$\,$\ref{thm:ME} for translation invariant states.

%
\begin{lem}
\label{lem:period}
Let  $a \in \NN$, $\ome$ be an $a \Znu$-invariant state and $\pot \in \PBI$.\\
\ $($\rm{1}$)$ The  mean entropy 
\begin{eqnarray}
\label{eq:aperS}
s(\ome)=   \lim_{n \to \infty} \left\{
\frac{S(\ome_{\Al(\Cna)})}{|\Cna|} \right\} 
\end{eqnarray}
exists. It is affine, weak$\ast$ upper semicontinuous
 in $\ome$ and translation invariant$:$
\begin{eqnarray}
\label{eq:Zurashi0}
 s(\ome)=s(\dshiftk(\ome)),\quad (k \in \Znu).
\end{eqnarray}
\ $($\rm{2}$)$ The 
mean energy  
\begin{eqnarray}
\label{eq:aperE}
e_{\pot}(\ome)=
 \lim_{n \to \infty} \left\{
\frac{(\ome\bigl( U(\Cna) \bigr)}{|\Cna|} \right\}
\end{eqnarray}
 exists. It is linear in $\pot$, bounded by $\potnor$,
 affine and weak$\ast$ continuous in $\ome$,
  and translation invariant$:$
\begin{eqnarray}
\label{eq:Zurashi1}
 e_{\pot}(\ome)=e_{\pot}(\dshiftk(\ome)), \quad (k \in \Znu).
\end{eqnarray}
\end{lem}
\proof\ We introduce a new lattice system $(\Al^{a},\, \Al^{a}(\I))$ 
 where the total algebra $\Al^{a}$ is equal to $\Al$ and 
 its local algebra is 
 $\Al^{a}(\I)\equiv\Al\bigl(\cup_{m \in \I}(\Cam)\bigr)$ for 
 each finite subset $\I$ of $\Znu$.

For this new
  system   $(\Al^{a},\, \{\Al^{a}(\I)\})$,
we assign its
 local Hamiltonian
\begin{eqnarray*} 
H^{a}(\I)\equiv H\bigl( \cup_{m \in \I}(\Cam) \bigr)
\end{eqnarray*}
to  each finite $\I$, where $H(\cdot)$ denotes a local Hamiltonian
of the original system  $(\Al,\, \{\AlI\})$.

If $\ome$ is  an 
 $a \Znu$-invariant state of the 
 system $(\Al,\, \{\AlI\})$, 
then it goes over to a translation 
invariant state of the new system
  $(\Al^{a},\, \{\Al^{a}(\I)\})$.

We denote  mean entropy and   mean energy of $\ome$
 for the system $(\Al^{a},\,\{\Al^{a}(\I)\})$ 
 by $s^{a}(\ome)$ and $e^{a}_{\pot}(\ome)$
 which are shown to  exist by  
 Theorem$\,$\ref{thm:MENT}
 and Theorem$\,$\ref{thm:ME}.

Because of the scale change,
we have 
\begin{eqnarray}
\label{eq:scale1}
s(\ome)&=&
 \lim_{n \to \infty} \frac{S(\ome_{\Cna})}{|\Cna|}
=   {|\Ca|}^{-1}  s^{a}(\ome), \\
\label{eq:scale2}
e_{\pot}(\ome)&=& 
\lim_{n \to \infty}\frac{(\ome\bigl( U(\Cna) \bigr)}{|\Cna|} 
={|\Ca|}^{-1} e_\pot^{a}(\ome).
\end{eqnarray}
Hence those properties of mean entropy and 
mean energy of translation invariant states given 
 in Theorem$\,$\ref{thm:MENT}
 and Theorem$\,$\ref{thm:ME} go over to those 
 for periodic states.

Now we show (\ref{eq:Zurashi0}) for any 
$a \Znu$-invariant  state $\ome$ 
and  any $k \in \Znu$.
Due to the $a\Znu$-invariance of $\ome$,  
we only have to show the assertion  for 
 any $k \in \Ca$.
%
For any $n \in \NN$, we have
\begin{eqnarray}
\label{eq:nshift0}
S\bigl( \dshiftk \ome |_{\Al({\Cna})} \bigr)=S\bigl( \ome|_{\Al(\Cnak)}\bigr),
\end{eqnarray}
 which is to be compared with 
$S\bigl( \ome|_{\Al(\Cna)} \bigr)$.

Since $k \in \Ca$, we have
\begin{eqnarray}
\Cnmoa+ a(1,\cdots,1)\subset \Cnak\subset \Cnpoa. 
\end{eqnarray}
By (\ref{eq:ENTlogn}), (\ref{eq:SACAR}), and the periodicity
 of $\ome$,
\begin{eqnarray*}
 S(\ome_{\Al(\Cnak)}) &\le&
S\bigl(\ome_{\Al(\Cnmoa)} \bigr)+\bigl\{|\Cna|-|\Cnmoa| \bigr\} \log 2,
\nonum \\
  S(\ome_{\Al(\Cnak)})&\ge& S(\ome_{\Al(\Cnpoa)})-
\bigl\{ |\Cnpoa|-|\Cna| \bigr\} \log 2.
\end{eqnarray*}

Due to 
\begin{eqnarray}
\label{eq:Zurelimit}
\lim_{n \to \infty}\frac{|\Cna|}{|\Cnmoa|}=1,\quad 
 \lim_{n \to \infty}\frac{|\Cna|}{|\Cnpoa|}=1,
\end{eqnarray} 
and (\ref{eq:nshift0}), 
 we obtain 
\begin{eqnarray*}
s(\dshiftk \ome)&=&\lim_{n \to \infty}
\frac{S\bigl(\ome_{\Al(\Cnak)} \bigr)}{|\Cna|}  \nonum  \\
&=&\lim_{n \to \infty} 
\frac{S\bigl(\ome_{\Al(\Cna)} \bigr)}{|\Cna|}=s(\ome),
\end{eqnarray*}
 which is the desired equality  (\ref{eq:Zurashi0}).

It remains to  show (\ref{eq:Zurashi1}).
Applying the inequality (\ref{eq:UIUJ}) to 
the pair  $\I=\bigl( \Cnmoa+ a(1,\cdots,1)\bigr)$,  
$\J=(\Cnak)\setminus \{\Cnmoa+ a(1,\cdots,1)\}$
  and to the pair  $\I=\bigl(\Cnmoa+ a(1,\cdots,1) \bigr)$,  
$\J=\Cna \setminus \{\Cnmoa+ a(1,\cdots,1)\}$,
 we obtain
 \begin{eqnarray*}
\Vert U(\Cna)-U(\Cnak) \Vert &\le& 
\Vert U(\Cna)-U( \I )\Vert 
+ \Vert U(\I  )  -U(\Cnak) \Vert \\ 
&\le& 2 \potnor \bigl\{|\Cna|-|\Cnmoa| \bigr\},
\end{eqnarray*}
where $\I=\bigl( \Cnmoa+ a(1,\cdots,1) \bigr)$.
Hence due to (\ref{eq:Zurelimit})
 and the periodicity of $\ome$,
\begin{eqnarray*}
e_{\pot}(\dshiftk \ome)&=&
\lim_{n \to \infty}
\frac{ \ome\bigl( U(\Cnak) \bigr) }{|\Cna|} \\
&=& \lim_{n \to \infty}
\frac{ \ome\bigl( U(\Cna) \bigr)}{|\Cna|}
=e_{\pot}(\ome),
\end{eqnarray*}
 which is the desired equality (\ref{eq:Zurashi1}).
\proofend
\ \\

Now we resume the proof of Theorem$\,$ \ref{thm:VAREQ}.\\
\underline{{\it{Proof of Theorem\ref{thm:VAREQ} }} 
({\it{continued}}) }.

Due to $\Theta$-evenness of the internal energy $\UI$ for 
every  finite $\I \subset \Znu$,   we have 
\begin{eqnarray}
\vpic \in \evenstateI.
\end{eqnarray}

Let $a \in \NN$. 
For distinct  $m \in \Znu$,
 $\bigl\{\camb\bigr\}$ are mutually disjoint and their union for 
 all $m \in \Znu$ is $\Znu$.

 We apply Theorem$\,$\ref{thm:EXT} to the local 
 Gibbs states $\vpccm\in  \evenstatecam$, $m \in \Znu$
 and obtain an even product state of $\Al$,
 which we denote by $\vpac$.

For any $k \in \Znu$, $\dshift_{ak} \vpac=\vpac$
 by the uniqueness of the 
 product state with the same component states.
Thus $\vpac$ is an $a\Znu$-invariant state. 
 
By using  $\vpac$
 we construct an averaged state $\vpachat$
which is translation invariant as follows:
\begin{eqnarray}
\label{eq:AVG}
\vpachat \equiv        
  \sum_{ m  \in  \Ca}
\frac{\dshiftm \vpac }{|\Ca|} \in 
\invstate.
\end{eqnarray}

We now show (\ref{eq:VAREQ2}) by taking $\rhon=\vpnchat$.
By affine dependence of $s$ and $e_{\pot}$
 on the space of periodic states in Lemma$\,$
\ref{lem:period},
\begin{eqnarray*}
s(\vpachat)&=&{|\Ca|^{-1}} \sum_{m \in \Ca}  
s(\dshiftm\vpac),  \\
e_{\pot}(\vpachat)&=&
 {|\Ca|^{-1}}  \sum_{m \in \Ca}  e_{\pot}(\dshiftm\vpac).
\end{eqnarray*}
Due to (\ref{eq:Zurashi0}) and (\ref{eq:Zurashi1}),
they imply 
\begin{eqnarray}
\label{eq:saverage}
s(\vpachat)&=&
s(\vpac),  \\
\label{eq:eaverage}
e_{\pot}(\vpachat)&=&e_{\pot}(\vpac).
\end{eqnarray}
By (\ref{eq:saverage}),
we have
\begin{eqnarray}
\label{eq:svpachat}
s(\vpachat)&=&s(\vpac)=\frac{1}{|\Ca|}S(\vpcCa) \nonum  \\
 &=& \frac{1}{|\Ca|} 
\Bigl\{ \log\Tr_{\Ca} \bigl(\ebuCa \bigr)+ \beta \vpac\bigl( U(\Ca)\bigr) \Bigr\},
\end{eqnarray}
 where the last 
equality is given by the substitution of an explicit form of the 
density matrix of the local Gibbs state $\vpcCa$ in Definition 
 \ref{def:LocalGibbs}.

In order to show (\ref{eq:VAREQ2}),
 we now compare 
  $\epot(\vpac)$ with $\frac{1}{|\Ca|} \vpac \bigl( \UCa  \bigr)$
 in (\ref{eq:svpachat}).
Let $k \in \NN$ and 
consider the following division of $\Cka$
 as a disjoint  union of translates of $\Ca$:
\begin{eqnarray}
\label{eq:division1}
\Cka=\bigcup_{m \in \Ck }(\Cam).
\end{eqnarray}
 We give  the lexicographic ordering  for  elements in  $\Ck$ and 
set
\begin{eqnarray*}
\Ckaum\equiv \bigcup _{m^{\prime}<m  }( \Camprime)
\end{eqnarray*}
 for $m \in \Ck$.
For any $k \in \NN$, 
\begin{eqnarray*}
U(\Cka)-\sum_{m\in \Ck  }U(\Cam) 
=\sum_{m\in \Ck  } E_{\{\Cka  \setminus {\Ckaum  } \}} W(\Cam) .
\end{eqnarray*}
By  $\Vert E \Vert \le 1$  and the translation covariance 
($\pot$-f) of the potential $\pot$, we obtain 
\begin{eqnarray}
\frac{1}{|\Cka|}
\Vert U(\Cka)-\sum_{m\in \Ck  }U(\Cam) \Vert
&\le& \frac{1}{|\Cka|} \bigl(
|\Ck|\cdot 
 \Vert W(\Ca) \Vert \bigr) \nonum\\
&=& \frac{\Vert W(\Ca) \Vert  }{|\Ca|}. 
\end{eqnarray}
Therefore, by (\ref{eq:SKILL}), 
 there exists $a_{0} \in \NN$ for any $\eps>0$ such that for all $a>a_{0}$
\begin{eqnarray}
\left\Vert  \frac{1}{|\Cka|} \Bigl\{ 
\UCka-\sum_{m\in \Ck }U(\Ca+am) \Bigr\} \right\Vert
< \eps.
\end{eqnarray}
Note that the above $a_{0}$ can be taken independent of $k\in \NN$.
For any $a \in \NN$,
\begin{eqnarray*}
\vpac \bigl( U(\Ca+am) \bigr)=\vpac\bigl( U(\Ca) \bigr),
\end{eqnarray*}
 for any $m \in \Znu$, 
due to the $a \Znu$-invariance of $\vpac$.
Therefore, we obtain 
\begin{eqnarray*}
\left|  \frac{1}{|\Cka|} 
\vpac \bigl(\UCka\bigr)
-\frac{1}{|\Ca|} \vpac \bigl( U(\Ca) \bigr)
 \right| 
< \eps,
\end{eqnarray*}
 for $a>a_{0}$.
By taking the limit $k \to \infty$, we have
\begin{eqnarray*}
\Bigl| \epot(\vpac) -\frac{1}{|\Ca|} \vpac \bigl( \UCa  \bigr)\Bigr|<\eps.
\end{eqnarray*}
 From this estimate, (\ref{eq:eaverage} ) and  (\ref{eq:svpachat} ), 
it follows that
\begin{eqnarray*}
\Bigl| s(\vpachat)-\beta \epot(\vpachat)
 -\frac{1}{|\Ca|} 
\log\Tr_{\Ca} \bigl(\ebuCa \bigr) \Bigl|< |\beta| \eps,
\end{eqnarray*}
 for all $a \ge a_{0}$.
 This proves (\ref{eq:VAREQ2}) for 
 $\rhon=\vpnchat$ in view of (\ref{eq:TRACEpress1}).
\proofend
\subsection{Variational Principle}
\label{subsec:VP}
\begin{df}
\label{df:solution}
Any translation invariant 
state $\vp$ 
 satisfying 
\begin{eqnarray}
\label{eq:VARSOL}
\Pbpot= s(\vp)-\beta\epot(\vp)
\end{eqnarray}
(namely, maximizing the functional $s-\beta \epot$)
 is called a solution of the $(\pot,\,\beta)$-variational principle
 (or a translation invariant equilibrium state for $\pot$
 at the inverse temperature $\beta$).
The set of all solutions 
 of the $(\pot,\,\beta)$-variational principle
is denoted by  $\equisetb$.
\begin{eqnarray}
\equisetb \equiv 
\left\{ \vp;\ \vp \in \invstate,
\ \Pbpot=s(\vp)-\beta \epotvp    
  \right\}.
\end{eqnarray}
\end{df}

\ \\
{\it{Remark 1.}} 
Since $\beta \epot(\vp)=e_{\beta \pot}(\vp)$, 
the condition $\vp \in \equisetb$ is equivalent to the condition that 
 $\vp$ is a solution of the 
 $(\beta\pot,\,1)$-variational principle, and hence 
 $\equisetb$  is a consistent notation.

\ \\
{\it{Remark 2.}} 
 In the usual physical convention, 
 the functional $s-\beta e_{\pot}$ is $-\beta$ times the free energy functional.\ \\
\begin{thm}
\label{thm:VAREQsolution}
For any $\pot \in \PBI$ and  $\beta \in \R$,
there exists a solution $\vp(\in \invstate)$ of 
$(\pot,\,\beta)$-variational principle, 
namely, 
\begin{eqnarray*}
\equisetb \ne \emptyset.
\end{eqnarray*}
\end{thm}
\proof\ 
 $\{\vpachat \}$ in the proof of 
Theorem\,\ref{thm:VAREQ}
has an accumulation point in $\invstate$  
by the weak$*$-compactness of $\invstate$.
 Let $\vp$ be any such accumulation point.
By the proof of Theorem\,\ref{thm:VAREQ},
the weak$\ast$ continuity of  $\epot$ 
 and the
weak$\ast$ upper semicontinuity 
 of  $s$   
in $\ome$, the state $\vp$ satisfies
\begin{eqnarray}
\Pbpot=\lim_{a\to\infty}\Bigl(s(\vpachat)-\beta \epot(\vpachat)\Bigr)
\le s(\vp)-\beta\epot(\vp).
\end{eqnarray}
By (\ref{eq:VARINE}), we obtain (\ref{eq:VARSOL})
\proofend

 Our Fermion algebra $\Al$ is not asymptotically abelian
 with respect to the lattice translations,  but if 
 $\omega$ is translation invariant state of $\Al$, 
 it is well known that the pair 
 $(\Al,\omega)$ is $\Znu$-abelian and that $\omega$ is 
 automatically even 
 (see, for example, Example 5.2.21 in  \cite{BRA2}).
 From this consideration and Theorem$\,$\ref{thm:VAREQ}, 
 we obtain the following 
 result,  which corresponds to  Theorem$\,$6.2.44 in \cite{BRA2}
  in the case of quantum spin lattice systems, by 
the same argument as for that theorem.
 
For a convex set $K$, we denote the set of 
extremal points of $K$
  by $\E(K)$.
\begin{pro}
\label{pro:EQUISET}
For  $\pot \in \PBI$
and $\beta \in \R$, 
$\equisetb$ is a simplex with $\E(\equisetb) \subset \E(\invstate)$
  and the unique barycentric
decomposition of each $\vp$ in $\equisetb$
  coincides with its unique  ergodic decomposition.
\end{pro}

\section{Equivalence  of Variational Principle and KMS Condition}
\label{sec:VPequalKMS} 
Among 5 steps for establishing the equivalence stated in the 
title
 (which are  described in $\S$$\,$\ref{sec:INTRO}),
  Step (1) ``KMS condition $\Rightarrow$ Gibbs condition"
  is obtained in Theorem$\,$\ref{thm:KtoG} 
 in $\S$$\,$\ref{subsec:KtoG}, 
 Step (4) ``dKMS condition on $\Alinfty$ $\Rightarrow$ 
dKMS condition on $\Ddelal$''
 is obtained in  Corollary$\,$\ref{cor:DKMS-KMS}, and 
  Step (5) ``dKMS condition on $\Ddelal$ 
  $\Rightarrow$ KMS condition''
 is  stated in Theorem\,\ref{thm:RoepstroffSewell}. 

 In this section, we complete the remaining two steps 
of proof by showing 
 Step (2) ``Gibbs condition $\Rightarrow$ Variational principle"
 in $\S$$\,$\ref{subsec:GtoV}
 and  Step (3) ``Variational principle 
$\Rightarrow$ dKMS condition on $\Alinfty$" in $\S$
 $\,$\ref{subsec:dKMSfVP}.
As a preparation for the latter, some tools of convex analysis 
 is gathered in $\S$$\,$\ref{subsec:CONVEX}.

\subsection{Variational Principle from Gibbs Condition}
\label{subsec:GtoV}
\begin{pro}
 \label{pro:GtoV}
 For $\pot \in \PBI$, 
each translation invariant state $\vp$ 
  satisfying $(\pot,\,\beta)$-Gibbs
  condition
 is  a solution of the $(\pot,\,\beta)$-variational principle. 
 \end{pro}
\proof
\ We follow the method of proof in {\cite{ARAKI74cmp}}. 
The Gibbs condition for $\vp$ implies 
 \begin{eqnarray}
 \vpbwstate \: {\Big|}_{\AlI}=
   \vpic
 \end{eqnarray}
for every finite subset $\I$, where $\vpIc$ is given by 
(\ref{eq:vpIcEQ}), and 
 $\vpbwstate$ denotes the normalization of $\vpbw$
  given by (\ref{eq:pertstate}).

By (\ref{eq:VARI3}) with $\ome$ replaced by $\vp$, we have 
 \begin{eqnarray}
\label{eq:GtoV1}
S(\vpic,\,\vpI) &=&
-\Shat(\vpI)+\beta \vp(\UI)+ \log \tau \bigl( \ebu\bigr)  \nonum \\
&=& -S(\vpI)+\beta \vp(\UI)+ \log \trI \bigl( \ebu\bigr).
\end{eqnarray}

Since relative entropy is non-negative 
and is monotone non-increasing  under 
restriction of states,  it follows that  
\begin{eqnarray*}
 0 \le S \bigl(\vpic,\, \vp_{\I} \bigr)\le 
S \bigl(\pertsw,\, \vp \bigr).
\end{eqnarray*}
By  (\ref{eq:pertstate}), (\ref{eq:pertstate3}) and 
 (\ref{eq:pertstate2}),
 we have   
\begin{eqnarray*}
S \bigl(\pertsw, \, \vp \bigr)=
 \log \bigr(\vpbw(\identitybf) \bigl)
-\vp\bigl(\beta \WI  \bigr)
 \le 2 \Vert \beta W_{\I} \Vert.
 \end{eqnarray*}
 From these estimates and  (\ref{eq:GtoV1}), 
  it follows that    
\begin{eqnarray*} 
  0 \le S(\vpic, \, \vpI)=
-S(\vpI)+\beta \vp(\UI)+ \log \trI \bigl( \ebu\bigr)
\le 2 \Vert  \beta W_{\I} \Vert. 
\end{eqnarray*}    
(Up to this point, the assumption of   translation invariance
  of $\vp$ is irrelevant.)

We now divide the above 
  inequality by $\vert \I \vert$
  and take the van Hove limit $\vH$.
Then by the translation invariance of $\vp$ and (\ref{eq:SKILL}), 
we obtain 
\begin{eqnarray*}
s(\vp)-\beta e_{\pot}(\vp)=P(\bpot),
\end{eqnarray*}
 which completes  the  proof.
 \proofend \\

Combining this proposition with Theorem$\,$\ref{thm:KtoG}, 
we immediately obtain the following. 
\begin{cor}
\label{cor:KtoV}
Let $\alt$ be a 
dynamics of $\Al$
 satisfying the Assumptions ${\rm{(II)}}$ and
 ${\rm{(IV)}}$ 
in $\S$$\,$\ref{sec:DYNAMICS} and 
$\pot$ be the  (translation covariant)
 standard potential uniquely corresponding to this $\alt$. 
If $\vp$ is a translation invariant  $(\alt,\,\beta)$-KMS state of $\Al$,
 then 
$\vp$ is a solution of the $(\pot,\,\beta)$-variational principle.
\end{cor}

We have now completed the proof of Theorem A.
%
\subsection{Some Tools of Convex Analysis}
\label{subsec:CONVEX}
 We use the pressure functional 
$\pot \in \PBI \mapsto \Ppot \in \R$,
 which is a norm continuous convex function on the Banach space 
 $\PBI$ due to Corollary$\,$\ref{cor:TRACEpress}.

A continuous linear functional $\al \in \PBI^{\ast}$
 (the dual of $\PBI$) is called a tangent of the functional
$P$ at $\pot \in \PBI$ if it satisfies  
\begin{eqnarray}
\label{eq:CONV1}
P(\pot+\potpsi)\ge P(\pot)+\alpha(\potpsi)
\end{eqnarray}
for all $\potpsi \in \PBI$.
%
%
%
\begin{pro}
\label{pro:tangent-energy}
For any solution $\vp$ of the  
 $(\pot,\, 1)$-variational principle, define
\begin{eqnarray}
\label{eq:CONV2}
\alvp (\potpsi) \equiv -e_{\potpsi}(\vp)
\end{eqnarray}
for all $\potpsi \in \PBI$.
Then $\alvp$ is a tangent of $\PBI$ at $\pot$.
\end{pro}
\proof\ By linear dependence (\ref{eq:MEprop1}) of $e_{\potpsi}$ on $\potpsi$,
 $\alvp$ is a linear functional on $\PBI$.
 Due to $|e_{\potpsi}(\vp)| \le \potpsinor$ given by 
(\ref{eq:MEprop3}), 
 we have $\alvp \in \PBI^{\ast}$.
Due to  the variational inequality (\ref{eq:VARINE}),
\begin{eqnarray*}
P(\pot+\potpsi) &\ge& s(\vp)-e_{\pot+\potpsi}(\vp) \nonum \\ 
 &=& s(\vp)-e_{\pot}(\vp)-e_{\potpsi}(\vp) \nonum \\ 
&=&\Ppot+\alvp(\potpsi)
\end{eqnarray*}
 for all $\potpsi \in \PBI$, 
 where the last equality is due to the assumption
 that $\vp$ is a solution  of the  
 $(\pot,\, 1)$-variational principle.
\proofend
(We will establish  the bijectivity 
 between solutions
 of the $(\pot,\,\beta)$-variational principle  
 and tangents of $P$ at $\beta\pot$ through (\ref{eq:CONV2})
in Theorem$\,$\ref{thm:tan-sol}.)

 Since $P(\pot+ k \potpsi)$ is a convex continuous function 
 of $k \in \R$ for any fixed $\pot, \potpsi \in \PBI$, 
there exist its right and left derivatives at $k=0$,
\begin{eqnarray*}
(D^{\pm}_{\potpsi} P)(\pot)=\lim_{k \to \pm 0}
\frac{ P(\pot+k \potpsi)-P(\pot) }{k}.
\end{eqnarray*}
By the convexity of $P$,
\begin{eqnarray*}
(\Dpsip P)(\pot)\ge (\Dpsim P)(\pot).
\end{eqnarray*}
If and only if  they coincide, $P(\pot+k \potpsi)$ is differentiable
 at $k=0$. Then we define
\begin{eqnarray}
\label{eq:LRbibun3}
(\Dpsi P)(\pot)=(\Dpsip P)(\pot)=(\Dpsim P)(\pot).
\end{eqnarray}
 The derivatives $(\Dpsipm P)(\pot)$ and hence 
$(\Dpsi P)(\pot)$ (when it exists) satisfy
\begin{eqnarray}
\label{eq:LRbibun4}
\bigl|(\Dpsiopm  P)(\pot)-(\Dpsitpm P)(\pot)\bigr|
&\le& \Vert\potpsio -\potpsit \Vert,\nonum \\
\bigl|(\Dpsio  P)(\pot)-(\Dpsit P)(\pot)\bigr|
&\le& \Vert\potpsio -\potpsit \Vert,
\end{eqnarray}
 as is shown by the following computation in the limit 
$k \to \pm 0$.
\begin{eqnarray*}
&&\left| \frac{\bigl\{ 
P(\pot+k\potpsio)-P(\pot)\bigr\}
-\bigl\{P(\pot+k\potpsit)-P(\pot) \bigr\}   }{k}
\right| \nonum \\
&=& \left|  \frac{P(\pot+k\potpsio)-P(\pot+k\potpsit)}{k} \right| \nonum \\
&\le& |k|^{-1}\Vert k(\potpsio-\potpsit) \Vert=\Vert \potpsio-\potpsit\Vert,
\end{eqnarray*}
 where (\ref{eq:TRACEpress2}) is used for  the inequality.
If (\ref{eq:LRbibun3}) holds for all $\potpsi$, 
then $P$ is said to be differentiable at $\pot$.
Let $\PBIdiff$ be the set of all $\pot \in \PBI$
 where $P$ is differentiable.
\begin{pro}
\label{pro:tangent-VP}
 If  $\pot \in \PBIdiff$,
\begin{eqnarray}
\alpot(\potpsi)=(\Dpsi P)(\pot),\quad (\potpsi \in \PBI),
\end{eqnarray}
 defines an $\alpot \in \PBI^{\ast}$
  which is the unique tangent of $P$
 at $\pot$. Then any solution $\vp$ 
of $(\pot,\,1)$-variational principle satisfies
\begin{eqnarray}
\label{eq:alpot-alvp}
\alpot(\potpsi)=\alvp(\potpsi),
\end{eqnarray}
for all $\potpsi \in \PBI$,
 where $\alvp$ is given by (\ref{eq:CONV2}).
\end{pro}
\proof\ 
 By Theorem$\,$\ref{thm:VAREQsolution},
 there is a solution $\vp$ of the $(\pot,\,1)$-variational 
 principle and, by Proposition$\,$\ref{pro:tangent-energy},
 $\alvp$ is  a tangent of $P$ at $\pot$.

Let $\alprim$ be any tangent of 
$P$ at $\pot \in \PBIdiff$. 
 We have for $k>0$
%
\begin{eqnarray*}
P(\pot+k \potpsi)&\ge&P(\pot)+k \alprim(\potpsi),\nonum\\
P(\pot-k \potpsi)&\ge&P(\pot)-k \alprim(\potpsi).
\end{eqnarray*}
Hence 
\begin{eqnarray*}
(\Dpsip  P)(\pot)&=&\lim_{k\to +0}
\frac{ P(\pot+k \potpsi)-P(\pot) }{k} \ge \alprim(\potpsi),\nonum \\
(\Dpsim  P)(\pot)&=&\lim_{k\to +0}
\frac{ P(\pot-k \potpsi)-P(\pot) }{(-k)} \le \alprim(\potpsi).
\end{eqnarray*}
By (\ref{eq:LRbibun3}) for $\pot \in \PBIdiff$, we obtain
\begin{eqnarray*}
\alprim(\potpsi)=(\Dpsi  P)(\pot).
\end{eqnarray*}
Then $\alprim$ is unique and 
 (\ref{eq:alpot-alvp}) holds.
\proofend
\begin{lem}
\label{lem:Alinfty-POT}
For each $A \in \Alinfty$ such that $A=A^{\ast}=\Theta(A)$, 
there exists $\potpsiA \in \PBIf$
 such that
\begin{eqnarray}
\label{eq:epsivp}
e_{\potpsiA}(\vp)=\vp(A)-\tau(A)
\end{eqnarray}
 for all translation invariant states $\vp$.
\end{lem}
\proof\ 
Let $A =A^{\ast}=\Theta(A)\in \AlI$ 
for some finite $\I$ and 
\begin{eqnarray*}
\Ah \equiv A-\tau(A)\identitybf \,(\in \AlI).
\end{eqnarray*}
Since $\EIc(\Ah)=\tau(\Ah)\identitybf=0$, there 
 exists  a unique decomposition 
%
%
\begin{eqnarray}
\label{eq:Ahdec1}
\Ah&=&\sum_{\J \subset \I \atop {\J \ne \emptyset}}\AJ,
\quad \AJ \in \AlJ, \\
\label{eq:Ahdec2}
\EK\bigl( \AJ \bigr)&=&0 \quad {\text{for}}\ \K \not\supset\J.
\end{eqnarray}
To show these formulae,
 let  
\begin{eqnarray}
\label{eq:Ahdec3}
\AJ=
\sum_{\K \subset \J}(-1)^{\vert\J\vert-\vert\K\vert} \EK(\Ah)
\end{eqnarray}
 for all non-empty $\J \subset \I$, 
 a formula in parallel with (\ref{eq:POTfrUI2}).
Then 
\begin{eqnarray}
\EJ(\Ah)=\sum_{\K \subset \J \atop {\K \ne \emptyset}}\AK
\end{eqnarray}
 for  $\J \subset \I$
by exactly the same computation 
 as Step 1 of the proof of  Lemma$\,$\ref{lem:POTfrHI}.
 (When $\J=\emptyset$, the right-hand side is interpreted as $0$ and 
 $E_{\emptyset}(\Ah)=0$.)
We have
\begin{eqnarray}
\label{eq:Ahdec4}
\AJ^{\ast}=\AJ=\Theta\bigl(\AJ\bigr)\in \AlJ,
\end{eqnarray}
because $\AJ$ is a real linear combination of 
  $\EK(\Ah)$, $\K \subset \J$, and all 
$\EK(\Ah)$ satisfy the same equation.
We note  that  Step 4 of Lemma$\,$\ref{lem:POTfrHI}
 uses only the following properties of 
$\UK$
\begin{eqnarray*}
U(\emptyset)=0,\ \tau\bigl(\UK\bigr)=0,\ \EK\bigl(\UJ\bigr)=\UK,
\end{eqnarray*}
for $\K \subset \J \subset \I$,
and that all of them  are satisfied also by  $\EK(\Ah)$.
Therefore,  
(\ref{eq:Ahdec2}) follows from the same argument as 
Step 4 of Lemma$\,$\ref{lem:POTfrHI}.

We now construct $\psiJ \in \PBIf$ for each 
 $\AJ$ in (\ref{eq:Ahdec1})
 such that 
\begin{eqnarray}
\label{eq:epsiJ}
e_{\psiJ}(\vp)=\vp\bigl(\AJ\bigr)
\end{eqnarray}
for all translation invariant states $\vp$.
Then by linear dependence of $e_{\potpsi}$
 on $\pot\in \PBI$, we obtain for 
$\potpsi=\sum_{\J \subset \I}\psiJ$ the desired relation 
(\ref{eq:epsivp}):
\begin{eqnarray*}
e_{\potpsi}(\vp)=\sum_{\J\subset \I}e_{\psiJ}(\vp)=\sum_{\J\subset\I}
\vp\bigl(\AJ \bigr)=\vp(\Ah)=\vp(A)-\tau(A).
\end{eqnarray*}

 We define a potential 
$\psiJ$ for each $\J \subset \I$, $\J\ne\emptyset$
by
\begin{eqnarray}
\label{eq:psiJ}
\psiJ(\J+m)&=&\shiftm\bigl( \AJ \bigr),\quad (m \in \Znu),\nonum \\
\psiJ(\K)&=&0 \ \ {\text{if}}\ \K 
\ {\text{is not a translate of}} \ \J.
\end{eqnarray}
Due to the property (\ref{eq:Ahdec4}) and (\ref{eq:Ahdec2}),
$\psiJ$ belongs to $\PBIf$.
 We compute
\begin{eqnarray*}
\frac{1}{|\Ca|}\vp\bigl( U_{\psiJ}(\Ca)\bigr)&=&
\frac{1}{|\Ca|}\vp\bigl( \sum \{ \psiJ(\J+m);\ \J+m\subset \Ca  \}\bigr)\nonum\\&=&\frac{N_a}{|\Ca|}\vp\bigl(\AJ \bigr),
\end{eqnarray*}
 where $N_a$ is the number of $m$ such that $\J+m\subset \Ca$.

 We now show that $\frac{N_a}{|\Ca|}\to 1$
 as $a \to \infty$.
 Since $\J+m\subset \Ca$ is equivalent to 
 $\J\subset \Ca-m$, $N_a$ is the same as $\laJ$ 
(the number of translates of $\Ca$ containing  $\J$).
By (\ref{eq:laIEQ}), 
\begin{eqnarray*}
\lim_{a \to \infty}\frac{N_a}{|\Ca|}=
\lim_{a \to \infty}\frac{\laJ}{|\Ca|}=1
\end{eqnarray*}
Hence
\begin{eqnarray*}
e_{\psiJ}(\vp)=\lim_{a \to \infty}\frac{1}{|\Ca|}
 \vp\bigl( U_{\psiJ}(\Ca) \bigr)=\vp\bigl(\AJ\bigr).
\end{eqnarray*}
\proofend
\begin{cor}
\label{cor:distinctsol}
If $\vp_{1}$ and $\vp_{2}$ are distinct solutions
 of $(\pot,\,1)$-variational principle for  $\pot \in \PBI$,
 then the corresponding tangent of $P$ at $\pot$
 are distinct, that is,  $\alvpone\ne\alvptwo$.
\end{cor}
\proof\ 
If $\vp_1\ne \vp_2$, there exists  an $A \in \Alinfty$
 such that $\vp_1(A)\ne\vp_2(A)$.
Let $\Apm=\frac{1}{2}\bigl(A\pm \Theta(A)\bigr)$.
 Then $A=\Ap+\Ami$.
 Since $\vp_1$ and $\vp_2$ are translation invariant,
  both of them are $\Theta$-even, and hence 
 $\vp_1(\Ami)=\vp_2(\Ami)=0$.
 Thus   $\vp_1(\Ap)\ne\vp_2(\Ap)$.
 So we may assume that $\Theta(A)=A$.
 Let $\Ah=\frac{1}{2}\bigl(A+\Aast\bigr),
\Af=\frac{1}{2i}\bigl(A-\Aast\bigr)$, 
 $A=\Ah+i\Af$.
Then either  $\vp_1(\Ah)\ne \vp_2(\Ah)$ or 
 $\vp_1(\Af)\ne \vp_2(\Af)$.
Since $\Ah^{\ast}=\Ah$ and $\Af^{\ast}=\Af$,
 we may assume $A=A^{\ast}=\Theta(A)$.
 Let $\potpsiA \in \PBIf$ be given as in Lemma$\,$\ref{lem:Alinfty-POT}
 for this $A \in \Alinfty$.
Then 
\begin{eqnarray*}
\alvpone(\potpsiA)&=&-e_{\potpsiA}(\vp_1)=-\vp_1(A)+\tau(A) \nonum\\
&\ne&-\vp_2(A)+\tau(A)
=-e_{\potpsiA}(\vp_2)= \alvptwo(\potpsiA).
\end{eqnarray*}
Hence $\alvpone \ne \alvptwo$.
\proofend
\begin{cor}
\label{cor:Uniquesol}
For $\pot \in \PBIdiff$, a solution of 
$(\pot,\,1)$-variational principle is unique.
\end{cor}
\proof
\ This follows from Proposition$\,$\ref{pro:tangent-VP} 
 and Corollary$\,$\ref{cor:distinctsol}.
\proofend

We will use the following result in the proof of 
Theorem$\,$\ref{thm:DKMSfromV}.
\begin{thm}
\label{thm:MAZUR}
$(${\rm{1}}$)$ The set $\PBIdiff$ of points of unique 
 tangent of $P$ 
is residual (an intersection of a countable
 number of dense open sets) and dense in $\PBI$.\\
$(${\rm{2}}$)$ For any $\pot \in \PBI$, 
any tangent of $P$ at $\pot$ 
 is contained in the weak$\ast$ closed convex hull 
 of the set $\Gamma(\pot)$ which is defined  by
\begin{eqnarray}
\label{eq:Gammapot}
\Gamma(\pot) \equiv \bigl\{  \alpha &\in& \PBI^{\ast};
\ {\text{there exists a net}}
\  {\pot_{\gamma}} \in \PBIdiff \   
{\text{such that }}
\Vert\pot_{\gamma}-\pot \Vert\to 0,\nonum \\
\ &&{\text{and }}\alpha_{\pot_{\gamma}}
\to \alpha \ 
{\text{in the weak}}\ast\  
{\text{topology of}} \  
\PBI^{\ast} \bigr\},
\end{eqnarray}
where $\alpha_{\pot_{\gamma}}$ is the unique tangent 
 of $P$ at $\pot_{\gamma}$.
\end{thm}
\proof\ (1) is Mazur's theorem \cite{MAZUR}.

 (2) is  Theorem 1 of \cite{LANROBINSON68-2} 
where the function $f$ is to be set $f(\potpsi)=P(\pot+\potpsi)$
 for our purpose.
The proof in \cite{LANROBINSON68-2} is 
 by  the Hahn-Banach
 theorem. (Separability of $\PBI$ given by Corollary$\,$
 \ref{cor:separable}
is needed for both 
(1) and (2).)
\proofend

We now show a bijective correspondence between solutions
 of the $(\pot,\,\beta)$-variational principle  
 and tangents of $P$ at $\beta\pot$.
 We first prove a lemma about stability 
 of solutions of the variational principle
 under the limiting procedure in  
(\ref{eq:Gammapot}).
\begin{lem}
\label{lem:StabilitySol}%
Let  $\{\potr\}$ 
 be a net in $\PBI$ and $\{\vpr\}$
 be a net consisting of a solution $\vpr$
 of the $(\potr,\,\betar)$-variational principle 
 for each index $\gamma$ such that 
\begin{eqnarray*}
\Vert \potr-\pot\Vert \to 0,\ (\pot \in \PBI),
\quad \betar\to \beta\in \R,\\
\vpr\to \vp \in \invstate\ 
{\text{in the weak\,$\ast$ topolgy}}   \ {\text{of}}\  \invstate.
\end{eqnarray*}
Then $\vp$ is a solution of the 
 $(\pot,\,\beta)$-variational principle.  
\end{lem}
\proof\ 
By the norm continuity (\ref{eq:TRACEpress2}) of $P$,
 the weak$\ast$ upper semicontinuity 
 of s (Theorem$\,$\ref{thm:MENT})
 and the continuous  dependence of $\epot(\vp)$
on $\pot$ in the norm topology (uniformly in $\vp$)
and on $\vp$ in the weak$\ast$ topology
(Theorem$\,$\ref{thm:ME}),
 we have 
\begin{eqnarray*}
\Pbpot&=&\lim_{\gamma}P(\betar \potr),\\
s(\vp)&\ge&\limsup_{\gamma}s(\vpr),\\
\epot(\vp)&=&\lim_{\gamma} e_{\potr}(\vpr).
\end{eqnarray*}
Since, $\vpr$ is a solution of the 
 $(\potr,\,\betar)$-variational principle, 
we have 
\begin{eqnarray*}
\Pbpotr= s(\vpr)-\betar\epotr(\vpr).
\end{eqnarray*}
Hence
\begin{eqnarray*}
\Pbpot\le  s(\vp)-\beta\epot(\vp).
\end{eqnarray*}
By the variational inequality (\ref{eq:VARINE}), 
 we have 
\begin{eqnarray*}
\Pbpot= s(\vp)-\beta\epot(\vp)
\end{eqnarray*}.
\proofend
\begin{thm}
\label{thm:tan-sol}
For any $\pot \in \PBI$ and $\beta \in \R$, 
there exists a 
 bijective affine map $\vp \mapsto \alvp$
 from the set $\equisetb$
to the set of all tangents of the functional
 $P$ at $\beta \pot$, given by 
\begin{eqnarray}
\alvp (\potpsi) = -e_{\potpsi}(\vp), \quad\potpsi \in \PBI.
\end{eqnarray}
\end{thm}
\proof\ 
  By
 {{\it Remark 1}} after Definition$\,$\ref{df:solution},
 all solutions of the 
$(\pot,\,\beta)$- and 
$(\beta\pot,\,1)$- variational principle coincide.
 Furthermore, if $\vp$ is a solution of the 
$(\pot,\,\beta)$-variational principle,
 then 
\begin{eqnarray*}
P(\beta\pot+\potpsi)&\ge& s(\vp)-e_{\beta \pot+\potpsi}(\vp)\nonum\\
&=& s(\vp)-\beta \epotvp-e_{\potpsi}(\vp)\nonum\\
&=&\Pbpot +\alvp (\potpsi).
\end{eqnarray*}
Namely $\alvp$ is a tangent of $P$ at $\beta \pot$,
 exactly the same statement as for a solution 
 $\vp$ of the $(\beta\pot,\,1)$-variational principle.
Therefore, it is enough to prove the case of $\beta=1$.

The map $\vp\mapsto \alvp$ is an affine  map 
 from the set of all solutions of 
$(\pot,\,1)$-variational principle
 into the set of all tangents of $P$ at $\pot$.
 The map is injective by Corollary$\,$\ref{cor:distinctsol}.
To show the surjectivity of the map, 
 let $\alpha$ be a tangent of $P$ at $\pot$.
By Theorem$\,$\ref{thm:MAZUR},
 there exists 
 a net $\potr\in \PBIdiff$
 such that 
$\Vert\potr-\pot \Vert\to 0$,
and $\alpha_{\potr}
\to \alpha$ in the weak$\ast$  topology of $\PBI^{\ast}$,
 where $\alpha_{\potr}$ is the unique tangent of $P$
 at $\potr$.
By Theorem$\,$\ref{thm:VAREQsolution}, there exists
 a solution $\vpr$ 
of the $(\potr,\,1)$-variational principle.
By Proposition$\,$\ref{pro:tangent-energy}, 
  $\alpha_{\vpr}$ is a tangent of $P$ at $\potr$
 and hence must coincide with the unique tangent 
  $\alpha_{\potr}$.
 Due to the weak$*$ compactness of $\invstate$,
there exists a subnet $\{\vp_{\gamma(\mu)}\}_{\mu}$
 which converges to some $\vp \in \invstate$.
By Lemma$\,$\ref{lem:StabilitySol} and by 
$\Vert\potrm-\pot \Vert\to 0$, 
$\vp$ must be a solution of the  
$(\pot,\,1)$-variational principle.
 Furthermore, for any $\potpsi \in \PBI$, we have
\begin{eqnarray*} 
\alvp (\potpsi) &=& -e_{\potpsi}(\vp) \nonum\\
&=&
  -\lim_{\mu}e_{\potpsi}(\vp_{\gamma(\mu)})
= -\lim_{\mu} \alpha_{\gamma(\mu)}(\potpsi)\\
&=&\alpha(\potpsi).
\end{eqnarray*}
Hence $\alpha=\alpha_{\vp}$ and the map 
$\vp \to \alpha_{\vp}$ is surjective.
 \proofend

\subsection{Differential KMS Condition from  Variational Principle}
\label{subsec:dKMSfVP}
In this subsection, we give a proof 
 for Step 3. 
\begin{thm}
\label{thm:DKMSfromV}
Let $\pot \in \PBI$
 and $\vp$ be a translation invariant state.
 If $\vp$ is a solution of  $(\pot,\,\beta)$-variational principle,
 then $\vp$ is a  $(\delpot,\,\beta$)-dKMS state, 
 where $\delpot\in \DB$ corresponds to $\pot$
 by the  bijective linear map of Corollary \ref{cor:DB-PB}.
\end{thm}
\begin{rem}
We note that this theorem holds for any $\pot \in \PBI$
 without any further assumption on $\pot$ and we do not need
 $\alt$. Note that the domain $D(\delpot)$ is $\Alinfty$
 by definition.
\end{rem}
First we present some estimate needed in the proof of this theorem
 in the form of the following lemma.
%
%
\begin{lem}
\label{lem:DKMSV2}
Let $\I$ and $\J$ be finite subsets of $\Znu$.
If $A \in \AlJ$, then 
\begin{eqnarray}
\label{eq:DKMSV2}
\Bigl\Vert [\UI,\, A] \Bigr\Vert \le 2 \potnor \cdot
\Vert A \Vert \cdot |\I \cap \J|. 
\end{eqnarray}
\end{lem}
\proof
\ Let $\Io$ be the complement of $\IaJ$ in $\I$. 
 Then $\Io \cap \J=\emptyset$ and hence $U(\Io)$ commutes
 with $A(\in \AlJ)$ due to $\UIo \in \AlIop \subset \AlJcommut$.
Since $\Io$ and $\IaJ$  are disjoint and have the  union $\I$,
 the following computation proves (\ref{eq:DKMSV2}).
\begin{eqnarray*}
\Bigl\Vert [\UI, \,A] \Bigr\Vert&=&
 \Bigl\Vert [ \UI-U(\Io), \,A] \Bigr \Vert \nonum \\
 &\le& 2 \Vert \UI-U(\Io)   \Vert \, \Vert A \Vert
\nonum \\
  &\le& 2 \potnor \cdot\Vert A \Vert \cdot |\IaJ|,
\end{eqnarray*}
 where the last inequality is due to (\ref{eq:UIUJ}).
\proofend \\
\underline{{{\it{Proof of Theorem$\,$\ref{thm:DKMSfromV}}}}} :\\
We note that $(\pot,\,\beta)$-variational principle and 
  $(\beta\pot,\,1)$-variational principle
 are the same and 
$(\delpot,\,\beta)$-dKMS condition and 
 $(\delpotb,\,1)$-dKMS condition are the same.
 By taking $\beta \pot$ as a new $\pot$, we only have to prove
 the case $\beta=1$.

Let $\vpachat$ be the  translation invariant state 
 defined by (\ref{eq:AVG}) in the proof of Theorem\,\ref{thm:VAREQ}.
 Let $\vp$ be any accumulation point of $\{\vpachat \}_{a \in \NN}$.
 Then this $\vp$ is 
   a solution of $(\pot,\,1)$-variational principle
 as shown in Theorem$\,$\ref{thm:VAREQsolution}.

For the moment, let us assume 
 $\pot \in \PBIdiff$ (the set of $\pot \in \PBI$
 where $P$ is differentiable, defined in 
$\S$$\,$\ref{subsec:CONVEX}).
Due to the assumption $\pot \in \PBIdiff$, 
 any accumulation 
 point of $\{\vpachat \}_{a\in \NN}$
coincides with  the unique solution $\vp$ 
of $(\pot,\,1)$-variational principle,
and hence 
\begin{eqnarray}
\label{eq:vpachatlim}
\lim_{a\to \infty}\vpachat=\vp.
\end{eqnarray}

We now prove that the above $\vp$
satisfies the conditions (C-1) and (C-2) of Definition$\,$
\ref{df:DKMS} for each $A \in \Alinfty$ by using (\ref{eq:vpachatlim}).

Let $A \in \AlI$ for a finite subset $\I$  of $\Znu$. 
 Suppose $\Ca-k \supset \I$ ($a\in \NN,\ k\in \Znu$).
Since $\dshiftk \vpac$ is the 
  $({\mbox{Ad}}\, e^{it U(\Ca-k)},\,1 )$-KMS state
 on $\Al(\Ca-k)$, 
we have 
\begin{eqnarray}
\label{eq:V-D2}
{\mbox{\bf{Re}}}\bigl( \dshiftk \vpac \bigr)
\Bigl( A^{\ast} \bigl[iU(\Ca-k),\; A\bigr] \Bigr)=0,
\end{eqnarray}
\begin{eqnarray}
\label{eq:V-D3}
 {\mbox{\bf{Im}}} (\dshiftk \vpac)
 \bigl( A^{\ast} [iU(\Ca-k),\; A] \bigr)
 \geq  S\Bigl(\dshiftk\vpac(A \Aast),\,\dshiftk\vpac(\Aast A)  \Bigr).
 \end{eqnarray}
Our strategy of the proof is to replace
$\dshiftk \vpac$ and $[iU(\Ca-k),\; A]$ 
by $\vp$ and $\delpot(A)$, respectively,  
by using an approximation argument. 

By (\ref{eq:EZlim}) for $\JlimZ$, there exists 
 a finite subset $\Jeps$ of $\Znu$
for any given $\varepsilon >0$
 such that 
\begin{eqnarray}
\label{eq:V-D7}
\Vert \HI -\EJ\bigl( \HI \bigr)\Vert <\epsilon,
\end{eqnarray}
 for all $\J \supset \Jeps$.

Let $b$ be sufficiently large so that 
there exists  a translate $\Cblo$ of $\Cb$ containing 
both $\I$ and $\Jeps$.

We will use 
 the following convenient expression for $\vpachat(\in \invstate)$
 which is equivalent to (\ref{eq:AVG}):
\begin{eqnarray}
\label{eq:AVG-2}
\vpachat=\dshiftl\vpachat=  \sum_{ m  \in  \Ca}
\frac    
{   \dshiftlm \vpac   }
{|\Ca|} = \sum_{ m  \in  (\Ca+l)}
\frac    
{   \dshiftm \vpac   }
{|\Ca|}, 
\end{eqnarray}
 for any $l \in \Znu$.
 We will take $l=\lo$.
We divide $\Calo$ into the  following two disjoint subsets when 
$a>b$:
\begin{eqnarray}
\label{eq:divide}
\Cone\equiv \Camb+\lo, \quad \Ctwo\equiv\bigl(\Calo\bigr)\setminus\Cone.
\end{eqnarray}
Then 
\begin{eqnarray}
\label{eq:V-D8}
\Ca-k\supset \Cblo\supset\I\cup\Jeps
\end{eqnarray}
 if $k \in \Cone$, while
\begin{eqnarray}
\label{eq:V-D6}
\frac{|\Ctwo|}{|\Ca|}=\Bigl(1-\frac{|\Camb|}{|\Ca|}\Bigr)
\rightarrow 0,
\end{eqnarray}
as $a \to \infty$.

For $k\in \Cone$, $A(\in \AlI)$ belongs to $\Al(\Ca-k)$
 due to $\I\subset \Ca-k$.
By using the general property of the conditional expectation,
we have 
\begin{eqnarray*}
i\bigl[U(\Ca-k),\; A\bigr]&=&
iE_{\Ca-k}\bigl(\bigl[H(\Ca-k),\; A\bigr]\bigr)=
iE_{\Ca-k}\bigl(\bigl[\HI,\; A\bigr]\bigr) \nonum\\
&=&i\bigl[E_{\Ca-k}\bigl(\HI\bigr),\; A\bigr].
\end{eqnarray*}
By (\ref{eq:V-D7}) for $\J=\Ca-k(\supset \Jeps)$, 
this implies 
\begin{eqnarray*}
\bigl\Vert i\bigl[\HI,\; A\bigr]-i\bigl[U(\Ca-k),\; A\bigr] \bigr\Vert 
 < 2 \varepsilon \Vert A \Vert.
\end{eqnarray*}
Noting that $\delpot(A)=i\bigl[\HI,\; A\bigr]$, we have 
\begin{eqnarray}
\label{eq:V-D11}
\bigl\Vert \delpot(A)-i\bigl[U(\Ca-k),\; A\bigr] \bigr\Vert 
 < 2 \varepsilon \Vert A \Vert.
\end{eqnarray}
It follows from (\ref{eq:V-D2}) and (\ref{eq:V-D11})
 that 
\begin{eqnarray}
\label{eq:V-D13}
\Bigl|  {\mbox{\bf{Re}}} \bigl( \dshiftk \vpac \bigr)
\Bigl( A^{\ast} \delpot(A) \Bigr)
 \Bigr| < 2 \varepsilon \Vert A \Vert^{2}
\end{eqnarray}
 for  $k \in \Cone$.
For $k \in \Ctwo$, we use the following obvious estimate.
\begin{eqnarray}
\label{eq:V-D14}
\Bigl|  {\mbox{\bf{Re}}} \bigl( \dshiftk \vpac \bigr)
\Bigl( A^{\ast} \delpot(A) \Bigr)
 \Bigr| <  \Vert \Aast\delpot(A) \Vert.
\end{eqnarray}
Substituting (\ref{eq:V-D13}) and (\ref{eq:V-D14})
into (\ref{eq:AVG-2}), 
 we obtain 
\begin{eqnarray*}
&&\Bigl|  {\mbox{\bf{Re}}} \vpachat
\Bigl( A^{\ast} \delpot(A) \Bigr)
 \Bigr| \nonum \\ 
&\le&
\Bigl|  {\mbox{\bf{Re}}} \Bigl( \sum_{k \in \Cone }
\frac{1}{|\Ca|}{\dshiftk}\vpac \Bigr)
\Bigl( A^{\ast} \delpot(A) \Bigr)
 \Bigr| +
\Bigl| {\mbox{\bf{Re}}} \Bigl( \sum_{k \in \Ctwo }
\frac{1}{|\Ca|}{\dshiftk}\vpac \Bigr)
\Bigl( A^{\ast} \delpot(A) \Bigr)
 \Bigr| \nonum \\
&\le& 2 \varepsilon \Vert A \Vert^{2}+
\frac{|\Ctwo|}{|\Ca|} \Vert \Aast \delpot(A) \Vert. 
\end{eqnarray*}
Taking the limit $a\to\infty$
 and using (\ref{eq:V-D6}),
 we obtain 
\begin{eqnarray*}
\Bigl|  {\mbox{\bf{Re}}} \vp
\Bigl( A^{\ast} \delpot(A) \Bigr)
 \Bigr|
\le
  2 \varepsilon \Vert A \Vert^{2}.
\end{eqnarray*}
Due to arbitrariness of $\varepsilon>0$, we obtain
\begin{eqnarray}
\label{eq:V-D18}
\Bigl|  {\mbox{\bf{Re}}} \vp
\Bigl( A^{\ast} \delpot(A) \Bigr)
 \Bigr|=0.
\end{eqnarray}
 Hence the condition (C-1) holds.

 By (\ref{eq:V-D3}) and  (\ref{eq:V-D11}),  we have
the following inequality for $k \in \Cone$,
\begin{eqnarray*}
 {\mbox{\bf{Im}}}
(\dshiftk \vpac)
 \bigl( A^{\ast} \delpot(A) \bigr)
 \geq  S\Bigl(\dshiftk\vpac(A \Aast),\,\dshiftk\vpac(\Aast A)  \Bigr)
-2 \varepsilon \Vert A \Vert^{2}.
 \end{eqnarray*}

For $k \in \Ctwo$, we use simply the following estimate.
\begin{eqnarray*}
  {\mbox{\bf{Im}}}
(\dshiftk \vpac)
 \bigl( A^{\ast} \delpot(A) \bigr)
 \geq  
-  \Vert A \delpot(A) \Vert. 
 \end{eqnarray*}
From  these inequalities, we obtain
\begin{eqnarray}
\label{eq:V-D22}
&& {\mbox{\bf{Im}}}
\vpachat\bigl( A^{\ast} \delpot(A) \bigr) \nonum \\ 
 &=& 
   {\mbox{\bf{Im}}} \Bigl( \sum_{k \in \Cone }
\frac{1}{|\Ca|}{\dshiftk}\vpac \Bigr)
\Bigl( A^{\ast} \delpot(A) \Bigr)
 + 
 {\mbox{\bf{Im}}} \Bigl( \sum_{k \in \Ctwo }
\frac{1}{|\Ca|}{\dshiftk}\vpac \Bigr)
\Bigl( A^{\ast} \delpot(A) \Bigr)\nonum \\
&\geq&
\frac{1}{|\Ca|} 
\sum_{k \in \Cone }
S\Bigl(\dshiftk\vpac(A \Aast),\,\dshiftk\vpac(\Aast A)  \Bigr)\nonum\\
&& \quad \quad \quad \quad 
-2 \frac{|\Cone|}{|\Ca|}\varepsilon \Vert A \Vert^{2}  
-\frac{|\Ctwo|}{|\Ca|}\Vert A \delpot(A) \Vert .
\end{eqnarray}

 Due to the estimate (\ref{eq:V-D6}), the last term
tends to $0$ as $a \to \infty$, while the second last term
 tends to $-2\varepsilon \Vert A \Vert^{2} $ as $a \to \infty$. 
Due to the convexity of $S(\cdot,\cdot)$ in 
two variables, the  first term on the right-hand side 
has the following lower bound:
\begin{eqnarray}
\label{eq:V-Dlower}
\frac{1}{|\Ca|} 
\sum_{k \in \Cone }
S\Bigl(\dshiftk \vpac(A \Aast),\,\dshiftk\vpac(\Aast A)  \Bigr) 
\geq
\frac{|\Cone|}{|\Ca|} 
S\Bigl(\vpachatp(A \Aast),\,\vpachatp(\Aast A)  \Bigr),
\end{eqnarray}
where 
$\vpachatp$ is a  state of $\Al$ defined by
\begin{eqnarray*}
\vpachatp(B) \equiv \frac{1}{|\Cone|} 
\sum_{k \in \Cone }
 \dshiftk\vpac(B), \quad B \in \Al. 
\end{eqnarray*}
The difference of the states $\vpachat$ and $\vpachatp$
 can be estimated as
\begin{eqnarray*}
\vpnchatp-\vpnchat&=&
\left(\frac{1}{|\Cone|}-\frac{1}{|\Ca|} \right)
\sum_{k \in \Cone } \dshiftk\vpac
-\frac{1}{|\Ca|}  \sum_{k \in \Ctwo} \dshiftk\vpac \nonum \\
&=&\frac{|\Ctwo|}{|\Ca|}\vpnchatp-
\frac{1}{|\Ca|}\sum_{k \in \Ctwo} \dshiftk \vpac.
\end{eqnarray*}
Hence
\begin{eqnarray*}
\Vert \vpachatp-\vpachat\Vert  \le 2\frac{|\Ctwo|}{|\Ca|},
\end{eqnarray*}
which tends to $0$ as $a \to \infty$
 by (\ref{eq:V-D6}).
We note 
\begin{eqnarray*}
\lim_{a}\vpachatp(A \Aast)&=&
\lim_{a}\vpachat(A \Aast)=\vp(A \Aast) \nonum \\
\lim_{a}\vpachatp( \Aast A)&=&
\lim_{a}\vpachat(\Aast A)=\vp(\Aast A).
\end{eqnarray*}
By the lower semi-continuity of $S(\cdot,\cdot)$, we obtain
 \begin{eqnarray}
\label{eq:V-D30}
\liminf_{a}S\Bigl(\vpachatp(A \Aast),\,\vpachatp(\Aast A)  \Bigr)
\geq  S\Bigl(\vp(A \Aast),\,\vp(\Aast A)  \Bigr). 
\end{eqnarray}
Combining the estimates (\ref{eq:V-D22}), 
(\ref{eq:V-Dlower}), (\ref{eq:V-D30}) as well as (\ref{eq:V-D6}),
we obtain 
 the following inequality in the limit $a \to \infty$.
\begin{eqnarray*}
 {\mbox{\bf{Im}}}
\vp\bigl( A^{\ast} \delpot(A) \bigr) \geq 
 S\Bigl(\vp(A \Aast),\,\vp(\Aast A)  \Bigr)
-2 \varepsilon \Vert A \Vert^{2}. 
\end{eqnarray*}
 Due to arbitrariness
 of $\varepsilon$, we have 
\begin{eqnarray*}
   {\mbox{\bf{Im}}}
\vp\bigl( A^{\ast} \delpot(A) \bigr) \geq 
 S\Bigl(\vp(A \Aast),\,\vp(\Aast A)  \Bigr),
\end{eqnarray*}
for $A \in \Alinfty$.
 Hence the condition (C-2) holds.

Thus, we have shown that $\vp$ satisfies the  
$(\delpot,\,1)$-dKMS condition
 if $\vp$ is the  (unique) solution  
of $(\pot,\,1)$-variational principle
when  $ \pot \in \PBIdiff$.

For general $\pot \in \PBI$, 
 we will use  the standard argument 
of the convex analysis in the same way as \cite{LANROBINSON68-2},
 or Theorem 6.2.42 in \cite{BRA2}.

By Theorem$\,$\ref{thm:MAZUR}, any solution 
 of the $(\pot,\,1)$-variational principle can be obtained 
 by successive use of the following procedures, 
 starting with the unique solution of 
 $\vpal$ of  $(\potal,\,1)$-variational principle
for $\potal\in \PBIdiff$.\\
(1) Weak$\ast$ limits of any converging nets $\vpal$
 such that $\Vert \potal -\pot\Vert\to 0$. \\
(2) Convex combinations of limits obtained in (1).\\ 
(3) Weak$\ast$ limits of a converging net of states  
 obtained in (2).\\

By Lemma$\,${\ref{lem:stableC}},
 the conditions (C-1) and (C-2) are  stable 
 under these procedures.
As we have already shown these conditions  for $\vpal$
 when $\potal$ belongs to $\PBIdiff$, 
the same holds for any  $\pot \in \PBI$.
\proofend
%
We have now shown Theorem B.  
\section{Use of Other Entropy in the Variational Equality}
\label{sec:DYN}
We now consider  the possibility
to replace the mean  entropy 
$s(\ome)$ in Theorem$\,$\ref{thm:VAREQ} by 
other entropy.
We take up 
the CNT entropy $\hws$ with respect to the lattice 
translation automorphism group $\shift$ as one example. 
 But readers will find that any other entropy 
  will do if it has 
  those basic properties of CNT entropy which are used in 
the proof of Theorem$\,$\ref{thm:VarCNT}.
Note that it is not known whether CNT entropy is equal to the
mean entropy or not so far, either in some general context or
in the present case.
\subsection{CNT-Entropy}
\label{subsec:CNT}
The CNT-entropy is introduced by Connes-Narnhofer-Thirring 
\cite{CNT} for a single automorphism and its invariant state,
and is extended by Hudetz \cite{HUDETZ} 
to the  multi-dimensional case of the group $\Znu$
generated by a finite number(=$\nu$) of commuting automorphisms.
We will use the latter extended version
 for the group of lattice translation automorphisms 
$\shiftm$ ($m\in \Znu)$.
 
For a positive integer $k$, we consider
a finite decomposition of a state $\ome$ in the state
 space $\state$:
 \begin{eqnarray}
\label{eq:omedec}
\omega=\sum_{\dexcomma} \omega_{\dex},
 \end{eqnarray} 
 where each $i(l)$ runs over a finite subset of $\NN$,
 $l=1,\cdots,k$, and $\omega_{\dex}$ is a nonzero 
positive linear functional 
 of $\Al$. 
For each fixed $l$ and $i(l)$, let 
\begin{eqnarray}
\label{eq:DECwilfix}
\omega^{l}_{i(l)} \equiv  \!\!\!\!\!
\sum_{ {\dexcomma} \atop { i(l):   {\footnotesize {\mbox {fixed} } } 
 }} \omega_{\dex}, 
\quad 
\omehatlil \equiv \frac{\omelil}{\omelil(\identitybf)}.
\end{eqnarray}
Let $\eta(x) \equiv -x \log x$ for $x>0$ and $\eta(0)=0$.
For  finite dimensional subalgebras 
$\Ali, \Alii, \ldots, \Alk$ of   $\Al$, the 
so-called 
algebraic entropy $H_{\omega}(\Alkdex)$ is defined by
\begin{eqnarray}
\namadefsec,
\end{eqnarray}
  where the supremum is taken over all  finite decompositions 
(\ref{eq:omedec}) of $\omega$ with a fixed $k$.

If  $\omega$ is $\shift$-invariant, 
  the following limit (denoted by $\hwsN$) is 
known to exist (as the infimum over $a$)
   for  any finite dimensional subalgebra $N \subset \Al$,
 \begin{eqnarray*}
\hwsN \equiv 
\lim_{a \to \infty} \frac{1}{|\Ca|} 
H_{\omega}\bigl( N, \cdots, \shift^{k}(N), 
\cdots \shift^{a-1,\cdots,a-1}(N)
 \bigr),
\end{eqnarray*} 
 where there are $|\Ca|$ arguments for 
$H_{\ome}(\cdots)$ 
and each of them 
 is $\shift^{k}(N)$, $k\in \Ca$.
Let  $N_{1} \subset N_{2} \subset \cdots\subset N_{n}\subset\cdots$
be an increasing  sequence of finite algebras 
such that  the norm closure $\overline{\cup_{n} N_{n}}$ is 
 equal to $\Al$. By a Kolmogorov-Sinai type theorem 
(Corollary V.4 in \cite{CNT}), the CNT-entropy $\hws$ is  given by
\begin{eqnarray}
\label{eq:KStype}
\hws= \lim_{n \to \infty}\hwsn.
 \end{eqnarray} 
\subsection{Variational Equality in Terms of CNT-Entropy}
\label{subsec:Comparison} 
Let $\Ji$,$\Jii,\ldots, \Jk$ be disjoint finite subsets of $\Znu$
 with their union $\J$.
From Lemma VIII.1 in \cite{CNT} it follows that 
\begin{eqnarray}
\label{eq:HsyouS}
\Hw\bigl(\AlJi, \AlJii, \ldots, \AlJk\bigr)  \le 
S\bigl(\omegaJ\bigr). 
\end{eqnarray}
When $\omega$ is an even `product state',
the equality holds as follows (the following simple 
proof is due to a referee). 
\begin{lem}
\label{lem:CNTvnProd}
Let  $\Ji$,$\Jii,\ldots, \Jk$ be disjoint finite subsets
 with their  union  $\J$.
Let $\omega$ be a $\Theta$-even state of $\Al$.
Assume  that $\omega$ has  the following  product property\,$:$
\begin{eqnarray}
\label{eq:productKATEI}
\omega(\Ah \Af \cdots \Ak B )=
\omega(\Ah)\omega(\Af) \cdots \omega(\Ak) \omega(B),
\end{eqnarray}
where $A_j$ is an arbitrary element in $\Al(\J_{j})$ 
$(j=1,\ldots,k)$ and $B$ is an arbitrary element in $\AlJc$.
Then
\begin{eqnarray}
\label{eq:productHw}
\Hw \bigl( \AlJkdex \bigr)
=S(\omega_{\J})=\sum_{l=1}^{k} S(\omegaJl),
\end{eqnarray}
and 
\begin{eqnarray}
\label{eq:productHw2}
\Hw \bigl( \AlJkdex \bigr)
=\sum_{l=1}^{k}\Hw \bigl( \AlJl \bigr).
\end{eqnarray}
\end{lem}
\proof
\ We define  
 \begin{eqnarray*}
\EJaiplu \equiv\frac{1}{2}\bigl( {\mbox{id}}+\ThetaJaihat \bigr).
\end{eqnarray*} 
Then $\EJiJkplu  \equiv \EJiplu\cdots \EJkplu$ is 
the conditional expectation 
 from $\Al$ onto $\AlJip\otimes\cdots \otimes \AlJkp\otimes \AlJc$.
Since $\ome$ is a product state for the tensor product
 $(\AlJip\otimes\cdots \otimes \AlJkp)\otimes \AlJc$, 
there exists an $\omega$-preserving conditional 
 expectation $E^{\prime}_{\ome}$ from 
$(\AlJip\otimes\cdots \otimes \AlJkp)\otimes \AlJc$
 onto $\AlJip\otimes\cdots \otimes \AlJkp$.
Hence 
\begin{eqnarray*}
\EJiJkpluome \equiv 
E^{\prime}_{\ome} 
\EJiJkplu
\end{eqnarray*} 
 is an $\ome$-preserving conditional expectation 
 from $\Al$ onto $\AlJip\otimes\cdots \otimes \AlJkp$.
Hence
\begin{eqnarray*}
&&\Hw \bigl( \AlJkpdex \bigr)\\
&=&H_{\omega|_{\AlJip\otimes\cdots \otimes \AlJkp}}
\bigl(\AlJkpdex\bigr)\\
&=&\sum_{l=1}^{k} S(\omegaJlp)
=S(\omega_{\J}).
\end{eqnarray*} 
On the other hand, 
\begin{eqnarray*}
\Hw \bigl( \AlJkpdex \bigr) 
&\le& \Hw \bigl( \AlJkdex \bigr)\\ 
&\le& S(\omega_{\J}).
\end{eqnarray*} 
\proofend
%
%
%
We are now in a position to give the 
 main theorem of this subsection.
\begin{thm}
\label{thm:VarCNT}
Assume the same conditions on $\pot$ as Theorem$\;$\ref{thm:VAREQ}.
Then 
\begin{eqnarray}
\label{eq:VarCNT}
 \Pbpot=\sup_{\omega \in   \invstate} 
[ \hwshift- \beta\epot(\omega)                  ],
\end{eqnarray} 
where $\hwshift$ is the CNT-entropy of $\ome$
 with respect to the lattice translation $\shift$.
\end{thm}
\proof
\ 
Based on Lemma\,\ref{lem:CNTvnProd}, the proof will go in 
the same as 
 the case of  quantum lattice systems \cite{MORIYAvariation}.
  Basic properties of 
the CNT-entropy to which we use  in the proof  
 are as follows.\\
\ (i) Covariance under an automorphism of $\Al$
(the adjoint action on states and conjugacy action 
on the shift).
\\ \ (ii) Scaling property under 
the scaling of the  automorphism group.\\ 
\ (iii) Concave dependence on states.

Due to (\ref{eq:HsyouS}),
 we have 
\begin{eqnarray}
\label{eq:hsyous}
\hwshift \le s(\ome),
\end{eqnarray}
for any translation invariant state $\ome$.
Hence the variational inequality (\ref{eq:VARINE})
 obviously holds when $s(\ome)$ is 
 replaced by $\hwshift$.

Due to Lemma$\;$\ref{lem:CNTvnProd} and the product property 
 of $\vpac$, the translation invariant 
state $\vpachat$ defined in (\ref{eq:AVG})
will play an identical role as in the proof of 
Theorem$\;$\ref{thm:VAREQ}.
Therefore the  sequence
\begin{eqnarray*}
 \{h_{\vpachat}(\shift)- e_{\pot}(\vpachat)\}
\end{eqnarray*}
tends to the supremum value 
 $P(\pot)$ of  the variational inequality  as  $a \to \infty$.
Hence the theorem  follows.
\proofend\\
\begin{rem}
(iii) is a  general property of CNT-entropy 
(see e.g. \cite{ST2000rev}) and is enough for the proof. 
But in the situation of the above proof, the affinity 
 holds due to the specific nature of the states
 to be considered. 
\end{rem}
\

The preceding result is the variational equality.
We are then interested in the variational principle. 
\begin{pro}
\label{pro:ifCNT}
Suppose 
that a translation invariant state $\vp$
 satisfies 
\begin{eqnarray}
\label{eq:ifCNT1}
 \Pbpot=\hvpshift- \beta\epot(\vp).
\end{eqnarray} 
Then $\vp$ is a solution of the 
 $(\pot,\,\beta)$-variational principle
 and 
\begin{eqnarray}
\label{eq:ifCNT2}
 \hvpshift=s(\vp).
\end{eqnarray} 
\end{pro}
\proof
\ By (\ref{eq:HsyouS}), we have 
\begin{eqnarray*}
 s(\vp)- \beta\epot(\vp)\ge \hvpshift- \beta\epot(\vp)=\Pbpot.
\end{eqnarray*} 
By the variational inequality (\ref{eq:VARINE}), we have 
\begin{eqnarray}
\label{eq:ifCNT3}
 s(\vp)- \beta\epot(\vp)=\Pbpot.
\end{eqnarray} 
Therefore $\vp$ is a 
 solution of the 
 $(\pot,\,\beta)$-variational principle. 
From (\ref{eq:ifCNT1}) and (\ref{eq:ifCNT3}),
 we obtain (\ref{eq:ifCNT2}).

\ \\
{\it{Remark 1.}} 
We have no result about the existence theorem for a 
solution of the variational principle 
(\ref{eq:ifCNT1}) in terms of the CNT-entropy 
for a general $\pot\in \PBI$ 
(like Theorem$\,$\ref{thm:VAREQsolution})
nor the
stability of solutions of such a variational principle 
 (like Lemma$\,$\ref{lem:StabilitySol}), 
the obstacle in applying the usual method being absence
 of any result about weak\,$\ast$ upper semicontinuity of $\hws$ in $\ome$.
 
In this sense, Proposition$\,$\ref{pro:ifCNT}
 is a superficial result, 
  and Theorem$\;$\ref{thm:VAREQ}
 is short of `the variational principle' in terms of the CNT-entropy.
See also the discussion in  Section 4 
 of \cite{MORIYAvariation}. 

%
%
\ \\
{\it{Remark 2.}} 
Although we have used CNT entropy throughout this section,
other entropy such as 
$\htwsig$ defined by Choda
\cite{CHODA} can be substituted into $\hwshift$,
yielding similar results.
\section{Discussion}%
\label{sec:DISCUSSION}
The following are some of remaining problems 
about equilibrium statistical mechanics
 of Fermion lattice systems which are not covered
 in this paper.

\ \\
1. \underline{\bf{Dynamics which does not commute with $\Theta$}}

Obviously, there is an inner one-parameter group of 
$\ast$-automorphisms which does not commute with $\Theta$.
Examples of outer dynamics not commuting with 
$\Theta$ can be constructed in the 
following way (suggested by one of referees).
  Let $\{\I_i\}_{i=l,2,\cdots}$ be a partition of the lattice
  $\Znu$ into mutually disjoint finite subsets $\I_i$
  and let $\J_j \equiv \cup_{i\le j}\I_i$.
  Choose a self-adjoint $b_i$ in $\Al(\Ii)_{-}$ for each $i$ and
  set $\pot(\J_i)\equiv v_{\J_{i-1}}b_i$ 
 where $\vJ$ is given by (\ref{eq:vIEQ}). By
  Theorem \ref{thm:SOUGOUI}(1), they mutually commute and
  $\pot(\J_i)\in \Al(\J_{i-1})^{\prime}$ for
  each $i$. Hence $\alt^{(i)}\equiv {\mbox{Ad}} e^{
  it \pot (\J_i)}$, $i=1,2,\cdots$, are mutually
  commuting dynamics of $\Al$, $\alt^{(i)}$ leaving
  elements of $\Al(\J_{i-1})$ invariant. Hence $\alt
  \equiv \prod_{i=1}^{\infty} \alt^{(i)}$
 gives a dynamics of $\Al$ satisfying
  $\Theta \alt={\alpha}_{-t}{\Theta}$.
 (Namely, its generator anticommutes with $\Theta$.) 
The corresponding potential is given
  by $\pot(\I)=0$ if $\I\ne \J_i$ for
  any $i$ and $\pot(\I)=\pot(\J_i)$ if 
$\I=\J_i$. This potential satisfies the
  standardness condition $(\pot$-$\rm{d})$ 
if each $b_i$ satisfies it for the
  set $\I_i$. By looking at the behavior of
   $\UnN^{\ast}{\alpha}_t(\UnN)=e^{-2it{\sum_{i=0}^{N}}\pot (\J_{n+i})}$ for 
$\UnN\equiv \prod_{i=0}^{N}$
  $v_{\I_{n+i}}$ as $n\to {\infty}$, the dynamics is seen
  to be outer unless $\sum_{i}\pot (\J_i)$
  is convergent.

\ \\
2. \underline{\bf{Broken  $\Theta$-invariance of equilibrium states}}

In connection with the Gibbs condition, 
 we have shown in $\S$$\,$\ref{pro:evenGibbsProduct} 
 that the perturbed state 
either by surface energy or by the local interaction 
energy
 satisfies the product property 
if and only if the equilibrium state is $\Theta$-invariant.
However, we do not know an example of an equilibrium 
state which is not $\Theta$-invariant.
Existence or non-existence of such a state seems 
 to be an important question. 
It seems to be closely related to the next problem 3.
 
Note that any translation invariant state is $\Theta$-invariant.
 So we need broken translation invariance of an equilibrium state   
 for its broken $\Theta$-invariance.\\
\ \\
3. \underline{\bf{Local Thermodynamical Stability (LTS)}}

In parallel with the case of quantum spin lattice system,
 one can formulate the local stability condition 
(\cite{ARAKISEWELL77}, 
\cite{SEWELL77}) for our Fermion lattice system.
However, there seems to be two choices of the outside system for a local algebra $\AlI$ ($\I$ finite).
(1)The commutant $\AlIcommut$. (2) $\AlIc$.
For the choice (1), all arguments in the case of quantum spin 
 lattice systems seem to go through for the 
Fermion lattice system leading to equivalence of LTS with the KMS
 condition under 
 our basic Assumptions (I), (II) and (III).

On the other hand, (2) seems to be physically 
correct choice, although we do not have  an equivalence proof for 
(2) so far.  

In this connection, the problem 2 is crucial. If all equilibrium 
 state is $\Theta$-invariant, then the choice (2) also seems
 to give the LTS which is equivalent to the KMS under our basic assumptions. 
A paper on this problem is forthcoming \cite{AMLTS}.

\ \\
4. \underline{\bf{Downstairs Equivalence}}

We may say that the dynamics $\alt$ is working upstairs
 while its generator is working downstairs.
In particular, 
our arena for the downstairs activity is $\Alinfty$.
The stair going upstairs seems to be 
not wide open. 
On the other hand, there seems to be a lot more 
room downstairs.
There, we have  established the one-to-one
 correspondence  between ($\Theta$-invariant) 
 derivations on $\Alinfty$
 and standard potentials.
 We have shown that the solution of the variational principle
 (described in terms of a translation covariant potential)
 satisfies the dKMS condition on $\Alinfty$
 (described in terms of the corresponding derivation).
 How about the converse.

 There is also the problem of equivalence of LTS condition 
 (in terms of a potential)
 and the dKMS condition on $\Alinfty$
 (in terms of the corresponding derivation)
 where the translation invariance is not needed.
   Some aspects of this problem will also be included 
 in the forthcoming paper \cite{AMLTS}.

\ \\
5. \underline{\bf{Equivalent Potentials}}

 We have introduced the notion of general potentials and 
 equivalence
 among them in $\S$$\,$\ref{subsec:GP}.
 Our theory is developed only for the unique standard potential
 among each equivalence class.
 Natural questions about general potentials arise.

Does the existence of the limits defining the 
pressure $\Pbpot$ and 
the mean energy 
$\epot(\vp)$ hold also for translation covariant general potentials
$\pot$? Assuming the existence, are the  $\Pbpot$ and 
$\epot(\vp)$ the same as those for the unique standard potential 
 $\pots$ equivalent to $\pot$? If they are different, how about the solution of their variational principle?
 
We give a partial answer to these questions.
\begin{pro}
\label{pro:EquivPot}
Let $\pot$ be a translation covariant potential 
(which satisfies $(\pot$-{\rm{a,b,c,e,f}}$)$
by definition) fulfilling the following additional condition$:$
 the surface energy 
\begin{eqnarray}
\label{eq:EquivPot1}
\WIpot= \lim_{\JlimZ} 
\sum_{\K} \bigl\{ 
{\pot(\K) ; \K \cap \I \ne  \emptyset, \;
                              \K \cap \Ic \ne \emptyset,\ \K\subset \J}
\bigr\}. 
\end{eqnarray}
satisfies
\begin{eqnarray}
\label{eq:Wkilldis}%
\lvH \frac{\Vert \WIpot \Vert }{|\I|}=0.
\end{eqnarray}

Let $\pots$ be the standard potential (in $\PBI$) which is equivalent
 to $\pot$. Then both  van Hove limits defining $\Pbpot$
 and $\epot(\ome)$ for all $\ome\in \invstate$
 exist  
 if and only if 
\begin{eqnarray}
\label{eq:CpotEQ}
\Cpot\equiv \lvH \frac{\tau\bigl(  \HIpot \bigr) }{|\I|}
\end{eqnarray}
exists.
 
If this is the case, then
the following relations hold
\begin{eqnarray}
\label{eq:PRESSgeneral}
\Pbpot&=& \lvH \frac{1}{|\I|} \log \trI \bigl(e^{-\beta\HI} \bigr)
 \nonum \\
&=&\lvH \frac{1}{|\I|} \log \trI  
\bigl(e^{-\beta\UI} \bigr)\nonum \\
&=&P(\beta\pots)-\beta \Cpot,
\end{eqnarray}

\begin{eqnarray}
\label{eq:MEgeneral}
\epo &=& \lvH \frac{1}{|\I|} \ome \bigl(\HI  \bigr) \nonum \\
&=& \lvH \frac{1}{|\I|} \ome \bigl(\UI  \bigr)\nonum \\
&=&e_{\pots}(\ome)+\Cpot.
\end{eqnarray}
Furthermore, 
 $(\pot,\,\beta$)- and $(\pots,\,\beta$)-
variational principle
 give the same set of solutions.
\end{pro}
\begin{rem}
If $\tau\bigl(\potI\bigr)=0$ for all $\I$, then 
(\ref{eq:CpotEQ}) exists and $\Cpot=0$.
Hence $\Pbpot=P(\beta\pots)$
 and $\epot(\ome)=e_{\pots}(\ome)$.
This can be achieved for any general potential
 $\pot$ by changing it to $\poto=\pot-\potzero$
 where $\potzero$ is a scalar-valued potential given by
\begin{eqnarray*}
\potzero(\I)=\tau\bigl( \potI \bigr)\identitybf.
\end{eqnarray*}
\end{rem}
\proof\ Since $\pot$ and $\pots$ are equivalent, we have
\begin{eqnarray*}
\HIpot-\HIpots \in \AlIcommut.
\end{eqnarray*}
Since $\HIpot-\HIpots$ is $\Theta$-even by ($\pot$-c)
 for $\pot$ and $\pots$, we have
\begin{eqnarray}
\label{eq:EquivPot3}
\HIpot-\HIpots \in \AlIcp.
\end{eqnarray}
Hence,
\begin{eqnarray*}
\UIpot-\UIpots&=&\EI\bigl( \UIpot-\UIpots  \bigr)\\
&=&
\EI\bigl( \HIpot-\HIpots  \bigr)- \EI\bigl( \WIpot-\WIpots  \bigr)\\
  &=& \tau\bigl( \HIpot-\HIpots  \bigr)-\EI(\WIpot-\WIpots),
\end{eqnarray*}
 due to (\ref{eq:EquivPot3}).
By $\tau(\HIpots)=0$ and $\EI(\WIpots)=0$ due to ($\pot$-d), 
we have
\begin{eqnarray*}
\UIpot-\UIpots=\tau\bigl( \HIpot  \bigr)-\EI(\WIpot).
\end{eqnarray*}
By (\ref{eq:Wkilldis}), we have
\begin{eqnarray*}
\lvH\frac{1}{|\I|}\Vert \UIpot-\UIpots-\tau\bigl( \HIpot  \bigr)  \Vert=0.
\end{eqnarray*}
Also by (\ref{eq:Wkilldis}), 
\begin{eqnarray*}
\lvH\frac{1}{|\I|}\Vert \HIpot-\UIpot  \Vert=0.
\end{eqnarray*}
Hence (\ref{eq:MEgeneral}) follows:
\begin{eqnarray*}
\lvH \frac{1}{|\I|}\ome\bigl( \HIpot\bigr)&=&
\lvH \frac{1}{|\I|}\ome\bigl( \UIpot\bigr)\nonum \\
&=&\lvH \frac{1}{|\I|}\ome\bigl( \UIpots\bigr)
+\lvH \frac{1}{|\I|}\tau\bigl( \HIpot\bigr)\\
&=&\epots+\lvH\frac{1}{|\I|}\tau\bigl( \HIpot \bigr).
\end{eqnarray*}
We also have 
\begin{eqnarray*}
\lvH \frac{1}{|\I|} \log \trI \bigl(e^{-\HI} \bigr) 
&=&\lvH \frac{1}{|\I|} \log \trI  \bigl(e^{-\UI} \bigr)\nonum \\
&=&P(\beta\pots)-
\beta \Bigl\{\lvH \frac{1}{|\I|}\tau\bigl(\HIpot \bigr)\Bigr\},
\end{eqnarray*}
 which shows (\ref{eq:PRESSgeneral}).
\proofend
\begin{rem}
Suppose that $\pot$
 satisfies 
 $(\pot$-\rm{a}$)$, $(\pot$-\rm{b}$)$, 
$(\pot$-\rm{c}$)$, $(\pot$-\rm{f}$)$ and
\begin{eqnarray}
\label{eq:EquivPot7}
\sum_{\I \ni 0}\Vert \potI \Vert<\infty.
\end{eqnarray}
Then it satisfies 
 $(\pot$-\rm{e}$)$
 automatically and  is a
  general potential.
  Furthermore, (\ref{eq:Wkilldis})
   is known to be satisfied 
   (the same proof as Lemma$\,$\ref{lem:SKILL}
    holds except for estimates (\ref{eq:sd8})
    (\ref{eq:sd9}), (\ref{eq:sd10}) and (\ref{eq:sd12})
     which follow from the absolute convergence 
     of  (\ref{eq:EquivPot7}) due to (\ref{eq:WIterm})) and 
 \begin{eqnarray}
\Cpot=
 \lvH \frac{\tau\bigl(  \HIpot \bigr) }{|\I|}
=
 \lvH \frac{\tau\bigl(  \UIpot \bigr) }{|\I|}
=
\epot(\tau)
\end{eqnarray}
is known to converge.
 (The same proof as Theorem$\,$\ref{thm:ME}
  holds except for a modification 
  of proof of some estimates 
   for Lemma$\,$\ref{lem:UHBUNK} on the basis of the absolute convergence
of (\ref{eq:EquivPot7}). See also e.g. Proposition 
6.2.39 of \cite{BRA2}.)
    
 Therefore 
 (\ref{eq:PRESSgeneral}) and 
   (\ref{eq:MEgeneral}) hold and the solutions
    of 
   $(\pot,\,\beta$)- and $(\pots,\,\beta$)-
variational principle
 coincide.
\end{rem}
%

%
\newpage
\appendix
\section{Appendix: Van Hove Limit}
\label{sec:VHL}
For the sake of mathematical precision, 
we present some digression about Van Hove limit.
\subsection{Van Hove Net}%
\label{subsec:VAN}%

We introduce mutually equivalent two types of conditions 
for the van Hove limit. First we start with  our  notation 
about   the shapes of regions of $\Znu$, 
  which will be used hereafter. 
 Recall that $\Ca$ is a cube of size $a$ given 
 by (\ref{eq:CaEQ}).
For a finite subset $\I$ of $\Znu$ and  $a \in \NN$, let
  $\nvhaIp$  be  the smallest number of translates 
  of $\Ca$\
  whose union covers $\I$, while  $\nvhaIm$ be  the largest
  number of mutually disjoint translates of $\Ca$
  that can be packed in $\I$. 

Let $\Br({n})$ be a closed ball
 in $\R^{\nu}(\supset \Znu)$
 with the center $n \in \Znu$ and 
the radius $r \in \R$.
Denote  the surface of $\I$  with a thickness $r(>0)$ by 
\begin{eqnarray}
\surf(\I) \equiv 
\Bigl\{n\in \I; \ \Br(\{n\})\cap \Ic\ne \emptyset \Bigr\}.
\end{eqnarray}

In what follows, we consider a net of finite subsets $\Ial$
 of $\Znu$ where
  the set of indices $\alpha$ is a directed set.
  Its partial ordering need not 
  have any relation with the set inclusion partial 
   ordering of $\Ial$.
\begin{lem}
\label{lem:VAN}
For a net of finite subsets $\Ial$ of $\Znu$, 
the following two conditions are 
 equivalent $\rm{:}$

 $($\it{1}$)$ For any $a \in \NN$, 
\begin{eqnarray}  
\label{eq:VH1}
\lim_{\alpha} \frac{\nvhaIam}{\nvhaIap}=1.
\end{eqnarray}

 $($\it{2}$)$ For any $r>0$, 
\begin{eqnarray}
\label{eq:VH2}
\limal\frac{1}{|\Ial|}\Bigl| \surf(\Ia)\Bigr|=0  
\end{eqnarray}
\end{lem}
\proof
\ ($\it{1}$) $\to$ ($\it{2}$):\\
 Let $\eps>0$ and $r>0$ be given.
Let $a \in \NN$ be sufficiently large so that
$a\ge 2r+1$ and
\begin{eqnarray*}
\epso\equiv 1- \left(\frac{ [a-2r]^{\nu} }{a^{\nu}} \right)
 < \frac{\epsilon}{2},
\end{eqnarray*}
 where  $[b]$ indicates the maximal integer not exceeding
 $b$.

By the condition ($\it{1}$), there exists
 an index $\alpha_{0}$ of the net $\{\Ial \}$
 such that, for $\alpha  \ge \alpha_{0}$, 
\begin{eqnarray*}
\epst\equiv 1-\frac{\nvhaIam}{\nvhaIap} <\frac{\eps}{2}.
\end{eqnarray*}
Let  $\Done,\ldots,\DN$,
 with $N=\nvhaIam$,   be
  mutually disjoint translates of $\Ca$
contained  in  $\Ia$.
 
Let $\Dpi$ be a translate  of $\C_{[a-2r]}$
 placed in  $\Di$ with a distance larger than $r$
 from the complement of $\Di$ in $\Znu$
 for each $i=1,\ldots,N$ which exists.
Then 
\begin{eqnarray*}
\frac{|\Dpi|}{|\Di|}=\Bigl( \frac{[a-2r]^{\nu}}{a^{\nu}} \Bigr)
 = 1 -\epso.
\end{eqnarray*}
Let $D$ be the union of $\Done,\ldots,\DN$ and $\Dp$
 be the union of $\Dpone,\ldots,\DpN$.
Then
\begin{eqnarray*}
\frac{|D\setminus\Dp|}{|D|}=1-\frac{|\Dp|}{|D|}=
1-(1-\epso)=\epso.
\end{eqnarray*}
Since $\nvhaIap$ translates of $\Ca$
 covers $\Ial$, we have %
\begin{eqnarray*}
|\Ial| \le \nvhaIap |\Ca|=\nvhaIap a^{\nu}.
\end{eqnarray*}
Hence
\begin{eqnarray*}
\frac{ |\Ial \setminus D|}{|\Ial|}=1- \frac{|D|}{|\Ial|}
\le 1-\frac{|D|}{\nvhaIap a^{\nu}} =1-\frac{\nvhaIam}{\nvhaIap}
=\epst.
\end{eqnarray*}
Due to $\Ial \supset D$, 
\begin{eqnarray*}
\frac{|D\setminus\Dp|}{|\Ial|}
\le  \frac{|D\setminus\Dp|}{|D|}=\epso.
\end{eqnarray*}
By construction, the distance between  $\Dpi$ and the complement of $\Di$
 (in $\Znu$) is larger than  $r$, and hence    
 the distance between  $\Dpi$ and the complement of $\Ial$
 is larger than $r$.
Thus, 
\begin{eqnarray*}
{\rm{surf}}_{r}(\Ial) 
\subset \Ial 
\setminus 
\Dp=(D \setminus \Dp)\cup
 (\Ial \setminus D).
\end{eqnarray*}
For $\alpha \ge \alpha_{0}$, we obtain 
\begin{eqnarray*}
\frac{ | {\rm{surf}}_{r}(\Ial)|}{| \Ial |} \le \epso+ \epst < \eps.
\end{eqnarray*}
Now $(\it{1}) \to (\it{2})$ is proved. 
\ \\

($\it{2}$) $\to$ ($\it{1}$):\\ 
Let $\eps>0$ and $a \in \NN$ be given.
Take $r>\sqrt{\nu}a$. 
  Let $\alpha_{0}$ be an index  of the net $\Ia$ 
 such that,   for $\alpha \ge \alpha_{0}$,
\begin{eqnarray*}
\frac{|\surfr(\Ial)|}{|\Ial|}<a^{-\nu} \eps.
\end{eqnarray*}

The translates $\Ca +a n$ of $\Ca$
 are disjoint for distinct $n \in \Znu$
  and their union over $n \in \Znu$ is $\Znu$.
Let $\Oa$ be the union of all those $\canb$ contained in $\Ial$
 and $N_{1}$ be their  number.
Let $\Oap$ be the union of all those $\canb$ which have non-empty
 intersections with both  $\Ial$
 and $(\Ial)^{c}$, and $N_{2}$ be their  number.
From the construction, the following estimates  follow
\begin{eqnarray*}
N_{1} \le \nvhaIam\le  \nvhaIap \le  N_{1}+N_{2}.
\end{eqnarray*}
Furthermore, since $\canb$ in $\Oap$
 contains a point  in  $\Ial$ as well as  a point in $(\Ial)^{c}$,
  and the distance of any two points in it 
 is at most $\sqrt{\nu}a<r$, 
 it has a non-empty intersection with $\Ial$, which is 
 contained  in $\surfr(\Ia)$.
Therefore,
\begin{eqnarray*}
|\surfr(\Ial)| &\ge& N_{2}=(N_{1}+N_{2})-N_{1} \nonum \\
&\ge& \nvhaIap -\nvhaIam.
\end{eqnarray*}
We have also
\begin{eqnarray*}
|\Ial| \le \nvhaIap |\Ca|=\nvhaIap a^{\nu}.
\end{eqnarray*}
Combining above estimates, we obtain  for $\al \ge \alpha_{0}$
\begin{eqnarray*}
0 &\le& 1-\frac{\nvhaIam}{\nvhaIap}
=\frac{\nvhaIap-\nvhaIam}{\nvhaIap} \nonum \\
&\le& \frac{|\surfr(\Ial)| a^{\nu} }{|\Ial| } \nonum \\
&<& \eps.
\end{eqnarray*}
Hence,  $(\it{2}) \to (\it{1})$ is now proved. 
\proofend\\
\begin{df}
If a net of finite subsets $\{\Ial \}$ satisfies
  the above condition $(\it{1})$ (or equivalently $(\it{2})$), then
 it is said to be a van Hove net (in $\Znu$).
\end{df}

We introduce the third  condition on 
  a net  of finite subsets $\Ial$ of $\Znu$:\\
({\it{3}}) For any finite subset $\I$ of $\Znu$, there exists an 
 index $\alpha_{\circ}$ such that $\Ial \supset \I$ for all $\al \ge 
\alpha_{\circ}$.
%
%
\begin{df}
  If a  net $\{\Ial \}$ (in $\Znu$)  
 satisfies the conditions  $(${\it{1}}$)$ (or equivalently  $(${\it{2}}$)$ )
and  $(${\it{3}}$)$, 
 then it is said to be  a  van Hove net tending to  $\Znu$.
\end{df}
\begin{rem}
The condition  ({\it{1}}) (or equivalently  ({\it{2}}) )
 does not imply the condition ({\it{3}}).
$\{\Cn \}_{n\in \NN}$ 
of (\ref{eq:CaEQ}) is  obviously a van Hove sequence.
But it does not cover the whole $\Znu$.
Hence it is not a  van Hove sequence  tending to  $\Znu$.
\end{rem}
\begin{lem}
\label{lem:NOTSTOP}
For any van Hove net and for any van Hove net 
 tending to $\Znu$, the directed set can not 
  have a maximal element.
\end{lem}
\proof\ 
Let $\{\Ial\}_{\alpha\in A}$
be a van Hove net where $A$ is a directed set of indices.
We show that
for any $\alc \in A$,
there exists $\alpr \in A$
 satisfying $\alpr \ge \alc$, $\alp \ne \alc$.
 
In fact, for a given $\alc$, there exist $a(\alc) \in \NN$ and $n \in \Znu$
such that 
\begin{eqnarray*}
\I_{\alc} \subset  \C_{a(\alc)-n},
\end{eqnarray*} 
and hence 
\begin{eqnarray*}
\nbf^{-}_{a(\alc)}(\I_{\alc})=0.
\end{eqnarray*}
On the other hand, 
for the above  $a(\alc) \in \NN$
there exists $\alone$ such that 
\begin{eqnarray*}
1-\frac{\nbf^{-}_{a(\alc)}(\Ial) }{\nbf^{+}_{a(\alc)}(\Ial)} < \frac{1}{2}
\end{eqnarray*}
for all $\al \ge \alone$, since $\{\Ial\}(\al \in A)$ is a van Hove net.

For any $\alp \in A$ satisfying both $\alp \ge \alone$
 and $\alp \ge \alc$, we have $\nbf^{-}_{a(\alc)}(\I_{\alp})\ne0$ due to
$ \alp \ge \alone$, and hence $\alp \ne \alc$.
We have  shown
 the existence of a desired $\alp$. 
 
 A van Hove net tending to $\Znu$ is a special case of a van Hove 
 net. Hence the assertion  for this case obviously follows.
\proofend
%
%
\subsection{Van Hove Limit}
Let $f(\I)$ be an $\R$-valued  function of 
finite subsets $\I$ of $\Znu$. 
We first show the following lemma which asserts  the independence
 of the limit on the choice of van Hove net (van Hove net tending to $\Znu$)
when  $f(\Ial)$ has 
 a limit for any van Hove net 
 (van Hove net tending to $\Znu$) $\{\Ial\}$.
\begin{lem}
\label{lem:COFINAL}
If $f(\Ial)$  has a limit 
 for any van Hove net $\{\Ial \}$,
 then its limit is independent of such 
a net.

If $f(\Ial)$  has a limit 
 for any van Hove net $\{\Ial \}$ tending to $\Znu$,
 then its limit is independent of such 
a net.
\end{lem}
\proof
\ Let $\{\Ialf\}_{\al \in A}$ and 
$\{\Ibls\}_{\beta \in B}$ 
 be two van Hove nets where 
 $A$ and $B$ are directed sets of indices. 
 We introduce a new index set
\begin{eqnarray*}
C\equiv\Bigl\{(\al, \beta, i)\,; \ \al \in A,\; \beta \in B,\; i=1,2  \Bigr\}
\end{eqnarray*}
 with the partial ordering 
\begin{eqnarray*}
(\al, \beta, i) \ge (\al^{\prime}, \beta^{\prime}, i^{\prime})
\end{eqnarray*}
 either if $\al > \alpha^{\prime}$ and $\beta>\beta^{\prime}$
 or if $\al=\al^{\prime}$, $\beta=\beta^{\prime}$
 and $i \ge i^{\prime}$.

For any  $(\al_{1}, \beta_{1}, i_{1}) \in C$ and 
  $(\al_{2}, \beta_{2}, i_{2}) \in C$, 
 there exist $\al \in A$
 and $\beta \in B$ such that 
 $\al > \al_{1}$, $\al > \al_{2}$,
 $\beta > \beta_{1}$,  $\beta > \beta_{2}$,
 because $A$ and $B$ are directed sets
  without maximal elements due to 
  Lemma$\,$\ref{lem:NOTSTOP}.
 Hence $(\al, \beta, 2) (\in C)$ obviously satisfies  
\begin{eqnarray*}
(\al, \beta, 2) 
 > (\al_{1}, \beta_{1}, i_{1}), \quad 
(\al, \beta, 2) 
> 
(\al_{2}, \beta_{2}, i_{2}). 
\end{eqnarray*}
So $C$ is a directed set.

Let 
\begin{eqnarray*}
\I_{(\al, \beta, i) }=
\left\{ 
\begin{array}{ll}
\Ialf & {\mbox{if}} \ \;i=1,\\
\Ibls & {\mbox{if}} \ \;i=2.
\end{array}
\right. 
\end{eqnarray*}
Since
$\{ \Ialf \}$ and $\{\Ibls \}$ 
 are van Hove nets, there exists 
 $\al_{\circ} \in A$ and $\beta_{\circ} \in B$
for any $d>0$ and $\epsilon >0$ such that 
\begin{eqnarray*}
\frac{|\surfd(\Ialf)|}{|\Ialf| } < \varepsilon 
 \quad \mbox{if}\ \al \ge \al_{\circ}
\nonum \\
\frac{|\surfd(\Ibls)|}{|\Ibls| } < \varepsilon 
 \quad \mbox{if}\ \beta \ge \beta_{\circ}.
\end{eqnarray*}
Set $\gamma_{\circ}\equiv 
(\al_{\circ}, \beta_{\circ}, 1)$.
 For any $\gamma=(\al, \beta, i) \ge \gamma_{\circ}$, 
 we have obviously 
$\al \ge \alpha_{\circ}$ and  $\beta \ge \beta_{\circ}$ by the definition
 of the ordering.
Hence, 
\begin{eqnarray*}
\frac{|\surfd(\Irl)|}{|\Irl| }
\le \mbox{{max}} \left\{ 
\frac{|\surfd(\Ialf)|}{|\Ialf| },\ 
\frac{|\surfd(\Ibls)|}{|\Ibls| } 
\right\}
 < \varepsilon. 
\end{eqnarray*}
Thus $\{\Irl\}_{\gamma \in C}$ is also a van Hove net.
If $\{ \Ialf \}$ and $\{\Ibls \}$ 
 are  van Hove nets tending to $\Znu$, then 
$\{\Irl\}$ is also a van Hove net tending to $\Znu$
 by its definition.

Since 
$\{\Irl\}_{\gamma \in C}$ is a van Hove net (van Hove net tending to $\Znu$),
$f$ has the following limit by the assumption on $f$
\begin{eqnarray*}
\finf=\lim_{\gamma} \{f(\Irl),\ \gamma \in C \}.
\end{eqnarray*}
Thus for any $\varepsilon$, there exists a 
$\gamma_{\circ}=(\alc, \blc, 1)$ or 
 $\gamma_{\circ}=(\alc, \blc, 2)$
such that 
\begin{eqnarray*}
|\finf-f(\Irl)| < \varepsilon
\end{eqnarray*}
for $\gamma \ge \rlc$.
This inequality holds  especially 
for $\gamma=(\al, \beta, 1) \ge \rlc$ 
 with $\al > \alc$ and $\beta > \beta_{\circ}$.
For this $\gamma$, $\Irl=\Ialf$, and hence
$f(\Irl)=f(\Ialf)$.
Thus we have
\begin{eqnarray*}
|\finf-f(\Ialf)| < \varepsilon
\end{eqnarray*}
 for  $\al >  \alc$.
Therefore, we obtain 
\begin{eqnarray*}
\finf=\lim_{\al}f(\Ialf).
\end{eqnarray*}
Similarly,
\begin{eqnarray*}
\finf=\lim_{\beta}f(\Ibls).
\end{eqnarray*}
Now we have shown that the  limit is  the same for
 $\{\Ialf\}_{\alpha \in A}$ and 
 $\{\Ibls\}_{\beta \in B}$.
Hence the independence of the limit on the choice of the net
follows. 
\proofend
\begin{df}
If $f(\Ial)$ has a limit  
 for any van Hove net $\{\Ial \}$,
 then $f(\I)$ is said to have the van Hove limit for large $\I$,
 and its limit is denoted by 
\begin{eqnarray}
\lvH f(\I).
\end{eqnarray}
If $f(\Ial)$ has a  limit 
 for any van Hove net $\{\Ial \}$ tending to $\Znu$,
 then $f(\I)$ is said to have the van Hove limit 
for $\I$ tending to $\Znu$, and its limit is denoted by 
\begin{eqnarray}
\lvHZ f(\I).
\end{eqnarray}
\end{df}

 In general, the first condition is stronger than the second.
If $f(\I)$ is translation invariant, however, the
 existence of the two limits are
 equivalent  as shown below.
\begin{lem}
\label{lem:TRANSVH}
If $f(\I)$ is translation invariant in the sense that
\begin{eqnarray*}
f(\I+n)=f(\I)
\end{eqnarray*}
 for any finite subset $\I$ of $\Znu$
and $n \in \Znu$, 
then $f(\I)$ has the van Hove limit for large $\I$
 if and only if $f$ has the van Hove limit for  $\I$ tending to $\Znu$.
\end{lem}
\proof
\ The only if part is obvious. 
Let  $\{\Ial\}_{\alpha \in A}$ be an arbitrary van Hove net.
Let $a(\al)$ be the largest integer $a$ such that 
 a translate of $\Ca$ is contained in $\Ial$.
Let $\Caal+n\subset \Ial$ and hence $\Caal \subset \Ial-n$.
Now we shift an approximate center of 
$\Caal$ to the origin of $\Znu$ and simultaneously  shift 
$\Ial-n$ by the same amount.
More precisely, $\Ial-n$ is shifted to 
\begin{eqnarray*}
\Ial^{\prime} \equiv \Ial-n-\left[\frac{a(\al)-1}{2}\right](1,\cdots,1).
\end{eqnarray*}
Obviously, 
\begin{eqnarray*}
\frac{|\surfd(\Ial^{\prime})|}{|\Ial^{\prime}|}= 
\frac{|\surfd(\Ial)|}{|\Ial| } 
\end{eqnarray*}
for all $d>0$ and $\al \in A$.

We show that this $\{\Ial^{\prime}\} (\al \in A)$
 is tending to $\Znu$.
 Let $\I$ be a finite subset of $\Znu$.
For sufficiently large integer $a$, 
$\I \subset \C_{a-[\frac{a-1}{2}]}$.
 For this $a$, there exists $\alone$
such that $\nbf^{-}_{a}(\I_{\al})>0$
 for $\al \ge \alone$.
Then $\aal\ge a$
 and 
\begin{eqnarray*}
\Ial^{\prime}\supset 
 \C_{\aal-[\frac{\aal-1}{2}]}\supset
\C_{a-[\frac{a-1}{2}]} \supset \I
\end{eqnarray*}
for $\al \ge \alone$.
Thus $\{\Ial^{\prime}\} (\al \in A)$
 is a van Hove net tending to $\Znu$.
Since $f$ is translation invariant, 
\begin{eqnarray*}
f(\Ial)=f(\Ial^{\prime}).
\end{eqnarray*}
By  the assumption that $f$ has  the van Hove limit tending to $\Znu$, 
$\lim_{\al}f(\Ial^{\prime})$ exists, and hence
 $\lim_{\al}f(\Ial)$ exists.
\proofend

\end{document}